\documentclass[12pt]{elsarticle}

\RequirePackage{lineno}

\usepackage[ddmmyyyy]{datetime}

\usepackage{array}
\usepackage{graphicx}
\usepackage{subfiles}
\usepackage{color}
\usepackage{amsfonts}
\usepackage{amssymb}
\usepackage{amsmath}
\usepackage{subfigure}
\usepackage{url}
\usepackage{units}
\usepackage[dvips]{epsfig}
\usepackage[footnotesize]{caption}  
\usepackage{rotating}
\usepackage[colorlinks=true,linkcolor=DarkBlue,citecolor=DarkBlue]{hyperref}

\usepackage{fancyhdr}

\makeatletter
\def\ps@pprintTitle{%
 \let\@oddhead\@empty
 \let\@evenhead\@empty
 \def\@oddfoot{}%
 \let\@evenfoot\@oddfoot}
\makeatother
\usepackage[percent]{overpic}

\begin{document}
\thispagestyle{empty}
%
%
%

%
%
%

\providecommand{\nSigmaOneYrParametricFullSyst}{XXXXXXXXXXXXXXXXXXXXXXXXXXXXXXXXXXXXXXXXXXXX}
\providecommand{\YrsToThreeSigmaSystLimited}{4}
\providecommand{\YrsToThreeSigmaLLH}{5}
\providecommand{\YrsToThreeSigmaLLHWithPriors}{4}

%
%
%


\providecommand{\OneStringIncrementalCost}{\$XXXM}
\providecommand{\TwentyStringIncrementalCost}{\$XXXM}
\providecommand{\FortyStringIncrementalCost}{\$XXXM}
\providecommand{\FortyStringFacilityNoContingencyNoForeignContrib}{\$65M}
\providecommand{\FortyStringFacilityContingencyAlone}{\$15M}
\providecommand{\FortyStringFacilityContingencyNoForeignContrib}{\$80M}
\providecommand{\FortyStringFacilityContingencyWithForeignContrib}{\$55M}
\providecommand{\FortyStringFacilityLikelyForeignContrib}{\$25M}

\providecommand{\FortyStringStandaloneNoContingencyNoForeignContrib}{\$85M}
\providecommand{\FortyStringStandaloneContingencyAlone}{\$15M}
\providecommand{\FortyStringStandaloneContingencyNoForeignContrib}{\$105M}
\providecommand{\FortyStringStandaloneContingencyWithForeignContrib}{\$80M}
\providecommand{\FortyStringStandaloneLikelyForeignContrib}{\$25M}

\providecommand{\TwentyStringDrillCost}{\$5M}
\providecommand{\TwentyStringDeploymentCost}{\$5M}
\providecommand{\TwentyStringInstrumentationCost}{\$24M}
\providecommand{\TwentyStringManagementCost}{\$5M}
\providecommand{\TwentyStringTotalCost}{\$39M}
\providecommand{\TwentyStringFuel}{146\,000 gal}

\providecommand{\TwentySixStringDrillCost}{\$5M}
\providecommand{\TwentySixStringDeploymentCost}{\$5M}
\providecommand{\TwentySixStringInstrumentationCost}{\$32M}
\providecommand{\TwentySixStringManagementCost}{\$5M}
\providecommand{\TwentySixStringTotalCost}{\$47M}
\providecommand{\TwentySixStringFuel}{190\,000 gal}

\providecommand{\TwentyStringForeign}{\$25M}

\providecommand{\SixtyStringIncrementalCost}{\$XXXM}

%
%
\providecommand{\dmOT}{\Delta m^2_{\rm 12}}
\providecommand{\dmTO}{\Delta m^2_{\rm 31}}
\providecommand{\dmsolar}{\Delta m^2_{\rm solar}}
\providecommand{\dmatm}{\Delta m^2_{\rm atm}}
\providecommand{\dmTT}{\Delta m^2_{\rm 32}}
\providecommand{\dmsq}[1]{\Delta m^{2}_{#1}}
\providecommand{\thatm}{\Delta m^2_{\rm atm}}
\providecommand{\thOT}{\theta_{\rm 13}}
\providecommand{\thTT}{\theta_{\rm 23}}
\providecommand{\sinsq}[1]{\sin^{2}(\theta_{#1})}
\providecommand{\sinsqOT}{\sin^{2}(\theta_{\rm 12})}
\providecommand{\sinsqsolar}{\sin^{2}(\theta_{\rm solar})}
\providecommand{\sinsqTT}{\sin^{2}(\theta_{\rm 23})}
\providecommand{\sinsqTTtrue}{\sin^{2}(\theta_{\rm 23}^{\rm true})}
\providecommand{\sinsqTwoTT}{\sin^{2}(2\theta_{\rm 23})}
\providecommand{\dcp}{\delta_{\rm CP}}

\providecommand{\gsim}{\gtrsim}
\providecommand{\lsim}{\lesssim}
\providecommand{\Enu}{\rm{E}_\nu}
\providecommand{\Emu}{\rm{E}_\mu}
\providecommand{\Ecasc}{\rm{E}_{\rm casc}}
\providecommand{\Lmu}{\rm{L}_\mu}
\providecommand{\Lnu}{\rm{L}_\nu}
\providecommand{\Thetamu}{\theta_\mu}
\providecommand{\cosTheta}{\cos{\theta}}
\providecommand{\cosThetamu}{\cos{\theta_\mu}}
\providecommand{\cosThetanu}{\cos{\theta_\nu}}
\providecommand{\Thetanu}{\theta_\nu}
\providecommand{\nue}{\nu_{\rm e}}
\providecommand{\numu}{\nu_\mu}
\providecommand{\nutau}{\nu_\tau}
\providecommand{\nubar}{\overline{\nu}}
\providecommand{\nuebar}{\overline{\nu}_{\rm e}}
\providecommand{\numubar}{\overline{\nu}_\mu}
\providecommand{\nutbar}{\overline{\nu}_\tau}
\providecommand{\nux}{\nu_{\rm x}}

\providecommand{\mares}{{M_A^{RES}}}
\providecommand{\maccqe}{M_A^{CCQE}}
\providecommand{\ahtby}{A^{BY}_{HT}}
\providecommand{\bhtby}{B^{BY}_{HT}}
\providecommand{\cvouby}{C^{BY}_{V1u}}
\providecommand{\cvtuby}{C^{BY}_{V2u}}

\providecommand{\ket}[1]{|#1\rangle}

\providecommand{\ue}[1]{|U_{e #1}|}
\providecommand{\umu}[1]{|U_{\mu #1}|}
\providecommand{\utau}[1]{|U_{\tau #1}|}
\providecommand{\uesq}[1]{|U_{e #1}|^{2}}
\providecommand{\umusq}[1]{|U_{\mu #1}|^{2}}
\providecommand{\utausq}[1]{|U_{\tau #1}|^{2}}

\providecommand{\Nch}{${\rm N}_{\rm ch}\,$}
\providecommand{\Ndir}{${\rm N}_{\rm dir}\,$}
\providecommand{\Nstr}{${\rm N}_{\rm str}\,$}
\providecommand{\Aeff}{${\rm A}_{\rm eff}\,$}
\providecommand{\Veff}{${\rm V}_{\rm eff}\,$}
\providecommand{\VeffNS}{${\rm V}_{\rm eff}$}
\providecommand{\Meff}{${\rm M}_{\rm eff}\,$}

\providecommand{\Cerenkov}{Cherenkov }
\providecommand{\deltaChiSq}{\Delta \chi^{2}}
\providecommand{\deltaChiSqBar}{\overline{\Delta \chi^{2}}}

\begin{frontmatter}

  \title{Letter of Intent:\\ The Precision IceCube Next Generation Upgrade (PINGU)\\}

\author[Adelaide]{{\bf The IceCube Gen2 Collaboration\footnote{Please send correspondence to analysis@icecube.wisc.edu}:} M.~G.~Aartsen} 
\author[Zeuthen]{M.~Ackermann} 
\author[Christchurch]{J.~Adams} 
\author[BrusselsLibre]{J.~A.~Aguilar} 
\author[Copenhagen]{M.~Ahlers} 
\author[StockholmOKC]{M.~Ahrens} 
\author[Geneva]{I.~Al~Samarai} 
\author[Erlangen]{D.~Altmann} 
\author[Marquette]{K.~Andeen} 
\author[PennPhys]{T.~Anderson} 
% ...
\address[Adelaide]{Department of Physics, University of Adelaide, Adelaide, 5005, Australia}
% ...
\address[BrusselsLibre]{Universit\'e Libre de Bruxelles, Science Faculty CP230, B-1050 Brussels, Belgium}
% ...
\address[Christchurch]{Dept.~of Physics and Astronomy, University of Canterbury, Private Bag 4800, Christchurch, New Zealand}
% ...
\address[Copenhagen]{Niels Bohr Institute, University of Copenhagen, DK-2100 Copenhagen, Denmark}
% ...
\address[Erlangen]{Erlangen Centre for Astroparticle Physics, Friedrich-Alexander-Universit\"at Erlangen-N\"urnberg, D-91058 Erlangen, Germany}
\address[Geneva]{D\'epartement de physique nucl\'eaire et corpusculaire, Universit\'e de Gen\`eve, CH-1211 Gen\`eve, Switzerland}
% ...
\address[Marquette]{Department of Physics, Marquette University, Milwaukee, WI, 53201, USA}
% ...
\address[StockholmOKC]{Oskar Klein Centre and Dept.~of Physics, Stockholm University, SE-10691 Stockholm, Sweden}
% ...
\address[PennPhys]{Dept.~of Physics, Pennsylvania State University, University Park, PA 16802, USA}
% ...
\address[Zeuthen]{DESY, D-15738 Zeuthen, Germany}


\begin{abstract}

  The Precision IceCube Next Generation Upgrade (PINGU) is a proposed
  low-energy in-fill array of the IceCube Neutrino Observatory.
  Leveraging technology proven with IceCube, PINGU will feature the
  world's largest effective volume for neutrinos at an energy
  threshold of a few GeV, improving the sensitivity to several aspects
  of neutrino oscillation physics at modest cost.  With its
  unprecedented statistical sample of low-energy atmospheric
  neutrinos, PINGU will have highly competitive sensitivity to $\numu$
  disappearance, the $\thTT$ octant, and maximal mixing, will make the
  world's best $\nutau$ appearance measurement, allowing a unique probe
  of the unitarity of the PMNS mixing matrix, and will be able to
  distinguish the neutrino mass ordering at $3\sigma$ significance
  with less than $\YrsToThreeSigmaSystLimited$~years of data.  PINGU
  can also extend the indirect search for solar WIMP dark matter complimentary to the on-going and planned direct dark matter experiments.  At
  the lower end of the energy range, PINGU may use neutrino tomography
  to directly probe the composition of the Earth's core.  With its
  increased module density, PINGU will improve IceCube's sensitivity
  to galactic supernova neutrino bursts and enable it to extract the
  neutrino energy spectral shape.

\end{abstract}


\begin{keyword}


neutrinos \sep
neutrino oscillations \sep
neutrino ordering \sep
neutrino hierarchy \sep
cosmic rays \sep
dark matter \sep
supernovae 

\end{keyword}

\end{frontmatter}



\tableofcontents
\clearpage


\resetlinenumber

\IfFileExists{NewCommands.tex}       {}       {}
\IfFileExists{../NewCommands.tex}    {}    {}
\IfFileExists{../../NewCommands.tex} {} {}

\graphicspath{{figures/}{Introduction/figures/}}

\section{Introduction}

Over the past decade, the South Pole has emerged as a site for
world-class astronomy, particle astrophysics and neutrino physics. The
Amundsen-Scott South Pole Station offers very special characteristics
--- the deep, clear ice below the surface and the dry air and clear
sky above. The glacial ice at the South Pole is \unit[2.8]{km} thick
and extremely clear~\cite{Aartsen:2013rt}, making possible neutrino
telescopes of unprecedented scale and sensitivity. The South Pole ice at depths below \unit[2100]{m} is not only
exceptionally clear but also extremely pure. The age of the ice at a
depth of \unit[2500]{m} is about \unit[100\,000]{years}, and
radioactive contaminants in the deep ice are in the range of
\unit[0.1--1]{$\times 10^{-12}$~g(Uranium or Thorium)/g} and
\unit[0.1--1 $\times
10^{-9}$]{g(Potassium)/g}~\cite{Cherwinka:2011ij}. The cold
environment greatly reduces thermionic electron noise in the
photomultipliers.  Thus, the South Pole provides a uniquely hospitable
environment to host large-scale detectors.  IceCube~\cite{Aartsen:2016nxy},
the world's largest neutrino detector, has been in operation
with 5160 optical sensors distributed on 86 strings (cables) since 2011,
transforming one gigaton of clear ice into a kilometer-scale \Cerenkov
detector.

\begin{figure}
   \begin{center}
     \includegraphics[width=12cm, angle=0]{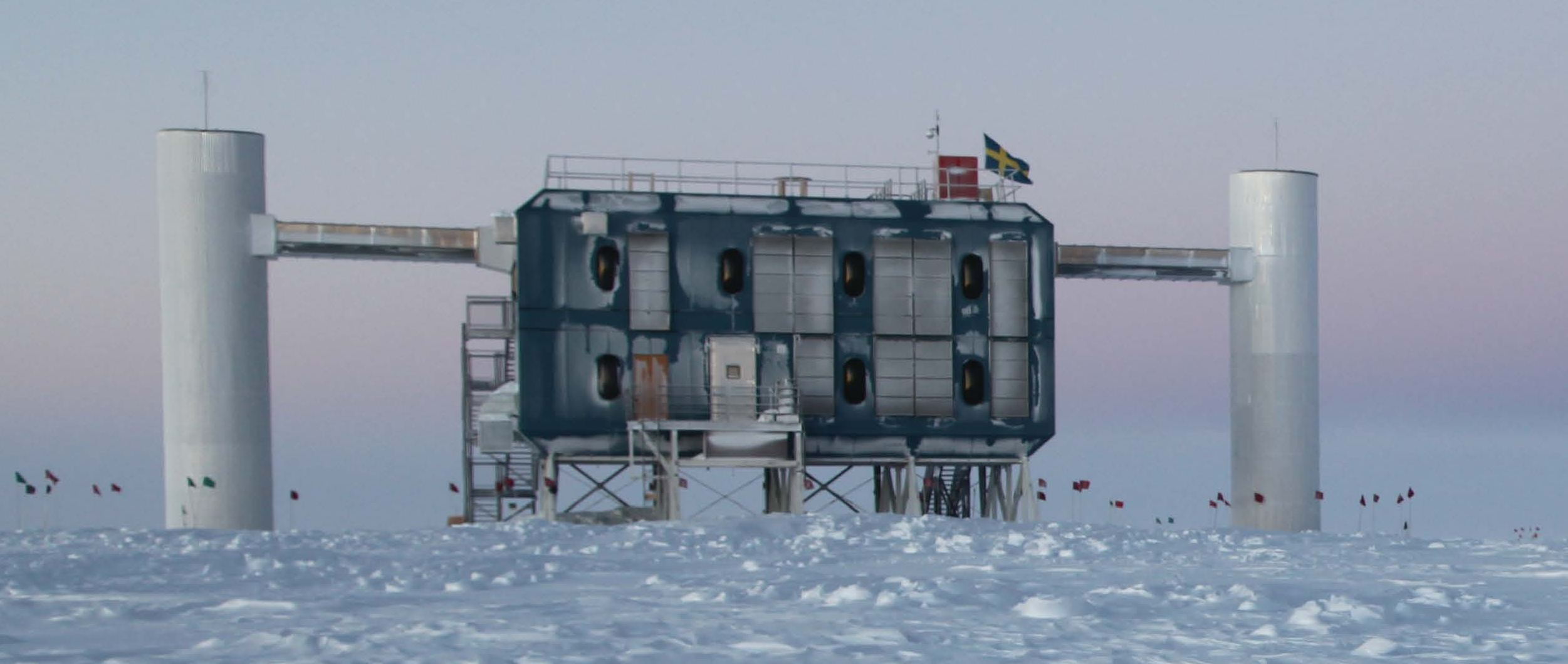}
   \end{center}
   \caption{The IceCube Laboratory building houses power,
     communications and data acquisition systems for IceCube and other
     experiments at the South Pole (photo by S. Lidstr\"om/NSF).}
   \label{fig:ICL}   
\end{figure}

Two smaller subarrays that were deployed along with IceCube -- IceTop
and DeepCore -- are key elements of the
facility. DeepCore~\cite{Aartsen:2016nxy} is the low-energy extension
of IceCube, located in the lower region of the detector's center,
which provides substantially increased sensitivity to neutrinos with
energies of approximately \unit[10--100]{GeV}.
The IceTop surface detector consists of 162 detector tanks, deployed
approximately in coincidence with IceCube strings, for measurements of
air showers above \unit[\~100]{TeV} and, therefore, may also act as a
veto against muons from these same events for the buried array.
IceCube was constructed with funding from the Major Research Equipment
and Facilities Construction (MREFC) program of the National Science
Foundation (NSF) \footnote{which supports the acquisition and
  construction of major research facilities and equipment that extend
  the boundaries of science, engineering, and technology} and a
roughly 10\% contribution from non-US sources.  The NSF's
Amundsen-Scott station provides excellent infrastructure for support
of IceCube's scientific activities, including the IceCube Laboratory
building (see Fig.~\ref{fig:ICL}) that houses power, communications,
and data acquisition systems.

\begin{figure}
   \begin{center}
     \includegraphics[width=10cm, angle=0]{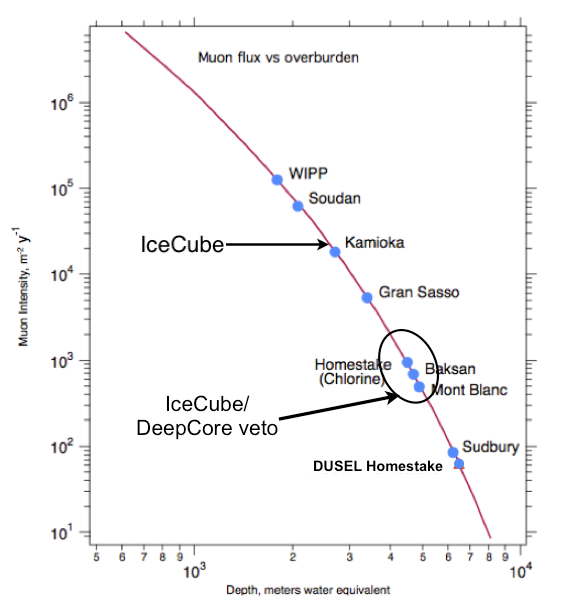}
   \end{center}
   \caption{Estimated muon rate in the deep ice after applying a veto
     based on a simple majority trigger (see, {\it e.g.,}~\cite{Mei:2005gm}).}
   \label{fig:MuonRate}   
\end{figure}

By using IceCube sensors to veto incoming muons, background rates due
to undetected muons in the deep detector can be reduced to levels
comparable to deep mines. Figure~\ref{fig:MuonRate} shows an estimate
of the muon rate after applying a downward-going muon veto based on a
simple majority trigger.

The discovery of high energy neutrinos of astrophysical origin by
IceCube~\cite{Aartsen:2013pza}, as well as competitive measurements of
neutrino oscillations~\cite{Aartsen:2013jza,NumuDisappearanceICRC2013}
and uniquely sensitive searches for dark matter with
DeepCore~\cite{Aartsen:2014yll}, have led the IceCube Collaboration to
begin investigating possible extensions of IceCube with improved
performance at both high and low energies.  The technological
solutions to drill and deploy instruments in the deep ice are proven,
the risks are small, and the costs are well understood. 

At high energies (\textgreater\unit[1]{TeV}), an expanded array in the
deep ice and an improved surface array for identifying air showers
that produce atmospheric neutrinos are under
consideration~\cite{Aartsen:2014njl}.  At low energies (around
\unit[10]{GeV}), the Precision IceCube Next Generation Upgrade (PINGU)
is proposed as an in-fill array for IceCube.  PINGU is designed to
study neutrino oscillations using atmospheric neutrinos that undergo
Mikheyev-Smirnov-Wolfenstein (MSW)~\cite{MSW-W,MSW-MS} and
parametric~\cite{Akhmedov,Petcov-NOLR} oscillations as they pass
through the Earth.

Using these neutrinos, PINGU will have enhanced sensitivity to several
important aspects of neutrino oscillation physics. It will be capable
of determining the neutrino mass ordering (NMO), making high precision
measurements of atmospheric muon neutrino disappearance and tau
neutrino appearance, studying whether or not the mixing angle $\thTT$
is maximal, and PINGU will also explore a lower mass range for Weakly
Interacting Massive Particles in searches for annihilation to
neutrinos in the terrestrial and solar cores, the galactic center,
dwarf spheroidals and other candidate astrophysical bodies.  The
detector will have enhanced sensitivity to very low energy neutrinos
from supernovae, and also has the potential to perform the first
neutrino-based tomographic probe of the Earth.

Together, the proposed detector enhancements at low and high energies
are denoted ``IceCube-Gen2.''  As a stepping stone to a full-scale
IceCube-Gen2 detector, a proposal to build IceCube-Gen2 Phase~1
(``Phase~1''), comprised of roughly one-quarter the strings envisioned
for the full PINGU array, has been submitted to the NSF and cognizant
funding agencies at collaborating institutions worldwide.  The goals
of Phase 1 are to perform the world's most stringent tests of the
unitarity of the PMNS neutrino mixing matrix in the tau sector, and to
improve the sensitivity of IceCube at both high and low energies
through better calibration of the optical properties of the ice and
the {\it in situ} response of IceCube photodetectors.  Better
calibrations will improve reconstruction of neutrino directions and
energies in both archived and future data, extending IceCube's
sensitivity to sources of high energy astrophysical neutrinos as well
as increasing the precision of neutrino oscillation measurements.

Our simulations of PINGU, informed by experience with data from
IceCube, and in particular DeepCore, indicate that our reconstruction
algorithms will provide sufficient angular and energy resolutions to
achieve the project goals. Known systematic uncertainties are found
to be sufficiently small to enable a measurement of neutrino
oscillation parameters with a few months to a few years of
data. These estimates are based on a conservative analysis in which
the detector geometry has yet to be fully optimized and a number of
possible improvements: such as the use of event elasticity to provide
some $\nu$-$\bar{\nu}$ separation, down-going atmospheric neutrinos to
provide an oscillation-free control sample, possible improvements in
event reconstructions, and better control and understanding of various
systematic errors,  not yet included, increasing our
confidence in this estimate.

PINGU will be composed of sensors of similar nature and size as the
IceCube DOMs, and installed in only two deployment seasons using the
same techniques and equipment.  The expertise developed in designing,
deploying, and operating IceCube means that the PINGU detector could
be deployed quickly and with well-understood and minimal
risk. Designed as an extension of IceCube, close integration of PINGU
with IceCube's online and offline systems will be straightforward,
enabling us to use the surrounding IceCube DOMs to provide a nearly
hermetic active veto against downward-going cosmic ray muons, the
chief background of all PINGU physics channels.  In addition, the cost
of both developing software systems and operating PINGU will be
incremental and thus considerably lower than normally expected for a
project of this scale.

This Letter of Intent (LoI), an update of our original LoI~\cite{LoI},
presents the detailed physics cases for the muon neutrino
disappearance, maximal mixing, tau neutrino appearance, neutrino mass
ordering, Earth tomography, supernova neutrino burst and WIMP dark
matter indirect detection measurements.  We describe the baseline
design of the PINGU detector and requirements for the hot water drill
and calibration devices, highlighting salient points of departure from
what was used in
IceCube~\cite{Achterberg:2006md,Collaboration:2011ym,Benson:2014ag}.
In~\ref{sec:gen2phase1}, we describe ``Gen2
Phase~1,'' the proposed first step for PINGU
and the future IceCube Gen2 array, highlighting the important physics
measurements attainable at both low and high neutrino energies with
its reduced string count.  Finally, we provide an estimate of the
schedule and cost for the design, construction, deployment and
operation of PINGU.

\clearpage

\resetlinenumber

\IfFileExists{NewCommands.tex}       {}       {}
\IfFileExists{../NewCommands.tex}    {}    {}
\IfFileExists{../../NewCommands.tex} {} {}

\graphicspath{{figures/}{DetectorDesignPredictedPerformance/figures/}}

\section{Detector Design and Predicted Performance}
\label{sec:DetectorDesignPredictedPerformance}

\subsection{Introduction}

The design of the PINGU detector closely follows that of the DeepCore
low-energy extension for IceCube. The current design consists of a
further in-fill of the central DeepCore volume using hardware following the design principles of the standard DeepCore Digital Optical Modules (DOMs).  The
additional modules will lower the neutrino detection threshold in
energy and significantly improve the sensitivity below \unit[20]{GeV}.


\subsection{Detector Geometries}
\label{sec:DetectorGeometries}

\begin{figure} [ht!]
   \begin{center}
      \includegraphics[width=14cm, angle=0]{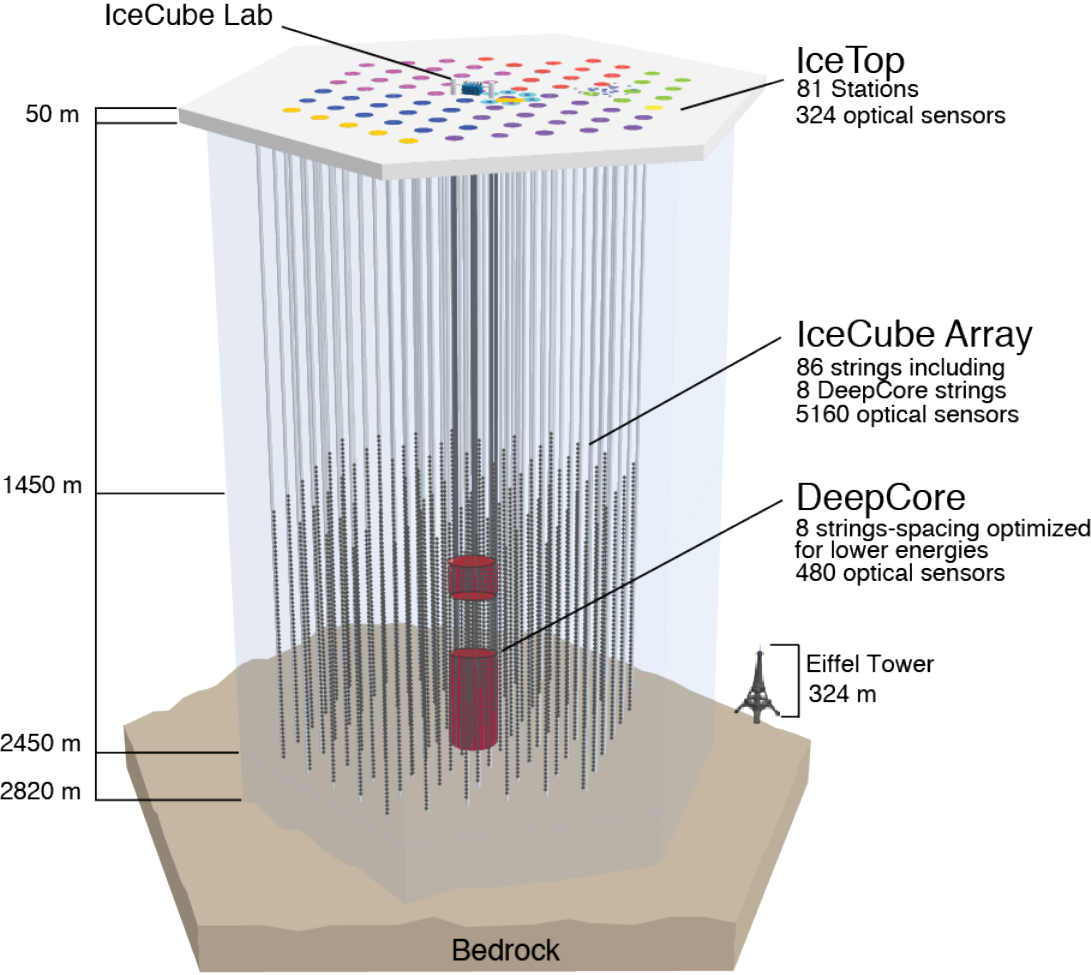}
   \end{center}
   \caption{Artist's rendering of the IceCube and DeepCore detectors.
     The PINGU detector strings would in-fill the existing DeepCore
     array at similar depths, with substantially closer vertical
     spacing between modules.  PINGU modules would occupy only the lower
     region of DeepCore indicated in the figure.}
   \label{fig:IceCubeDeepCoreDetector}
\end{figure}

An artist's rendering of the existing IceCube and DeepCore vertical
strings, with each string holding 60 DOMs, is shown in
Fig.~\ref{fig:IceCubeDeepCoreDetector}. The PINGU strings will be
deployed around the center of the existing DeepCore strings, with both
sub-arrays surrounded by the IceCube array.  This location maximizes
the sensitive volume for the PINGU detector while preserving the use
of the IceCube strings as a veto for incoming atmospheric muons.

Several geometries for PINGU have been studied, where the evaluation
metric between geometries was set by the sensitivity to the neutrino
mass ordering.  This particular metric was chosen since the
experimental signal has a small but measureable impact on the
oscillation probabilities.  Further, the geometry-dependent quantities
(the energy threshold, the angle and energy reconstruction resolution,
and the particle identification efficiency) affecting the determination of the mass
ordering are also common features affecting the majority
of the other analyses.

The first version of this LoI~\cite{LoI} used a
``$40\times60$'' baseline geometry composed of 40 strings with 60 DOMs per
string, all situated within the DeepCore fiducial volume.  Following a
subsequent period of optimization, the number of DOMs per string was
increased to 96 and the inter-string spacing was increased slightly.
This $40\times96$ geometry provided improved access to the lower-energy
neutrino signal and therefore to the physics topics discussed in this document.  Subsequent simulations using a similar
photocathode area with fewer strings (achieved by having more DOMs per
string) showed comparable physics reach at
reduced overall cost by virtue of one fewer drilling season and many
fewer holes, translating to lower on-ice personnel costs, less
required fuel, and other savings.

The studied geometries differ in the number of strings, their
positions, and (as detailed above) the number of DOMs per string.  A
summary of the geometries studied in the greatest detail is
shown in Table~\ref{Tab:geometries}.  The configuration used in the
previous version of this LoI is shown first for comparison. For
reference, the $40\times96$ geometry has undergone the most study, and
was the focus of PINGU presentations until March 2016.

The new baseline PINGU geometry features 26~strings and 192 DeepCore-style modules
per string.  This geometry delivers comparable physics sensitivities
at significantly reduced cost relative to the previous $40\times96$
geometry.  The anticipated installation schedule calls for eight
strings to be installed in the first deployment season and 18 in the
following season.  This schedule fits well with the knowledge gained
during the installation of IceCube and DeepCore, when up to 20 strings
were installed per season.  With each string populated with 192
modules, the photocathode in the ice is increased (30\% compared to the $40\times96$ geometry) and offsets the
reduced number of strings.  To model the physics impact of a challenging
drilling scenario, we have also studied a scaled-down version of the
new baseline geometry with only 20 strings deployed in two seasons.
This $20\times192$ geometry matches the total photocathode
area of the previous $40\times96$ geometry, but has a slightly larger inter-string spacing.

\begin{table}[h!]
   \begin{center}
   \begin{tabular}{|c|c|c|c|c|} \hline
   Configuration & Number of & Average Inter- & Number of   & Inter-DOM \\
                 & Strings   & string Spacing & DOMs/String & Spacing   \\ \hline\hline
    $40\times60$ (initial baseline) &    40    &      20~m      &     60      &   5~m     \\ \hline
    $40\times96$ &    40    &      22~m      &     96      &   3~m     \\ \hline
    $26\times192$ (new baseline) &    26     &      24~m      &     192      &   1.5~m     \\ \hline
    $20\times192$ &    20     &      30~m      &     192      &   1.5~m     \\ \hline
   \end{tabular}
   \end{center}
   \caption{Summary of the parameters for the studied PINGU detector geometries.  
     Shown are the initial LOI~\cite{LoI} 
     baseline $40\times60$ geometry, its enhanced $40\times96$ version with greater 
     deployed photocathode, the new baseline
     $26\times192$ geometry, and its scaled-back $20\times192$ version used to study
     the impact of a challenging drilling scenario.}
   \label{Tab:geometries}
\end{table}

More recently we have explored the physics potential of a geometry
with 6--7 PINGU baseline strings, deployed as ``Gen2 Phase~1'' (more
simply, ``Phase~1'') in one season.  The Phase~1 configuration is
effectively what the first year of PINGU deployment would provide.
Phase~1 is described in more detail in~\ref{sec:gen2phase1}.

\begin{figure}
	\begin{center}
	 	\includegraphics[scale=0.8]{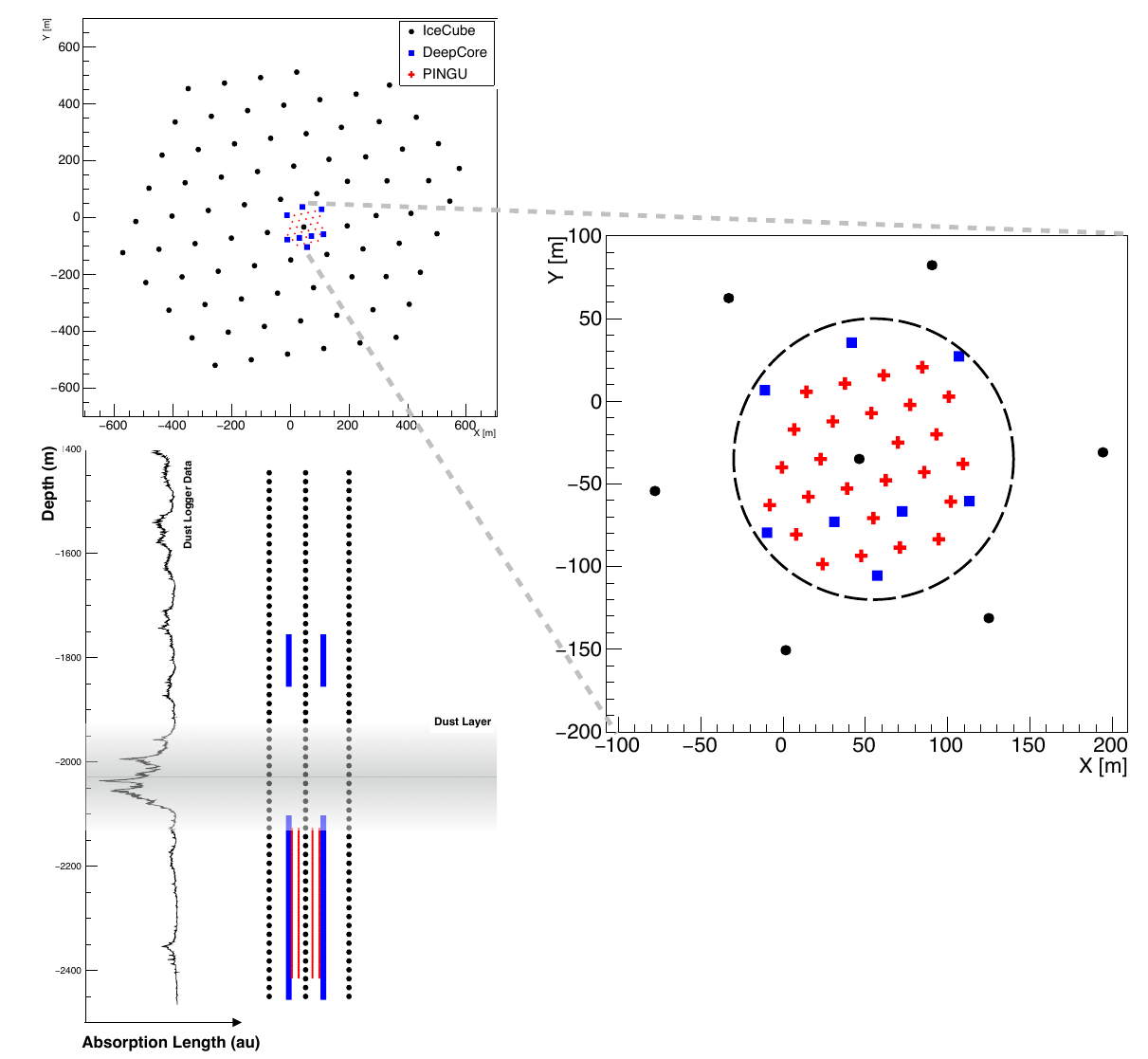}
	 \end{center}
         \caption{The left figure shows the overhead and side views of
           the new baseline PINGU geometry.  It also shows the
           surrounding IceCube and DeepCore strings, and vertical
           spacings for DeepCore and PINGU modules.  In the side view
           only some of the strings are shown for clarity.  The
           leftmost graph along the side of the figure delineates the
           dust concentration in the ice and shows that PINGU occupies
           the clearest ice.  Contained in the right figure is a
           zoom-in of the new baseline PINGU geometry, showing only
           the strings used in the PINGU triggering algorithm as well
           as the fiducial radius of PINGU (dashed line).}
	\label{fig:pingu_geos}
\end{figure}

A more detailed view of the IceCube, DeepCore, and future PINGU
strings for the new baseline geometry is shown in
Fig.~\ref{fig:pingu_geos}.  The top figure shows the complete IceCube
detector, with an average \unit[125]{m} spacing between strings (shown
in black circles) and a footprint comprising roughly one square
kilometer.  The 8 DeepCore strings, with average string spacings of
approximately \unit[70]{m}, are located in the center of the IceCube
geometry and shown in blue squares.  These are co-situated with the 26
PINGU strings (shown in red circles) having an average string spacing
of \unit[24]{m}.  The vertical module distribution is also shown in the
bottom left of Fig.~\ref{fig:pingu_geos}, showing the decrease in
module spacing from IceCube (\unit[17]{m}) to DeepCore (\unit[7]{m})
to PINGU (\unit[1.5]{m}).  This plot also shows the position of the
``dust layer,'' a region of ice with a much higher concentration of
particulate matter than the surrounding areas.  The DeepCore DOM
positions show the location of the ``plug'' of 10 DOMs above this
layer while the remaining 50 DOMs are located below.  The veto plug
provides additional information about events which enter the dust
layer since in that region their photons are largely lost due to the
increased absorption.  The PINGU modules would be positioned below the
dust layer, and inset slightly with respect to the DeepCore DOMs to
facilitate the veto of the cosmic ray muon background.

The final, zoomed-in portion of Fig.~\ref{fig:pingu_geos} shows a more
detailed overhead view of the center of the IceCube detector.  The
figure shows all strings that are used in the PINGU trigger,
including the 15 strings comprising DeepCore.  The dashed line
indicates the approximate fiducial region of PINGU, a radius of
\unit[85]{m} around the center of the detector and a length of
\unit[290]{m}, for a total fiducial mass of roughly \unit[6]{Mton}.

\begin{figure}
  \centering
  \subfigure[${\rm M}_{\rm eff}$($\numu$ CC)]{
  	\begin{overpic}[scale=1.3]{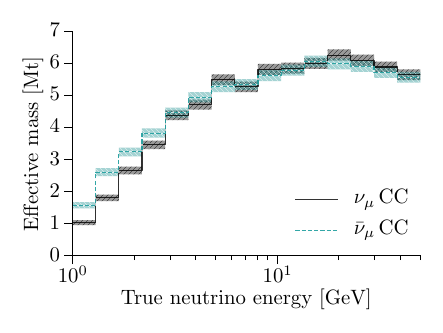}
		\put(50,40){\large \textcolor{red}{Preliminary}}
	\end{overpic}

    \label{fig:VeffNumu}
  }
  \subfigure[${\rm M}_{\rm eff}$($\nu_{all}$ NC)]{
  	\begin{overpic}[scale=1.3]{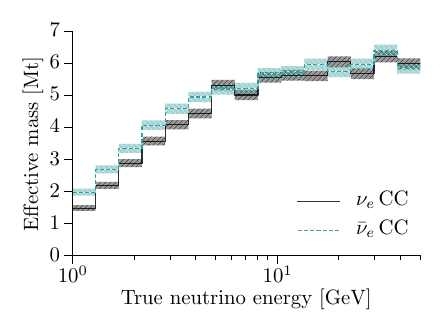}
	\put(30,60){\large \textcolor{red}{Preliminary}}
	\end{overpic}

    \label{fig:VeffNue}
  }
  \caption{Effective mass for (a) CC~$\numu$ and (b) CC~$\nue$
    interactions in the new baseline PINGU geometry as a function of
    neutrino energy.  Only events passing the final event selection
    criteria described in this Section are included in the histograms,
    causing a slight turnover in the effective mass from
    CC~$\nu_{\mu}$ events producing a muon that is not fully contained
    in the PINGU fiducial volume.}
   \label{fig:pingu_effvol}
\end{figure}

One of the metrics used to determine the efficiency of the detector
to incoming neutrinos is the ``effective mass'' of the geometry.  This
value is calculated using Monte Carlo simulation at analysis level and
is defined as
\begin{equation}
  M_{\rm eff} = M_{\rm gen} \times \frac{ N^\nu_{\rm reco} }{ N^\nu_{\rm gen} },
\end{equation}
where $M_{\rm gen}$ is the mass of ice in which the simulated
neutrinos interact, $N^{\nu}_{\rm reco}$ is the number of simulated
neutrinos passing all analysis selection criteria, and $N^{\nu}_{\rm
  gen}$ is the total number of neutrinos generated.  The final
selection criteria used for this calculation are described in the
following parts of this Section, with the results of this calculation
shown as a function of the true neutrino energy for CC~$\numu$
interactions in Fig.~\ref{fig:VeffNumu} and for CC~$\nue$ interactions
in Fig.~\ref{fig:VeffNue}.

\subsection{Detector Hardware and Simulation}
\label{sec:simulation}

Anticipated improvements to the modules, described in
Sec.~\ref{sec:NewPhotonDetectionTechnologies}, are designed to provide
more useful single photoelectron (SPE) information compared to that
collected by DeepCore DOMs.  However, we have conservatively built
upon the existing IceCube and DeepCore Monte Carlo software to produce
the PINGU simulated data (see Sec.~\ref{sec:MCSoftware} for full details).  To simulate low-energy events, the GENIE
neutrino generator~\cite{Andreopoulos:2009rq} is used to model the
neutrino interactions.  Following the interaction,
GEANT4~\cite{Agostinelli:2002hh} is used to track the produced
particles as they travel through the ice (including full simulation of
all secondaries produced at the interaction vertex).  The \Cerenkov
photons produced by these particles are tracked individually using
CLSim, a GPU-based software similar to that described
in~\cite{Chirkin:2013tma} which properly treats the position-dependent
photon scattering and absorption in IceCube (see
Sec.~\ref{sec:SoftwareLightPropagation} for more details).  For more
information on the simulation please see Sec.~\ref{sec:MCSoftware}.

In addition to the particle and photon tracking, the simulation
software also generates the non-physics-related signals, {\it i.e.},
instrumental noise.  The model for these noise hits has changed since
the first version of this Letter of Intent in that improvements have
been made to the method by which the noise is
simulated~\cite{LarsonMasters}.  The current noise generation method
matches the data measured using the IceCube detector better than the
previous version, largely due to the addition of intra-DOM correlated
noise.  The resulting increase in noise hits has affected the
performance of the event reconstruction which will be discussed in
detail next.

%

\subsection{Event Reconstruction}
\label{sec:EventReconstruction}

The most advanced PINGU reconstruction employs likelihood methods at
the single photon level for energy and direction estimation, as well
as neutrino flavor identification, building on the algorithms used for
IceCube~\cite{Aartsen:2013vja}.  Since the scattering length of
\Cerenkov photons in the deep Antarctic ice is approximately 20--30\%
of the absorption length, we are in an intermediate regime
between free-streaming photons and diffusive propagation and therefore
rely on numerical descriptions of light propagation through the ice.
The expected detector responses are computed and tabulated for a
variety of event topologies in the detector, in addition to different
depths and angular orientations.  These tables are then fit with
splines to reduce numerical instabilities from the binning and ensure
a smooth parametrization~\cite{Chirkin:icrc2013}.
To reconstruct an event, all DOM readouts
are subdivided in time, and a Poisson likelihood is calculated for the
contents of each time bin for all DOMs in PINGU, DeepCore, and
IceCube, comparing a reconstruction hypothesis to the data. The
interaction hypothesis is adjusted, and the process is repeated until
the hypothesis with the maximum likelihood is found.

PINGU is designed to observe neutrinos with energies as low as a few
GeV.  At this energy-scale, most or all of the secondary particles and
\Cerenkov photons created in a detected neutrino interaction will be
contained within the detector volume.  This is a significant change
from the most frequent events observed in the IceCube detector --
high-energy through-going muons from cosmic ray air showers.
Furthermore, at neutrino energies below roughly \unit[100]{GeV}, the
hadronic shower at the interaction vertex can contribute a significant
fraction of all the \Cerenkov photons detected in the event and must
be considered in the event reconstruction.  However, stochastic
processes such as bremsstrahlung are much more rare than at high
energies, and muon tracks produced by the CC~$\numu$ events most
relevant for PINGU analyses are in the minimum-ionizing regime.

This change in the anticipated signal necessitated a new
reconstruction strategy appropriate for the low energy events relevant
for the study of neutrino oscillations.  A simultaneous global
likelihood fit is performed using all eight event parameters: the
interaction vertex position and time, the zenith and azimuthal angles,
the energy of the cascade at the vertex, and the length of the
daughter muon (primarily for CC~$\numu$ events).  For CC~$\numu$
events, the angles are those of the emerging muon, while for all other
events, the angles are those of the cascade at the interaction vertex.
In principle, since the muon and the hadronic shower at the CC~$\numu$
interaction vertex are not perfectly aligned, their relative
directions could be treated as independent parameters in the fit, but
this is not done in these analyses.  It is possible to extract the
measured event inelasticity from the fit and use that information to
improve the sensitivity to the neutrino oscillation
parameters~\cite{Ribordy:2013xea}, but for the analyses covered in
this letter we do not yet take advantage of this.  Use of the event
inelasticity and complete interaction kinematics will be explored in
future refinements of the analyses presented here.

In order to successfully fit all eight parameters described above, we
use the nested sampling algorithm MultiNest~\cite{MultiNest_1}.  This
reconstruction was applied to fully simulated PINGU events,
independent of interaction type, for the geometries shown in
Table~\ref{Tab:geometries}. In order to remove events close to the
boundaries of the sampling space and ensure good reconstruction
quality, the reconstructed vertex of each event was required to be
contained within the PINGU fiducial volume.


%
%
\begin{figure}
	\begin{center}
	 	\begin{overpic}[scale=0.5]{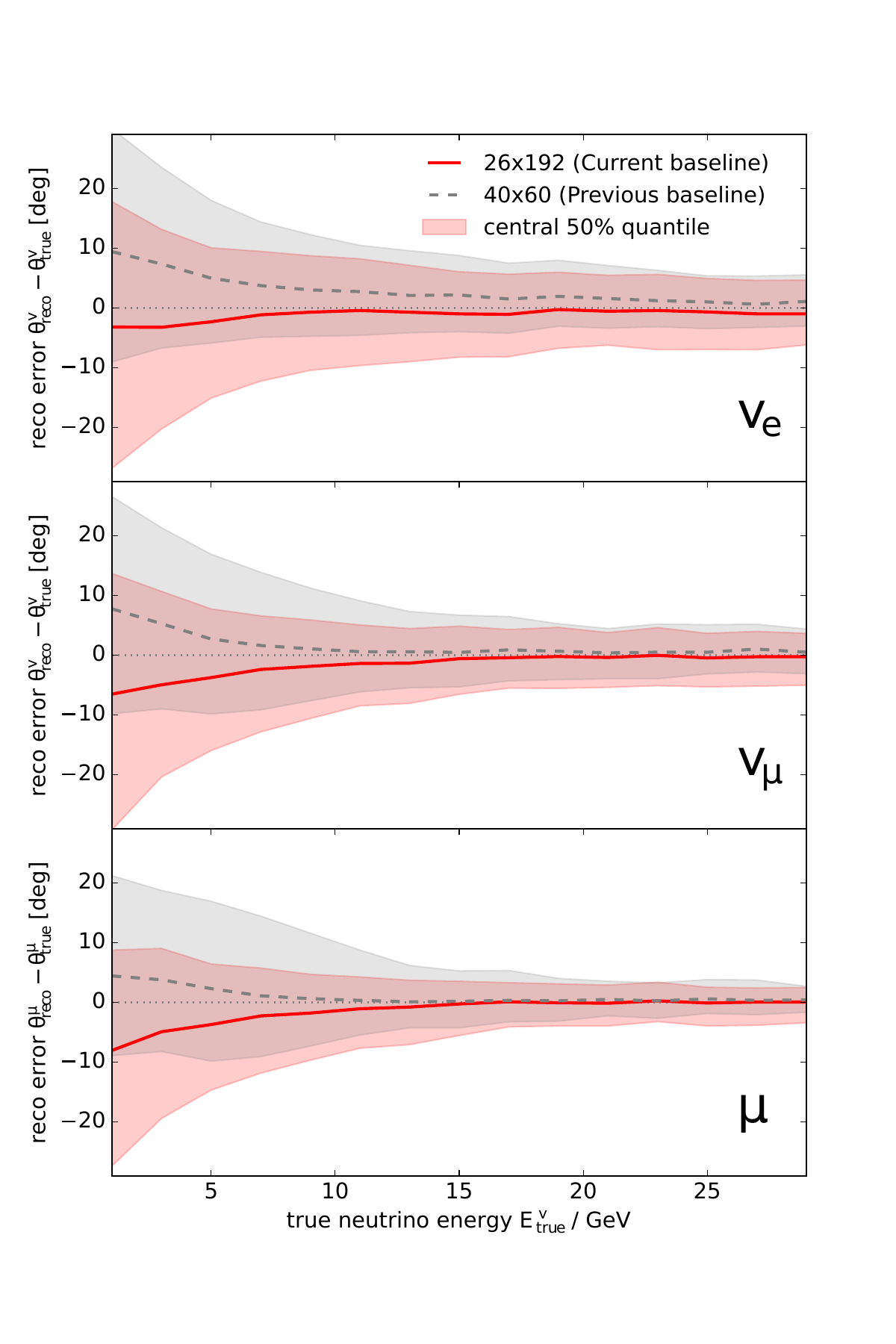}
		\put(35,15){\large \textcolor{red}{Preliminary}}
		\end{overpic}
	 \end{center}
         \caption{Zenith angle accuracy for events used in our
	    atmospheric-neutrino-based analyses.  The lines show the
	    median of the distribution and the shaded bands indicate the central
	    50\% of events.  The top and center plots show the resolution
	    of the reconstructed zenith angle with respect to the incoming neutrino for
	    $\nue$ and $\numu$ events, respectively.  The bottom plot
	    shows the zenith angle resolution with respect to the outgoing
	    muon for $\numu$ CC events.  Values are shown for the current baseline
	    (26$\times$192) and the previous baseline (40$\times$60) in all cases.}
	\label{fig:zenithRecoRes}
\end{figure}

\begin{figure}
	\begin{center}
	 	\begin{overpic}[scale=0.55]{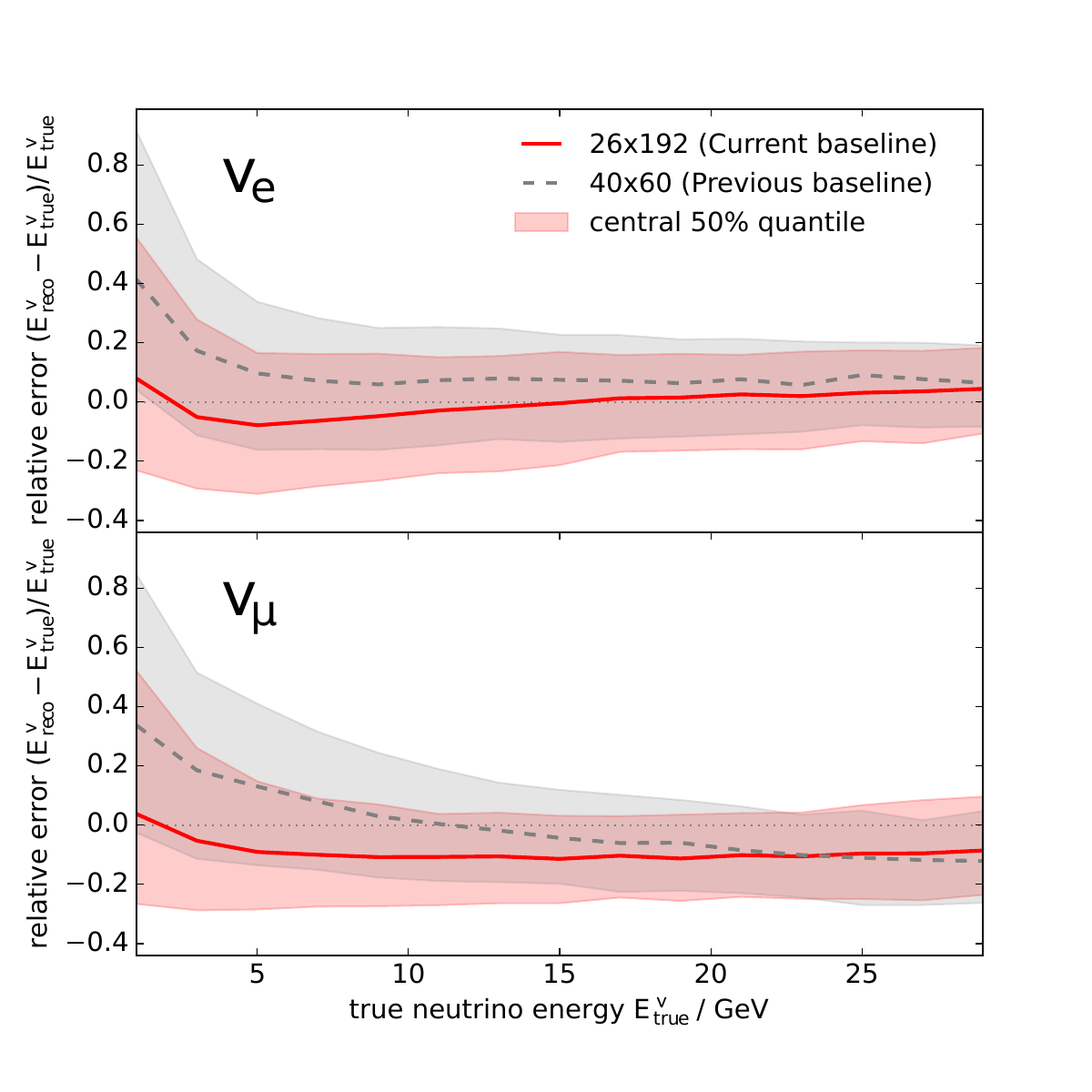}
		\put(60,45){\large \textcolor{red}{Preliminary}}
		\end{overpic}
	 \end{center}
         \caption{The relative reconstructed energy resolution for events used in our
         		atmospheric-neutrino-based analyses.  The lines show the median of
			the distribution and the shaded bands indicate the central 50\% of events.
			The top plot shows the resolution of the reconstructed energy for $\nue$
			events while the bottom is for $\numu$, both relative to the energy of the
			incoming neutrino. Values are shown for the current baseline (26$\times$192)
			and the previous baseline (40$\times$60) in all cases.}
	\label{fig:energyRecoRes}
\end{figure}

Since the previous version of this Letter of Intent~\cite{LoI}, we
have updated the expected detector response spline table with newer
parameterizations of the ice model surrounding the detector and with
the better simulation of noise described above. Both of these changes
affected the accuracy of the reconstruction with respect to energy.
To permit a direct comparison of the effect of the change of geometry,
we have shown results from both the previous and current baselines in
Figs.~\ref{fig:zenithRecoRes} and~\ref{fig:energyRecoRes}.  An
important note is that the cylindrical radius used for containment has
changed from the previous baseline (where it was \unit[75]{m}) to the
new baseline (where it is \unit[85]{m}), meaning that more events are
available for analysis with the latter.

\begin{figure}
   \begin{center}
            \begin{overpic}[width=1.0\textwidth]{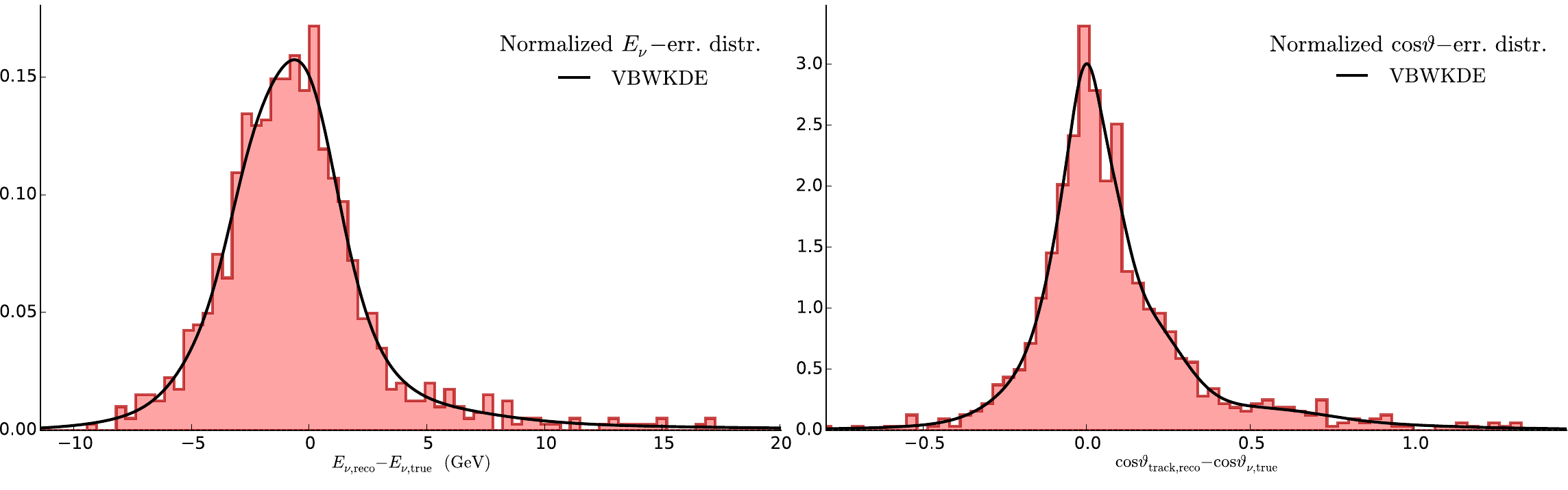}
	\put(30,10){\large \textcolor{red}{Preliminary}}
	\put(80,10){\large \textcolor{red}{Preliminary}}
	\end{overpic}   

  \end{center}
  \vspace{-2em}
  \caption{Resolutions for energy (left) and cosine-zenith (right) for
    CC $\numu$ and $\numubar$ events simulated with neutrino energies
    from \unit[9.46--10.59]{GeV}.  Plotted are histograms of the
    individual events' reconstruction errors overlaid by lines
    showing the variable-bandwidth kernel density estimates of the
    underlying distributions.}
  \label{fig:resolutions_vbwkde_numu_cc_bin20}
\end{figure}

It should also be noted that in several of our analyses described in
Sec.~\ref{sec:NeutrinoOscillations} the information in
Figs.~\ref{fig:zenithRecoRes} and~\ref{fig:energyRecoRes} is not used
directly.  Instead, the distributions--created using the full PINGU
simulation and reconstruction--are parameterized as functions of
energy.  This parametrization was done using a variable-bandwidth
kernel density estimate (VBW KDE), based
on~\cite{botev2010,abramson1982,10.2307/2345597}, of the distributions
of the reconstruction errors for several energy bins.  Example energy
and cosine-zenith error distributions and the computed VBW KDE
resolutions are shown in
Fig.~\ref{fig:resolutions_vbwkde_numu_cc_bin20}.


The reconstruction algorithm described above successfully reconstructs
about 90\% of atmospheric neutrino events that satisfy a loose trigger
criterion (three modules hit in spatial and temporal coincidence) and
whose true vertex is contained within the PINGU fiducial volume; these
events produce enough photoelectrons to fully constrain the fit.  We
do not attempt to recover the remaining 10\% of events although that
could be done by re-running the fit with a different random selection
of starting points.


\subsection{Event Selection}
\label{sec:CosmicRayMuonBackgroundRejection}

Downward-going atmospheric muons dominate the event rate in the detector,
outnumbering atmospheric neutrinos by a factor of roughly $10^6$:$1$
at the initial trigger level. Criteria must be put in place
to reject this background at high efficiency. The published analysis
of showering events induced by atmospheric
neutrinos~\cite{Aartsen:2012uu} in the IceCube/DeepCore detector
attained an atmospheric muon rejection factor of over $10^7$ by
vetoing events with ``early'' light (detected before the light in the
fiducial volume) in the surrounding IceCube modules. Furthermore, the
improved reconstruction performance for low energy events described in
Sec.~\ref{sec:EventReconstruction} also provides the ability to reject
any remaining downward-going events based on the reconstructed
direction (at the cost of rejecting downward-going neutrinos). We
therefore anticipate minimal contamination from this particular source
of background in PINGU, especially when restricting the analysis to
up-going events only.

In order to select events we reconstruct well, we require events to
consist of at least 8 optical modules that have detected light. We
also require the reconstructed interaction vertices of the events
(described in Section~\ref{sec:EventReconstruction}) to be within the
more densely instrumented PINGU fiducial volume, which is defined as a
cylinder with radius \unit[85]{m} and height \unit[320]{m} that runs
vertically between depths of \unit[2450]{m} and \unit[2130]{m} from
the surface as shown in Figure~\ref{fig:pingu_geos}. We do not require
full containment of muon tracks produced in $\nu_{\mu}$ charged
current interactions; depending on incident angle, muon tracks will be
contained in the IceCube volume up to energies of $E_{\mu}\sim
\mathcal{O}(\unit[100]{GeV})$.

Downward-going atmospheric muons dominate the event rate in the
detector, outnumbering atmospheric neutrinos by a factor of roughly
$10^{6}$ at the initial trigger level. Event selection criteria must
be put in place to reject this background at high efficiency. Current
measurements of muon neutrino disappearance in the IceCube/DeepCore
detector~\cite{ref:JasonNeutrinoProceedings} attain an atmospheric
muon rejection factor of more than $10^{8}$ by vetoing events with
`early' light in the surrounding IceCube modules (i.e. light detected
before the light in the DeepCore fiducial volume), highly elongated
events without a visible shower characteristic of a neutrino-nucleon
interactions, and events with hits too widely spread in time, which
are associated with muons traversing the full detector. The current
PINGU analysis applies similar criteria to account for the impact of
these selection criteria on the neutrino signals.

In current IceCube/DeepCore measurements, sideband data are used to
model the residual atmospheric muon contamination, and neutrinos from
the full sky (downward-going as well as upward-going) are used to help
constrain systematic uncertainties. However, it is not currently
possible to simulate enough atmospheric muons to make a full
estimation of the impact of atmospheric muons in the PINGU analyses,
nor to use data to make such an estimate. We have therefore decided to
remove all downgoing events in the analyses presented here. Once PINGU
data is available we are confident the current methods used to
estimate such background contamination from data in IceCube/DeepCore
will continue working and will allow us to perform full-sky analyses
in PINGU, which will aid in controlling flux systematics as discussed
in~\ref{sec:FluxSysts}.

With the cuts described above and the requirement of events to be
reconstructed as upgoing, we do not anticipate contamination from
atmospheric muons in PINGU, given that the similar techniques used in
IceCube/DeepCore analyses have already led to very small
contaminations from this source in the upgoing region. We will
therefore neglect such atmospheric muon contamination in the physics
studies presented here. A summary of the expected events rates after
the full event selection is given in
Table~\ref{tab:FinalSelectedEventRates}.

\begin{table}
\centering
\begin{tabular}{|c|c|}
\hline
Interaction type & Events per year \\
\hline
$\nu_{\tau} + \overline{\nu}_{\tau}$ CC & 2,800\\
$\nu_{\mu} + \overline{\nu}_{\mu}$ CC & 32,600\\
$\nu_{e} + \overline{\nu}_{e}$ CC & 25,400\\
$\nu+\overline{\nu}$ NC & 7,400\\
\hline
\end{tabular}
\caption{Annual event rates expected in the PINGU detector after the full event selection and including the effects of neutrino oscillation (assuming the normal mass ordering).}
\label{tab:FinalSelectedEventRates}
\end{table}

\subsection{Particle ID}
\label{sec:ParticleID}


Neutrino events at the energies relevant for PINGU analyses fall into
two channels: track-like events with an associated muon from CC
$\nu_\mu$ interactions and cascade-like events coming from
CC~$\nu_{e,\tau}$ as well as all neutral current (NC)
interactions. The ability to separate track-like and cascade-like
events benefits many PINGU analyses as we expect different flavors of
neutrinos to carry different pertinent information. In the context of
the neutrino oscillation measurements described in detail in
Sec.~\ref{sec:NeutrinoOscillations}, $\numu$ (classified mostly as
track-like) and $\nue$ (classified mostly as cascade-like) undergo
effects that are induced at different energies and different
baselines, and the differences in those signatures affect the
oscillation measurements.

\begin{figure}
	\begin{center}
		\begin{overpic}[width=0.4\textwidth]{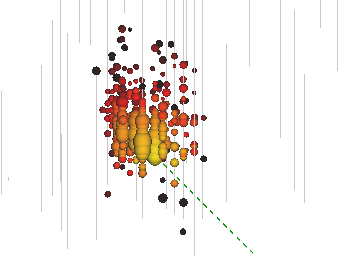}
			\put(50,65){\large \textcolor{red}{Preliminary}}
		\end{overpic}
		\hspace{0.05\textwidth}
		\begin{overpic}[width=0.4\textwidth]{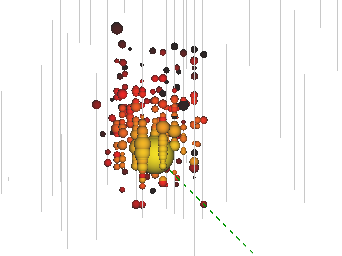}
			\put(50,65){\large \textcolor{red}{Preliminary}}
		\end{overpic}   
	\end{center}
	\vspace{-1em}
	\caption{Event displays of a CC~$\numu$ (left) and a CC~$\nue$
          (right) event.  The spheres indicate the DOMs which recorded
          photons where the total amount of charge is indicated by the
          size of the sphere. The color
          indicates the time when the DOM observed the first photon,
          while the dashed line shows the true neutrino direction
          direction.  In both panels are shown \unit[12]{GeV} CC~$\nu$ producing a
          \unit[10]{GeV} lepton ($\mu$ or $e$) crossing the detector leaving
          several groupings of hits on consecutive strings in a short
          time interval.
          In both cases the interaction vertex and direction is identical to make the
          comparison between events easier. We can distinguish the CC~$\numu$ and CC~$\nue$ events
          by comparing if the charge is extended in the diagonal (which happens in CC~$\numu$ events)
          or more concentrated around the vertex (which indicates no $\mu$ is present in the event).}
	\label{fig:PID:event_display}
\end{figure}

\begin{figure} [h!]
	\begin{center}
		\begin{overpic}[scale=0.5]{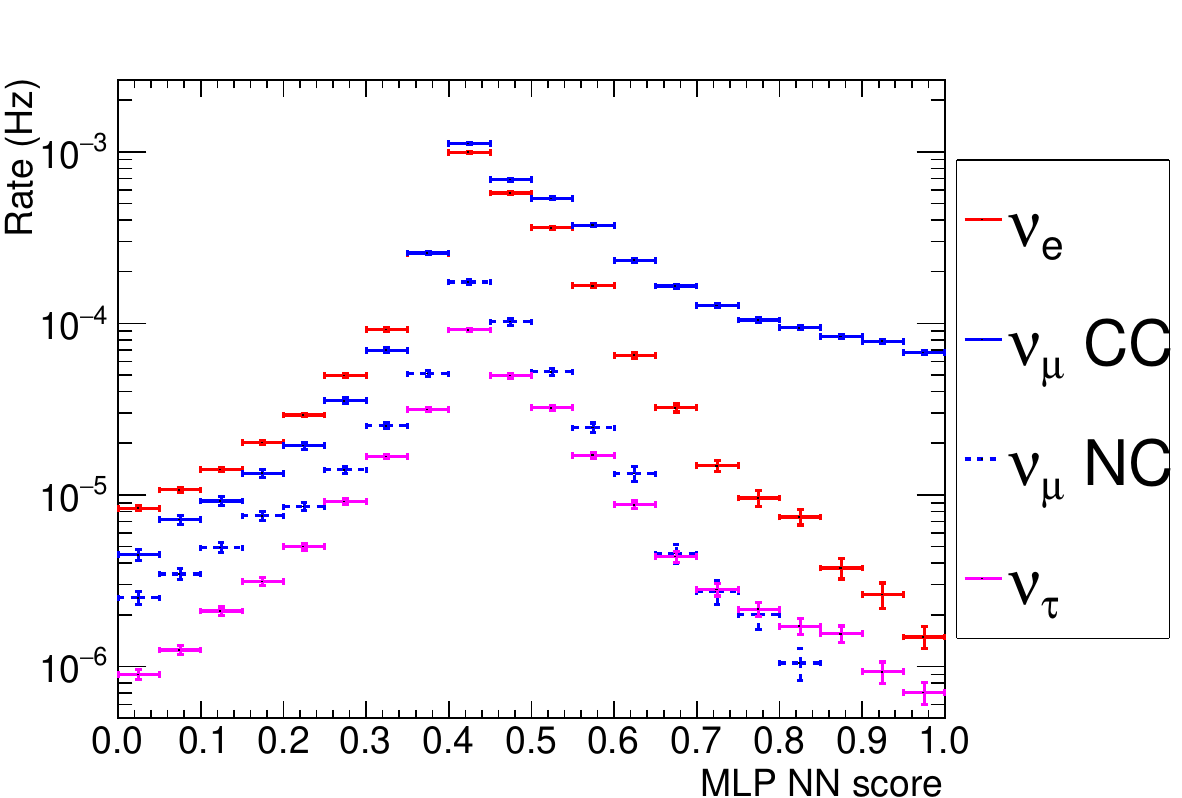}
			\put(30,15){\large \textcolor{red}{Preliminary}}
			\put(60,55){Track-like}
			\put(10,55){Cascade-like}
		\end{overpic}   
	\end{center}
	\vspace{-1em}
	\caption{Output of the MLP neural network classifier for the
          new baseline geometry.  Coefficients close to 1 represent
          more track-like events, while values closer to 0 correspond
          to events being more cascade-like. The large number of
          events around 0.4 typically will contain mainly lower energy
          events which are harder to classify.  The TMVA MLP method
          with the BFGS algorithm was used for training and Bayesian
          regulators, with 600 training cycles, N+5 hidden layers,
          ``tanh'' as the function type for neuron activation, and
          an overtraining test performed every 5 epochs~\cite{Hocker:2007ht}.  }
	\label{fig:PID:classifier}
\end{figure}

\begin{figure} [h!]
   \begin{center}
            \begin{overpic}[scale=0.6]{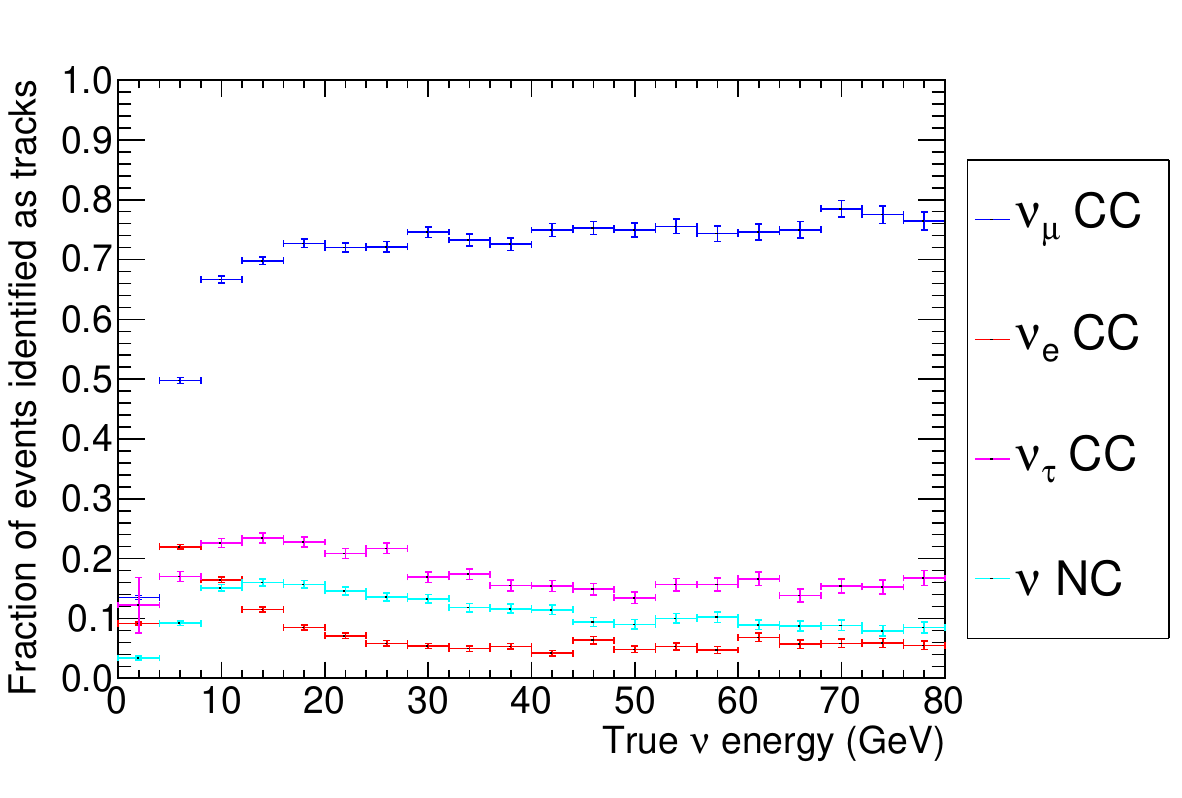}
	\put(50,55){\large \textcolor{red}{Preliminary}}
	\end{overpic}   

  \end{center}
  \vspace{-2em}
  \caption{The fraction of events identified as track-like in the new
    baseline geometry as a function of true neutrino energy for each
    neutrino flavor and interaction type (CC or NC) using the
    TMVA~\cite{Hocker:2007ht} MLP method described in
    Fig.~\ref{fig:PID:classifier} and a score of $0.55$.}
  \label{fig:PID}
\end{figure}

The central goal of classifying the event as track-like or
cascade-like is the identification of the presence (or absence) of a muon
track.  Figure~\ref{fig:PID:event_display} shows an example of a
track-like and a cascade-like event in PINGU.
To classify events as track-like or cascade-like we have trained a
``multilayer perceptron'' (MLP) neural network (NN) using the TMVA
toolkit~\cite{Hocker:2007ht} with variables based on the output of the
event reconstruction described in Sec.~\ref{sec:EventReconstruction}.
These include variables such as the reconstructed muon length, the
fraction of the energy reconstructed in the muon to total energy
reconstructed, and the difference in the log-likelihood between the
best fit and the fit forcing the assumption that there was no track
present.
Another class of variables that provides information related to the
classification of events is the timing of the hits.  Since muons
travel at roughly the speed of light $c$ (while emitting \Cerenkov
radiation), but the \Cerenkov photons emitted during hadronic and
electromagnetic cascades travel at $c/n$ (where $n$ is the refractive
index of the ice), the light emitted in events with muons will be
detected earlier than for those with only cascades.
Figure~\ref{fig:PID:classifier} shows the output of the algorithm
described above, and Fig.~\ref{fig:PID} shows the separation achieved
between track-like and cascade-like events used in the oscillation
analysis based on the use of the MLP neural network with a score of
$0.55$.  This creates a relatively pure CC~$\numu$ track-like sample,
while mixing all neutrino flavors in the cascade channel.

\clearpage

\resetlinenumber

\IfFileExists{NewCommands.tex}       {}       {}
\IfFileExists{../NewCommands.tex}    {}    {}
\IfFileExists{../../NewCommands.tex} {} {}

\graphicspath{{figures/}{AtmosphericNeutOsc/figures/}}

\section{Atmospheric Neutrino Oscillations}
\label{sec:NeutrinoOscillations}

\subsection{Introduction}
\label{sec:NeutrinoOscillationsIntroduction}
\begin{figure}[ht!]
    \centering
    \includegraphics[width=12cm, angle=0]{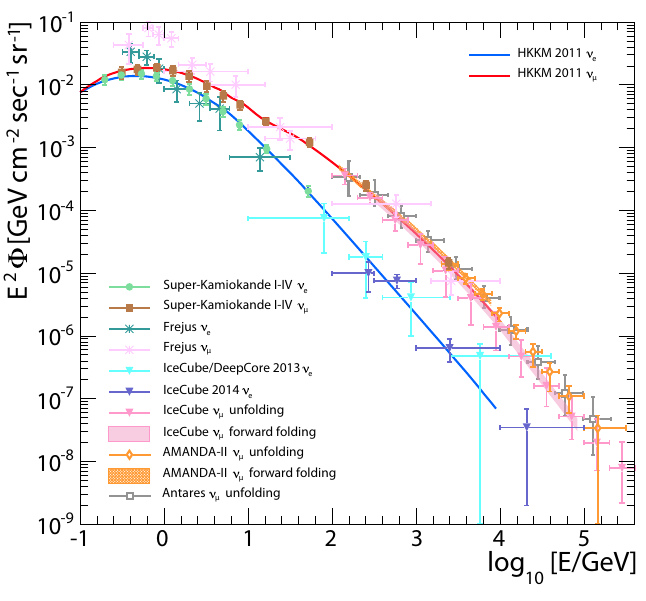}       
    \caption{Energy spectra of the atmospheric $\nue$ and $\numu$ fluxes
      measured by Super-Kamiokande~\cite{Richard:2015aua}, Frejus~\cite{Daum:1994bf},
      AMANDA-II~\cite{Abbasi:2008ih,Abbasi:2010qv}, Antares~\cite{Adrian-Martinez:2013bqq} and
      IceCube~\cite{Aartsen:2012uu,Aartsen:2014muf,Abbasi:2010ie,Abbasi:2011ui}. The
      solid lines show the prediction from the HKKM'11 model~\cite{Honda:2011nf}
      including the effects of oscillations. Figure adapted
      from~\cite{Richard:2015aua}. \label{fig:atmoFlux}}
\end{figure}

The atmosphere of the Earth is one of a number of sources that can be
used to provide a large sample of neutrinos~\cite{PDG_review,
  Honda:2006qj}.  A constant flux of cosmic rays (primarily protons
but with a component of light nuclei) interacts in the upper
atmosphere, producing a shower of hadrons.  The majority of these are
charged pions, which decay quickly into a muon and a muon-type
antineutrino (or an antimuon and muon-type neutrino).  The muon can
then decay into an electron, a muon neutrino and an electron-type
antineutrino, with a similar chain for the antimuon decays. As shown in
Fig.~\ref{fig:atmoFlux}, these
fluxes are described to first order as a power law spanning many
orders of magnitude in energy from roughly 10~GeV to beyond the TeV
scale.

The propagation of neutrinos through vacuum (in their mass eigenstates
and natural units) can be described by plane wave solutions of the
form~\cite{PDG_review}

\begin{equation}
\mid \nu_j(t) \rangle = e^{-i(E_jt-\vec{p_j}\cdot\vec{x})} \mid \nu_j(0) \rangle,
\label{eq:propagation1}
\end{equation}

in which $j \in \{1,2,3\}$, $t$ is the propagation time, $E_j$ and
$\vec{p_j}$ are the energy and momentum of the neutrinos and $\vec{x}$
is the position relative to the starting position.  All neutrinos
observed have energies of at least 1~GeV, $\mid \vec{p_i}\mid \gg
m_i$, and therefore the ultrarelativistic version of the momentum can
be used, and $t$ can be approximated by the distance travelled $L$.
These modifications to Eq.~\ref{eq:propagation1} produce the following
equation:

\begin{equation}
\mid \nu_j(L) \rangle = e^{-im_j^2L/2E} \mid \nu_j(0) \rangle.
\label{eq:propagation2}
\end{equation}

Using Eq.~\ref{eq:propagation2}, the probability of a neutrino of
flavor $\alpha$ being observed as flavor $\beta$ after propagation
can be expressed as
\begin{equation}
	\begin{split}
          P_{\alpha\rightarrow\beta} & = \mid \langle\nu_{\beta}\mid\nu_{\alpha}(t)\rangle \mid^2 \\
          & =  \Bigl| \sum_i U_{\alpha j}^* U_{\beta i} e^{-im_j^2L/2E} \Bigr|^2 \\
          & = \sum_{j,k} U_{\alpha j}^* U_{\beta j} U_{\alpha k}
          U_{\beta k}^* e^{-i\Delta m_{jk}^2L/2E},
	\end{split}
\label{eq:prob1}
\end{equation}
where $U_{\alpha j}$ are the elements of the PMNS matrix, as shown
in Eq.~\ref{eq:PMNS}, and $\Delta m^2_{jk} = m^2_j - m^2_k$ are the
mass squared differences.  If we assume there are only three families
of neutrinos it follows that the $\mathrm{U}_{\mathrm{PMNS}}$ written
in Eq.~\ref{eq:IndividualPMNS} is a unitary matrix and can be
parametrized using three mixing angles ($\thOT$, $\thTT$ and
$\theta_{\rm 12}$) and a Charge-Parity (CP)-violating phase $\delta$:
\begin{eqnarray}
\label{eq:PMNS}
\left( \begin{array}{c}
\nue\\
\numu\\
\nutau\\
\end{array} \right)
&=& {\Large \textrm{U}}_{\mathrm{PMNS}}
\left(
\begin{array}{c}
\nu_1\\
\nu_2\\
\nu_3\\
\end{array} \right)\\
\label{eq:IndividualPMNS}
&=&
\left(
\begin{array}{c c c}
U_{e1} & U_{e2} & U_{e3}\\
U_{\mu1} & U_{\mu2} & U_{\mu3}\\
U_{\tau1} & U_{\tau2} & U_{\tau3}\\
\end{array} \right)
\left(
\begin{array}{c}
\nu_1\\
\nu_2\\
\nu_3\\
\end{array} \right).
\end{eqnarray}

The mixing angles and mass-squared differences that describe
oscillations in the neutrino sector have been measured with good
precision through the efforts of a variety of experiments
worldwide~\cite{PDG_review}.  IceCube, with its DeepCore extension,
has demonstrated~\cite{Aartsen:2014yll} the ability to measure the
``atmospheric'' mixing parameters $\thTT$ and $\dmTT$ and, as analysis
techniques are refined and systematic uncertainties are better
understood, is on track to produce results that are competitive with
those of world-leading experiments.  The remaining unknowns in the
leptonic sector include the nature of the neutrino (Dirac or
Majorana), the extent to which CP symmetry may be violated in the
sector, determination of the presence of a potential sterile neutrino,
and the ordering of the mass eigenstates.

\begin{figure}[ht!]
    \centering
    \includegraphics[width=7cm, angle=0]{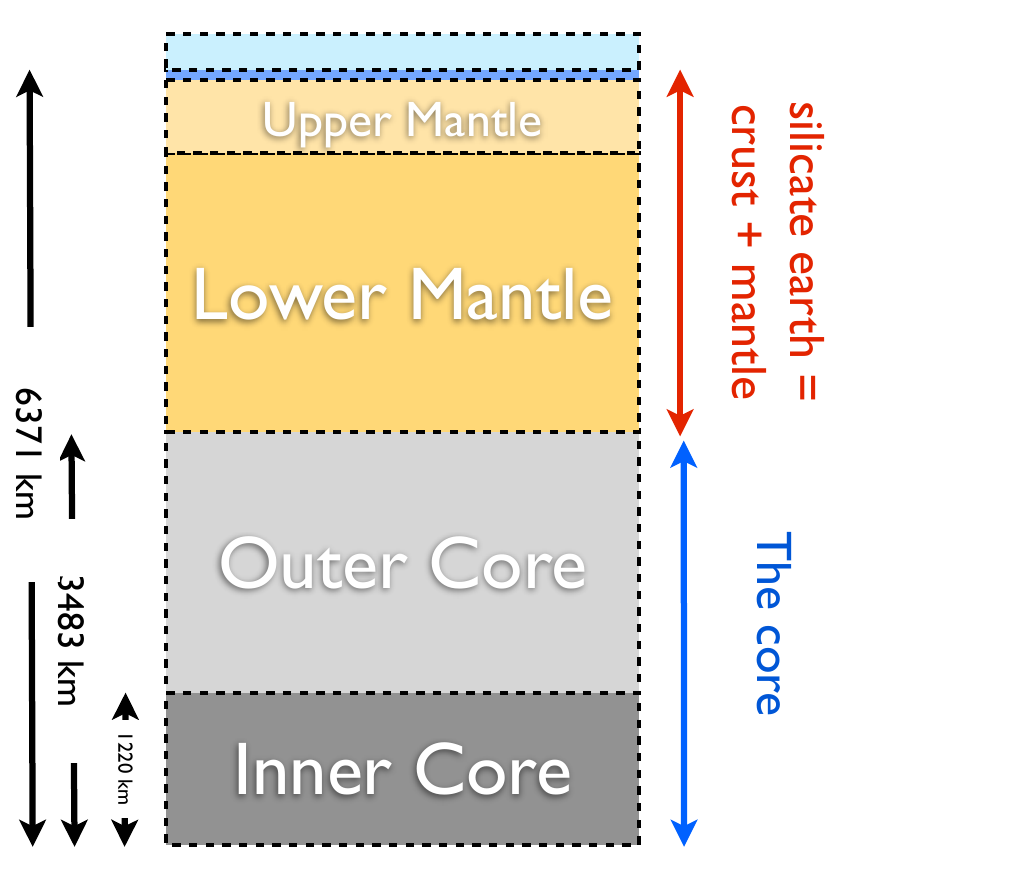}       
    \includegraphics[width=7cm, angle=0]{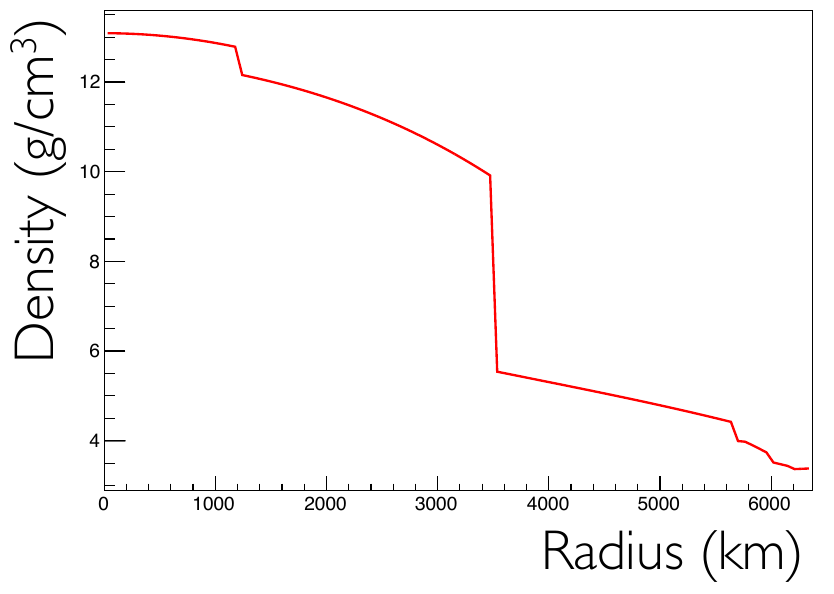}
    \caption{The Preliminary Reference Earth Model
      (PREM)~\cite{Dziewonski:1981xy} model's structure of the Earth
      and density as a function of radius $r$ from the center, showing
      the large change in density at $r \simeq \unit[3\,500]{km}$.
      The PREM is used to map the Earth's interior mass density and to
      account for matter effects in neutrino oscillations.}
          \label{fig:earth_density}
\end{figure}

In addition to vacuum oscillations there are two distinct physical
effects that play a role in the neutrino flavor propagation through
the Earth.  The first is the MSW effect~\cite{MSW-W,MSW-MS} that
enhances the oscillation probability for $\numu \rightarrow \nue$ or
$\numubar \rightarrow \nuebar$ (depending on the neutrino mass
ordering), which is strongly dependent on the matter density for all
path lengths through the Earth. The second effect arises from the
density transition at the Earth's mantle-core interface (see
Fig.~\ref{fig:earth_density}) where neutrinos passing through this
interface can undergo ``parametric enhancement'', which is also called
``neutrino oscillation length resonance'', of their oscillation
probability~\cite{Akhmedov,Petcov-NOLR}, further enhancing neutrino or
anti-neutrino oscillation probabilities depending on the ordering
as in the MSW effect.

\begin{figure}[ht!]
    \centering
    \includegraphics[width=7.2cm, angle=0]{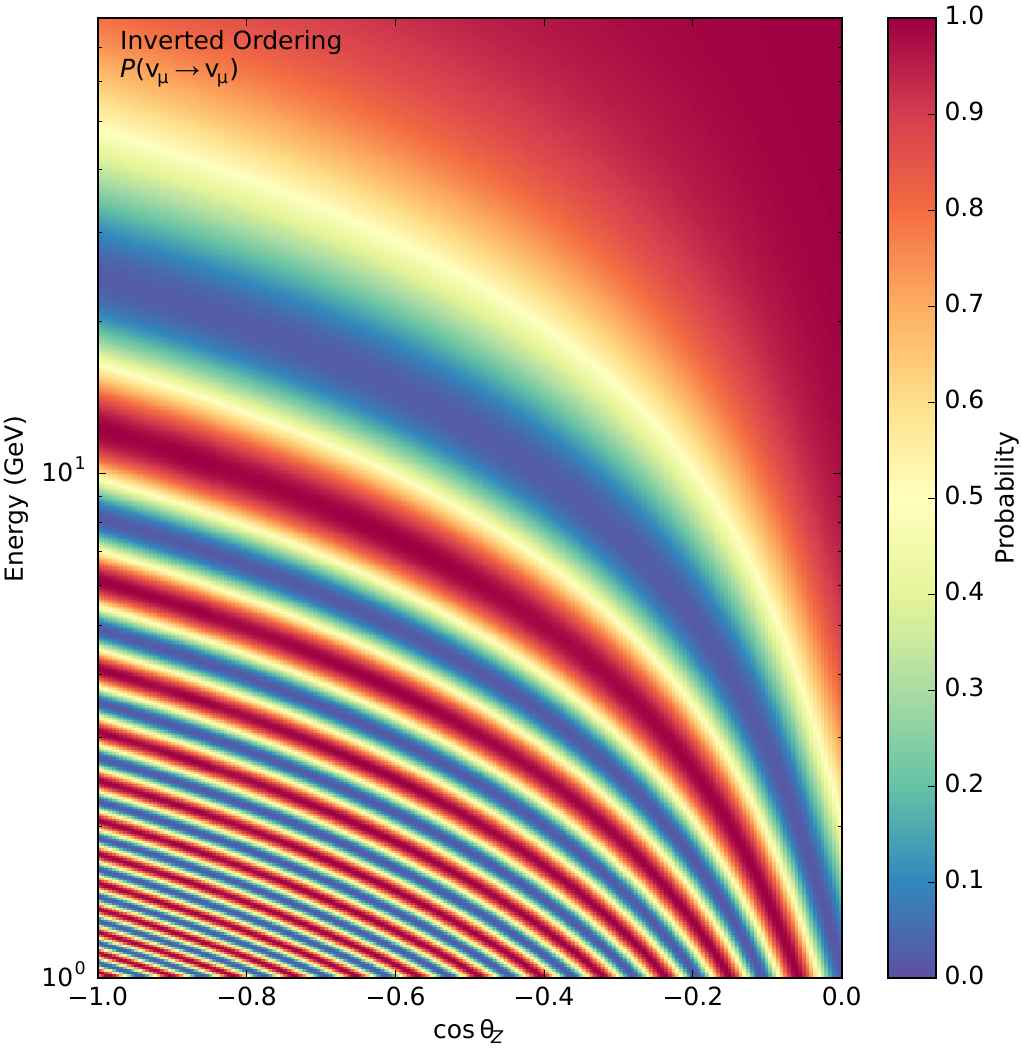}
    \includegraphics[width=7.2cm, angle=0]{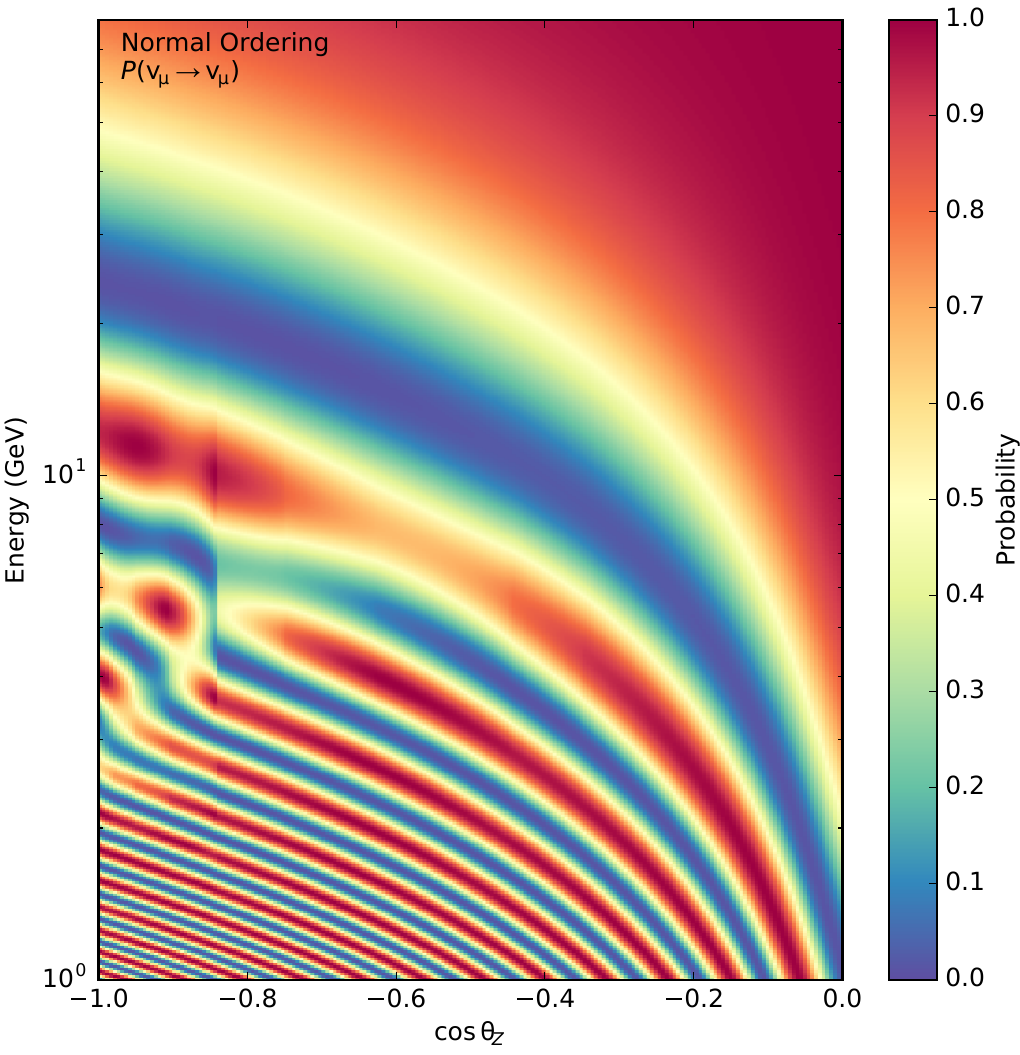}
    \caption{Muon neutrino survival probability after traveling
      through the Earth, as a function of the true neutrino energy and
      cosine of the true zenith angle, shown for the inverted ordering
      (left) and normal ordering (right). A path directly through
      the center of the Earth corresponds to $\cos{\theta}=-1$.  The
      survival probabilities for antineutrinos in a given ordering are
      essentially the same as those for neutrinos under the opposite
      ordering.}
    \label{fig:oscillograms}
\end{figure}

The effects of these additional modifications to the oscillation
probabilities can be seen in the ``oscillograms'' that represent the
probability of oscillation of the neutrino across a range of travel
distances and energies.  Examples of these oscillograms are shown in
Fig.~\ref{fig:oscillograms}, which displays the probability that a
neutrino created as a muon-type is detected as a muon-type
($P_{\numu\rightarrow\numu})$.    Since these plots show the 
oscillation probability for neutrinos, the resonance conditions are
met in the normal ordering case.  For anti-neutrinos, the conditions
are met for the inverted ordering.



\subsubsection{Sample Generation and Event Selection}
\label{sec:atmo_event_sel}

In order to investigate the sensitivity of PINGU to the various
neutrino oscillation parameters, a large sample of simulated neutrino
events is required.  This sample is generated with the Monte Carlo
software previously used for IceCube and DeepCore, using the GENIE and
GEANT4 packages as described in Sec.~\ref{sec:simulation}.  The entire
sample of these events is then reconstructed as described in
Sec.~\ref{sec:EventReconstruction} and Particle Identification (PID)
is applied to it as described in Sec.~\ref{sec:ParticleID}.


Although we conservatively reject downward-going atmospheric neutrinos
using the reconstructed event direction, in the future the subset of
these events that start in the PINGU fiducial volume could be retained
and used as an un-oscillated dataset for normalization purposes, with
the benefit of further understanding and constraining various
systematics.

\subsubsection{Overview of Analyses}
\label{sec:atmo_analysis}

While we have employed several different high-level analysis
techniques, many of the analyses described below use a common
framework and all use the same general methodology.  A detailed
discussion of the statistical approach is provided
in~\ref{sec:NMOAnalysisTechnique}.  In the following sections we will
highlight where a particular analysis deviates from this common
approach.


The analysis methods developed begin by binning the data based on the
reconstructed energy, zenith angle and particle identification.  The
current analysis uses 20 linearly spaced bins in the cosine of the
zenith angle for $-1 < \cos \theta_{\nu} < 0$ and 39 logarithmically
spaced bins in energy for $1 < \Enu < 80~GeV$, and also 2 bins for
particle identification referred to as track-like and cascade-like.
The content of these bins is derived from the previously-described
simulations, and primarily depends on the neutrino flux, the
oscillation probabilities, the cross-sections, and the effects of the
detector.  These values also depend on nuisance parameters that can be
adjusted to account for systematic effects
(cf. Table~\ref{Tab:Systematics}).

In the following sections, we present analyses that use atmospheric
neutrinos in the energy range of \unit[1--80]{GeV} with fluxes as
predicted by~\cite{Honda:2015fha}.  The neutrinos are tracked through
the Earth using a full three-flavor formalism~\cite{prob3++} including
matter effects based on the standard ``PREM'' model of the
Earth~\cite{Dziewonski:1981xy}.  Our predicted measurement of the
neutrino oscillation probabilities is given in
Sec.~\ref{sec:MuonNeutrinoDisappearance}, followed by our
sensitivities to maximal mixing in Sec.~\ref{sec:MaximalMixing} and
the neutrino mass ordering in Sec.~\ref{sec:NeutrinoMassHierarchy},
and finally our predicted measurement of tau neutrino appearance and
PMNS unitarity in Sec.~\ref{sec:TauNeutrinoAppearance}.  Unless
otherwise noted, all results presented below were fully simulated and
reconstructed with the new baseline PINGU geometry ($26\times192$).

The first analysis method, referred to as the ``LLR'' (log-likelihood
ratio) method, is primarily used for the neutrino mass ordering
analysis (see Sec.~\ref{sec:NeutrinoMassHierarchy}), in which two
possibilities for the ordering need to be compared (as opposed to the study of a
continuous parameter space).  This method begins with a ``template''
histogram generated using chosen values of the parameters being
studied. A pseudo-experiment is then created by making a
copy of the template histogram and applying Poisson fluctuations to
the contents of each bin. Then, for both the orderings being studied,
systematic uncertainties are varied in order to find those parameter
values that yield the template that maximizes the likelihood of
observing the pseudo-experiment under each hypothesis. This process is
repeated for many pseudo-experiments, building up a distribution of
the log-likelihood ratio with respect to the parameters studied.  A
more detailed description of this method is presented
in~\ref{sec:LLRAnalysis}.  A variation of this method is also used in
Secs.~\ref{sec:TauNeutrinoAppearance} and~\ref{sec:tomography}.

In the second, ``$\deltaChiSqBar$'' method, we make the Asimov
assumption, i.e. that the mean sensitivity of the experiment is well
represented by the significance obtained from the mean experimental
outcomes, removing the need for individual pseudo-experiments.  This
translates to a reduction of ${\cal{O}} (10^4)$ in processing time for
the neutrino mass ordering analysis.\footnote{The overline in
  ``$\deltaChiSqBar$" indicates that no statistical fluctuations are
  applied, cf.~\ref{sec:AsimovAnalysis}.} The significance for the NMO
determination follows from the two $\deltaChiSq$ values separating the
mean experimental outcomes under the different hypotheses, using a
Gaussian approximation for the distribution of $\deltaChiSq$, as
detailed in~\ref{sec:AsimovAnalysis}. In other analyses using this method,
described in Secs.~\ref{sec:MuonNeutrinoDisappearance}
and~\ref{sec:MaximalMixing}, Wilks theorem~\cite{wilks1938} is used to
convert the $\deltaChiSq$ to significance.

Note that for the analyses in the following two sections,
\ref{sec:MuonNeutrinoDisappearance} and \ref{sec:MaximalMixing},
we have included a nuisance parameter that explicitly accounts for
a systematic error in the overall efficiency of the individual optical modules,
assumed to be known with $10\%$ uncertainty. Wherever this parameter
is present, the uncertainty on the energy scale is reduced from $10\%$ to
$0.5\%$.


\resetlinenumber

\IfFileExists{NewCommands.tex}       {}       {}
\IfFileExists{../NewCommands.tex}    {}    {}
\IfFileExists{../../NewCommands.tex} {} {}

\graphicspath{{figures/}{AtmosphericNeutOsc/NumuDisappearance/figures/}}

\subsection{Muon Neutrino Disappearance}
\label{sec:MuonNeutrinoDisappearance}

The recent atmospheric neutrino oscillation results from
IceCube/DeepCore~\cite{Aartsen:2014yll} have demonstrated the physics
potential of a neutrino detector in the ice at the South Pole.  The
combination of the large flux of neutrinos incident on the detector,
along with their range of path lengths and energies, provides a highly
suitable dataset for analysis of the neutrino oscillation parameters.
This dataset will be increased with the PINGU detector and the
reconstructions upgraded, significantly improving the final result.
The method and projected results will be discussed here.

The determination of the atmospheric neutrino parameters relies on the
disappearance of muon-type neutrinos; following their travel through
the Earth, the flux of these neutrinos will be reduced.  The
probability of muon neutrino survival during propagation through the
Earth is shown in Eq.~\ref{eq:pmumu_full} in the limit that $|\Delta
m^2_{32}| \gg |\Delta m^2_{21}|$ and ignoring matter effects:
\begin{equation}
P(\nu_{\mu}\rightarrow\nu_{\mu}) \simeq 1 -
4\cos^2(\theta_{13})\sin^2(\theta_{23})[1-\cos^2(\theta_{13}) \times
  \sin^2(\theta_{23})] \sin^2(1.27\Delta m^2_{32} L/E_{\nu}).
\label{eq:pmumu_full}
\end{equation}
In Eq.~\ref{eq:pmumu_full} the atmospheric mixing angle and square of
the mass difference ($\thTT$ and $\dmTT$) can be determined using the
range of distances ($L$[km]) and energies ($\Enu$[GeV]) available in
the data set.  The remaining mixing angle ($\theta_{13}$) is treated
as a nuisance parameter with central value and uncertainty given by
best-fit values (from~\cite{NuFIT20}).

There have been many improvements to the analysis following the work
presented in~\cite{Aartsen:2014yll}: The inclusion of additional
photodetectors in the PINGU geometry lowers the energy threshold for
neutrino detection (see Sec.~\ref{sec:EventReconstruction});
the use of more sophisticated reconstruction algorithms retains a much
larger fraction of the events, resulting in a much larger data set at
all energies; the use of flavor identification (see
Sec.~\ref{sec:EventReconstruction}) and cascade-like events lends
added statistical power to the analysis and sharpens the features to
which it is sensitive.  These increases combine to significantly
improve the sensitivity of PINGU to the neutrino oscillation
parameters.

\subsubsection{Event Selection and Reconstruction}
\label{sec:NuMuAppearance:EvtSelectionReco}

The sample used for this study contains neutrinos of all flavors, and
uses the previously described PID algorithm to separate track-like and
cascade-like events.  The majority of the analyses in the previous
version of this document~\cite{LoI} used only the track-like events in
the final analysis, but newer analysis methods have shown that the
inclusion of the cascade-like events can serve to constrain the
non-atmospheric mixing parameters, producing an improvement in the
results.


\subsubsection{Analysis Method}
\label{sec:NuMuAppearance:LikelihoodAnalysis}

In order to determine the oscillation parameter constraints obtained
for a given NMO and corresponding fiducial model, we make use of the
Asimov approach described in~\ref{sec:AsimovAnalysis}. Given the
Asimov dataset for the true ordering, we perform a fine scan in the
$(\sinsqTT, \dmTO)$ plane, and evaluate $\deltaChiSqBar$, defined in
analogy with Eq.~\ref{Eq:AsimovChi2}, but minimized only over the
parameters within the true ordering and
over the remaining systematic parameters, including $\thOT$ which is
allowed to vary around the global fit to the
data in~\cite{NuFIT20}.  The solar neutrino oscillation
parameters $\theta_{12}$ and $\dmOT$ are fixed to their global best
values from the same fit since PINGU is insensitive to
them. Similarly, $\dcp$ is fixed to zero. We then report the
Confidence Level (C.L.) at which any pair of $(\sinsqTT, \dmTO)$
values can be excluded by assuming a $\chi^2$ distribution with 2 d.o.f..

\begin{figure}[htb]
  \centering
  \begin{overpic}[width=0.6\textwidth]{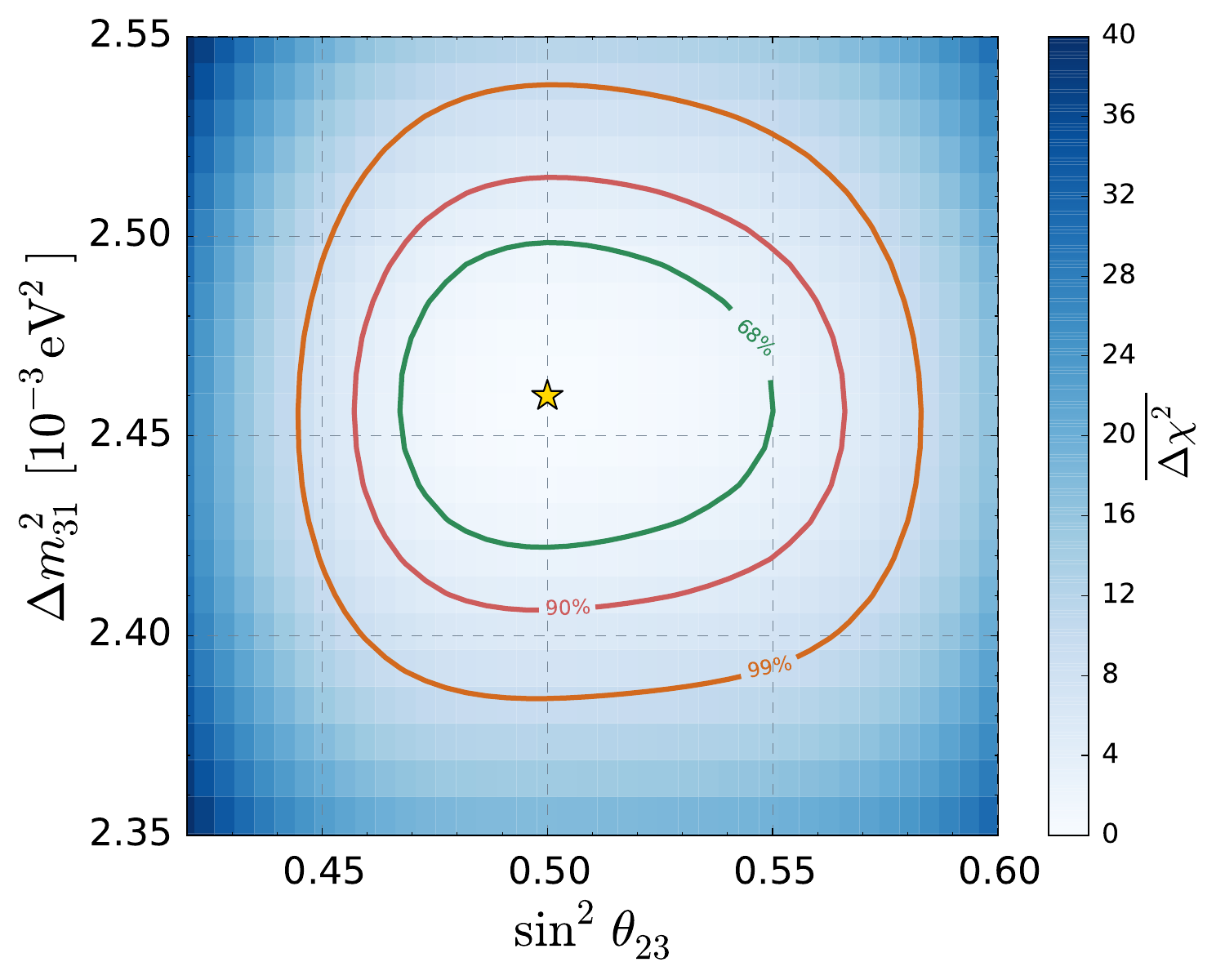}
  \put(52,15){\large \textcolor{red}{Preliminary}}
  \end{overpic}
  \caption{$\overline{\Delta \chi^2}$ profile and various confidence
    regions for the atmospheric oscillation parameters $\dmTO,
    \sinsqTT$ for maximal mixing and normal ordering, assuming a
    four-year exposure time.}
  \label{fig:numu_disappearance_analysis}
\end{figure}

An example of the process is presented in
Fig.~\ref{fig:numu_disappearance_analysis} in which the
$\deltaChiSqBar$ space is shown for the oscillation parameters with
the assumption of maximal mixing within the normal ordering. The three
contours shown here correspond to the $68\%$, $90\%$ and $99\%$
C.L. regions.

\subsubsection{Results}
\label{sec:NumuDisappearance:Results}

\begin{figure}[htb]
  \centering
  \begin{overpic}[width=0.55\textwidth]{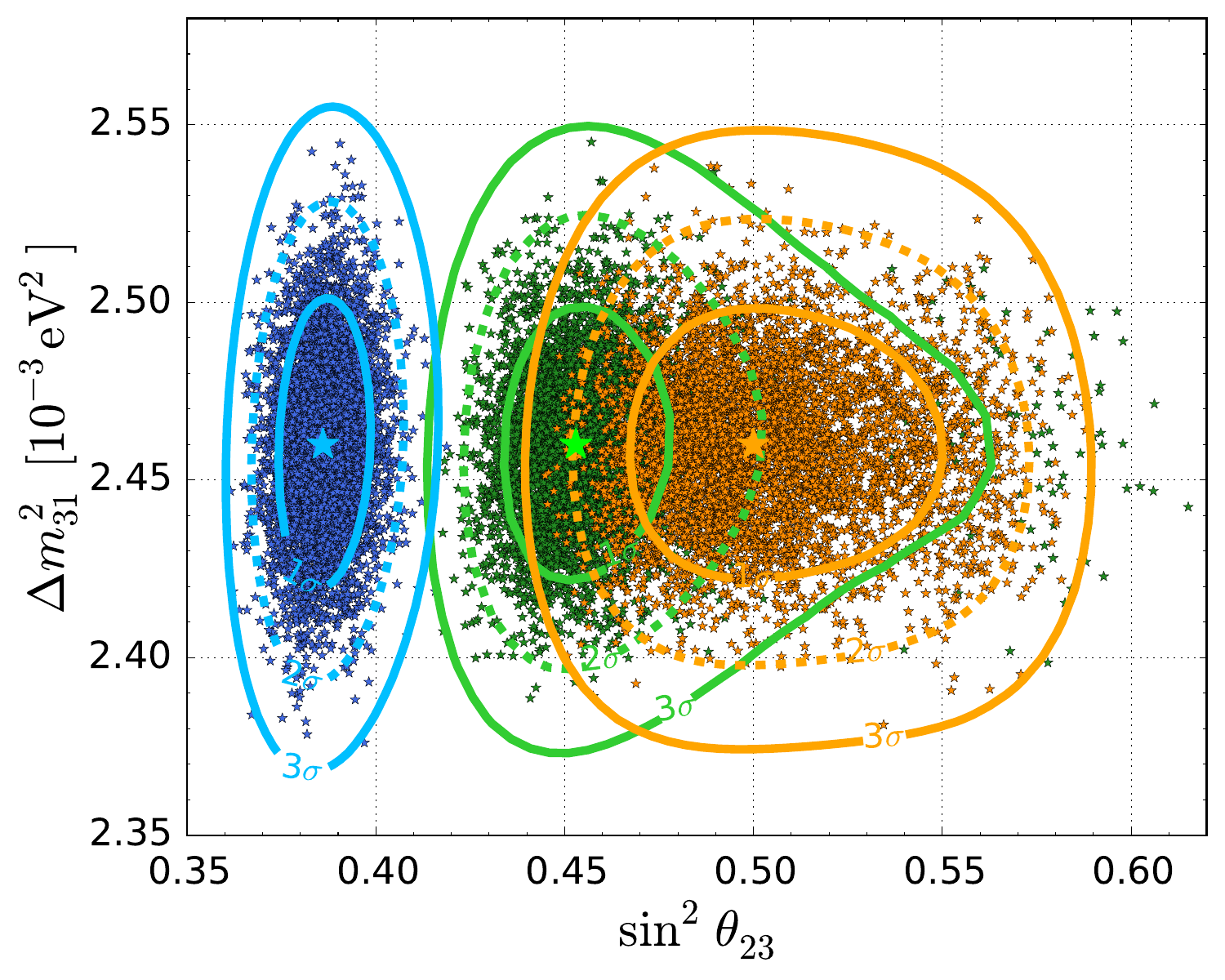}
      \put(66,72){\large \textcolor{red}{Preliminary}}
  \end{overpic}
  \caption{Normal ordering confidence regions under a $\chi^2$
    approximation for the atmospheric oscillation parameters obtained
    from injecting the respective Asimov dataset, superimposed on the
    best fit points from around 5\,000 pseudo-experiments,
    using three different true values for $\sinsqTT$.  The blue points
    and lines represent the Fogli 2012 input~\cite{PhysRevD.86.013012}
    for $\thTT$, the green points and lines represent the NuFit 2014
    inputs~\cite{NuFIT20}, and the orange lines and points represent
    the maximal mixing case.}
  \label{fig:PseudoExpVsScanResults}
\end{figure}

The analysis described above has been performed for a detector
livetime of four years. Since the precision with which $\thTT$ can be
measured depends strongly on its true value, we report the confidence
regions for three different values of $\sinsqTT$: maximal mixing and
two values taken from recent global best fits. In addition to
performing scans using the Asimov method, we have generated around
5\,000 pseudo-experiments for each true value of $\sinsqTT$ and fit
each experiment by optimizing all nine free parameters simultaneously in
order to verify the coverages of the different contours obtained under
the $\chi^2$ approximation. The outcomes of the pseudo-experiments,
projected into the $(\sinsqTT,\dmTO)$ plane, are superimposed on the
scan results in Fig.~\ref{fig:PseudoExpVsScanResults}. Identification
of the experimental trials that are contained within a certain contour
allows for a comparison between the expected and actual coverages,
based on the Asimov assumption and obtained from pseudo-experiments,
respectively. In testing the Asimov assumptions with many generated
experiments, we find that the contours are conservative all the way up
to a C.L. of around $99\%$. For an expected coverage of $90\%$, we
find true coverages that are on the order of $3\%$ larger.  This is
more thoroughly discussed in~\ref{sec:NMOAnalysisTechnique}.


\begin{figure}[htb]
  \centering
  \subfigure[Normal neutrino mass ordering assumed.]{
   	\begin{overpic}[scale=0.3]{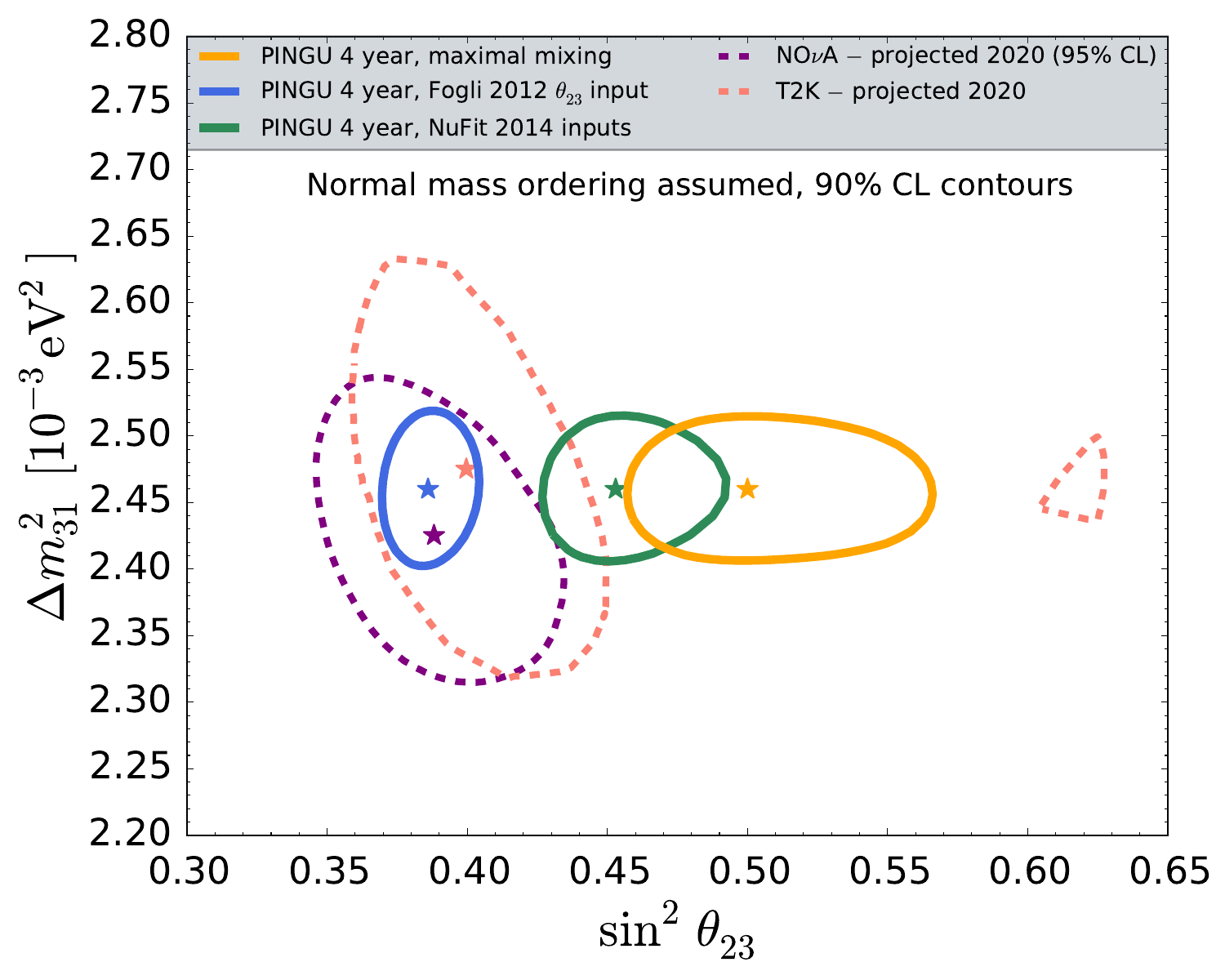}
	  \put(60,15){\large \textcolor{red}{Preliminary}}
 	 \end{overpic}
	}
   \subfigure[Inverted neutrino mass ordering assumed.]{
   	\begin{overpic}[scale=0.3]{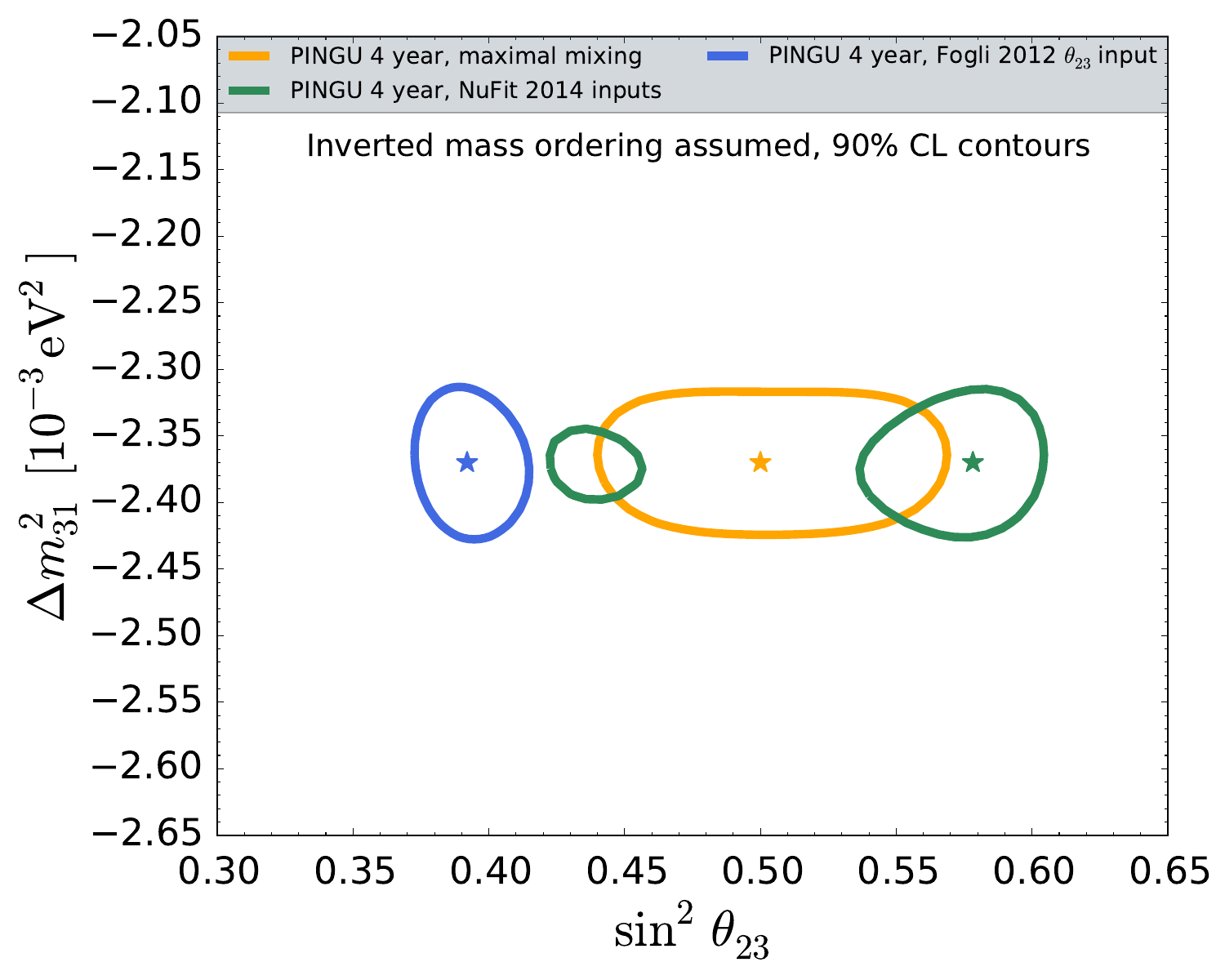}
	  \put(60,15){\large \textcolor{red}{Preliminary}}
 	 \end{overpic}	
	}
  \caption{The atmospheric neutrino oscillation contours are shown
    under assumptions of both the normal and inverted orderings.  Both
    orderings show the effect of the use of different inputs for $\theta_{23}$;
   maximal mixing, Fogli 2012~\cite{PhysRevD.86.013012} and NuFit
    2014~\cite{NuFIT20}. The normal ordering assumption includes
    projected contours from NOvA~\cite{NOvA} and
    T2K~\cite{Abe:2014tzr}.}
  \label{fig:numu_disappearance_llhscan}
\end{figure}

In Fig.~\ref{fig:numu_disappearance_llhscan} we display the $90\%$
confidence regions for the atmospheric oscillation parameters for the
different injected values of $\sinsqTT$, assuming correctly identified
normal mass ordering in the left panel and inverted in the right.
The precision of the $\sinsqTT$ measurement improves
as the true value moves away from maximal ($\sinsqTT=0.5$),
while the measurement of $\dmTO$ remains unaffected. 
In the inverted ordering case, a first octant solution is still allowed
at the $90\%$ C.L. if the true value of $\sinsqTT$ is at its current global best
fit. Projected constraints in the normal ordering case from the
running NOvA and T2K experiments are also included in the left
panel for comparison.

\clearpage
\resetlinenumber

\IfFileExists{NewCommands.tex}       {}       {}
\IfFileExists{../NewCommands.tex}    {}    {}
\IfFileExists{../../NewCommands.tex} {} {}

\graphicspath{{figures/}{AtmosphericNeutOsc/MaximalMixing/figures/}}

\subsection{Sensitivity to the Octant of $\thTT$ and Maximal Mixing}
\label{sec:MaximalMixing}

One of the open questions in neutrino physics is that of the
near-maximal value of $\thTT$. In this section we examine the
prospects for excluding maximal mixing if $\thTT$ differs from maximal
( {\it i.e.}, $\thTT\neq45^\circ$) and, if so, of determining its
octant, {\it i.e.}, whether $\thTT<45^\circ$ or $\thTT>45^\circ$. The
analysis follows the NMO sensitivity studies; see
Table~\ref{Tab:Systematics} (Sec.~\ref{sec:NMHAnalysisSystematics}),
for a list of the oscillation, flux, cross-section and
detector-related systematics that are taken into account. In order to
determine the confidence level at which a certain hypothesis (in this
case a value of $\thTT$) can be rejected, we make use of the Asimov
dataset, and convert the resulting $\deltaChiSqBar$ from any given fit
to an experimental outcome into a statistical significance by assuming
a $\chi^2$ distribution with one degree of freedom.

\begin{figure}[t]
	\begin{center}{
	\begin{overpic}[scale=.55]{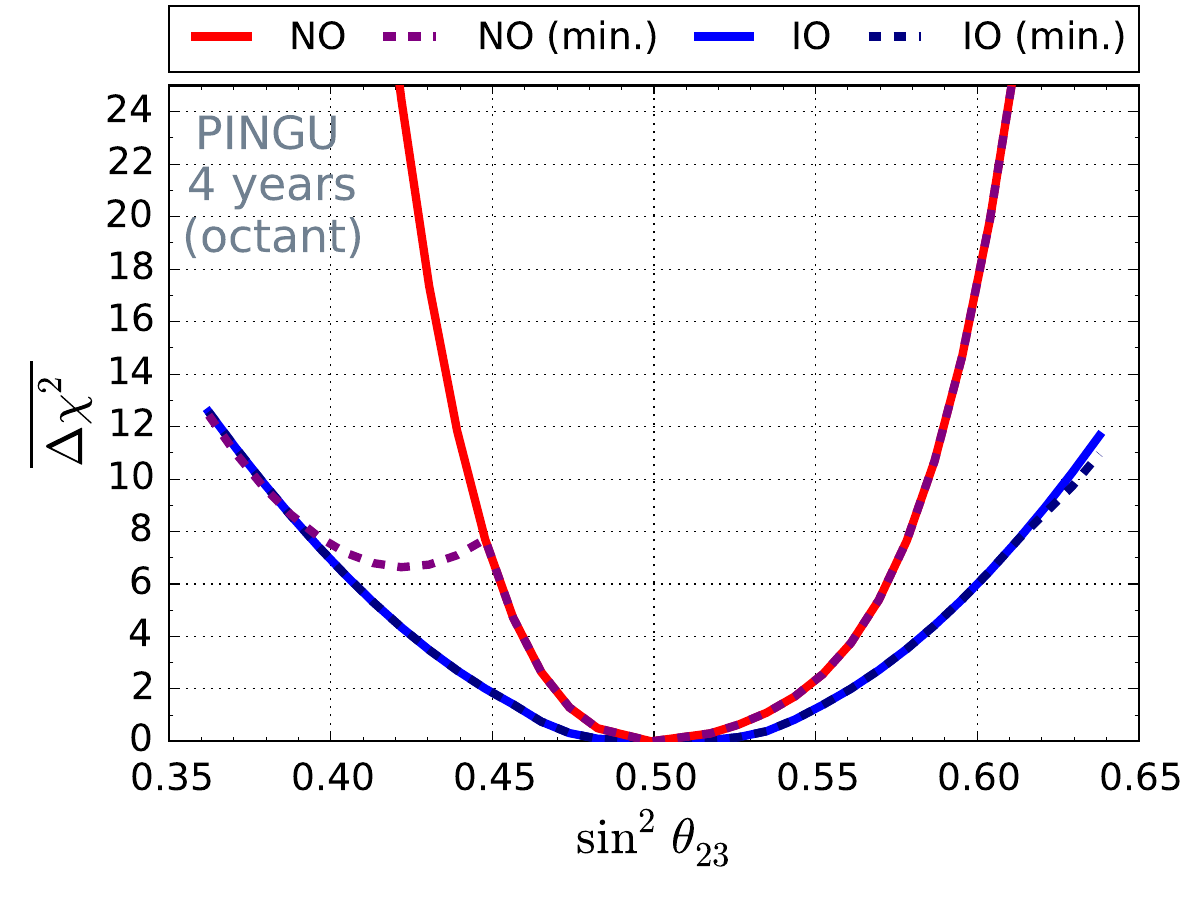}
		\put(45,60){\large \textcolor{red}{Preliminary}}
	\end{overpic}
	}
	\end{center}
	\caption{PINGU sensitivity to the octant of $\thTT$ assuming
          four years of exposure time, as a function of $\sin^2 \thTT$
          and depending on the true neutrino mass ordering.  For the
          solid curves, the test ordering is kept fixed at the truth,
          while for the dashed curves we treat it as a free parameter
          in the fit.}
	\label{fig:4yrOctantSensitivity}
\end{figure}

Figure~\ref{fig:4yrOctantSensitivity} shows the projected sensitivity
of PINGU to the octant of $\thTT$ for an assumed exposure time of four
years as a function of the true value of $\sinsqTTtrue$ (where
$\sinsqTTtrue = 0.5$ corresponds to maximal mixing). The solid line
labeled ``NO" assumes that the true neutrino mass ordering is normal,
while that labelled ``IO" refers to data generated assuming the
inverted ordering. Due to the existence of the ordering-octant
degeneracy (see~\ref{sec:AsimovAnalysis} and especially the right
panel of Fig.~\ref{fig:theta23fitvstrue}), we show two additional
lines which are obtained by minimizing over not only the regular set
of continuous nuisance parameters, but also over the ordering
itself. All $\deltaChiSqBar$ values result from restricting $\thTT$ to
the wrong octant in the fitting procedure. Note that for some values
of $\sin^2(\theta_{23}^\mathrm{true})$ and combinations of true and
tested mass ordering, there might be a second minimum in the wrong
octant apart from $\sin^2(\theta_{23}^\mathrm{test})=0.5$, which we
have ensured will be captured correctly in the case in which it
provides a better fit to the data.

If the ordering is inverted, and whether it is correctly identified or not, four years 
of data taking with PINGU will allow the octant to be determined at more than
$3\sigma$ C.L. if $\sin^2\thTT \textless 0.38$ or $\sin^2\thTT \textgreater
0.62$. If the ordering is normal and correctly identified, $\deltaChiSqBar$ increases
much faster as $\thTT$ deviates from maximal mixing, and a $3\sigma$
octant determination becomes possible for $\sin^2\thTT \textless 0.44$ or
$\sin^2\thTT \textgreater 0.58$.

If the ordering is inverted, we find that testing for a wrong octant
solution within the NO has no effect on the octant sensitivity for
$\sin^2 \thTT \textless 0.62$, and only very slightly reduces it above. The
latter region is where the ordering-octant degeneracy is largest. A
much more significant reduction in $\deltaChiSqBar$ occurs for the
true NO case if the test ordering is optimized, but only for
$\sin^2 \thTT \textless 0.45$. The second octant can then only be
excluded at the $3\sigma$ level for $\sin^2 \thTT \textless 0.38$.

\begin{figure}[ht]
	\begin{center}{
	\begin{overpic}[scale=.5]{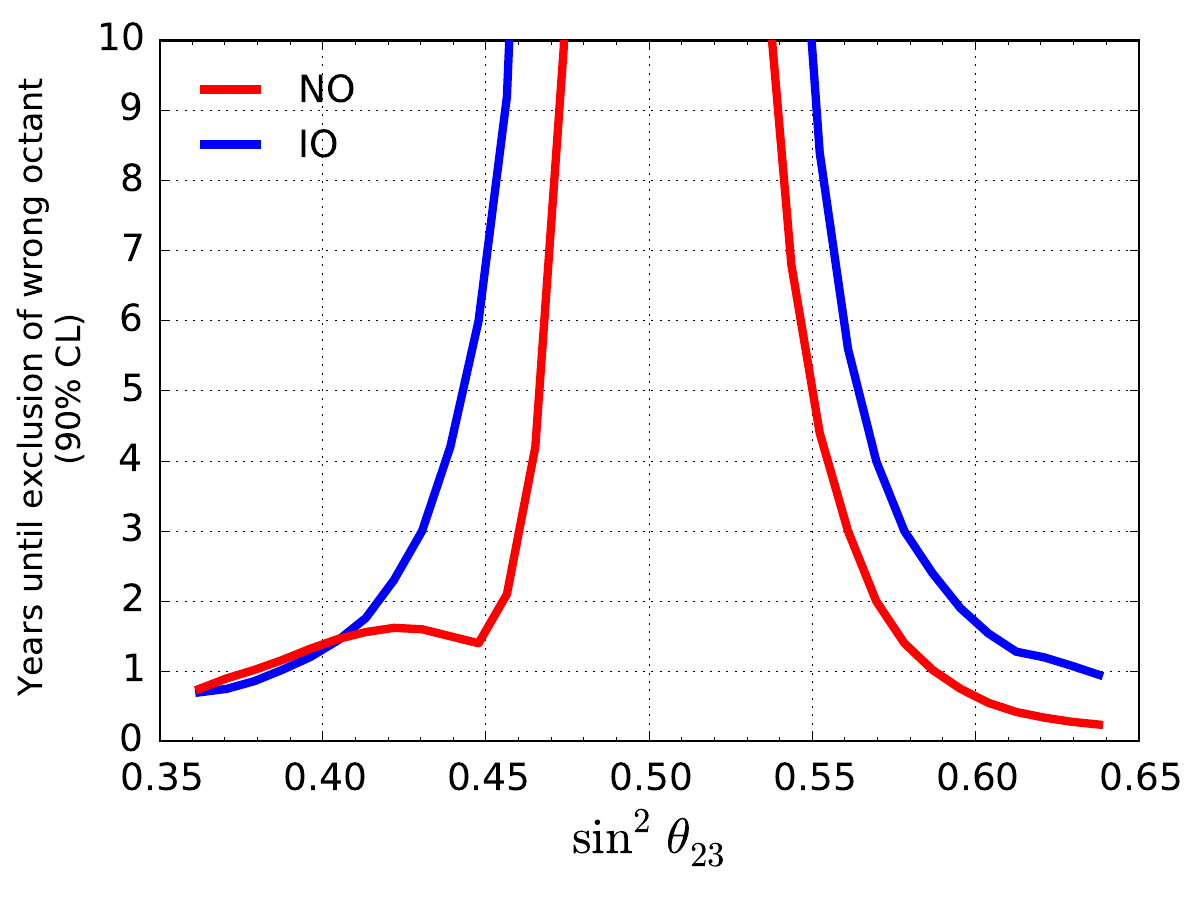}
		\put(45,20){\large \textcolor{red}{Preliminary}}
	\end{overpic}
	}
	\end{center}
	\caption{Exposure time in years required for PINGU to exclude
          the wrong octant of $\thTT$ at $90\%$ C.L., as a function of
          $\sin^2 \thTT$ and the true neutrino mass
          ordering. $\deltaChiSqBar$ is also minimized over the test
          ordering.}
	\label{fig:TimeUntilOctantExclusion}
\end{figure}

In addition to fixing the exposure time as in
Fig.~\ref{fig:4yrOctantSensitivity}, we have investigated how many
years of exposure time it would require in order for PINGU to make a
$90\%$ C.L. measurement of the octant, again depending on the true NMO
and $\sin^2(\theta_{23}^\mathrm{true})$. This effectively probes the
scaling of $\deltaChiSqBar$ with time. In
Fig.~\ref{fig:TimeUntilOctantExclusion} the results of this study are
presented; minimization over the ordering is included in both
curves. The very flat behavior of the value of $\deltaChiSqBar$ at
roughly 0 when $\thTT$ is close to maximal observed in
Fig.~\ref{fig:4yrOctantSensitivity} finds its counterpart here in the
very steep rise in exposure time as maximal mixing is approached from
both sides. The ``shoulder" in the NO case for $\sin^2 \thTT \textless
0.45$ and the barely visible bump at $\sin^2 \thTT \textgreater 0.62$ for IO
again correspond to the regions where misidentification of the NMO degrades
the octant sensitivity and leads to an increase in the exposure time
required to determine it.  PINGU is projected to make a $90\%$
C.L. determination of the octant of $\thTT$ within less than a year if
the NO is true and either $\sin^2 \thTT \textless 0.38$ or $\sin^2 \thTT
\textgreater 0.59$, or if the IO is true and either $\sin^2 \thTT \textless
0.39$ or $\sin^2 \thTT \textgreater 0.63$. A five-year measurement at the
$90\%$ C.L. is in reach for NO and $\sin^2 \thTT \textless 0.47$ or
$\sin^2 \thTT \textgreater 0.55$, or for IO and $\sin^2 \thTT \textless 0.44$
or $\sin^2 \thTT \textgreater 0.56$.

\begin{figure}[ht]
	\begin{center}{
	\begin{overpic}[scale=.5]{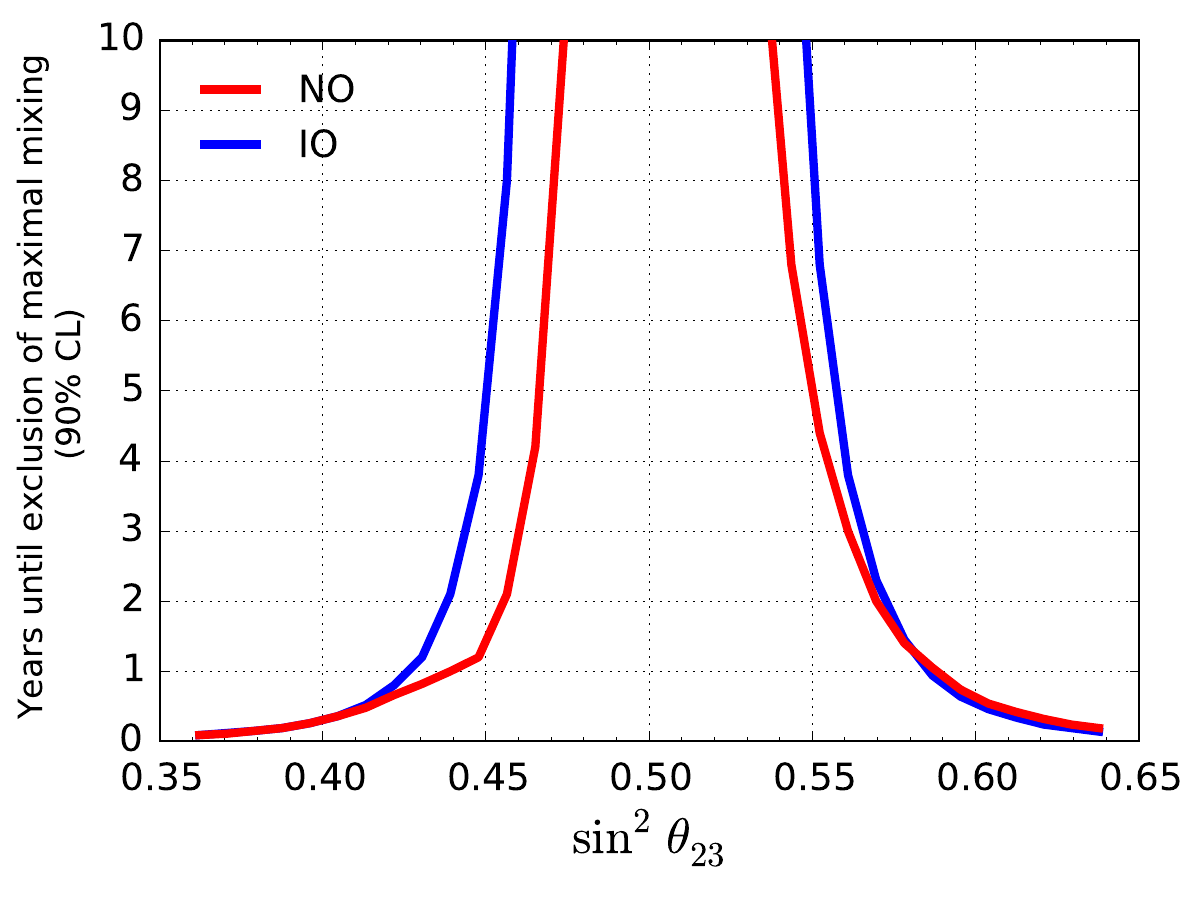}
		\put(45,20){\large \textcolor{red}{Preliminary}}
	\end{overpic}
	}
	\end{center}
	\caption{Exposure time in years required for PINGU to exclude
          maximal mixing at $90\%$ C.L., as a function of $\sin^2
          \thTT$ and the true neutrino mass ordering. $\deltaChiSqBar$
          is also minimized over the test ordering.}
	\label{fig:TimeUntilMaxMixExclusion}
\end{figure}

If we instead examine the time it takes until a non-maximal $\thTT$ is
established at $90\%$ C.L. assuming it is non-maximal
(Fig.~\ref{fig:TimeUntilMaxMixExclusion}), the overall trend is quite
similar. Since the best wrong-octant fit of $\thTT$ is
never a worse match to the data than maximal mixing, the exposure time
that is required to exclude a maximal $\thTT$ at a given C.L. provides
a lower bound on the time to exclude the wrong octant at the same C.L.
Uncertainty of the NMO only has minor impact on the measurement for NO
and $\sin^2 \thTT \textless 0.45$, where assuming maximal mixing in the
wrong IO yields better fits than it does in the true NO.
Minimization over the test ordering does not impact the sensitivity to
maximal mixing at all if the IO is true.

A non-maximal $\thTT$ can be established at $90\%$ C.L. with a
one-year exposure time if the ordering is normal and $\sin^2 \thTT
\textless 0.44$ or $\sin^2 \thTT \textgreater 0.59$, or if the ordering is
inverted and $\sin^2 \thTT \textless 0.43$ or $\sin^2 \thTT \textgreater
0.58$. Non-maximal mixing can be determined in less than five years in
the NO case for $\sin^2 \thTT \textless 0.47$ or $\sin^2 \thTT \textgreater
0.55$, and in the IO case for $\sin^2 \thTT \textless 0.45$ or $\sin^2
\thTT \textgreater 0.56$.

\clearpage
\resetlinenumber

\IfFileExists{NewCommands.tex}       {}       {}
\IfFileExists{../NewCommands.tex}    {}    {}
\IfFileExists{../../NewCommands.tex} {} {}

\graphicspath{{figures/}{AtmosphericNeutOsc/Hierarchy/figures/}}

\subsection{Neutrino Mass Ordering}
\label{sec:NeutrinoMassHierarchy}

\begin{figure}
  \centering
  \subfigure[$\numu$ CC events.]{
   	\begin{overpic}[scale=0.33]{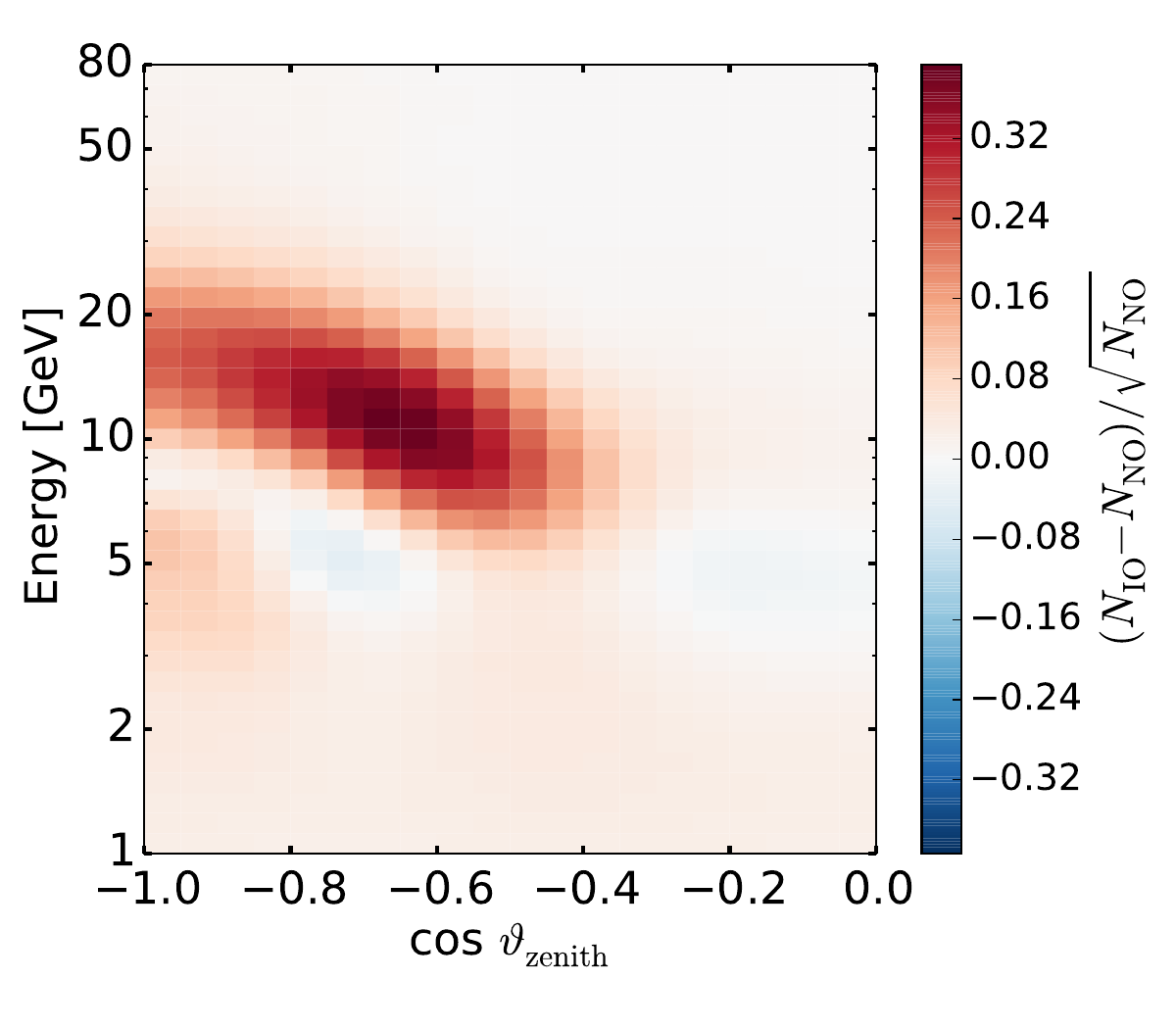}
	\put(34,72){\large \textcolor{red}{Preliminary}}
	\end{overpic}
        \label{fig:DistinguishabilityAfterReconstructionTracks}
  }
  \subfigure[$\nue$ CC events.]{
   	\begin{overpic}[scale=0.33]{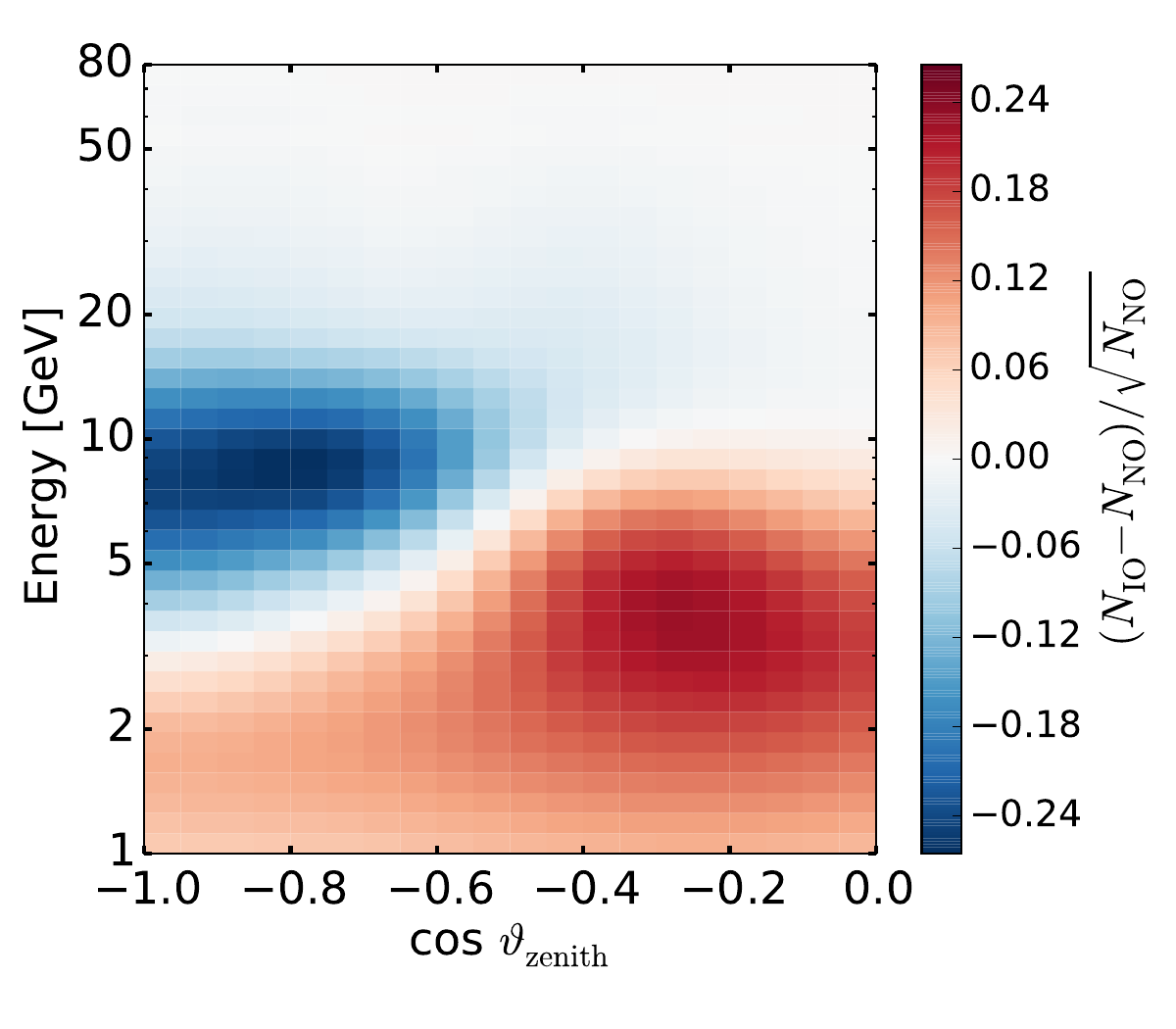}
	\put(34,72){\large \textcolor{red}{Preliminary}}
	\end{overpic}
    \label{fig:DistinguishabilityAfterReconstructionNue}
  }
  \subfigure[$\nutau$ CC events.]{
   	\begin{overpic}[scale=0.33]{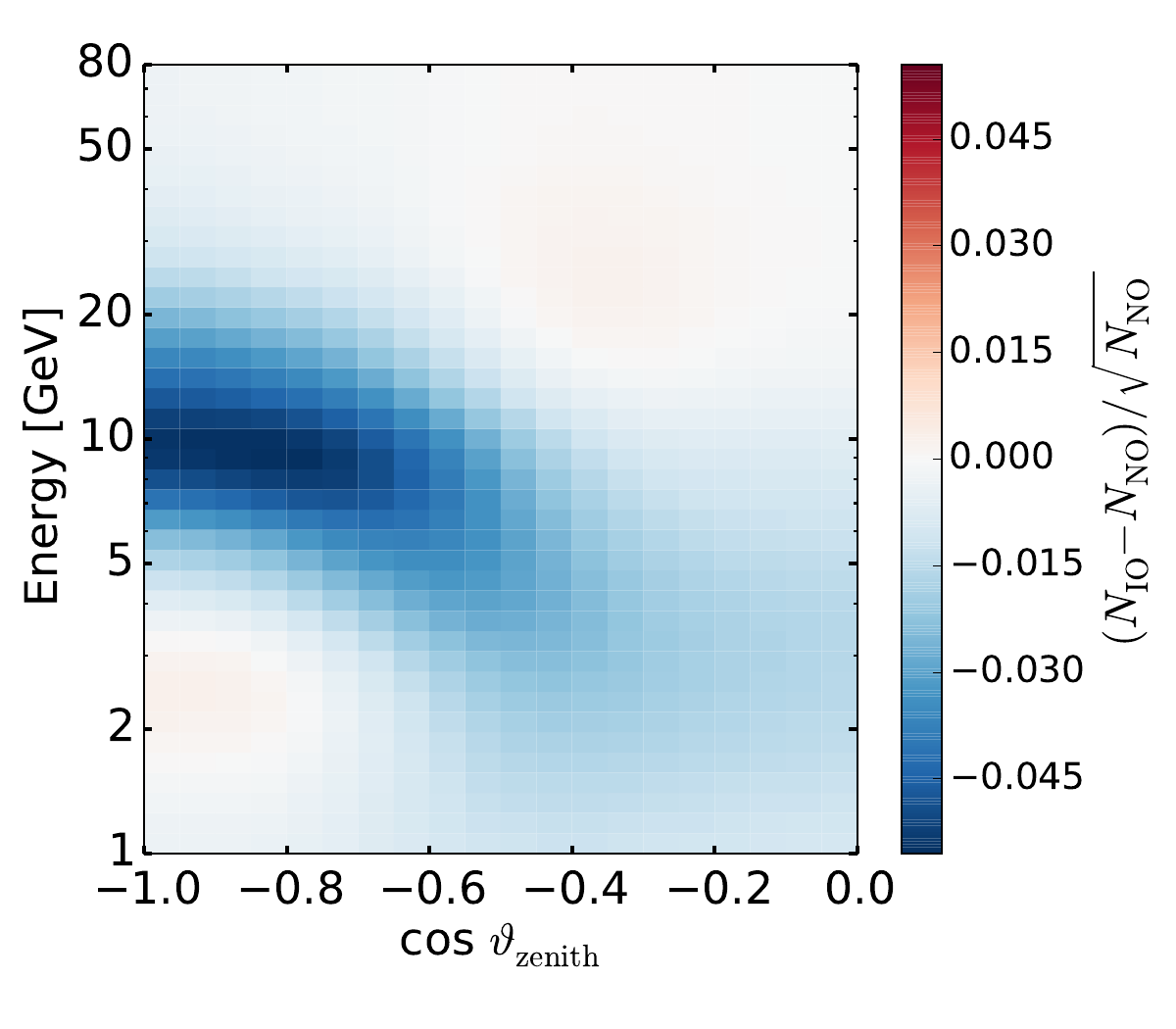}
	\put(33,72){\large \textcolor{red}{Preliminary}}
	\end{overpic}
        \label{fig:DistinguishabilityAfterReconstructioNutau}
  }
  
  \caption{Distinguishability metric as defined in~\cite{ARS} for one
    year of simulated PINGU data, with parametrized reconstruction
    resolutions as described in Sec.~\ref{sec:atmo_analysis}.  The
    square-root of a quadratic sum over the bins in each plot gives an
    estimate of the number of $\sigma$ separating the two orderings.
    For illustrative purposes we have assumed perfect flavor ID.
    The top left figure (a) shows track-like events from CC~$\numu$
    interactions.  The top right figure (b) shows CC~$\nue$ events and
    the bottom figure (c) shows CC~$\nutau$.}
  \label{fig:Distinguishability}
\end{figure}

Data from solar neutrino measurements~\cite{ref:SNOPaper} have shown
that $m(\nu_2) > m(\nu_1)$, but the position of $\nu_3$ in the
ordering of these masses is, as yet, unknown.  The sensitivity of the
PINGU detector to neutrinos with energies in the range of roughly
$5$--$15$~GeV allows this ordering to be determined.
Figure~\ref{fig:Distinguishability} shows three distinguishability
plots similar to those described in Ref.~\cite{ARS}.  To illustrate
the individual contributions to the ordering signal, three neutrino
flavors are shown separately under the assumption of perfect particle
identification (PID).  These plots identify regions in which the
number of events expected for the NO is greater than that expected for
the IO (blue regions) and vice-versa (red regions).  While this metric
is helpful for highlighting the regions of interest in the
energy-angle space from which useful information may be extracted, it
provides only an approximation of the PINGU sensitivity to the NMO
without systematics.  More detailed simulations and analysis methods
are required to determine the actual PINGU sensitivity, as discussed
below.

Compared to the first version of this document~\cite{LoI}, we have
incorporated several improvements into our NMO analysis.  In addition
to changing the geometry of the detector (see
Sec.~\ref{sec:DetectorDesignPredictedPerformance}), we have also
included a detailed simulation of the intrinsic noise.  We use a
detector livetime of four years which matches the study of atmospheric
$\numu$ disappearance (Sec.~\ref{sec:MuonNeutrinoDisappearance}).
Continuing to employ a full event reconstruction and particle
identification to discriminate track-like and cascade-like events, we
have studied the impact of numerous additional systematic
uncertainties. These have been partly implemented in the
log-likelihood ratio (LLR) analysis, as will be described.  We have
also changed the choice of the wrong ordering hypothesis to be the one
that most resembles the true ordering Asimov template, {\it i.e.},
that which maximizes the confusion between orderings (see
Sec.~\ref{sec:LLRAnalysis} for more details).

\subsubsection{Systematics}
\label{sec:NMHAnalysisSystematics}

Numerous sources of systematic errors have been identified and their
impact on the PINGU sensitivity to the determination of the NMO has been
quantified.  For convenience, we have categorized these uncertainties
into three subsets: neutrino oscillation parameters, neutrino flux and
cross section, and detector modeling.  In what follows we describe
and quantify each of these subsets of systematics, and indicate
possible ways to better constrain them to reduce their impact on the
final significance.  It should be noted that the calibration systems
installed with the PINGU detector (see Sec.~\ref{sec:Calibration})
will reduce the magnitude of the detector-related systematics.

The specific systematic uncertainties we have studied are listed in
Table~\ref{Tab:Systematics}. Those implemented as nuisance parameters
in the analysis procedures (see Sec.~\ref{sec:atmo_analysis}) are
indicated with an asterisk, with central values and priors indicated
for most parameters in the table. The uncertainties related to the
atmospheric flux and cross-sections were investigated only using the
$\deltaChiSqBar$ approach and are not included in the results of the
LLR analysis due to the computational time required to study them.
\begin{table}
\begin{center}
   \begin{tabular}{lccl} \hline \noalign{\vskip 1mm}
                  & LLR & $\overline{\Delta \chiˆ2}$ & \\
                  & $\ast$ & $\ast$ & $\Delta m^2_{31} = 2.46\times 10^{-3}\,{\rm eV}^2,\; -2.37\times 10^{-3}\,{\rm eV}^2$~\cite{NuFIT20} \\
      Oscillation & $\ast$ & $\ast$ & $\theta_{23} = 42.3^\circ,\;49.5^\circ$~\cite{NuFIT20} \\
                  & $\ast$ & $\ast$ & $\theta_{13} = 8.5^\circ \pm 0.2^\circ$~\cite{NuFIT20} \\
                  &  & $\dag$ & $\dcp = 0^\circ$ \\ \hline
                  & $\ast$ & $\ast$ & Event rate = nominal \\
                  & $\ast$ & $\ast$ & $\nue/\numu$ flux ratio = nominal $\pm\;3\%$~\cite{Barr2006} \\
  Flux \&         & $\ast$ & $\ast$ & $\nu/\nubar$ flux ratio = nominal $\pm\;10\%$~\cite{Barr2006} \\
  Cross Section   & $\ast$ & $\ast$ & Atmospheric spectral index = nominal $\pm\;0.05$~\cite{Barr2006} \\
                  &        & $\dag$ & Air-shower interactions~\cite{Barr2006} \\
                  &        & $\dag$ & Neutrino cross-section (see Sec.~\ref{sec:GENIESysts}) \\ \hline
                  & $\ast$ & $\ast$ & Energy scale $= 1.0 \pm 10\%$ $(\dag\, \pm 0.5\%)$ \\
      Detector    &   &  \dag & Individual module efficiency = nominal $\pm 10\%$ \\
                  &   &   & Ice properties \\ \hline
   \end{tabular}
   \caption{List of uncertainties studied in the NMO analysis, with nominal
     values and Gaussian priors (where applicable) shown.
     Asterisks indicate those systematics that were implemented as
     nuisance parameters in the analysis
     procedures.  Daggers indicate systematics used only in specific 
     studies and not fully incorporated into the main analysis.
     \label{Tab:Systematics}}
\end{center}
\end{table}
The oscillation parameters to which PINGU is most sensitive are
$\thTT$ and $\dmTO$.  We present significances with these included as
nuisance parameters in the ordering analysis, both with (LLR analysis
only) and without priors. The other oscillation parameters allowed to
vary in the analysis are $\thOT$ and $\dcp$, but they have smaller
impact on the overall sensitivity (the implementation of $\dcp$ as a
systematic is discussed in Sec.~\ref{sec:NMHdeltaCP}). The remaining
oscillation parameters ($\dmOT$, $\theta_{12}$) were also considered
but they have negligible impact and are held fixed. The injected true
values of these parameters are taken from the global best fit
analysis~\cite{NuFIT20}, and the $1\sigma$ error is used as a prior on
$\thOT$. The analysis exhibits a strong degeneracy between the octant
of $\thTT$ and the ordering in most of the allowed parameter space of
$\thTT$, as discussed in Sec.~\ref{sec:NMOAnalysisTechnique} (and, {\it
  e.g.}, in~\cite{Capozzi:2015bxa}), and this is accounted for in the
analysis procedure.

The flux uncertainties are also treated as nuisance parameters.  For
the atmospheric neutrino flux ratios and spectral index we use central
values and $1\sigma$ uncertainties from~\cite{Barr2006} as the nominal
values and priors, respectively. The overall atmospheric flux
normalization is allowed to vary without a prior in order to account
not only for the atmospheric flux normalization, but also any
uncertainties in the overall detector rate normalization.  The
systematic impact of numerous parameters describing neutrino
interaction cross-sections were studied using the
GENIE~\cite{Andreopoulos:2009rq} neutrino event generator, full
details of which are available in Secs.~\ref{sec:MCSoftware}
and~\ref{sec:GENIESysts}. 

Detector-related systematics introduce uncertainties in the visible
event energy as determined by our reconstruction algorithms.  These
uncertainties could arise from mis-calibrated module detection
efficiencies or imperfections in our modeling of the optical
properties of the ice in which PINGU will be deployed.  A complete
study of these effects requires a significant amount of simulated
data. Instead of implementing these uncertainties as
nuisance parameters, we evaluated the NMO significance for a small
number of independent datasets, each of which featured a variation of
either the overall module efficiency or the ice property model.  As a
proxy for other possible uncertainties, we allow the reconstructed
visible energy to vary globally as a nuisance parameter, with a
conservative prior of 10\% guided by estimations from IceCube
calibrations with {\it in situ} light sources (see
Sec.~\ref{sec:Calibration} for more details).  None of these
uncertainties had a strong impact on the NMO significance.

In the energy range of relevance to the
NMO measurement, neutrino interactions with matter are dominated
by deep inelastic scattering (DIS). Since DIS uncertainties are
relatively modest, the overall impact of these uncertainties was found
to be small.  The impact of the neutrino cross section systematics on the NMO
determination have also been studied using the $\deltaChiSqBar$
analysis method but with the addition of six more systematic
parameters listed in Table~\ref{tab:geniesyst}.  The impact of these systematics on the NMO
measurement is found to be very small.  More details on these
systematics and how they were implemented can be found
in Sec.~\ref{sec:GENIESysts}. 

\renewcommand{\arraystretch}{1.3}
\begin{table}
\centering

\begin{tabular}{c c c}\hline \hline
Name & nominal value  & uncertainty (\%) \\\hline
$\maccqe$	& 0.99 & $-15,+25$ \\
$\mares$	& 1.120 & $\pm 20$ \\
$\ahtby$	& 0.538 &  $\pm 25$ \\
$\bhtby$	& 0.305 &  $\pm 25$ \\
$\cvouby$	& 0.291 &  $\pm 30$ \\
$\cvtuby$	& 0.189 &  $\pm 30$ \\
\hline \hline
\end{tabular}

\caption{List of parameters and their associated uncertainties based
  on GENIE\label{tab:geniesyst}. Nominal values of the parameters are taken from
  Appendix B of the GENIE User Manual. See Sec.~\ref{sec:MCSoftware} for more details. }
\end{table}

\renewcommand{\arraystretch}{1}

To quantitatively assess the impact of each of these groupings of
systematics, each was varied independently of the other
groups. Table~\ref{Tab:SystematicsImpact} indicates the ``impact'' of
these different groups of systematics by comparing the 4-year PINGU
sensitivity with each set of systematics to the theoretical limit
where statistical uncertainties dominate (all the systematics are
known to arbitrary precision). Clearly, the oscillation parameter
uncertainties are the dominant contributors to the decrease in
significance compared to the ``no systematics'' case. The other two
groups shown have only a moderate effect when included by themselves.

\begin{table}
\begin{center}
   \begin{tabular}{c|cc|cc} \hline  \noalign{\vskip 1mm}
                       & \multicolumn{2}{|c|}{LLR}               & \multicolumn{2}{c}{$\deltaChiSqBar$}   \\ \hline
      Systematic       & 4~yr $n_\sigma$ (NO) & 4~yr $n_\sigma$ (IO) & 4~yr $n_\sigma$ (NO) & 4~yr $\sigma$ (IO)\\ \hline
      None             	& 5.5		& 5.5		&  5.5		& 5.5 \\
      Flux only        	& 5.2		& 5.0		&  5.1		& 5.1 \\
      Detector only    	& 4.3		& 4.4		&  4.5		& 4.5 \\
      Oscillation only 	& 3.4, 3.5 	& 2.9, 3.1	&  3.4		& 3.0 \\
      All              	& 2.9, 3.3	& 2.6, 2.9	&  2.8		& 2.6\\
   \end{tabular}
   \caption{ Summary of the systematic errors and their impacts on the
     estimated four-year significance of the mass ordering
     measurement, for both the LLR and $\deltaChiSqBar$ analysis
     methods and for the cases in which the normal and inverted
     orderings are true. Where two significances are given, the first
     value is for the case where NuFit~2.0~\cite{NuFIT20} priors are
     not used in the fit and the second is for where they are used
     (see Sec.~\ref{sec:NMOAnalysisTechnique} for
     details). \label{Tab:SystematicsImpact}}
\end{center}
\end{table}

\subsubsection{Results}
\label{sec:NMHAnalysisResults}

\begin{figure}
  \centering
  \subfigure[Normal neutrino mass ordering assumed.]{
    \begin{overpic}[scale=0.3]{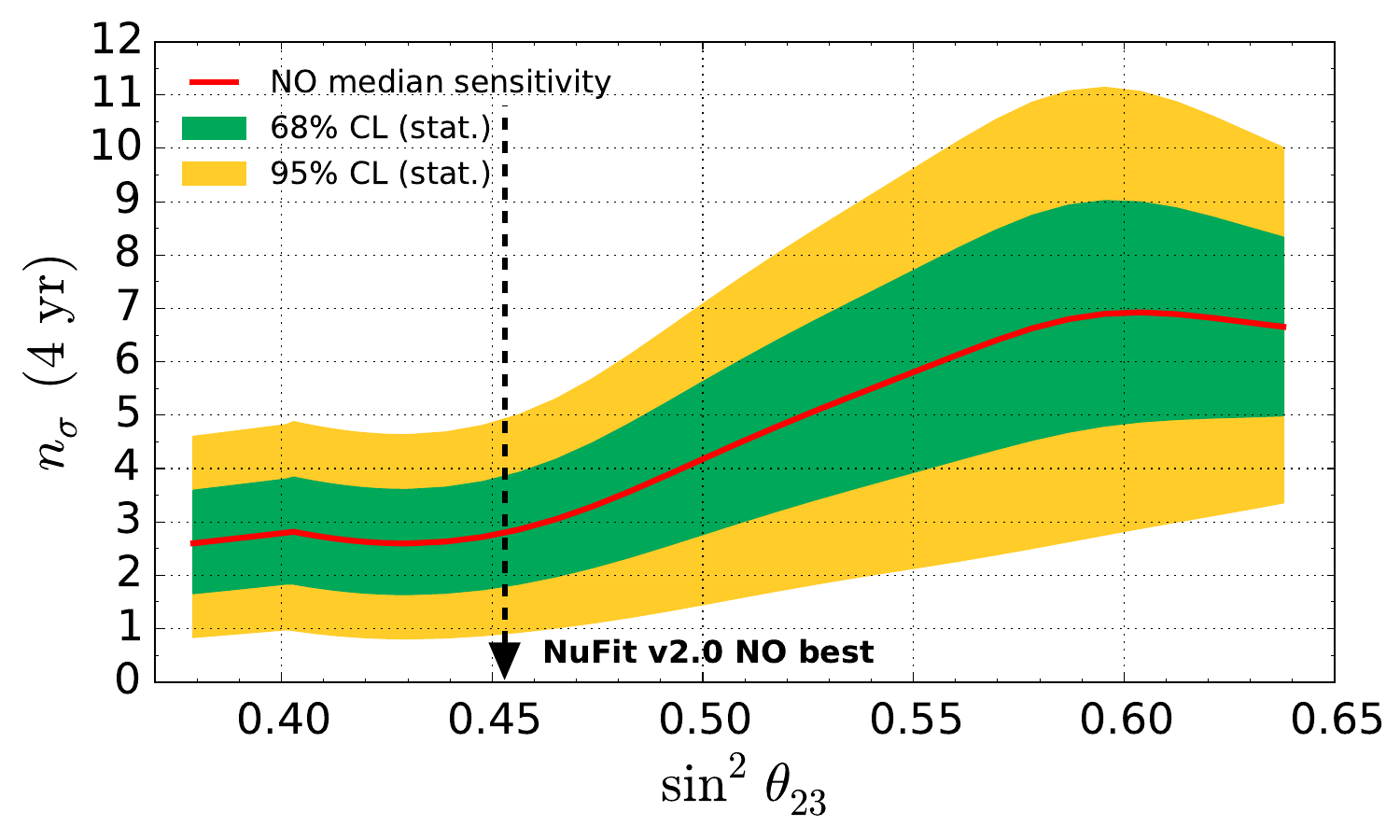}
      \put(65,12.6){\textcolor{red}{Preliminary}}
    \end{overpic}
  }
  \subfigure[Inverted neutrino mass ordering assumed.]{
    \begin{overpic}[scale=0.3]{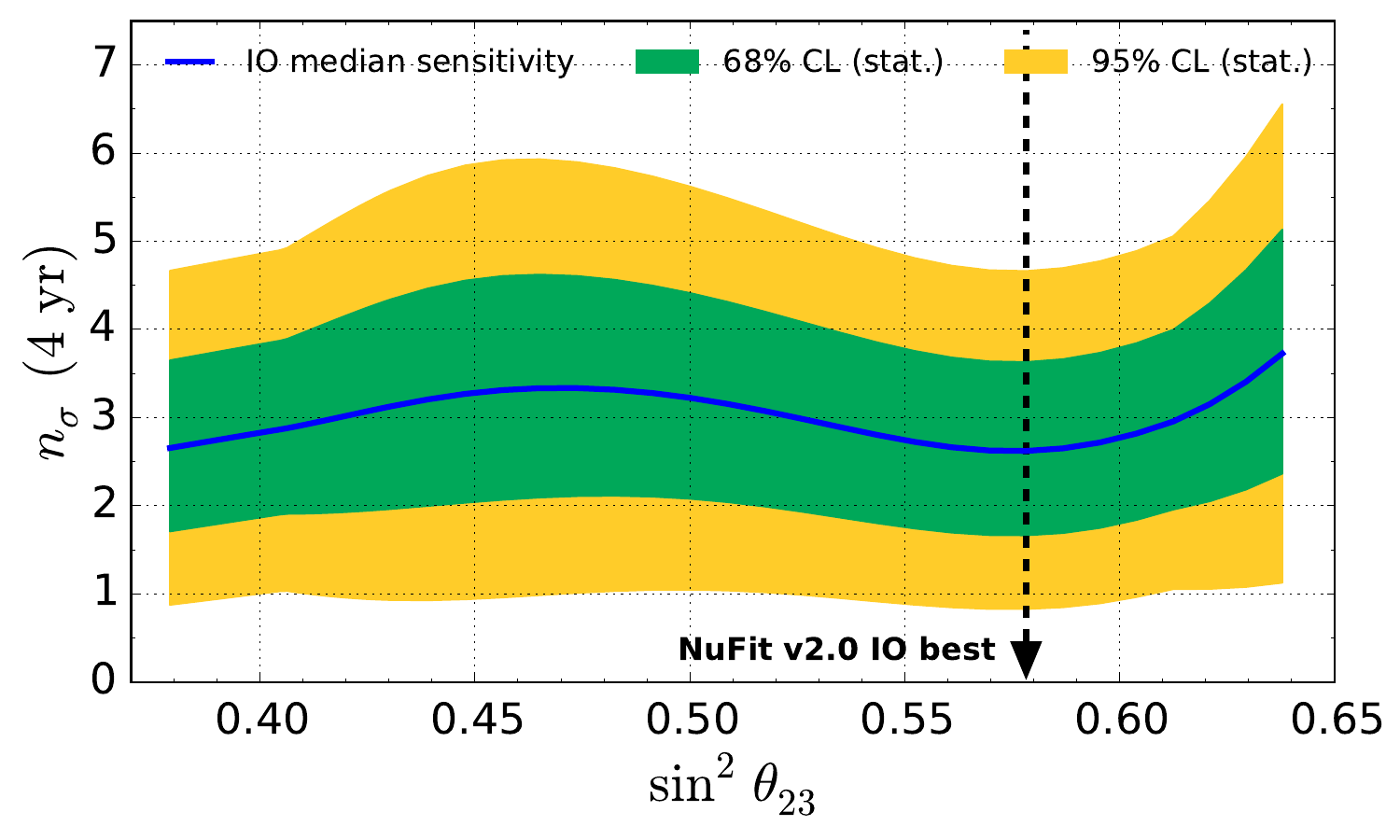}
      \put(14,12.6){\textcolor{red}{Preliminary}}
    \end{overpic}
  }
  \caption{Expected significance with which the neutrino mass ordering
    will be determined using four years of data, as a function of the true
    value of $\sinsqTT$.  Solid red (NO) and blue (IO) lines show median
    significances, while the green and yellow bands indicate the range
    of significances obtained in 68\% and 95\% of hypothetical
    experiments.  The significance for determining the ordering when
    the true ordering is inverted is relatively insensitive to
    $\thTT$, while for the normal ordering large values of $\thTT$ are
    advantageous.  The range shown corresponds roughly to the current
    $3\sigma$ allowed region of $\thTT$; the global best-fit values from the
    NuFit group~\cite{NuFIT20} for both orderings are indicated by
    black arrows.}
  \label{fig:SigmaVsThetaTT}
\end{figure}

\begin{figure} [ht!]
   \begin{center}
   \includegraphics[scale=0.6]{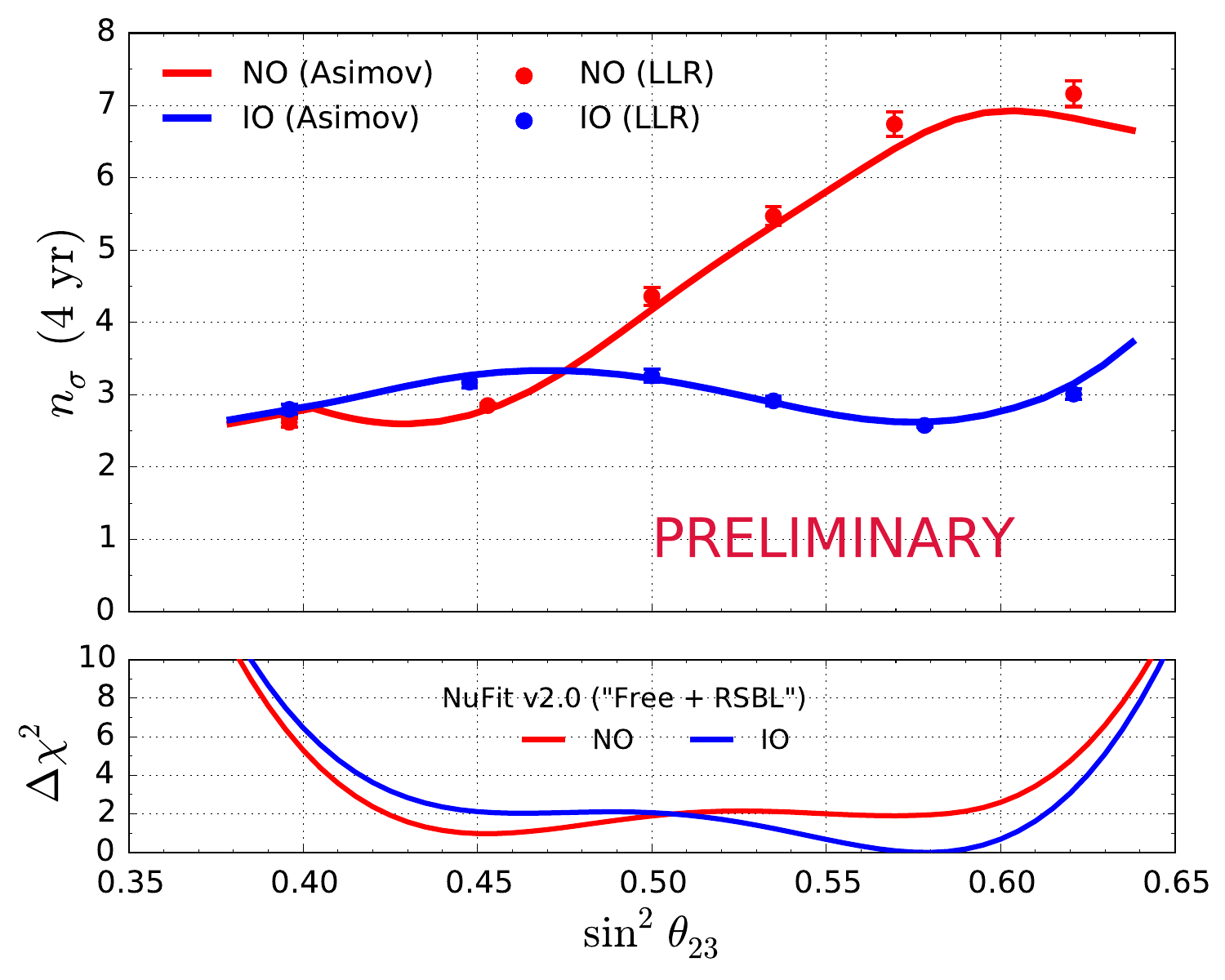}
   \end{center}
   \caption{\textit{Upper panel:} Projected four-year significance of
     the neutrino mass-ordering determination as a function of
     $\sinsqTT$ for the cases where the true mass ordering is normal
     (red) and inverted (blue).  The significance for determining the
     ordering when the true ordering is inverted is relatively
     insensitive to $\thTT$, while for the normal ordering large
     values of $\thTT$ are advantageous.  The NuFIT~2.0 $\sinsqTT$
     values used here (0.451 for NO and 0.576 for IO) produce almost
     the lowest significance possible in each ordering, and thus
     correspond to nearly the most conservative values in the entire
     range.  \textit{Lower panel:} Constraints on $\sinsqTT$ from
     NuFIT~2.0~\cite{NuFIT20} under the assumption of both normal and
     inverted ordering.}
   \label{fig:SigmaVsTheta23}
\end{figure}

Several numerical studies have been performed to quantify the ability
of PINGU to determine the NMO with all systematics considered.  Using
global best fit values~\cite{NuFIT20} for the oscillation parameters,
PINGU will determine the ordering with a significance of 3$\sigma$ in
roughly~\YrsToThreeSigmaSystLimited~years.  This significance depends
quite strongly on the actual value of $\thTT$, which is not well known
(see, {\it e.g.,}~\cite{Gonzalez-Garcia:2015qrr}
and~\cite{Marrone:2015nip}).  The expectation
of~\YrsToThreeSigmaSystLimited~years to reach 3$\sigma$ significance
is conservative in the sense that PINGU's sensitivity to the NMO would
be greater in almost any region of the allowed parameter space of
$\thTT$ other than the assumed global best fit, as shown in
Figs.~\ref{fig:SigmaVsThetaTT} and~\ref{fig:SigmaVsTheta23}.

In addition, the NMO sensitivity of PINGU as a function of time has
been calculated assuming the same global best fit oscillation
parameters, with and without world-average priors from~\cite{NuFIT20}
on $\thTT$ and $\dmTO$.  The technical description of how the priors
were applied, as well as the plot of NMO sensitivity vs. time, are
presented in Sec.~\ref{sec:NMOAnalysisTechnique}.  Finally, it is
worth noting that the time required to achieve a significant result is
shortened by roughly 6~months if the partially deployed PINGU detector
data, as well as roughly 10~years of DeepCore data available by then,
are included in the analysis.

%
%

\subsubsection{Impact of the Charge-Parity Phase $\dcp$}
\label{sec:NMHdeltaCP}

\begin{figure} [t!]
   \begin{center}

   \includegraphics[scale=0.6]{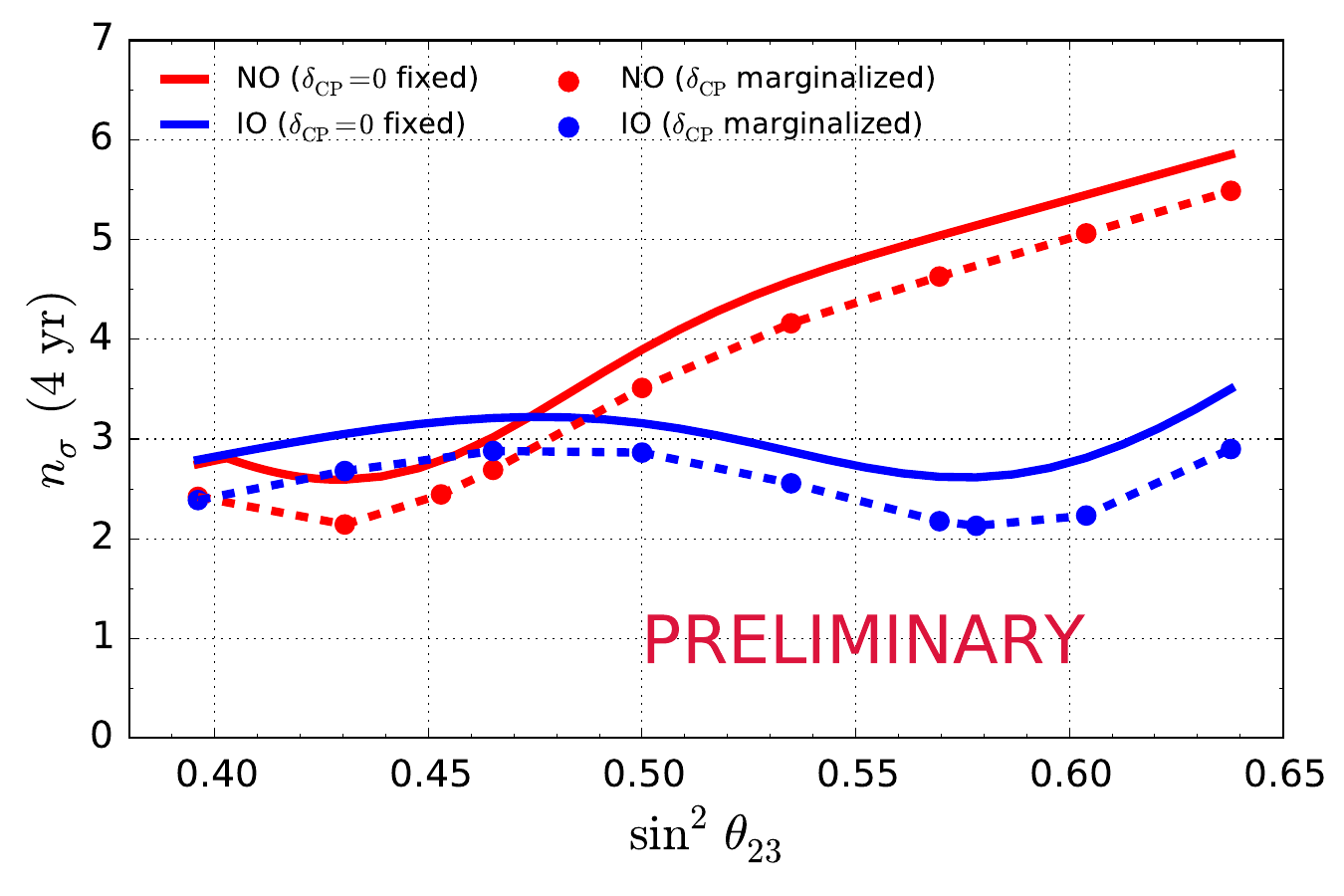}
   \end{center}
   \caption{Projected reduction of the four-year significance of the
     NMO determination as a function of $\sin^2 \thTT$ when $\dcp$ is
     included as a nuisance parameter, compared to our nominal case
     where $\dcp=0$ and $\dcp$ is not minimized over,
     cf. Fig.~\ref{fig:SigmaVsTheta23}.  Note that the dashed lines
     are only present in order to guide the eye. See text for details.
   }
   \label{fig:DeltaCPSigmaReductionVsTheta23}
\end{figure}

Using the $\deltaChiSqBar$ method we have investigated the effect of
taking into account $\dcp$ as a nuisance parameter in our NMO
analysis. This was included following all other systematics considered
since $\dcp$ represents a highly nonlinear parameter. Making sure that
it is correctly treated by the minimizer would therefore require a
very large computational effort. Instead, for a given true ordering
hypothesis, we opted to vary the injected $\dcp$ in steps of
$45^\circ$ between $0^\circ$ and $360^\circ$ and finely scan $\dcp$ in
the opposite-ordering fit, simultaneously minimizing over our default
set of eight systematics at each step. As before, we have taken care
to find the correct solution for $ \thTT$, by searching for a minimum
in both octants for each fixed value of the hypothesized
$\dcp$. Figure~\ref{fig:DeltaCPSigmaReductionVsTheta23} shows the
resulting reduction of the projected NMO sensitivity as a function of
$\sin^2 \thTT$ for four years of exposure time, compared to the
reference projection from Fig.~\ref{fig:SigmaVsTheta23}. Note that we
show the maximum impact on the sensitivity here, i.e. $\dcp$ is
allowed to take any value. Also, the absolute significances in this
figure deviate from those in the previous figures since here we make
the approximation that the absolute values of the means of the two
test statistic distributions (one for each NMO hypothesis) are the
same; see~\ref{sec:AsimovAnalysis} for details.  We find that $\dcp$
has a non-negligible influence on the significance of the NMO
measurement, leading to a decrease of about $\sim 10\%-20\%$. It does
not obscure the NMO signal, however, independent of whether the
ordering is inverted or normal, and independent of the octant of
$\thTT$.

\subsubsection{Conclusions}
\label{sec:NMHConclusions}

We have performed a detailed study of the expected sensitivity of
PINGU to the neutrino mass ordering.  The sensitivity estimates are
based on detailed Monte Carlo simulations of the detector and full event
reconstructions, and the statistical methods used were validated
against a full likelihood analysis for a more limited range
of systematics.  Our estimates are also in agreement with
external studies~\cite{Blennow:2013oma,Winter:2013ema,Ge:2013zua,Ge:2013ffa}.

We find that most detector-related systematics investigated so far,
with the exception of the energy calibration scale, play a smaller
role than physics-related systematics.  The latter arise from
uncertainties in measured oscillation parameters, some of which PINGU
itself will be able to measure.  We also find that including
non-$\numu$-CC events in the final sample with simple particle ID
greatly improves the significance of the measurement.

Using the $\overline{\Delta \chi^{2}}$ technique with
parameterizations based on fully simulated and fully reconstructed
events of all flavors, incorporating a wide array of detector- and
physics-related systematics, and using an initial form of particle ID,
we estimate that PINGU will be able to determine the neutrino mass
ordering to $\sim 3\sigma$ with \YrsToThreeSigmaLLHWithPriors~years of
data. However, this significance may very well be higher if the true
value of $\theta_{23}$ differs from the current world average.

There are a number of future improvements that will further enhance
the NMO significance.  The most prominent involve refinements in
reconstruction, flavor identification, and the use of the
reconstructed inelasticity of the neutrino event, which is a weak
$\nu/\bar{\nu}$ discriminator.  More sophisticated particle ID will
enable us to better exploit the distinct patterns of $\numu$ events
relative to those of $\nue$ and $\nutau$.  The use of the inelasticity
would help us distinguish neutrinos from antineutrinos on a
statistical basis and could provide a 20--50\% increase in
significance~\cite{Ribordy:2013xea}.  In addition, further detector
geometry optimization, improved event selection efficiency and more
accurate event reconstruction will also improve the significance.

\clearpage
\resetlinenumber

\IfFileExists{NewCommands.tex}       {}       {}
\IfFileExists{../NewCommands.tex}    {}    {}
\IfFileExists{../../NewCommands.tex} {} {}

\graphicspath{{figures/}{AtmosphericNeutOsc/NutauAppearance/figures/}}

\subsection{Tau Neutrino Appearance}
\label{sec:TauNeutrinoAppearance}

Following the IceCube-DeepCore detector's success in measuring $\numu$
disappearance~\cite{Aartsen:2014yll}, a $\nutau$ appearance analysis
was started and is still underway at the time of this writing.  Tau
neutrinos are not present in the conventional atmospheric neutrino
flux but can appear in significant quantities due to $\numu
\rightarrow \nutau$ oscillations.  Generally speaking, the appearance
measurement faces two challenges: the $\nutau$ must have a minimum
energy of \unit[3.5]{GeV} to undergo a charged-current (CC)
interaction, and the expected signal rate from this interaction is low
compared to the signal from other neutrino interactions.  In the PINGU
detector, the increased photocathode density will mainly improve the
separation of the tau-type interactions from the muon-type. Better
energy and zenith angle resolutions will also enable improved
measurements of the regions that are richest in the oscillated
$\nutau$ flux.

In the ``three neutrino framework'' discussed previously, the
two clear channels to measure $\utau{3}$ directly are $\nutau \rightarrow
\nutau$ and $\numu \rightarrow \nutau$. The $\nutau \rightarrow \nutau$
channel probes $\utau{3}$ directly, but requires an experimentally 
challenging (and hitherto unrealized) high statistics $\nutau$ beam and long
baseline setup. A more feasible experimental channel is $\numu
\rightarrow \nutau$, which probes a combination of $U_{\mu 3}$ and
$U_{\tau 3}$, where any degeneracy between $\umu{3}$ and $\utau{3}$ can
be broken by either external constraints on $\umu{3}$ or by
simultaneously measuring $\numu \rightarrow \numu$ and $\numu
\rightarrow \nutau$.

Going beyond the three-neutrino-family framework, {\it i.e.,} assuming
${\textrm{U}}_{\mathrm{PMNS}}$ is not unitary, means that the
summation in Eq.~\ref{eq:prob1} cannot be simplified as discussed above.
In that case, a long-baseline $\numu \rightarrow \nutau$ measurement
probes a combination of $U_{\tau i}$ and $U_{\mu i}$ elements, which
can then be compared to combinations of $U_{e i}$ and $U_{\mu i}$
obtained from $\numu \rightarrow \nue$ and $\numu \rightarrow \numu$
measurements at similar $L/E$ values. For simplicity, in the following
discussion where comparisons to cases outside the
three-neutrino-family framework are made, the Standard Model
expectation to measure $\utausq{3} \approx 1/2$ is used as equivalent
to $P(\numu \rightarrow \nue) + P(\numu \rightarrow \numu) + P(\numu
\rightarrow \nutau) = 1$.

A measured value of $\utausq{3} \approx \frac{1}{2}$ would:
\begin{itemize}
\item{ Strengthen the three-active-neutrino interpretation, and}
\item{ Confirm unitarity of the third mass eigenstate.}
\end{itemize}

 A measured value of $\utausq{3} \not \approx \frac{1}{2}$ would:
 \begin{itemize}
 \item{ Provide a distinct signature of new physics,}
 \item{ Further motivate explorations of possible Non-Standard
     Interactions (NSI)~\cite{Antusch:2006vwa, Antusch:2008tz, Meloni:2009cg}, and}
 \item{ Offer an additional experimental anomaly including a possible sterile
     neutrino(s) explanation via
     $\uesq{3}+\umusq{3}+\utausq{3}+{|U_{{\rm sterile}\,3}|^{2}}=1$.}
\end{itemize}

\subsubsection{Event Selection and Reconstruction}
\label{sec:TauNeutrinoAppearance:EvtSelectionReco}

The analysis for measuring $\nutau$ appearance in PINGU follows the
prescription described in Sec.~\ref{sec:atmo_event_sel} for
the study of atmospheric neutrino parameters with a notable difference
related to the particle identification.

\begin{figure}
	\begin{overpic}[scale=0.40]{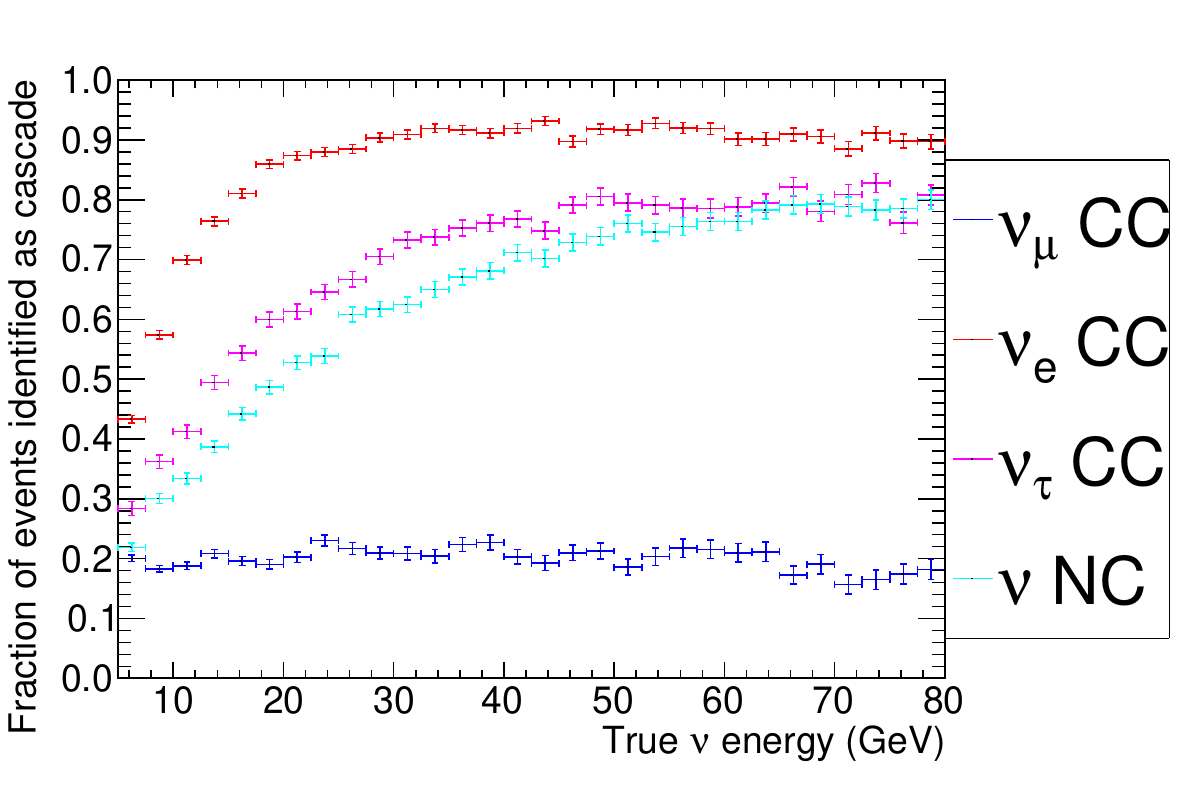}
		\put(40,30){\large \textcolor{red}{Preliminary}}
	\end{overpic}
	\begin{overpic}[scale=0.40]{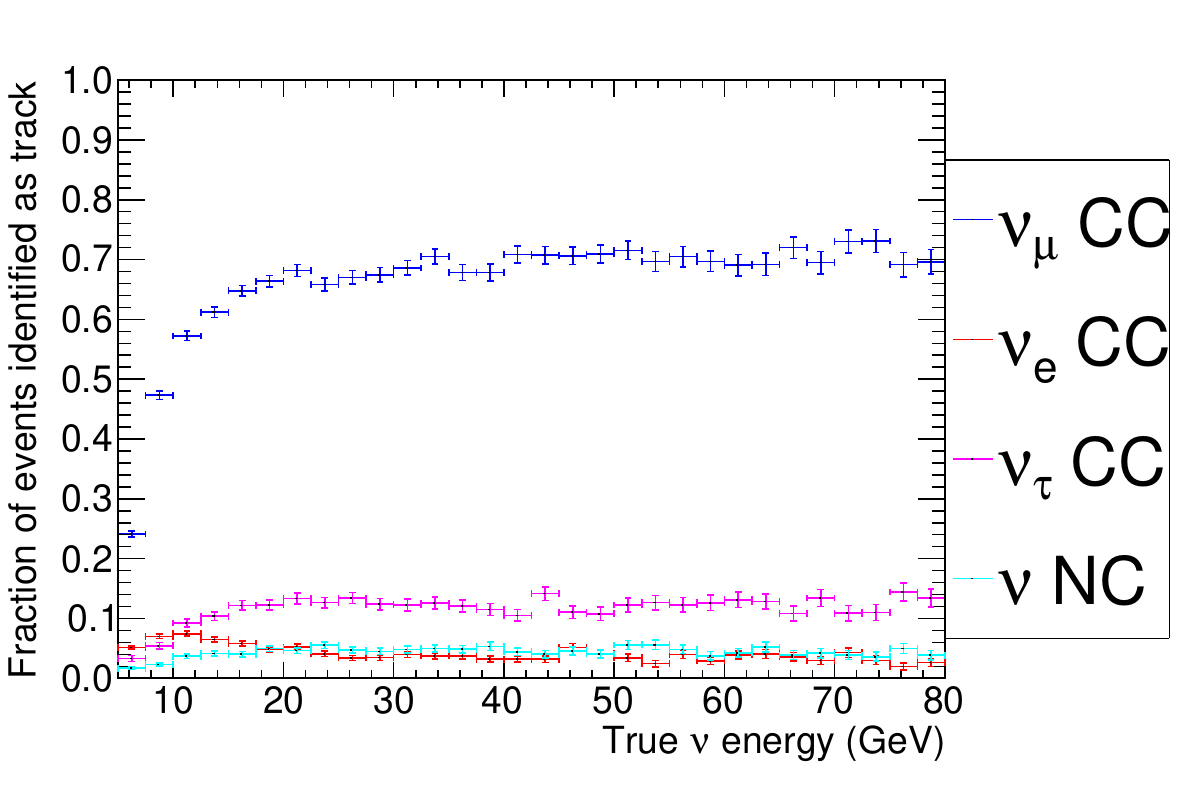}
		\put(40,30){\large \textcolor{red}{Preliminary}}
	\end{overpic}
	\caption{Particle identification efficiency of the samples
          used for the $\nutau$ appearance analysis.  On the left is
          shown the efficiency to identify the event as a cascade in
          the ``purer cascade PID'' region. The efficiency to identify
          the event as a track in the ``purer track PID'' control
          sample is shown on the right.  }
	\label{Fig:TauNeutrinoAppearance:PID}
\end{figure}

The Particle IDentification (PID) used for this analysis utilizes the
same training that is applied for all analyses, described in
Sec.~\ref{sec:ParticleID}.  The division of the sample into track-like
and cascade-like events is accomplished by requiring the output score
of the PID calculation to be more cascade-like (MVA score $< 0.424$),
the efficiency of which is shown in
Fig.~\ref{Fig:TauNeutrinoAppearance:PID}.

\begin{figure}
	\begin{center}
		\begin{overpic}[scale=0.50]{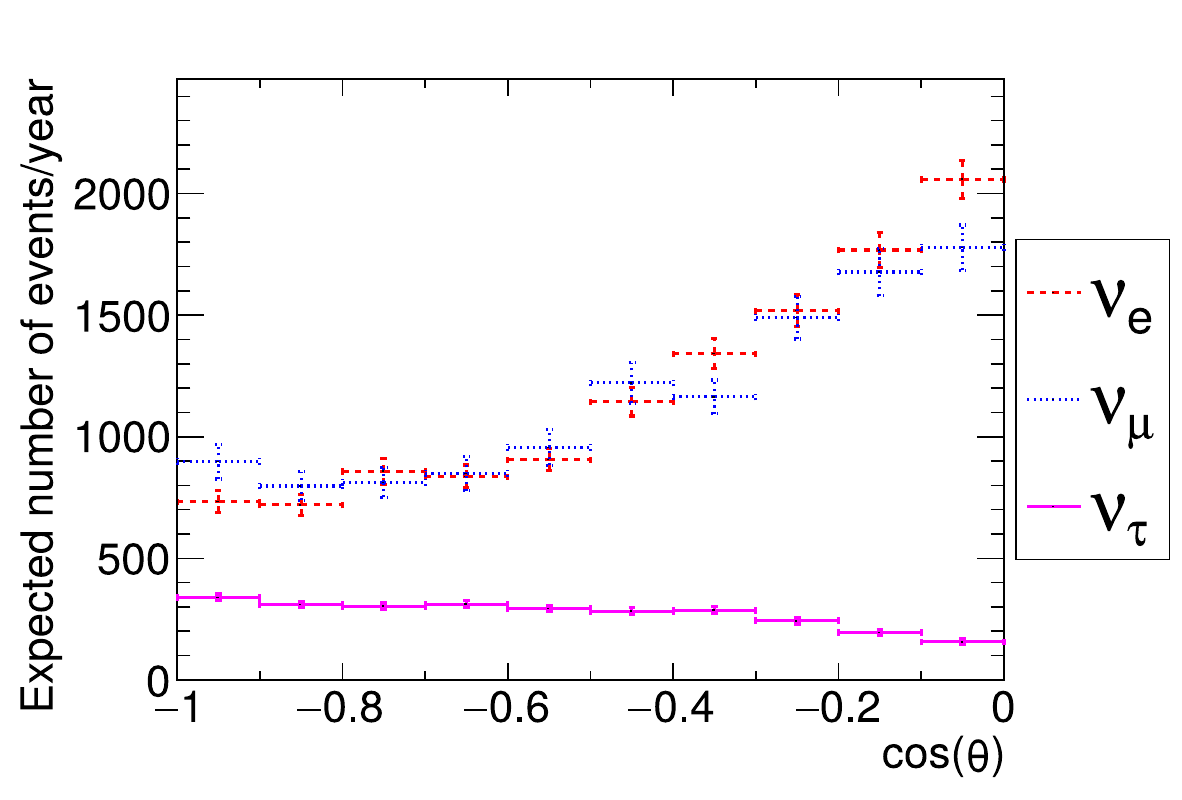}
			\put(30,50){\large \textcolor{red}{Preliminary}}
		\end{overpic}
	\end{center}
	\caption{The expected number of neutrino events per year per
          flavor as a function of the reconstructed $\nu$ incidence
          angle after regular NMO event selection and the ``purer''
          cascade PID cut.}
	\label{Fig:TauNeutrinoAppearance:FlavorCosZen}
\end{figure}

The goal of applying a stricter PID criterion is to reject the
mis-identified low-energy $\numu$~CC events that would reduce the
ability to distinguish $\nutau$ events. Some events that are not in
the ``purer cascade'' category are selected in a ``purer track''
sample that is used as a control
sample. Table~\ref{Tab:TauNeutrinoAppearance:EvtSelectionReco:rates}
shows expected rates for signal and background in each sample. Even
though there is still a larger fraction of $\nue$ and $\numu$,
primarily made up of charged-current interactions in the sample, we
can further improve the measurement given that $\nue$ and $\numu$ will
pile up around the horizon ($\cos\theta_{\mathrm{zenith}}=0$), while
$\nutau$ from $\numu \rightarrow \nutau$ oscillations will mostly be
found close to the up-going region, as shown in
Fig.~\ref{Fig:TauNeutrinoAppearance:FlavorCosZen}. Note that in this
analysis we consider explicitly the ``sterile neutrino'' scenario
where oscillation to sterile neutrinos will impact both the
charged-current and neutral-current (NC) interactions, as sterile
neutrinos by definition do not interact with the $Z^0$ boson. In the
case of NSI we could imagine a different effective change to
charged-current and neutral-current events, however this is currently
not considered in this analysis.


\begin{table}[h!]
	\begin{center}
		\begin{tabular}{r|c|c}
			& \multicolumn{2}{c}{Exp. number of events/year/$10^3$} \\
                        & Purer cascade PID & Purer track PID \\ \hline
			Background: $\nue+\numu$ & 23.7    & 12.0 \\
			CC & 20.7    & 11.7 \\
			NC &  3.0    &  0.3 \\
			\hline
			Signal: $\nutau$          &  2.7    &  0.4\\
			CC &  1.7    &  0.3\\
			NC &  1.0    &  0.1\\
		\end{tabular}
	\end{center}
	\caption{Rates for signal and background on the cascade and track samples used for this
          analysis. Breakdown in CC and NC events for the signal and background samples are also
          shown.}
	\label{Tab:TauNeutrinoAppearance:EvtSelectionReco:rates}
\end{table}

\begin{figure}
	\begin{center}
		\begin{overpic}[scale=0.50]{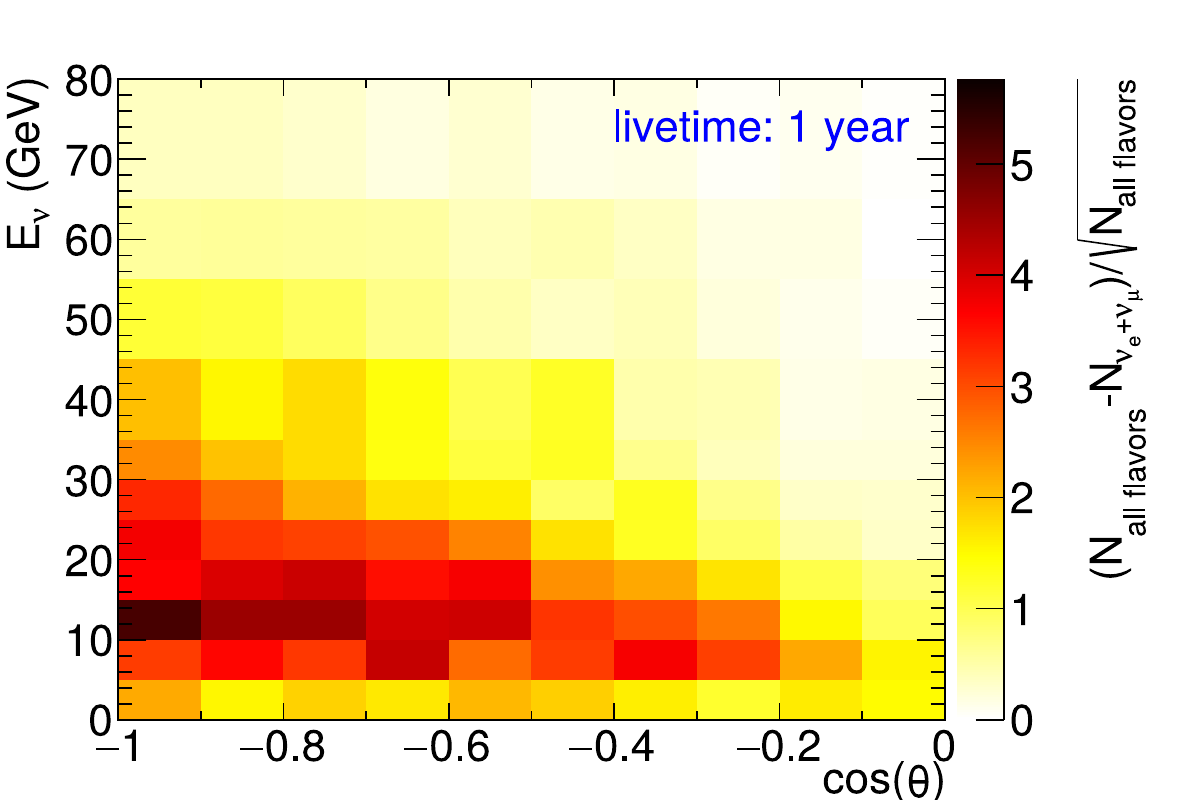}
			\put(50,50){\large \textcolor{blue}{Preliminary}}
		\end{overpic}
	\end{center}
	\caption{The distinguishability metric, as defined
          in~\cite{ARS}, for one year of simulated PINGU data. The
          quadratic sum of the absolute values in each bin gives an
          estimate of the number of $\sigma$ separating the
          no-$\nutau$ appearance hypothesis from the standard
          oscillation picture with $\nutau$ appearance, without
          accounting for any systematic uncertainties.}
	\label{Fig:TauNeutrinoAppearance:EvtSelectionReco:distinguishability}
\end{figure}

To better evaluate the potential of the measurement we construct a
distinguishability metric shown in
Fig.~\ref{Fig:TauNeutrinoAppearance:EvtSelectionReco:distinguishability},
similar to the one described in Sec.~\ref{sec:NeutrinoOscillations} (and
in~\cite{ARS}), that shows where the expected $\nutau$ can be best
distinguished. In that figure, the color scale indicates the
significance of a given bin in the final analysis (in the absence of
systematic effects) when trying to distinguish the case without
$\nutau$ appearance. It is important to note that the peak of the
$\nutau$ appearance signal is at a lower energy than the first
oscillation maximum (around 25~GeV for neutrinos traversing the
Earth). This difference is due to the undetected outgoing neutrinos
produced either at the $\tau$ decay for $\nutau$~CC interactions or at
the vertex of the NC interaction.  These secondary neutrinos carry
away part of the interacting $\nutau$ energy, resulting in event
reconstruction at lower energy.

\begin{figure}
	\begin{center}
		\begin{overpic}[scale=0.50]{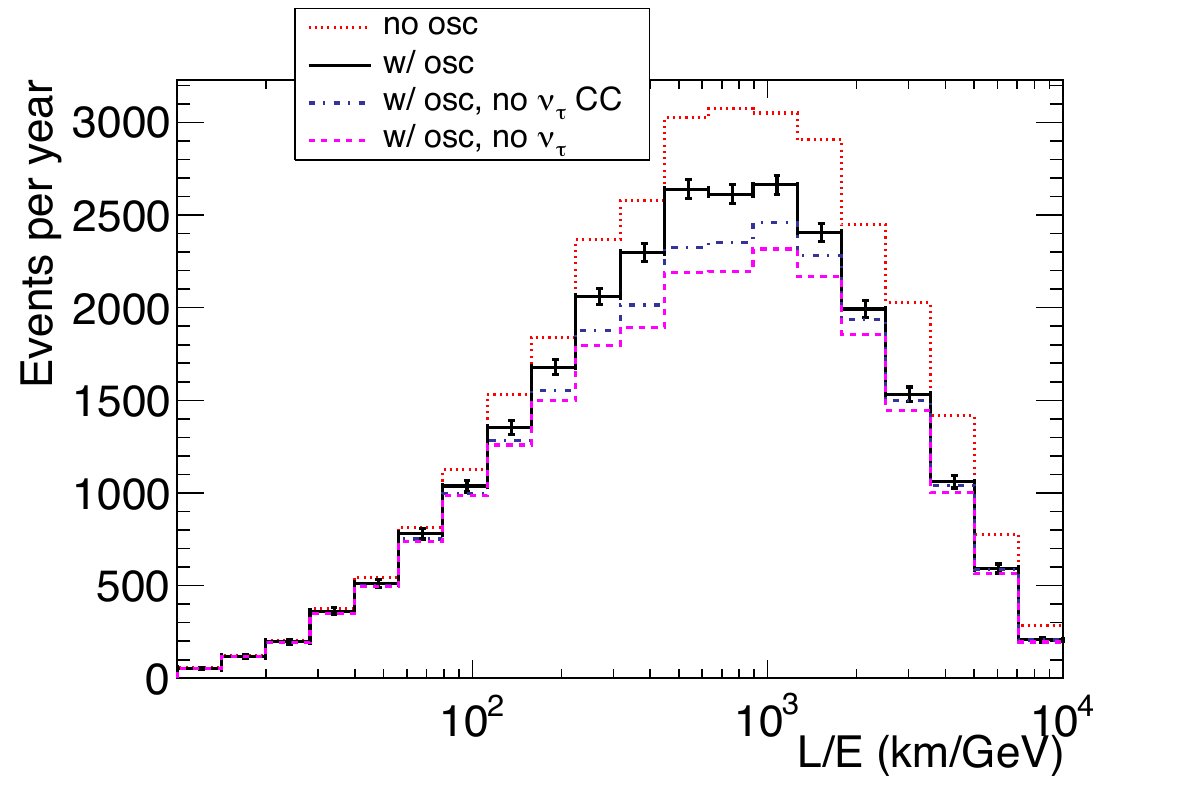}
			\put(50,15){\large \textcolor{red}{Preliminary}}
		\end{overpic}
	\end{center}
	\caption{Expected number of neutrino events per year as a
          function of reconstructed L/E after regular event selection
          and for the ``purer'' cascade PID cut. Four different
          scenarios are shown: no neutrino oscillations occur (red
          dotted line), regular three-flavor neutrino oscillations (black
          line) shown with statistical error bars, neutrino
          oscillations with $\nu_\mu \rightarrow \nu_s$ (magenta
          dashed line) and neutrino oscillations with the assumption
          that the $\nu_\tau$~CC vanish while the $\nu$~NC
          interactions do not vanish (blue dot-dashed line).}
	\label{Fig:LoverE_tau}
\end{figure}

Another way to visualize the potential to measure the $\nutau$
component is to look at the expected distribution of events as a
function of the ratio of neutrino pathlength and energy (L/E), on
which the neutrino oscillations depend directly as shown in
Eq.~\ref{eq:prob1}. This is illustrated in Fig.~\ref{Fig:LoverE_tau},
where the expected number of events for different scenarios is
shown. A clear difference is observed between the usual three-flavor
oscillations and the scenarios where oscillation-induced $\nutau$ do
not appear. In the figure, the case where only $\nutau$ CC events
disappear is also shown for reference.

\subsubsection{Analysis method}
\label{sec:TauNeutrinoAppearance:LikelihoodAnalysis}

The analysis method measures the $\nutau$ normalization, defined as
the ratio of the number of measured tau neutrinos to the number
expected in the standard three-neutrino paradigm.  The methodology is
similar to that described in~\ref{sec:LLRAnalysis}, but the
pseudo-data set generation is handled differently.  The main difference here 
is that the pseudo-data generation is marginalized over the systematics.

To exemplify the procedure described above we consider the first trial
run for the $\nutau$ normalization at the nominal value for one month
of livetime. First the expected distribution of events for a
realization of the systematic errors was calculated from the PINGU
simulation; for this example: \unit[$\Delta m_{31}^2 = 2.52\times 10^{-3}$]{eV$^2$}
[+1.01$\sigma$], $\sin^2\theta_{23}=0.491$ [-0.16$\sigma$], overall
normalization of 1.11 [+0.72$\sigma$], $\nue/\numu$ flux normalization
of 0.96 [-2.00$\sigma$], $\bar{\nu}/\nu$ normalization of 0.94
[-0.40$\sigma$], spectral index of 0.05 above nominal [+1.00$\sigma$],
and energy scale reduced by 2.5\% [-0.25$\sigma$]. This expectation is
then scaled to the equivalent number of events after \unit[1]{month} of data,
and a Poisson fluctuation is applied to each bin in the
$(\cosThetanu, \Enu)$ histogram to create the pseudo-data
template. The pseudo-data is then used to estimate which $\nutau$
normalization is better represented by the given histogram while
marginalizing over all systematic parameters; for this example:
$\nutau$ normalization of 0.89, \unit[$\Delta m_{31}^2 = 2.54\times
10^{-3}$]{eV$^2$} [+1.20$\sigma$], $\sin^2\theta_{23}=0.520$
[+0.35$\sigma$], overall normalization of 1.13 [+0.85$\sigma$],
$\nue/\numu$ flux normalization of 1.01 [+0.27$\sigma$],
$\bar{\nu}/\nu$ normalization of 0.97 [-0.17$\sigma$], spectral index
of 0.038 above nominal [+0.75$\sigma$], and energy scale reduced by
4.8\% [-0.48$\sigma$].

\begin{figure}
	\begin{center}
		\begin{overpic}[scale=0.50]{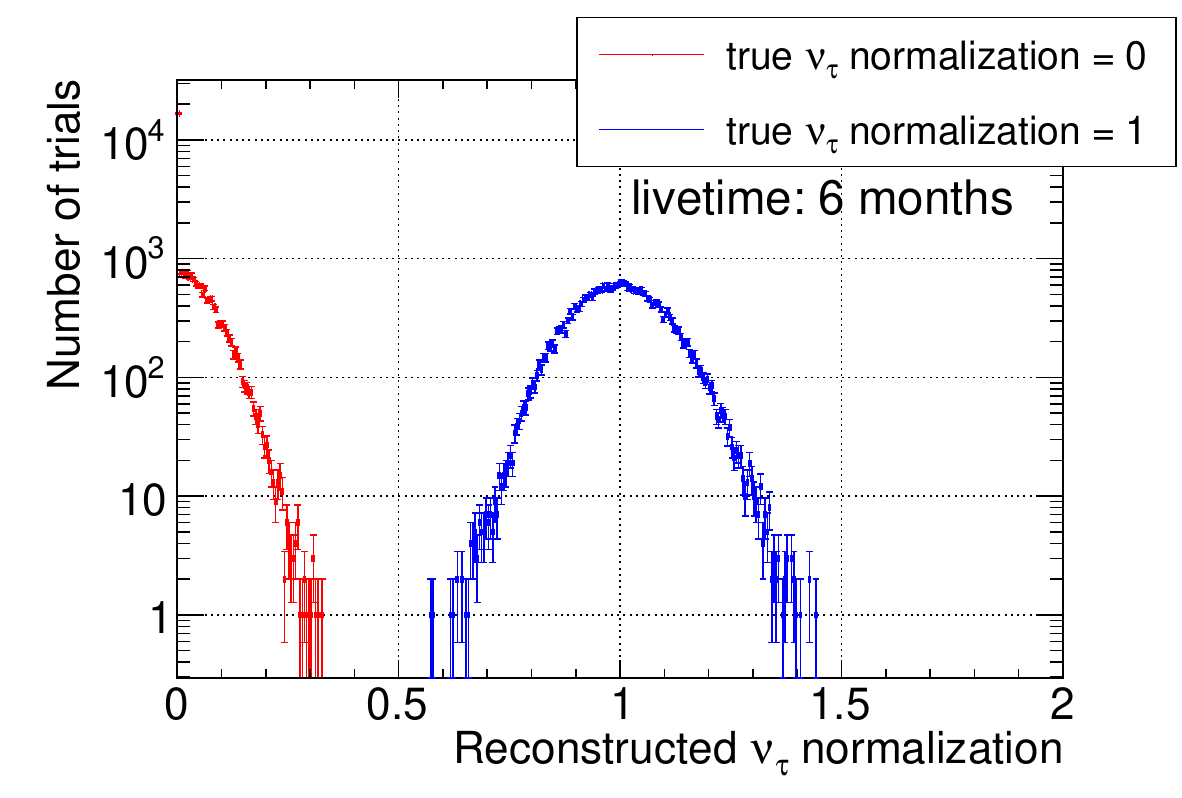}
			\put(20,50){\large \textcolor{red}{Preliminary}}
		\end{overpic}
	\end{center}
	\caption{Resulting distribution of reconstructed $\nutau$
          normalization obtained using the procedure described in the
          text.  An $x$-axis value of 0 means no $\nutau$ appearance
          while a value of 1 corresponds to $\nutau$ appearance, as
          expected in the standard oscillation scenario.  Under the
          assumptions of Gaussian distributions and standard
          oscillations, the median significance to exclude no $\nutau$
          appearance is roughly 13$\sigma$ after six months of
          livetime.}
	\label{Fig:TauNeutrinoAppearance:reco_nutau_normalization}
\end{figure}

The procedure detailed above was run with 30\,000 generated pseudo-data
trials for several different $\nutau$ normalizations (ranging between
0 and 1).  By comparing the spread of the reconstructed
$\nutau$ normalizations among those cases we obtain the sensitivity to
distinguish between different true $\nutau$ normalizations, as shown
in Fig.~\ref{Fig:TauNeutrinoAppearance:reco_nutau_normalization}.

\subsubsection{Systematic uncertainties}
\label{sec:TauNeutrinoAppearance:SystematicUncertainties}

We consider systematic uncertainties due to the overall normalization
of the neutrino flux, the $\nue$ flux in relation to the $\numu$ flux,
the $\bar{\nu}$ rate in relation to the $\nu$ rate, the atmospheric
oscillation parameters ($\Delta m^2_{31}$ and $\theta_{23}$), and
the spectral index and energy scale. All considered systematics are
listed in Table~\ref{Tab:TauNeutrinoAppearance:ListOfSystematics}
along with the priors used for either generating trials or for both
the trial generation and the fitting.

\begin{table}[h!]
	\begin{center}
		\begin{tabular}{l|c}
			Systematics & Gaussian prior \\ \hline
			$\Delta m^2_{31}$$^\dag$ & \unit[$(2.42\pm0.10) \times 10^{-3}$]{eV$^2$} \\
			$\sin^2(\theta_{23})$$^\dag$ & $0.500\pm0.055$ \\
			Normalization$^\dag$ & $\pm15\%$ \\
			$\nue/\numu$ flux normalization & $\pm2\%$ \\
			$\bar{\nu}/\nu$ normalization & $\pm15\%$  \\
			Spectral index of $\numu$ flux & $\pm0.05$ \\
			Energy scale         & $\pm10\%$ \\

		\end{tabular}
	\end{center}
	\caption{List of systematics and priors used for this analysis. For systematics
	marked with a $^\dag$ the priors listed are only used in the generation of trials.}
	\label{Tab:TauNeutrinoAppearance:ListOfSystematics}
\end{table}


These priors differ slightly from those used in other analyses described in this letter because
of either a different implementation of degenerate parameters (as is
the case for the spectral index of the $\numu$ flux and effective area
energy dependence).
One particular instance where the prior is changed is the $\nue/\numu$
flux normalization, where we assume a tighter prior based
on~\cite{Honda:2006qj}.  The sampling distribution of the oscillation
parameters was centered around maximal mixing for simplicity, and for
both parameters used larger Gaussian errors than the current best fit
values.  We have also tested the impact of using either the DOM
efficiency parametrization as done in
Sec.~\ref{sec:MuonNeutrinoDisappearance} or the energy scale
systematic, and found both yield very similar results, so we have
decided to only use the energy scale systematic in this analysis with
a 10\% prior, as is done in Sec.~\ref{sec:NeutrinoMassHierarchy}.

Additional systematic effects considered in other PINGU neutrino
oscillation analyses but not in this one, such as the uncertainty
related to the atmospheric neutrino flux model investigated in
Sec.~\ref{sec:NeutrinoMassHierarchy}, are not expected to
significantly impact the result.



\subsubsection{Results}
\label{sec:TauNeutrinoAppearance:Results}

\begin{figure}
	\begin{center}
		\begin{overpic}[scale=0.50]{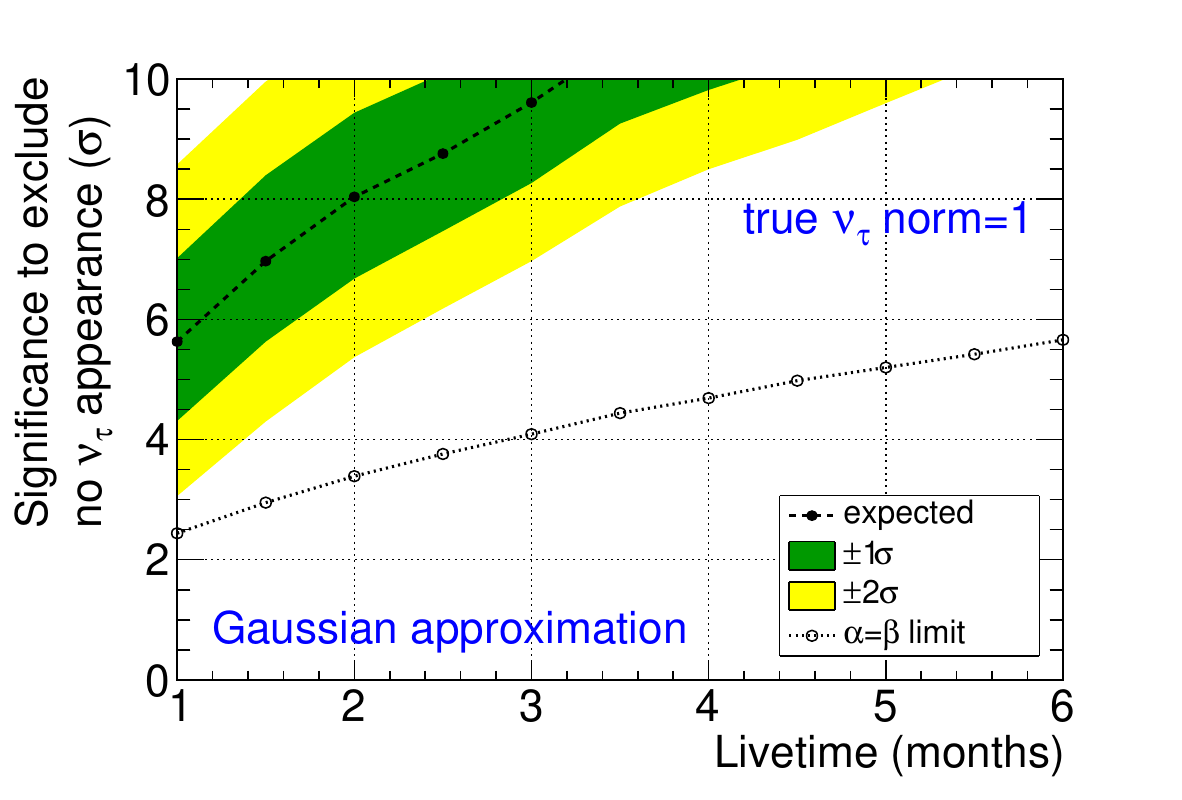}
			\put(62,54){\large \textcolor{red}{Preliminary}}
		\end{overpic}
	\end{center}
	\caption{Significance to exclude no~$\nutau$ appearance
          assuming the expected $\nutau$ appearance from the standard
          three-flavor $\nu$ oscillation. The expected result along
          with the $\pm 1\sigma$ and $\pm 2\sigma$ regions are shown,
          as well as the significance at which the type-I error
          ($\alpha$), that is incorrectly rejecting a true hypothesis,
          has the same probability as the type-II error ($\beta$),
          that is incorrectly accepting a false hypothesis.}
	\label{Fig:TauNeutrinoAppearance:ExclusionToNoNuTau}
\end{figure}

Performing this analysis for several different detector livetimes, we
expect to reach $5\sigma$ exclusion of the no-$\nutau$-appearance
hypothesis with one month of data, as shown in
Fig.~\ref{Fig:TauNeutrinoAppearance:ExclusionToNoNuTau}. The effect of
statistical fluctuations in the measurement are shown by the colored
bands.

\begin{figure}
  \begin{center}
    \begin{overpic}[scale=0.5]{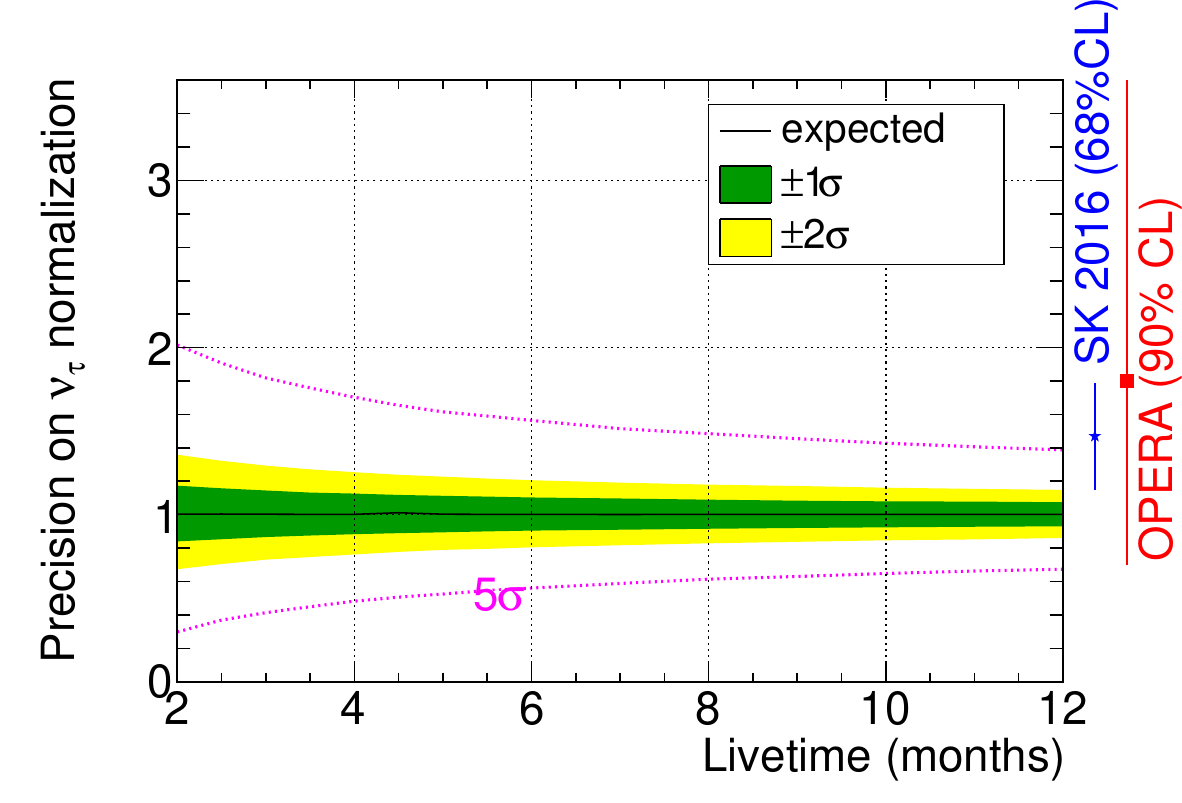}
      \put(26,54){\large \textcolor{red}{Preliminary}}
    \end{overpic}
  \end{center}
  \caption{Precision with which the rate of $\nutau$ appearance can be
    measured, in terms of the PMNS expected rate, as a function of
    exposure (in months).  The true value is assumed to be 1.0 (the
    standard expectation) for illustration.  The expected
    $\pm 1\sigma$ and $\pm 2\sigma$ regions and $\pm 5\sigma$ limits
    are shown, as well as measurements by
    Super-Kamiokande~\cite{Abe:2012jj,MoriyamaNu2016} and OPERA~\cite{Agafonova:2015jxn}.  
  }
  \label{Fig:TauNeutrinoAppearance:NormalizationUncertainty}
\end{figure}

With one year of data taking, and assuming the expected rate of
$\nutau$ appearance from standard three-flavor oscillations, the
precision of the $\nutau$ normalization measurement should be better
than 10\%, as shown in
Fig.~\ref{Fig:TauNeutrinoAppearance:NormalizationUncertainty}. The
current analysis has a $\mathcal{O}(1\%)$ bias in the determination of
the true $\nutau$ normalization that shrinks with time and is much
smaller than the uncertainty in the parameter determination.


\begin{figure}
	\begin{center}
		\begin{overpic}[scale=0.50]{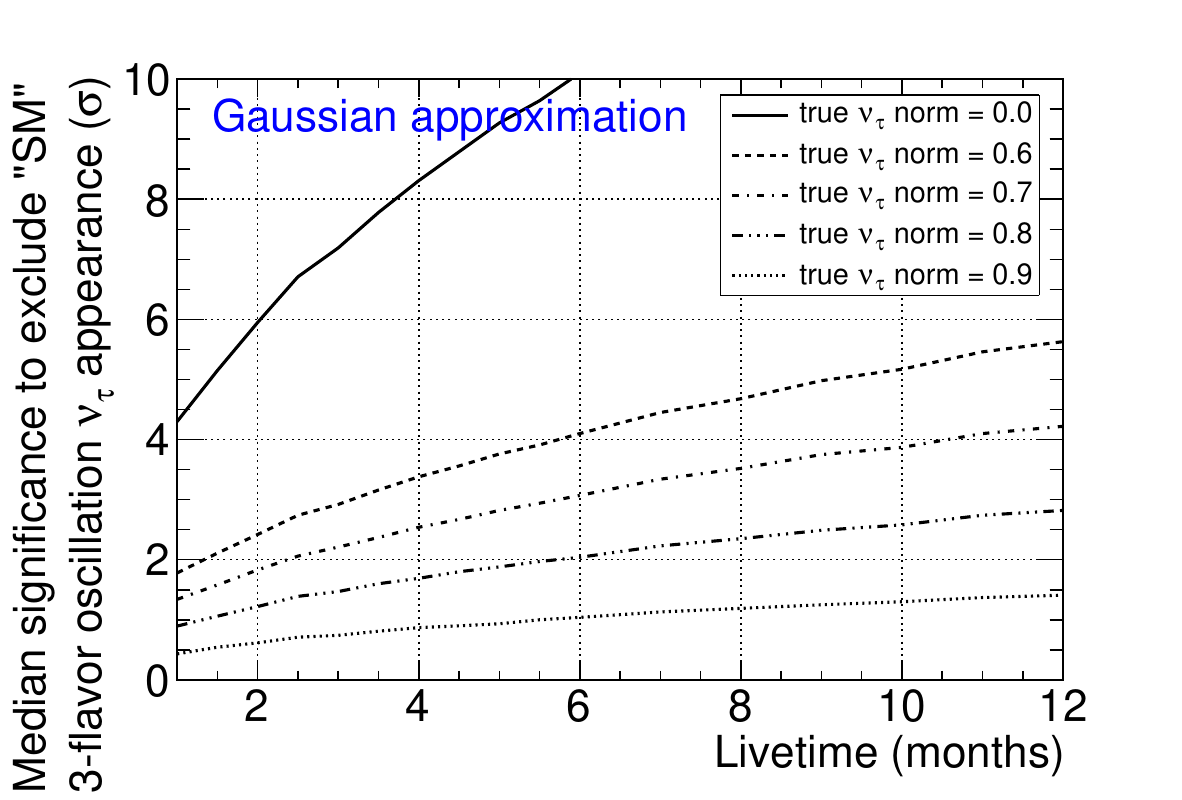}
			\put(32,42){\large \textcolor{red}{Preliminary}}
		\end{overpic}
	\end{center}
	\caption{Significance to exclude the usual three-flavor $\nu$
          oscillation for different true $\nutau$ normalizations for
          the median case. For estimations above $3\sigma$ a Gaussian
          approximation is used.}
	\label{Fig:TauNeutrinoAppearance:SignificanceExclude1}
\end{figure}

Alternatively, instead of assuming the $\nutau$ normalization is the
one given by three-flavor oscillations, we can estimate the level at
which we can distinguish different true values of the $\nutau$
normalization from the three-flavor oscillation prediction. As shown
in Fig.~\ref{Fig:TauNeutrinoAppearance:ExclusionToNoNuTau}, excluding
standard $\nutau$ appearance in the case where there is none can be
reached with a little bit more than one month of livetime. With a year
of data the true $\nutau$ normalizations can be distinguished at the
$5\sigma$ level for a true $\nutau$ normalization of 0.6, and at the
$3\sigma$ level for a true $\nutau$ normalization of 0.8, as shown in
Fig.~\ref{Fig:TauNeutrinoAppearance:SignificanceExclude1}.

The expected median significances from
Figs.~\ref{Fig:TauNeutrinoAppearance:ExclusionToNoNuTau}
and~\ref{Fig:TauNeutrinoAppearance:SignificanceExclude1} for distinguishing the case where there is no $\nutau$ appearance are
different because in each figure a different assumption is made.  Each case 
is therefore defined by a different expected rate of events
that will produce slightly different required livetimes to
exclude the other case at the same significance.

\subsubsection{Comparison to other measurements}
\label{sec:TauNeutrinoAppearance:ComparisonToOther}

Recently OPERA reported finding a 5\textsuperscript{th} $\nutau$
candidate event~\cite{Agafonova:2015jxn} from a $\numu$ beam produced
at CERN, increasing their significance for having observed $\nutau$
appearance to 5.1$\sigma$.  Super-Kamiokande has also published
evidence of $\nutau$ appearance at 3.8$\sigma$~\cite{Abe:2012jj} and,
more recently, announced evidence at the 4.6$\sigma$
level~\cite{MoriyamaNu2016}, using atmospheric neutrinos. With these
measurements it is clear that there is $\nutau$ appearance occurring in conjunction with $\numu$ disappearance.

It is important to increase the precision on the determination of the
rate of $\nutau$ production in order to verify the unitarity of the
neutrino mixing matrix and to test for NSI. While OPERA may find
additional events in the remaining portion of their data and
Super-Kamiokande may update their results using additional years of
data, the current precision will not increase signficantly; OPERA
has stopped taking data and Super-Kamiokande's newly announced result was
based on data from 1996 until 2016 with an exposure of 22.5~kton over
roughly 5\,000 days while their published result was based on data from 1996
until 2008 with a 12-yr exposure of 22.5~kton over 2\,806 days. Therefore by the time PINGU is expected to start taking data Super-Kamiokande is expected to have increased their sample size by about 50\%.


\subsubsection{Conclusion}
\label{sec:TauNeutrinoAppearance:Conclusion}

PINGU's dense instrumentation and the resulting improvement of the
reconstruction quality for low-energy events will allow PINGU to
measure $\nutau$ appearance with improved accuracy. We expect to be able to exclude the ``no $\nutau$
appearance'' scenario at $5\sigma$ signficance with \unit[1]{month} of
data, and have a better than 10\% precision on the $\nutau$ appearance
normalization after the first year of data. Also with the first year
of data we would be able to distinguish a
true $\nutau$ normalization below 0.6 at a level of $5\sigma$.

\clearpage


\clearpage

\resetlinenumber

\IfFileExists{NewCommands.tex}       {}       {}
\IfFileExists{../NewCommands.tex}    {}    {}
\IfFileExists{../../NewCommands.tex} {} {}

\graphicspath{{figures/}{NonAtmospheric/figures/}}

\section{Other Neutrino--Based Analyses}
\label{sec:Non-Atmo}

In addition to neutrino oscillation measurements using atmospheric
neutrinos, PINGU has the ability to make several other scientifically
compelling measurements.  These measurements may use other
neutrino sources in addition to the atmosphere and are detailed below.

\resetlinenumber

\IfFileExists{NewCommands.tex}       {}       {}
\IfFileExists{../NewCommands.tex}    {}    {}
\IfFileExists{../../NewCommands.tex} {} {}

\graphicspath{{figures/}{NonAtmospheric/Tomography/figures/}}

\subsection{Neutrino Tomography}
\label{sec:tomography}
\subsubsection{Motivation}

Seismological data based on primary (compressional) p-waves and
secondary (shear) s-waves have been used in the construction of the
preliminary reference Earth model (PREM)~\cite{Dziewonski:1981xy} that
describes the Earth matter density profile.  While the matter density
is well known through these measurements, the chemical composition of
the interior of the Earth has not yet been measured. Neutrino tomography provides the
first, and possibly the only, way to directly probe the Earth's
composition. By exploiting the dependence of neutrino oscillation
probabilities on the electron density, PINGU, with its large
atmospheric neutrino sample, could be the first experiment to be able
to use neutrino tomography to begin distinguishing between proposed
Earth-core composition models.  Systematic uncertainties associated
with the unknown true values of oscillation parameters are reduced by
the observation of neutrinos that do not pass through the Earth's
core.

\subsubsection{Introduction and Earth Composition Models}

The Earth's geomagnetic field was discovered at the end of 16th
century~\cite{Gilbert1600}.  Since then many models have been
proposed to explain its origin. The dynamo model is the leading explanation of the field. It implies that the Earth contains a
conducting fluid that is convecting~\cite{Buffett16062000}. A
measurement of the composition of the Earth's interior may help
resolve the longstanding mystery of the origin of the geomagnetic
field and further advance our understanding of the Earth.

Direct sampling of the interior of the Earth is limited by the reach
of drills, which have only penetrated down to a depth of about
12~km~\cite{Popov1999345}. Coarse information about deeper regions of
the Earth can be obtained by eruption entrainment
sampling~\cite{Hofmann1997}, and the interior of the Earth has mainly
been studied using seismic waves.
 
The inner Earth consists of two distinct parts: ``silicate Earth'', and
the core (see Fig.~\ref{fig:earth_density} for a schematic view). The
silicate Earth consists of the crust and lower and upper mantles while
the core is divided into the outer and inner regions. Boundaries between
these layers are known with uncertainties on their
positions smaller than $10$~km ~\cite{2003TrGeo...2..547M}.  The
outer core was determined to be liquid from the absence of detected
s-waves~\cite{Oldham01021906}.  The PREM matter density structure was
developed by combining astronomic-geodetic parameters, free
oscillation frequencies, and seismic wave
velocity~\cite{Dziewonski:1981xy}. The uncertainty on the average
density of the lower mantle is estimated to be less than
0.7\%~\cite{Resovsky2002}, and that of the inner core as less than
0.5\%~\cite{Smylie1992}. Interestingly, mantle densities could also be
measured with PINGU itself with a few percent precision~\cite{Winter:2015zwx}.

It is believed that the bulk chemical composition of the Earth is the
same as the composition of the ``Ivuna'' type carbonaceous chondritic
meteorites (CI)~\cite{McDonough1995}. The crust and upper mantle
contain less iron, nickel, and sulfur than CI chondrites, implying
the core region should contain more iron and nickel.  By comparing
high pressure experimental data and seismological velocity profiles,
the inner and outer core are presumed to be mostly made of iron with
some additional light elements~\cite{Birch1952}.

The outer core is assumed to have a combination of
thermal and compositional convection. Without the compositional
convection, it is difficult to maintain the
geo-dynamo~\cite{Fearn1981}. Measuring the composition of the Earth's
core is therefore expected to lead to understanding the
geo-dynamo model. In addition, scenarios of the Earth's formation depend
on the core composition models~\cite{Allegre1995}.  There is little doubt in the interpretation that the outer core is
composed of liquid iron alloyed with nickel and some light elements,
but the content and type of the light elements are still uncertain
because of the limitations of the observational data. From
high-pressure experimental constraints, possible candidates for the
light elements are hydrogen, carbon, oxygen, silicon, and
sulfur~\cite{Li2007}.  While, among these, hydrogen is the least
understood in the context of Earth's chemical composition, liquid iron
alloyed with 1\% hydrogen by weight could also explain the outer core
density~\cite{Narygina2011}.  A variety of other outer core
compositions have been
proposed~\cite{2003TrGeo...2..547M,Allegre200149,Huang2011}.


\subsubsection{Methodology}

The Earth composition study with PINGU is made feasible by the
fact that neutrino oscillations depend on the electron density,
$N_e$~\cite{Rott:2015kwa}. The electron density is related to the mass
density via the factor $Y=Z/A$ (the proton to nucleon ratio), weighted
by the relative elemental abundances. As an example iron has a value
of $Y=0.466$, while lighter elements have values closer to $Y=0.5$,
and hydrogen has a very distinctive value of $Y=1.0$.

\begin{figure} [tbp]
  \centering
   \begin{overpic}[scale=0.35]{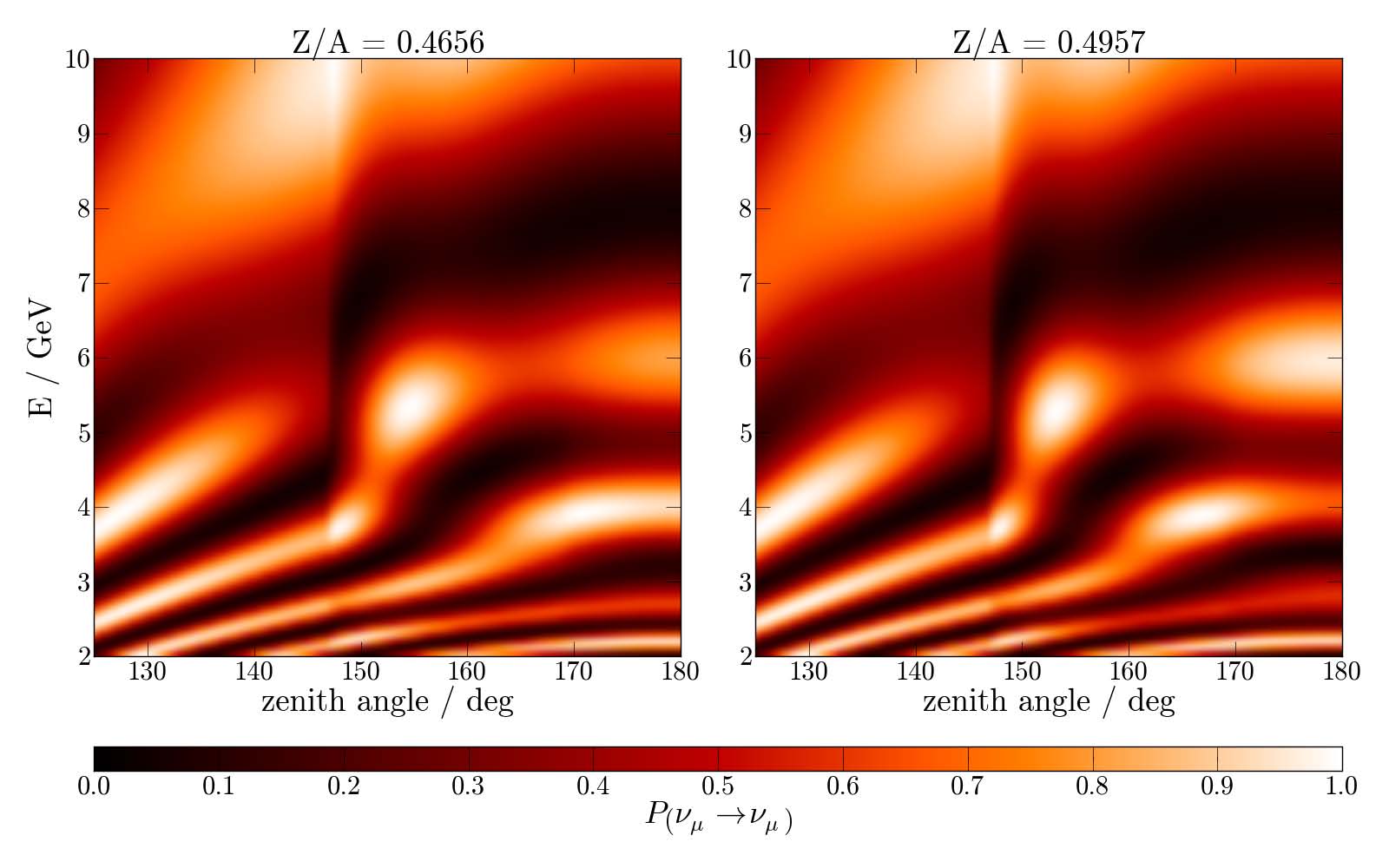}	
    \put(10,50){\textcolor{white}{Preliminary}}
    \put(60,50){\textcolor{white}{Preliminary}}
  \end{overpic}
  \begin{overpic}[scale=0.55]{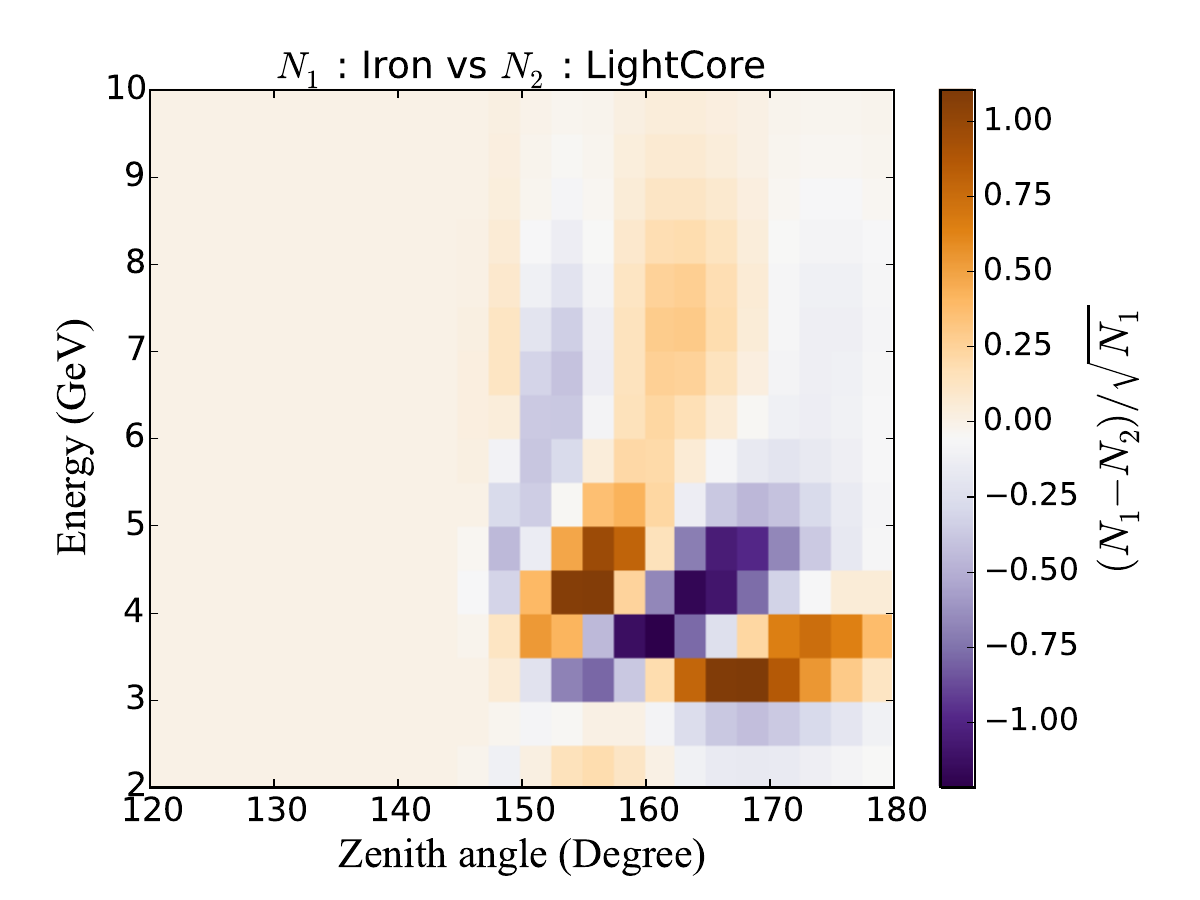}
    \put(15,45){\large\textcolor{red}{Preliminary}}
  \end{overpic}
  \caption{(Top) The impact of a changed core composition on the
    muon-neutrino survival probabilities is demonstrated by comparing
    the left figure (pure iron core, $Z/A=0.4656$) with the 
    right (iron mixed with lighter elements, $Z/A=0.4957$). (Bottom)
    ``Distinguishability metric'' plot showing the ratio between the
    expected difference in number of events and their statistical
    uncertainty assuming one year of data with 40\% electron neutrino
    contamination.  Events are binned by their true neutrino energy
    and direction.}
   \label{fig:tomo_oscillograph}
\end{figure}

The effect of a different core composition (or electron density) on the
survival probability for muon-type neutrinos is illustrated in
Fig.~\ref{fig:tomo_oscillograph}, with the ``distinguishability
metric'' plot included for these cases.  Modifications are visible for
the up-going neutrino events that cross the inner or outer core,
corresponding to zenith angles greater than $\theta_\nu=168^\circ$
($\cos{\theta_{\nu}}=-0.98$) for the inner core while the outer core
extends from the inner core to $\theta_\nu=147^\circ$
($\cos{\theta_{\nu}}=-0.84$). The most relevant energy range is
between about 2~GeV and 6~GeV. Good neutrino energy and zenith angle
resolutions in this energy range are essential for the success of a
neutrino tomography measurement.

\subsubsection{Analysis Method}

%

The tomography analysis presented here uses the same atmospheric
neutrino sample discussed in Sec.~\ref{sec:NeutrinoOscillations},
leading to approximately 30\,000 upward-going $\numu$ CC events per
year.  Roughly 50\% of these will have energies between 2 and 6~GeV.
The reconstruction and PID (see Sec.~\ref{sec:EventReconstruction})
proceed in the same manner as previously discussed.

For the sensitivity calculation, a likelihood ratio test as described
in Sec.~\ref{sec:NeutrinoOscillations} is performed for different core
composition models. For each bin, the likelihood for a given core
composition is calculated. Systematic uncertainties are treated as
nuisance parameters in the fitting procedure. These uncertainties
include the atmospheric spectral index, the muon-to-electron neutrino
flux ratio, the neutrino-to-antineutrino flux ratio, an overall rate
normalization, the energy scale, and the neutrino oscillation
parameters $\Delta m^2_{31}$, $\theta_{23}$, and $\theta_{13}$.
Priors are given in Table~\ref{Tab:Systematics}.
The likelihood is computed for a given $Z/A$ ratio with respect to the
Asimov dataset obtained for a pure-iron outer core.

\begin{figure} [tbp]
   \begin{center}
    	\begin{overpic}[scale=0.33]{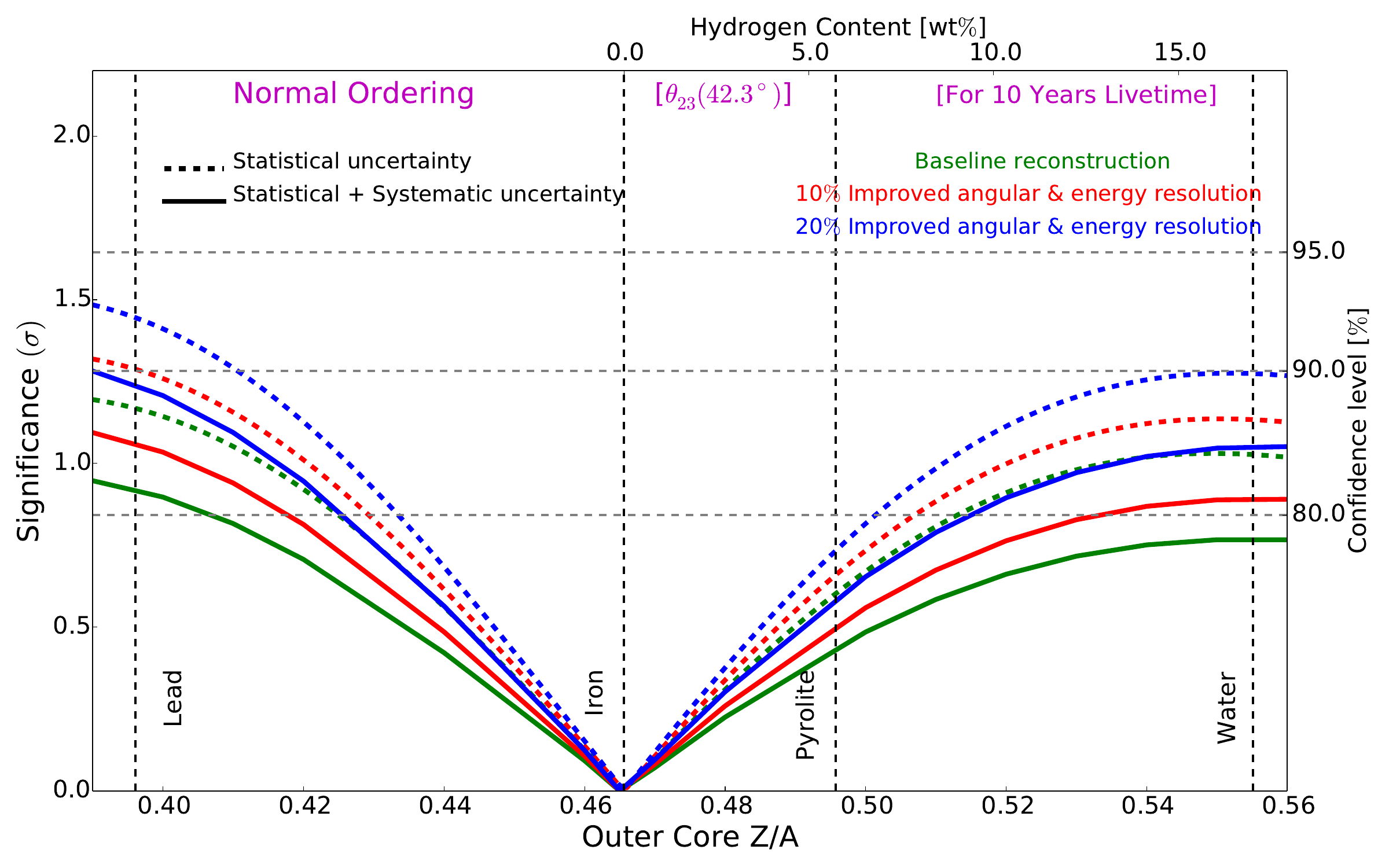}
	\put(68,10){\large \textcolor{red}{Preliminary}}
	\end{overpic}
	
	\vspace{0.6cm}	
	
	\begin{overpic}[scale=0.33]{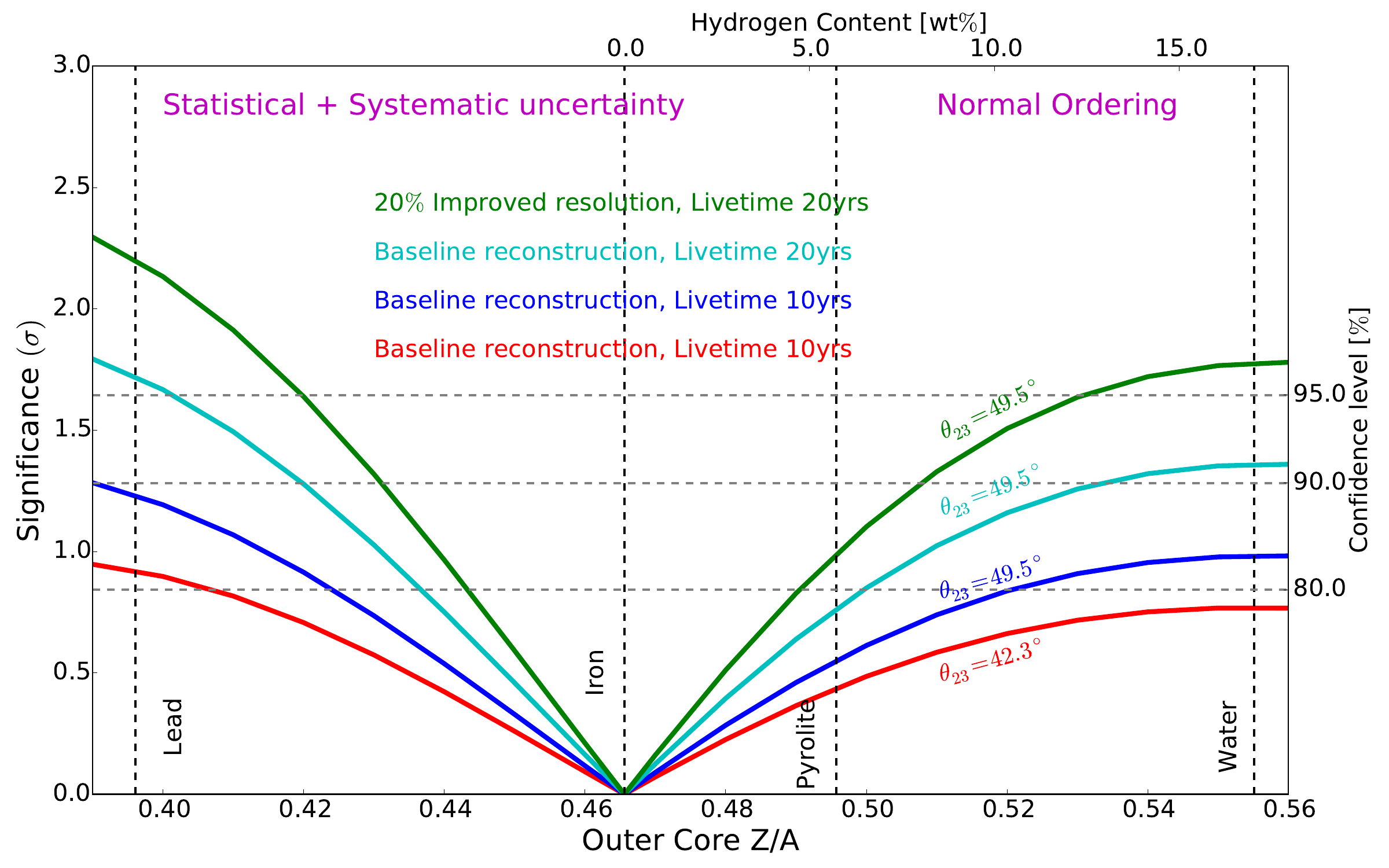}
	\put(68,10){\large \textcolor{red}{Preliminary}}
	\end{overpic}
   \end{center}
   \caption{Significances obtained from the mean of $10^5$ likelihood
     scans over proton to nucleon ratio $Z/A$ of the inner and outer
     Earth core with 10~years of PINGU data. The input value assuming
     a pure iron Earth core is reproduced correctly. The bottom plot
     shows how sensitivity improves for $\theta_{23}$ in the second
     octant, with more livetime, and with improved reconstructions.}
   \label{fig:tomography_exclusion}
\end{figure}

Scans over the electron density of the Earth's outer core, i.e. the
proton-to-nucleon ratio $Z/A$ can be seen in
Fig.~\ref{fig:tomography_exclusion}.  The top panel assumes normal
ordering to be true and ten years of livetime with $\theta_{23}$ in
the first octant.  The bottom panel shows how the sensitivity changes
for $\theta_{23}$ in the second octant and with more livetime and
improved reconstruction. The significance of excluding a given $Z/A$
is shown when the true value of $Z/A$=0.4656 corresponds to that of
pure iron.  The top axis shows the equivalent hydrogen content by
weight in an iron core. For each point, the nuisance parameters have
been fit so that the likelihood is maximal.  From a comparison of
dashed and solid lines, one can see that the systematic uncertainties
reduce the significances of the measurement by roughly 20\%.  For the
baseline reconstruction, and if $\theta_{23}$ lies in the second
octant, a true electron density of $Z/A$=0.4656 for the outer Earth
core can be constrained with an accuracy of $\pm 13\%$ ($1\sigma$)
after ten years of livetime, assuming statistical uncertainties
only.

The systematic uncertainties dominating this measurement are those on
the neutrino-to-antineutrino ratio, the overall rate normalization, and
the energy scale.
In the studies presented in sections \ref{sec:MuonNeutrinoDisappearance} and
\ref{sec:MaximalMixing}, the energy scale uncertainty was reduced from 10\% to 0.5\%
and a new systematic covering the direct optical efficiency was added with a 10\% uncertainty.
In both cases the new systematics improved the precision of the measurements.  While
this verification has not yet been studied in the context of the presented measurement, it is expected the
effect of changing these systematics to either have no impact on the result or increase
the significance to distinguish between models.
No significant dependence on the inner core
composition is observed.

Improvements in the event reconstruction performance are expected over
the lifetime of the detector.  Figure~\ref{fig:tomography_exclusion}
shows the impact of further improved reconstructions,
simultaneously improving the energy and angular resolution by $10\%$
and $20\%$, respectively.

\begin{figure} [tb]
   \begin{center}
   \begin{overpic}[scale=0.45]{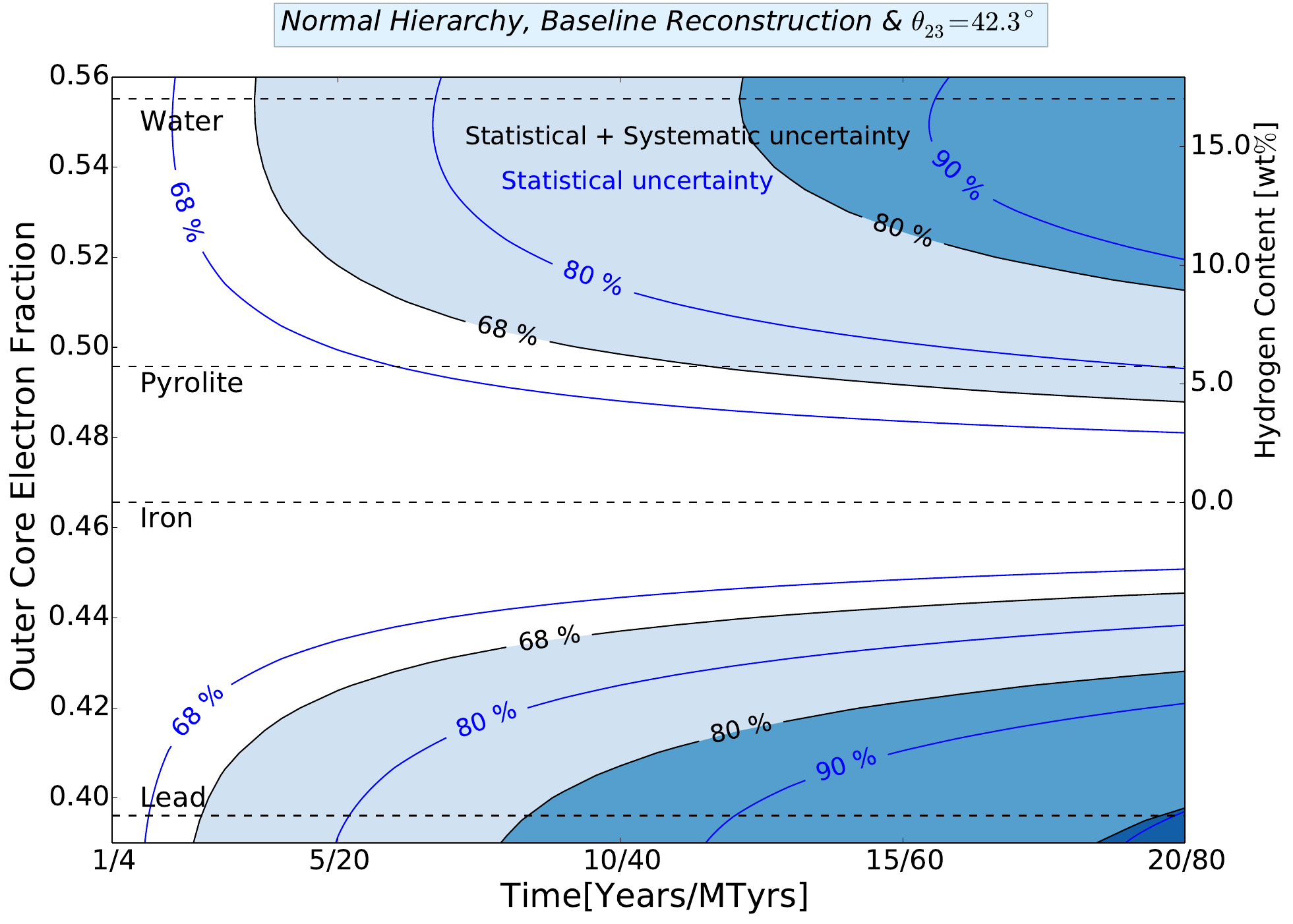}
        \put(50,28){\large\textcolor{red}{Preliminary}}
   \end{overpic}
   \end{center}
   \caption{Contour plot of exclusion levels with respect to a pure iron core 
     as function of livetime using the LLR method.}
   \label{fig:tomography_excl_time}
\end{figure}

Figure~\ref{fig:tomography_excl_time} shows the exclusion level
vs. time for a pure iron core ($Z/A=0.4656$) relative to other
compositions (the top panel of Fig.~\ref{fig:tomography_exclusion}
corresponds to a vertical slice of Fig.~\ref{fig:tomography_excl_time}
at ten years).  A mantle-like pyrolite core ($Z/A=0.4957$) can be
distinguished with more than $68\%$ confidence with systematic
uncertainties in 10~years or 40~megaton-years (as the average effective
mass of PINGU is $\sim$4 megaton in the energy range 2~GeV to 6~GeV)
assuming normal neutrino mass ordering and $\theta_{23}$ in the first
octant.  Lead could be excluded at $80\%$ confidence after
8~years. The right axis of Fig.~\ref{fig:tomography_excl_time} shows that the hydrogen content in the outer
core could be constrained with 68$\%$ confidence to less than $5\%$ by
weight after 10~years.

The tomography measurement strongly depends on the neutrino mass
ordering. If the ordering is inverted, the possibility to exclude one
model with respect to another drops by 50\%. Under the normal
ordering, resonant matter oscillations~\cite{MSW-W,MSW-MS,Akhmedov,Petcov-NOLR}
occur only for neutrinos; under the
inverted ordering, they occur only for antineutrinos. In the energy
range of interest, the cross section for the interaction of neutrinos
with matter is roughly twice that of
antineutrinos~\cite{RevModPhys.84.1307}.  This decreases the relative
number of signal events in the inverted ordering case, leading to a
drop in significance.  The octant of $\theta_{23}$ also has a significant
impact on the measurement where assuming $\theta_{23}$ is in the second octant
improves sensitivities by about 25\%.

\subsubsection{Conclusions}
\label{sec:TomographyConclusions}

Resonant neutrino oscillations inside the Earth (driven by the relatively large
magnitude of the $\theta_{13}$ mixing angle) can be used to probe the composition
of the Earth's core. Since these oscillations are sensitive to the electron
density, with good knowledge of the mass density from seismic
measurements one can derive the proton-to-nucleon ratio $Z/A$ inside
the Earth.  In particular, the composition of the liquid outer core
can be tested. PINGU could be the first experiment able to verify the
Earth has an iron-like core.  In this measurement, knowledge of the
neutrino mass ordering is critical for understanding the composition
of the core.  However, measurement of the mass ordering is not
strongly dependent on the core composition since the ordering
measurement also uses neutrinos that cross only the mantle. These
neutrinos can be used to fix the oscillation parameters and minimize
the induced systematics.  The uncertainty on $\theta_{13}$ is small
enough to have negligible impact on the Earth model determination.  PINGU
thus offers the best and possibly only way to test geophysical models
of the composition of the Earth's core by measuring the matter-induced
neutrino oscillations of atmospheric neutrinos.

Further improvements to the Earth tomography analysis are possible
with an improved event selection, potentially based on high-level
reconstructions described in Sec.~\ref{sec:EventReconstruction}. These
improvements would yield higher statistics in the region of interest,
and allow a better identification of matter-induced oscillation
effects by adding inelasticity measurements to the likelihood fit.

\clearpage
\resetlinenumber

\IfFileExists{NewCommands.tex}       {}       {}
\IfFileExists{../NewCommands.tex}    {}    {}
\IfFileExists{../../NewCommands.tex} {} {}

\graphicspath{{figures/}{NonAtmospheric/Supernova/figures/}}

\subsection{Supernova Neutrinos}
\label{sec:supernova}

As some stars approach the end of their life, they collapse to a
neutron star.  The majority of the energy radiated away during this
collapse is in the form of neutrinos of all flavors, emitted in a
span of tens of seconds.  These neutrinos have energies much lower
than those discussed in the previous sections; between \unit[10]{MeV}
and \unit[30]{MeV} instead of $\mathcal{O}$(\unit[1]{GeV}).

Due to the significantly lower energy of the neutrinos from
supernovae, the detection method is different from those detailed
previously.  Whereas the atmospheric neutrino analyses use the light
produced by hadrons and leptons following charged or neutral current
interactions, supernova neutrinos are detected after an inverse beta
decay interaction.  The positron produced in this interaction carries
away the majority of the energy from the neutrino, and it produces
\Cerenkov light along a short track (roughly \unit[0.6]{cm/MeV} and
\unit[325]{photons/cm}~\cite{2011JPhCS.309a2029K}).

The lower energy of the neutrinos arriving in the detector from a
supernova collapse also rules out reconstructing the initial neutrino
direction in IceCube.  As such, studies of supernova neutrinos search
for a short-term, coherent increase in the rate of signals in all the
DOMs in the detector.   The
noise rates of individual DOMs are roughly \unit[500]{Hz}, which can
be cut approximately in half by applying an artificial deadtime,
eliminating most of the correlated noise observed at low temperatures.

\subsubsection{IceCube Sensitivity to Supernova Neutrino Bursts}

If a core-collapse supernova explosion were to happen in our galaxy,
IceCube would provide the world's most precise neutrino rate
determination, with a significance larger than 20 standard deviations
for distances up to \unit[30]{kpc}~\cite{2011JPhCS.309a2029K}. 
The supernova detection significance deteriorates significantly at the
positions of the Magellanic Clouds and beyond because the intrinsic
DOM noise becomes dominant while the supernova neutrino fluence drops
with the inverse distance squared. Furthermore, individual neutrino
interactions are not detected, meaning that the energy and type cannot
be determined for each neutrino.  However, it has been
shown~\cite{ICRCSN} that the average neutrino energy can in principle
be determined to better than 30\% accuracy for distances less than
\unit[10]{kpc}. Recently, the IceCube data acquisition system has been
upgraded to store the timestamps of all hits.  With this information
at hand, one can deduce the neutrino energy from a comparison of the
rate at which a single neutrino interaction deposits light in just one
module vs.\ two (or more) neighboring modules.  In IceCube, however,
the coincidence rates (with a suitably tight time window) are only on
the order of a few per thousand.

Due to the increased sensor density both vertically and horizontally,
the coincidence rates in PINGU will be much higher, of order 1\%.
This will lead to a much improved determination of the average
neutrino energies, particularly at larger distances, and will allow
the extraction of information concerning the neutrino energy spectral shape. In
addition, some improvement in the detection significance for supernova
explosions in dwarf galaxies orbiting our Milky Way may be achieved by
employing coincidences in a short interval, $\Delta t$.
Note that the significance for a supernova detection with $n_k$ signal
events and optical module noise $r$ roughly scales like $n_k\cdot
(\Delta t\cdot r)^{-k/2}$ for a $k$-fold coincidence.

\subsubsection{Monte Carlo Simulation}

Two simulations were conducted to independently characterize the
sensitivity of the PINGU detector. In the first simulation, the
effective volume is evaluated for various coincidence conditions for
each detector configuration, namely IceCube, DeepCore, and PINGU with
an older, truncated geometry of 20 strings of 60 standard IceCube DOMs
operating with 4$\pi$ sensitivity, leading to a conservative
result. In the second simulation, the significance of the collective
rate deviation is determined with respect to supernova distance for
the IceCube detector in comparison to its subsequent generation
extensions, DeepCore and PINGU.  This simulation considers the
$40\times96$ and $20\times192$ PINGU geometries.  For supernova
neutrinos, we expect that a $26\times192$ PINGU geometry
will provide minor improvements over the $40\times96$ geometry.

The first simulation was conducted and cross checked with two
independent GEANT4-based~\cite{Agostinelli:2002hh} codes for various
detector configurations (IceCube and DeepCore, plus the PINGU
$20\times192$ geometry). The light yield is found to roughly scale
with the absorption length in the ice.  Uncertainties connected to
varying the optical properties of the ice as a function of depth were
assessed.

Since it is important to reject atmospheric muons when applying tight
coincidence criteria, atmospheric muon simulations were also 
produced. These simulations were generated following the Gaisser flux
model~\cite{Gaisser}, to which a simple and efficient muon rejection
algorithm was applied. A second atmospheric muon background set was
simulated as a cross check, using standard IceCube simulation tools
and the CORSIKA air shower generator~\cite{Heck:1998vt}. These studies
demonstrated that the distance reach and average energy determination
are not strongly affected by the presence of the remaining atmospheric
muon background. Some improvements to these results should be possible
by using the realistic description of the detector, particularly by
including the outer detector layers that act as a veto.

The detector sensitivity can be characterized by the effective volume
for positrons per optical module $V_{\rm eff}(e^+)=N_{\rm
  detected}/n({\rm e}^+)$, where $N_{\rm detected}$ is the number of
detected hits in each optical module for a given coincidence
condition. The energy-dependent density of positrons from neutrino
interactions in the ice is denoted by $n({\rm
  e}^+)$. Table~\ref{tab:effectiveVolumes} summarizes these effective
volumes, assuming an average neutrino energy $ \langle
E_{\bar\nu_{\rm{e}}}\rangle=12.6$~MeV.  The numbers are provided for
IceCube, DeepCore and the truncated $20\times60$ PINGU geometry,
imposing various coincidence criteria. Clearly, PINGU provides
substantially larger coincidence probabilities than IceCube.

\begin{table*}[htbp]
	\centering
		\begin{tabular}{l | l | l | l  }
		\hline
		Detector         & single          & nearest   & triple       \\
		                 & hit             & neighbor  & coincidence  \\
		\hline
                        IceCube            &   583 m$^3$ &  0.6 m$^3$  &  0.0002 m$^3$  \\
                        DeepCore           &   767 m$^3$ &  2.7 m$^3$  &  0.03 m$^3$    \\      
                        PINGU ($20\times60$) &   912 m$^3$ &  4.4 m$^3$  &  0.11 m$^3$    \\
	           \hline     
	           \end{tabular}
                   \caption{Effective volumes for positrons $V_{\rm
                       eff}(e^+)$ per optical module for various
                     coincidence conditions in IceCube, DeepCore and
                     the truncated PINGU $20\times60$ geometry, assuming $
                     \langle E_{\bar\nu_{\rm{e}}}\rangle=12.6$~MeV.}
	\label{tab:effectiveVolumes}
\end{table*}

\begin{figure*}[ht]
  \begin{center}
   	\begin{overpic}[scale=0.30]{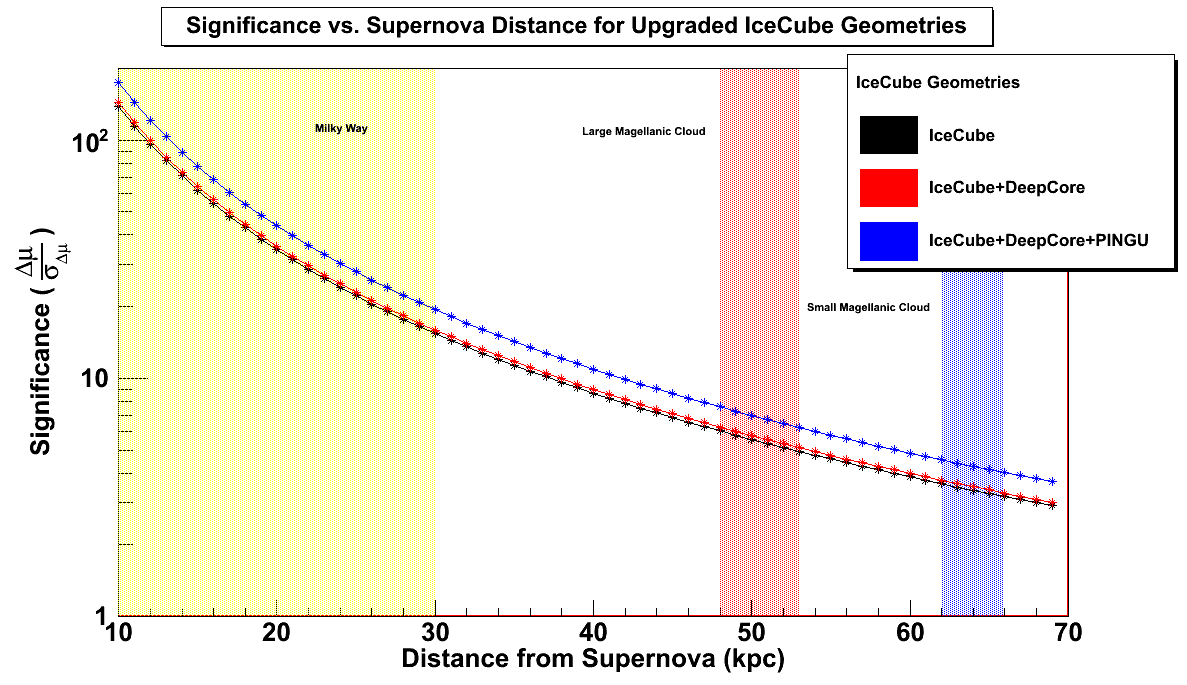}
	\put(35,40){\large \textcolor{red}{Preliminary}}
	\end{overpic}

   \end{center}
   \caption{Significance as a function of distance for an O-Ne-Mg,
     8.8M$_{\odot}$ supernova~\cite{garching}.  The sensitivity of
     the $40\times96$ PINGU geometry is compared to the standard
     IceCube geometries. For reference, the positions of the Milky Way
     and the Magellanic Clouds are indicated. Note that a trial factor
     is not included.}
	\label{fig:baselinesensitivity}
\end{figure*}

In the second study, low energy supernova neutrinos were injected in
each detector geometry (IceCube, DeepCore, and the $40\times96$ PINGU
geometry) with levels responding to an O-Ne-Mg, 8.8M$_{\odot}$
supernova~\cite{garching} at 10~kpc.  The collective hit rate
throughout the entire detector is compared to the expected
non-Poissonian noise levels, generated with the detector noise
model. The significance, calculated as the collective rate excess per
DOM, $\Delta \mu$, divided by its measured uncertainty $\sigma_{\Delta
  \mu}$, is approximately proportional to the square root of the total
number of optical modules within the detector (see
Fig.~\ref{fig:baselinesensitivity}). For statistically independent DOM
rates, it should be distributed according to a Gaussian with unit
width\footnote{In practice, statistically correlated hits introduced
  by atmospheric muons cannot be fully removed, leading to a widening
  of the distribution.}, such that its value corresponds to the number
of standard deviations from the null hypothesis of no signal.

\subsubsection{Determination of the average neutrino energy}

The methods described above search for supernovae by examining the
collective hit rate within the detector.  This method provides a large
target but very little information is obtained from the events.  A
second method considers coincident hits from events occurring very
close to the strings~\cite{Demiroers:2011am}.  This method reduces the
effective volume of the detector, but may provide more information
about the events themselves, and the average neutrino energy could be
determined in IceCube with MeV resolution for a supernova at 10 kpc.
The energy dependence becomes stronger when the optical modules are
closer together yielding a higher rate of coincidences. It has
previously been demonstrated that sub-MeV resolutions for dense core
detectors~\cite{ICRCSN,Varenna-Ribordy} can be achieved.

\begin{figure}[tb]
\begin{center}
\includegraphics[width=0.99\textwidth]{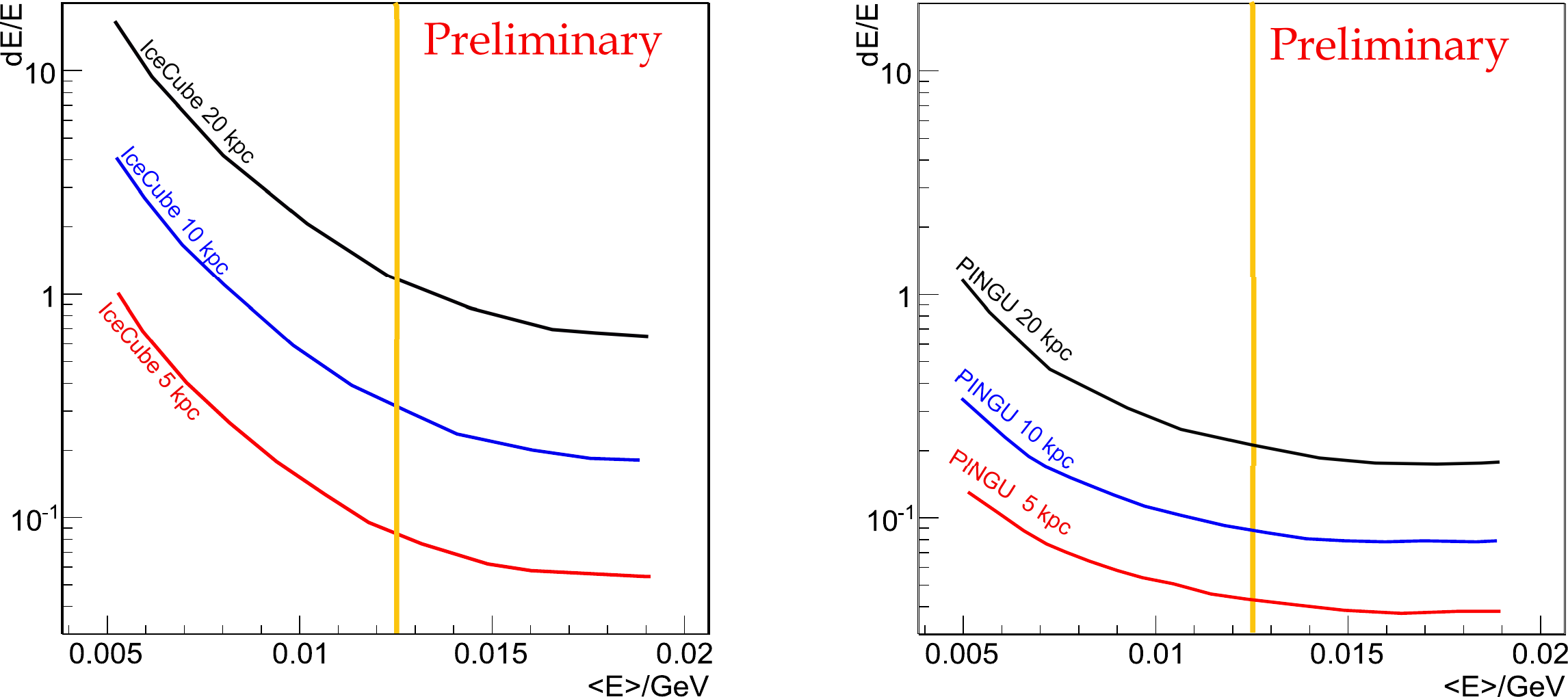}
\end{center}
\caption{Comparison of the precision on the determination of the
  average neutrino energy in (left) IceCube and (right) the
  truncated PINGU $20\times60$ geometry for a
  supernova at 20 (black), 10 (blue) and 5 (red) kpc distance. }
	\label{fig:averageEnergy}
\end{figure}

The relative precision of the energy determination is depicted in
Fig.~\ref{fig:averageEnergy} as a function of distance and average
energy. To be most conservative, flux, energy, and spectral shape were
taken from the collapse of an O-Ne-Mg 8.8$M_\odot$ progenitor star,
the lowest mass progenitor known to undergo a core collapse. We also
assume the conservative case of the truncated  ($20\times60$) PINGU
geometry.

\subsubsection{Providing a measure of the spectral shape}

Different coincident hit modes (e.g. multiple hits on single DOMs or
between neighbors) show distinct differences in the neutrino energy
dependence~\cite{Bruijn:2013ibl}. These can be used to characterize
the spectral shape of the neutrino emission, in addition to the
determination of the average energy.  Assuming a neutrino spectral
shape parametrized with three parameters (the luminosity $L_\nu$, the
average neutrino energy $\langle E_{\bar\nu_{\rm{e}}}\rangle$ and a
shape parameter $\alpha$~\cite{Keil}) and using single, double and
triple coincident hit modes in a $\chi^2$ fit, $ \langle
E_{\bar\nu_{\rm{e}}}\rangle$ and $\alpha$ can be extracted
simultaneously.

\begin{figure*}[ht]
\begin{center}
\subfigure{
\begin{overpic}[scale=0.37]{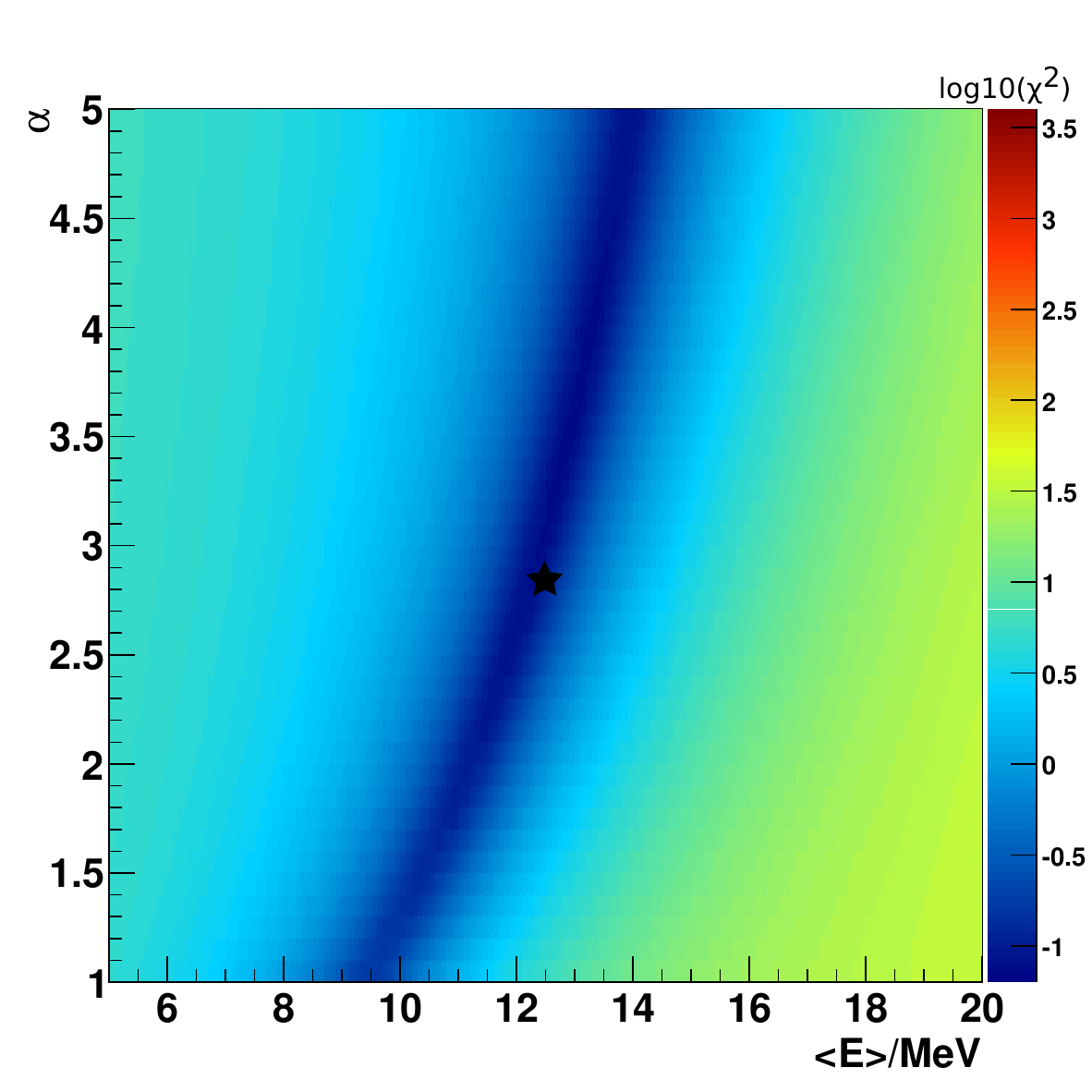}
	\put(15,80){\large \textcolor{black}{Preliminary}}
	\end{overpic}
}	
\subfigure{
\begin{overpic}[scale=0.37]{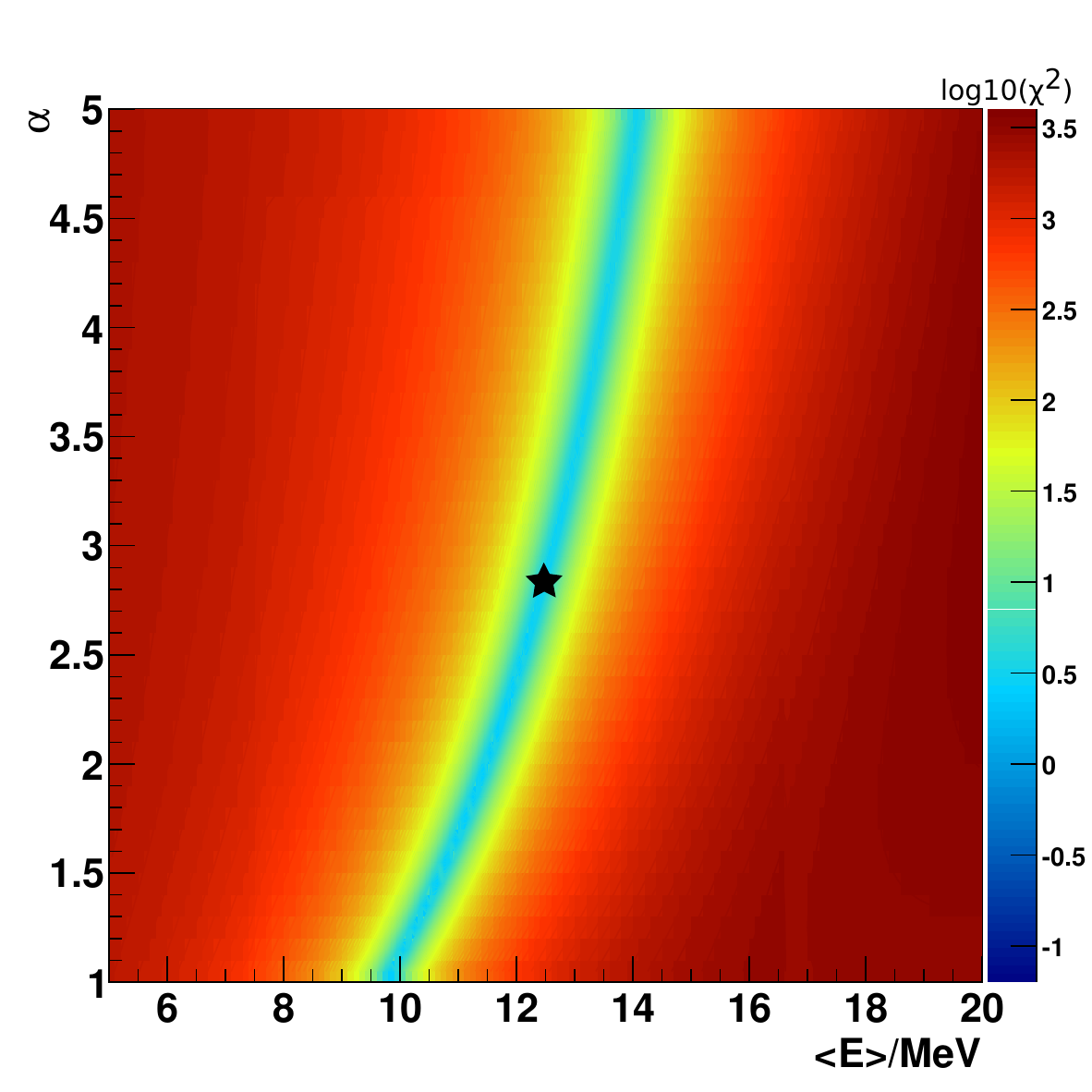}
	\put(15,80){\large \textcolor{black}{Preliminary}}
	\end{overpic}
	}
\end{center}
\caption{Combined determination of the average neutrino energy and the
  spectral shape parameter $\alpha$ for IceCube (left) and PINGU
  (right) using a $\chi^2$ method. The input value ($\langle
  E_{\bar\nu_{\rm{e}}}\rangle=12.6$ MeV, $\alpha=2.84$) is denoted by
  a star. }
	\label{fig:ShapeEnergy}
\end{figure*}

A scan of the parameter space for a simulated supernova at 10~kpc
distance is shown in Fig.~\ref{fig:ShapeEnergy} as function of average
energy and $\alpha$ for IceCube and PINGU. The color scale indicates
the log$_{10}$($\chi^2$) value for the parameter values tested. Note
that $\alpha$ and $\langle E_{\bar\nu_{\rm{e}}}\rangle$ are almost
degenerate.  The increased rates at higher hit modes in PINGU lead to
a substantially smaller uncertainty on the parameters.

\subsubsection{Conclusions and outlook}

PINGU will enhance IceCube's ability to detect and interpret very low
energy ($\Enu \sim 15$~MeV) neutrino bursts from supernova
explosions. Taking IceCube and DeepCore together with the $40\times96$
PINGU geometry, the galactic supernova sensitivity will increase by
approximately $\sqrt 2$ when compared to IceCube alone. The greatest
benefit comes from the improved determination of the average supernova
neutrino energy provided by PINGU's closer module spacing.  The
average energy will also be measured roughly a factor of five better
than with IceCube. As shown with the truncated $20\times60$ geometry,
PINGU would also allow us to determine the spectral shape, once the
average neutrino energy is known.  A PINGU geometry
with significant increase in the module density will improve this result.

Lowering the PMT noise rates will be an important ingredient for the
detection of supernovae in neighboring galaxies in an envisaged low
energy, multi-megaton future detector succeeding PINGU. Very low noise
rates could be achieved by collecting light with wavelength shifters
read out with highly sensitive PMTs with a small cathode
area. Extending the accepted wavelength range down to 200~nm by using
fused quartz windows would increase the detection efficiency by a
factor of approximately 2.7 (weighted for the optical module
sensitivity and the $1/\lambda^2$ dependence of the \Cerenkov photon
flux).  The PINGU deployment may allow us to test these technical
improvements by deploying prototype modules as discussed in
Sec.~\ref{sec:WOMs}.

\clearpage
\resetlinenumber

\IfFileExists{NewCommands.tex}       {}       {}
\IfFileExists{../NewCommands.tex}    {}    {}
\IfFileExists{../../NewCommands.tex} {} {}

\graphicspath{{figures/}{NonAtmospheric/WIMPs/figures/}}

\subsection{Dark Matter}
\label{sec:WIMPDarkMatter}

\subsubsection{Motivation}
\label{sec:WIMPDarkMatterMotivation}

Observational evidence indicating the existence of dark matter can be
obtained at a wide variety of scales, from the motion of stars to
imprints on the cosmic microwave background (CMB)~\cite{Zwicky:1933gu,
  Komatsu:2010fb, Bertone:2004pz}. Many dark matter candidates could
be indirectly detected through the observation of particles created in
dark matter self-annihilations.  The search for neutrinos created as
part of these annihilations is of particular interest, since neutrinos
can be used to probe the properties of the dark matter, including its
self-annihilation and nucleon scattering cross-sections.  One of the
attractive candidates for this dark matter is the Weakly Interacting
Massive Particle (WIMP, denoted by $\chi$).  This candidate arises
naturally in many theories beyond the Standard Model of Particle
Physics developed to explain the origin of electroweak symmetry
breaking and solve the gauge hierarchy
problem~\cite{Bertone:2004pz}. Although we will use the terms ``WIMP''
and ``dark matter candidate'' interchangeably in the rest of the
section, the reader should note that there are other scenarios that
provide viable non-WIMP dark matter
candidates~\cite{Bertone:2010,Kolb:2001} with similar signatures in
neutrino telescopes.

Given that they interact gravitationally, WIMPs could be captured in
the Sun (after scattering off nuclei), accumulate, and self-annihilate
producing a flux of neutrinos. Under the assumption of equilibrium
between WIMP capture rate ($\Gamma_C$) and annihilation rate
($\Gamma_A$) in the Sun, $\Gamma_A$ depends only on the total
scattering cross-section. Since the Sun is primarily a proton target,
strong constraints can be derived on the spin-dependent WIMP-proton
scattering cross-section ($\sigma_{p,SD}$) by measuring a neutrino flux
from the Sun.  IceCube has set the world's best limits on this process
for WIMP masses above 50~GeV~\cite{Aartsen:2012kia}.

Searches for dark matter annihilation signals in the Milky Way can be
used to test the thermal average of the WIMP annihilation rate, which
is the product of the annihilation cross-section and the relative
velocity of WIMPs averaged over the velocity distribution, $\langle
\sigma_A v \rangle$. IceCube has set tight constraints on $\langle
\sigma_A v \rangle$ with searches for signals from the Galactic
halo~\cite{Abbasi:2011eq}, Galactic Center~\cite{Abbasi:2012ws}, and
dwarf spheroidal galaxies and clusters of
galaxies~\cite{Aartsen:2013dxa,IceCube:2011ae}. These results improved
upon theoretical predictions~\cite{Yuksel:2007ac,Dasgupta:2012bd}.

The lowered energy threshold of PINGU will allow the enhancement of
current searches to test WIMP masses that
are below the IceCube detection threshold. WIMP scenarios motivated by
DAMA's annual modulation signal~\cite{Savage:2008er} and
isospin-violating scenarios~\cite{Feng:2011vu} would also be testable.

\subsubsection{Limit Calculation}
\label{sec:WIMPDarkMatterSolarLimit}

The primary WIMP annihilation spectrum is model-dependent, hence we
consider two extreme benchmark scenarios, where WIMPs annihilate
exclusively to $\tau^+\tau^-$ and $b\bar{b}$.  The first case results
in a ``hard'' neutrino spectrum, while the second case leads to a
``soft'' spectrum with lower average neutrino energy. One can expect
that a general model with a mix of various annihilation channels will
be bracketed by these two extrema.

\subsubsection{Experience from DeepCore}
\label{sec:WIMPDarkMatterSolarWIMPsLessonsFromDeepCore}

The most recent IceCube Solar WIMP analysis~\cite{Aartsen:2012kia}
searched for signal neutrinos originating from potential WIMP masses
ranging from 20~GeV to 5~TeV. Within this mass-range, signal events
can have very different event topologies in the detector. To
accommodate all expected event topologies within one single analysis,
the full dataset is split into three independent, non-overlapping
event selections that were later combined.  The first division is into
two seasonal data streams, ``summer'' and ``winter'', when the Sun is
above and below the horizon at the South Pole, respectively. The
winter dataset is then further divided into a low-energy and
high-energy sample. During the summer period, downward-going events
are selected that interact inside DeepCore to reduce the atmospheric muon background. The winter high-energy
event selection has no particular track-containment requirement,
selecting upward-going muon tracks. The low-energy counterpart is
focused on upward-going starting or fully-contained neutrino-induced
muon tracks inside DeepCore.

The inclusion of DeepCore thus allowed the search to be extended to
the austral summer when the Sun is above the horizon and expanded the
WIMP mass reach from 50~GeV to 20~GeV, which was not accessible to
previous IceCube searches~\cite{Aartsen:2012kia}.  With its lower
energy threshold, PINGU will allow further extension to even lower
WIMP masses.


\subsubsection{Solar WIMPs}
\label{sec:WIMPDarkMatterSolarWIMPs}

Building on the experience gained with DeepCore analyses, we utilize
the lowered energy detection threshold of PINGU to significantly
improve the sensitivity for WIMP masses in the range between 5 and
50~GeV.  We perform a straightforward event-based Monte Carlo (MC)
study, using the $26\times192$ PINGU geometry (see
Table~\ref{Tab:geometries}).  
The datasets are identical to those used in the atmospheric neutrino
studies, described in Sec.~\ref{sec:atmo_event_sel}.  In contrast to
previous IceCube WIMP searches where we relied solely on the track
channel produced by muon-type neutrino CC interactions, here we
include all neutrino flavor channels taking advantage of PINGU's
directional reconstruction capability for cascades.  PINGU will also
offer the potential to use energy spectral
information~\cite{Rott:2011fh}, however to be conservative it is not
used for the calculation of the sensitivity. This study focuses on the
low-mass WIMP range and considers only events that interact inside the
PINGU volume.

\begin{figure}[h]
\begin{center}
\begin{overpic}[width=0.49\textwidth]{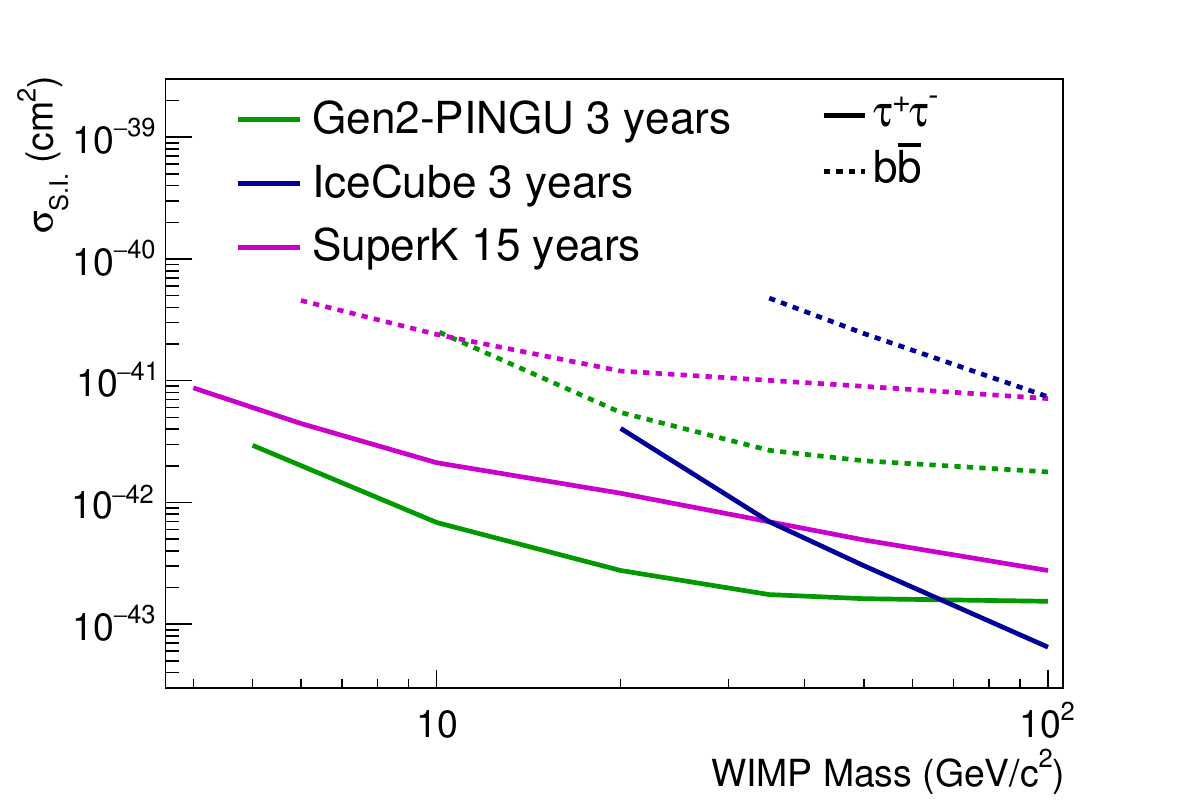}
     \put(18,12){\large \textcolor{red}{Preliminary}}
   \end{overpic}
\begin{overpic}[width=0.49\textwidth]{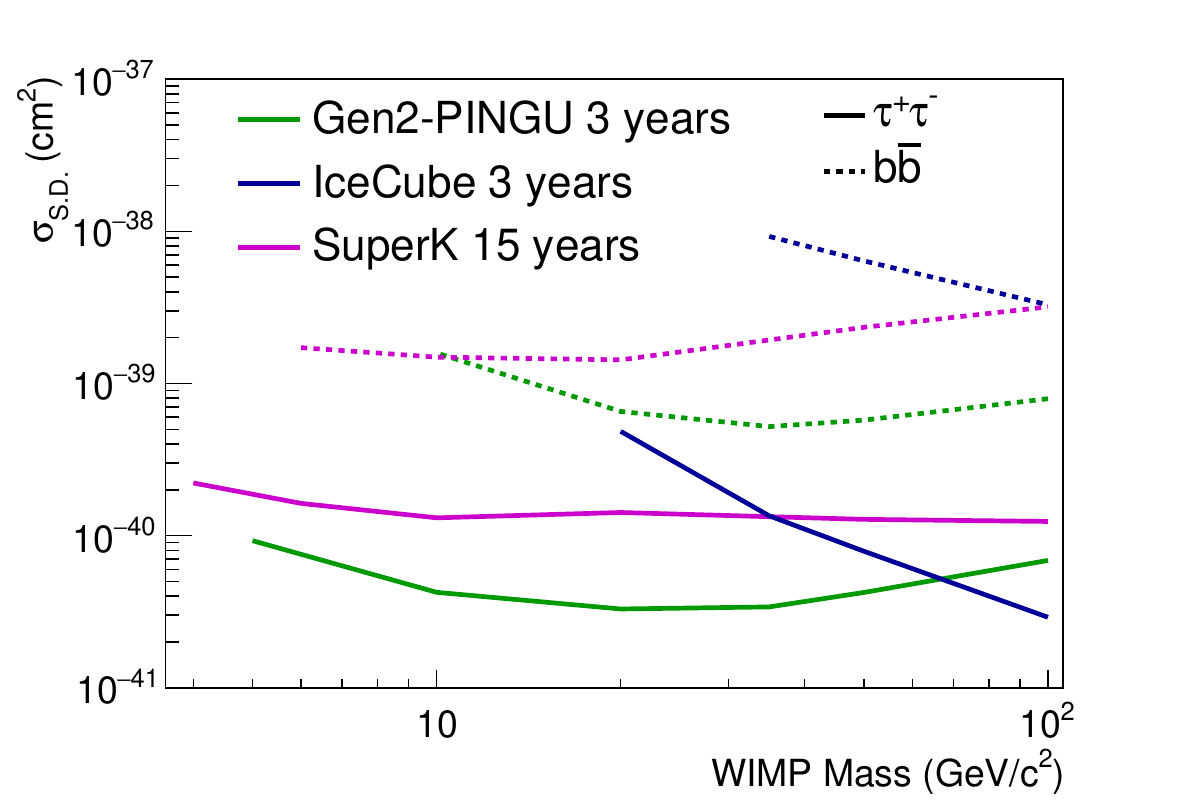}
     \put(18,12){\large \textcolor{red}{Preliminary}}
   \end{overpic}
   \end{center}
   \caption{Sensitivity of the $26\times192$ PINGU geometry to
     $\sigma_{p,SI}$ (left figure) and $\sigma_{p,SD}$ (right figure)
     for hard (solid lines) and soft (dashed lines) annihilation
     channels over a range of WIMP masses for a livetime of three
     years.  The sensitivities are compared
     to the present IceCube limits~\cite{Aartsen:2012kia} and limits
     from Super-K~\cite{Choi:2015ara}.}
\label{fig:PINGU_SolarWIMP_sens}
\end{figure}

In accordance with the assumptions in Sec.~\ref{sec:atmo_event_sel},
the down-going atmospheric muon background is assumed to be reduced to
a negligibly small level, so that the background consists only of
atmospheric muon neutrinos, which are generated following the flux
model in~\cite{Honda:2015fha}. The background at final analysis level
from solar atmospheric neutrinos (those neutrinos generated in the
atmosphere of the Sun) has previously been calculated to be of order
one event in 317~days of data~\cite{Aartsen:2012kia}, and is
consequently not included.

All signal simulations are made with DarkSUSY~\cite{Gondolo:2004sc}
and WimpSim~\cite{Edsjoe:2013ws}, which describe the capture and
annihilation of WIMPs inside the Sun and the consequent production and
propagation of neutrinos from the core of the Sun to the detector,
including three-flavor oscillations and matter effects. The expected
neutrino signal flux from WIMP annihilations in the Sun is calculated
by weighting the simulated neutrino events with the WIMP channel
dependent neutrino spectra obtained from DarkSUSY. The
zenith-dependent Sun position throughout the year is taken into
account.


In the absence of signal, we estimate the 90\% C.L. median upper limit
$\mu_{\mathrm{s}}^{90}$ on the number of signal events, given
$n_{\mathrm{bg}}$, following~\cite{Rolke:2005tr}. The limit on
$\mu_{\mathrm{s}}^{90}$ is compared to $n_{\mathrm{s}}$, which was
calculated for $\Gamma_A=1$~s$^{-1}$, to derive the 90\% C.L. limit on
$\Gamma_A$ for each WIMP model. Limits on $\Gamma_A$ are then
converted into limits on the spin-dependent ($\sigma_{p,SD}$) and
spin-independent ($\sigma_{p,SI}$) WIMP-proton scattering
cross-sections using the method
from~\cite{Edsjoe:2013ws}. Figure~\ref{fig:PINGU_SolarWIMP_sens} shows
the calculated sensitivities as a function of WIMP mass for the soft
(dashed) and hard (solid) annihilation channels obtained in this
study.  Additional ideas to
further advance this search are discussed below.


\subsubsection{Galactic Center WIMPs}
\label{sec:WIMPDarkMatterGalacticCenterWIMPs}

The expected differential neutrino flux from dark matter annihilations
at the Galactic Center (GC) is proportional to the differential
neutrino multiplicity per annihilation, $\frac{dN_{\nu}}{dE}$, and the
line of sight integral over the square of the dark matter density at
an angle of $\psi$ from the GC, $J(\psi)$~\cite{Yuksel:2007ac}.

Searches for dark matter self-annihilations in the GC have been
performed by IceCube using the NFW dark matter density
profile~\cite{Navarro:1995iw} as a benchmark. The first GC analysis
used the partially instrumented 40-string IceCube detector, probing
$\left<\sigma_A v\right>$ for WIMP masses down to 100~GeV for several
annihilation channels~\cite{Abbasi:2012ws}. The two most recent GC
analyses use the almost-completed IceCube detector (comprised of 79
strings), including DeepCore~\cite{Aartsen:2013ae}.  The sensitivity
to $\left<\sigma_A v\right>$ in the GC was improved by up to four
orders of magnitude for WIMP masses of 100~GeV ($\tau^+\tau^-$)
compared to the previous IceCube 40-string analysis. For PINGU an even
lower energy threshold in the sub-10~GeV region is expected.

The GC, in the southern hemisphere at $-29^\circ$ declination, is a
challenging target for IceCube due to the high rate of down-going
atmospheric muon background events.  Veto methods against atmospheric
muons~\cite{Aartsen:2013ae}, developed for the most recent
IceCube 79-string GC analysis, significantly improved the
sensitivities of analyses focused on low-energy events in the southern
hemisphere. This turns IceCube into an efficient 4$\pi$ detector for
indirect dark matter searches.  These methods reject muon background
while retaining low energy starting events inside a fiducial region,
and make it feasible to achieve an event selection with an adequate
neutrino purity for PINGU.

The sensitivity to $\left<\sigma_A v\right>$ for a GC analysis for
WIMP masses between 5 and 30~GeV with PINGU is derived in a
straightforward event-based MC analysis. This analysis is very similar
to the study described in Sec.~\ref{sec:WIMPDarkMatterSolarWIMPs}. In
contrast to the solar WIMP analysis, where a perfectly efficient
atmospheric muon veto was assumed, we use the level of atmospheric
muon background in the IceCube 79-string GC analysis (neutrino purity
of 10\% at final analysis level).  This provides a conservative
sensitivity estimate for PINGU.  As for the solar WIMPs, we assume a
live-time of one year, the $40\times96$ PINGU detector geometry, and WIMP
annihilations into neutrinos for this study. The resulting neutrino
line spectrum yields approximately equal numbers of neutrinos and
anti-neutrinos of all flavors after annihilation and neutrino
oscillations.  We consider muons produced through CC interactions as
the only signal-detection channel for this conservative study
(all-flavor prospects are discussed below). The number of signal and
background events are calculated in a cone around the GC with a
half-opening angle of 10$^\circ$ by weighting the simulated neutrino
events according to the expected signal and atmospheric neutrino
background flux respectively. The number of expected atmospheric muon
background events ($n_{\mathrm{bg},\mu}$) are added to the atmospheric
neutrino background ($n_{\mathrm{bg},\nu}$) with expected purity to
determine the total background ($n_{\mathrm{bg}}$).

\begin{figure}[t!]
\centering
\begin{overpic}[scale=0.25]{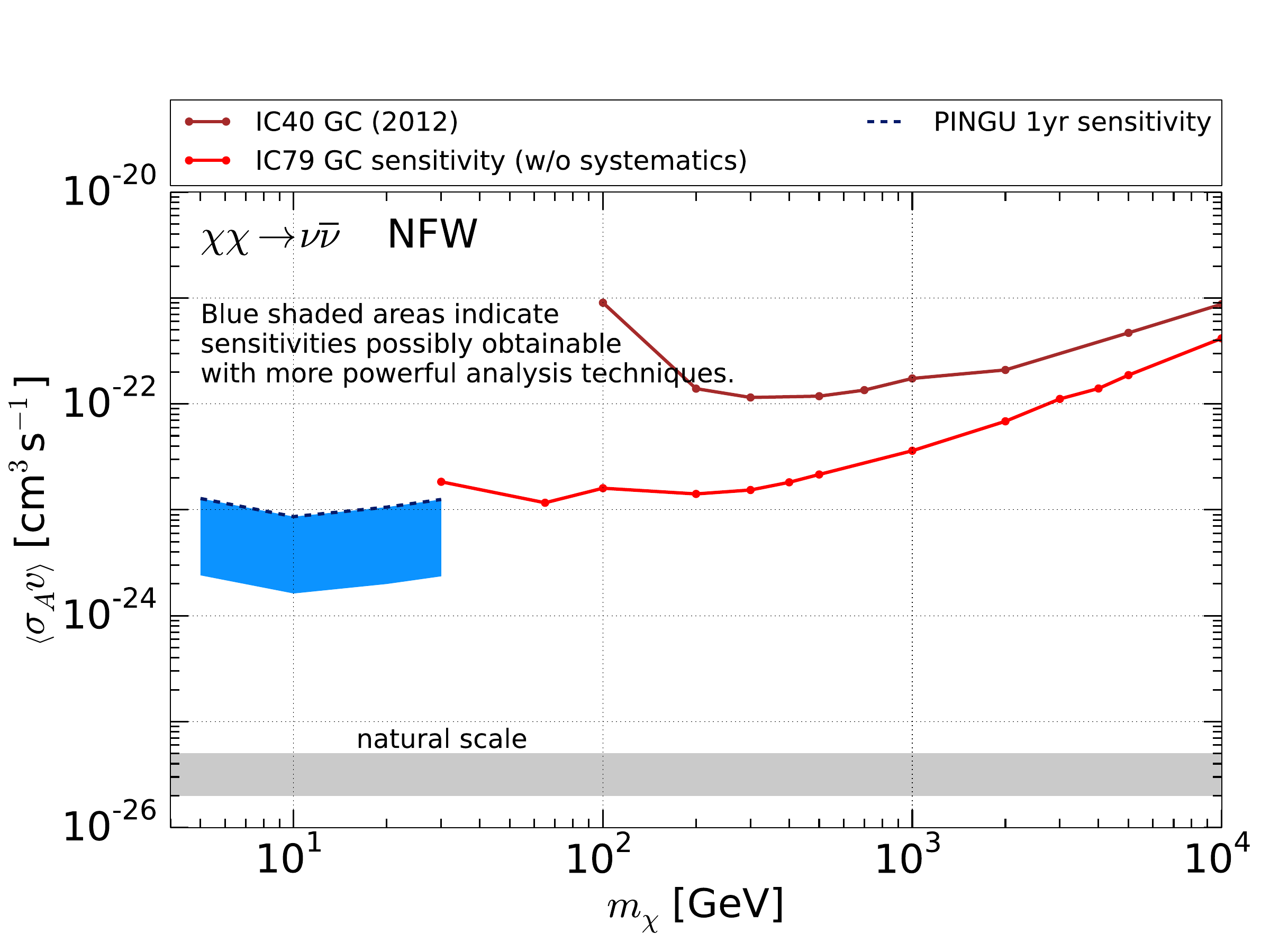}
     \put(55,40){\large \textcolor{red}{Preliminary}}
   \end{overpic}
   \caption{Sensitivity of the $40\times96$ PINGU geometry to
     $\left<\sigma_A v\right>$ for direct annihilations into neutrinos
     over a range of WIMP masses for a live-time of one year. In light of the results presented in
Sec.~\ref{sec:NeutrinoOscillations}, for this signature we expect the
$40\times96$ and $26\times192$ geometries to perform comparably well. The
     sensitivities are compared to limits from the IceCube 40-string
     analysis~\cite{Abbasi:2012ws} and the sensitivities of the most
     recent IceCube 79-string GC analyses~\cite{Aartsen:2013ae}. }
\label{fig:PINGU_GC_sens}
\vspace{-1.\baselineskip}
\end{figure}

Using the statistical method as described in
Sec.~\ref{sec:WIMPDarkMatterSolarWIMPs}, we derive the 90\% C.L. limit
on $\left<\sigma_A v\right>$ for each WIMP model.
Figure~\ref{fig:PINGU_GC_sens} shows the calculated sensitivities as a
function of WIMP mass for the direct neutrino annihilation channel
obtained in this study.  The blue shaded areas indicate the expected
range of sensitivity which could be obtained with improved analysis
techniques.  Here, the expected range of sensitivities is driven by
the remaining contribution of atmospheric muons to
$n_{\mathrm{bg}}$. Consequently, the most optimistic range of the blue
shaded area assumes a more efficient atmospheric muon veto (perfect
muon veto in the limit).

\subsubsection{Outlook and conclusions}
\label{sec:WIMPDarkMatterOutlookAndConclusions}

With one year of data, PINGU will be able to test direct detection
claims of signals in the spin-independent region using dark matter
annihilations in the Sun with hard neutrino spectra. The search for
neutrinos from dark matter annihilations in the Sun will provide an
independent method to test the longstanding anomalous annual
modulation signal from
DAMA/LIBRA~\cite{Bernabei:2010mq, Savage:2008er}. PINGU will also probe
a region of parameter space that is intrinsically difficult to access
in direct detection and has often resulted in excess events in
searches, such as CoGeNT~\cite{Aalseth:2010vx, Aalseth:2011wp}, and the
silicon data from CDMS-II~\cite{Agnese:2013rvf} (CDMS-Si), as well as
being able to test for dark matter annihilations near the Galactic
Center.

The PINGU sensitivities discussed here are based on standard analysis
methods utilized in IceCube, which have been kept intentionally
simple. The results presented are therefore conservative,
since PINGU offers the possibility of applying analysis methodologies
that go beyond what we have used thus far in IceCube and
DeepCore. These include a more precise definition of contained and
partially-contained events, a better energy estimation from the
measured track length allowing us to use spectral information, and
particle identification, each of which can potentially confer greater
sensitivity to dark matter.

\clearpage

\clearpage

\resetlinenumber

\IfFileExists{NewCommands.tex}       {}       {}
\IfFileExists{../NewCommands.tex}    {}    {}
\IfFileExists{../../NewCommands.tex} {} {}

\graphicspath{{figures/}{Hardware/figures/}}

\section{Instrumentation}
\label{sec:Instrumentation}

\subsection{Introduction}
\label{sec:InstrumentationIntroduction}

In this section we present initial design considerations for PINGU
sensors, down-hole cables and the drill.  Experience gained with
IceCube is leveraged in many instances to reduce design effort and
mitigate risk.  At this stage in the design process, several possible
sensor ideas are being pursued.  While simulations based on IceCube
DOMs enable us to perform detector studies using existing IceCube
Software infrastructure, improved sensor designs are expected to
provide better performance at lower cost.
We note the cable and drill designs are not strongly dependent on
the sensor design.

\subsection{Sensor Configuration}

As described in Sec.~\ref{sec:DetectorGeometries}, an initial
detector design consists of 26 strings with 4992 sensors (PINGU
Digital Optical Modules, or PDOMs) deployed within the DeepCore
section of the IceCube volume (Fig.~\ref{fig:pingu_geos}), at depths
of 2150-2450~m.  Following the design of the IceCube/DeepCore
DOM~\cite{Abbasi:2009domdaq}, each module houses a 10-inch
diameter high-quantum-efficiency PMT along with electronics to
operate the PMT, send digitized signals to the surface, and perform
calibration tasks.

\subsection{Optical Sensor Design}
\label{sec:OptSensDesign}

\begin{figure}[tb]
\begin{center}
   $\begin{array}{cc}
    \includegraphics[width=2in]{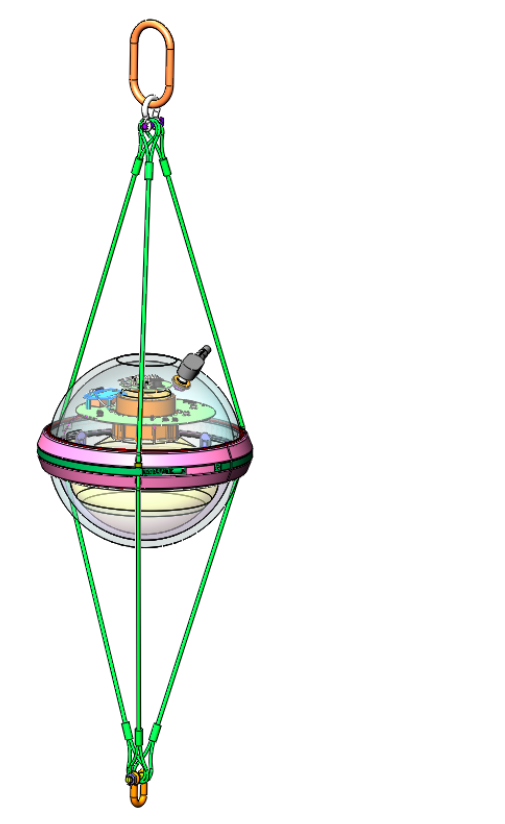} &
    \includegraphics[width=3in]{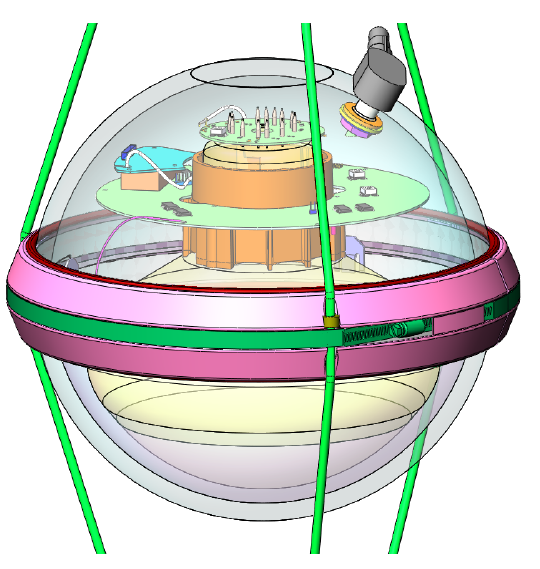}
    \end{array}$
   \caption{PINGU Digital Optical Module (PDOM).}
   \label{fig:PDOMAppearance}
\end{center}
\end{figure}

Figure~\ref{fig:PDOMAppearance} shows a sketch of the PDOM, whose
external components (glass sphere, cable penetrator, connector, etc.)
are essentially identical to that of the IceCube DOM.  Internal
components (PMT, optical gel, etc.) would also be similar.  IceCube
experience demonstrates the robustness of the DOM design: of the 5160
IceCube DOMs deployed at depth, 98.4\% were operable after this
critical phase, with only 0.4\% failing in subsequent long-term
operation.

\begin{figure}[tb]
   \begin{center}
      \includegraphics[width=6in]{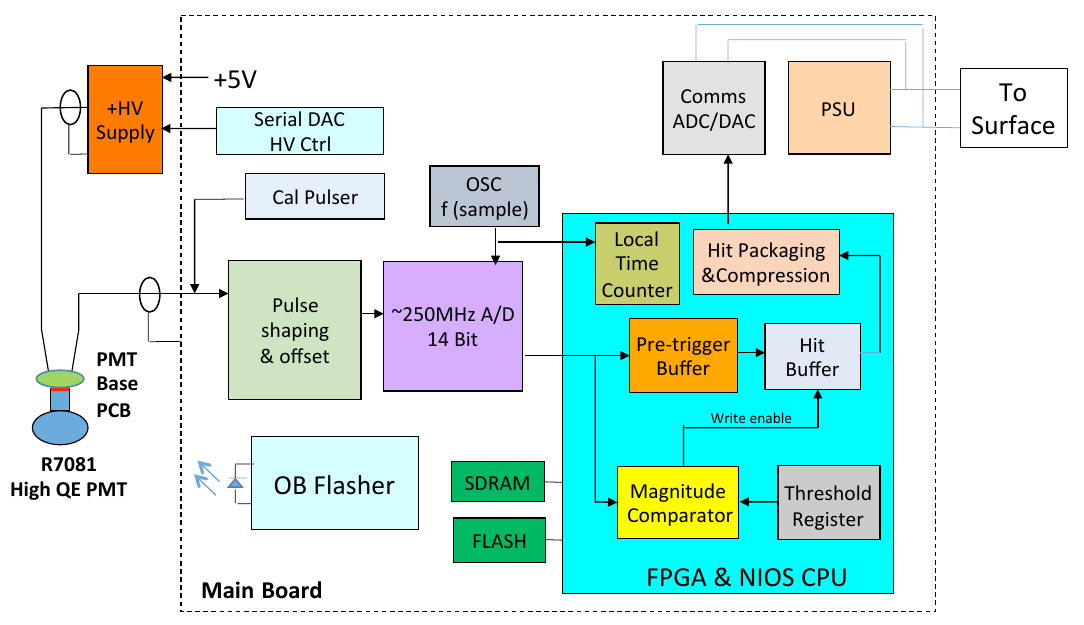}
      \caption{Functional block diagram for data acquisition in the
        PDOM, focusing on the Main Board.}
      \label{fig:PDOMFunctionalBlockDiagram}
   \end{center}
\end{figure}

As shown in Fig.~\ref{fig:PDOMFunctionalBlockDiagram}, the PDOM has
a single ``Main Board'' containing electronics responsible for
waveform digitization, control, and communication with the surface.
Modules on this board reproduce most of the functionality of the
original DOM Main Board, which cannot be identically recreated because
several of the key parts are obsolete and unavailable.  The new design
is simplified and uses modern electronic technology to reduce
cost, power and board space.  The previous multi-channel analog
circuitry and custom digitizers are replaced by a single commercial
ADC chip, with triggering and data processing handled by a more
capable FPGA-based processor.  All triggered waveforms are sent to
the surface after suitable compression, allowing the new design to
omit complex circuitry and {\it in situ} interconnections previously
used to detect local coincidences between modules.  As now found in IceCube modules, PMT
waveforms will be digitized and deconvolved into individual
photoelectron signals as needed; the new design will allow the
deconvolution to be done before transmission of data to the surface.


Simplification of the Main Board's design will reduce component count, complexity, and overall cost.  The previous design also included
an analog delay board to allow for triggering the custom waveform
digitizers after detection of a PMT pulse.  This is not needed in the
new design because the continuously-digitized PMT output is buffered
in the FPGA while threshold crossings are detected digitally.

New PDOM firmware is being written as part of the Main Board redesign.
The code preserves the original structure and communications protocols
as much as possible, ensuring a high level of compatibility with
existing modules in the IceCube data acquisition system.

\begin{figure}
  \centering
  \includegraphics[width=1.0\textwidth]{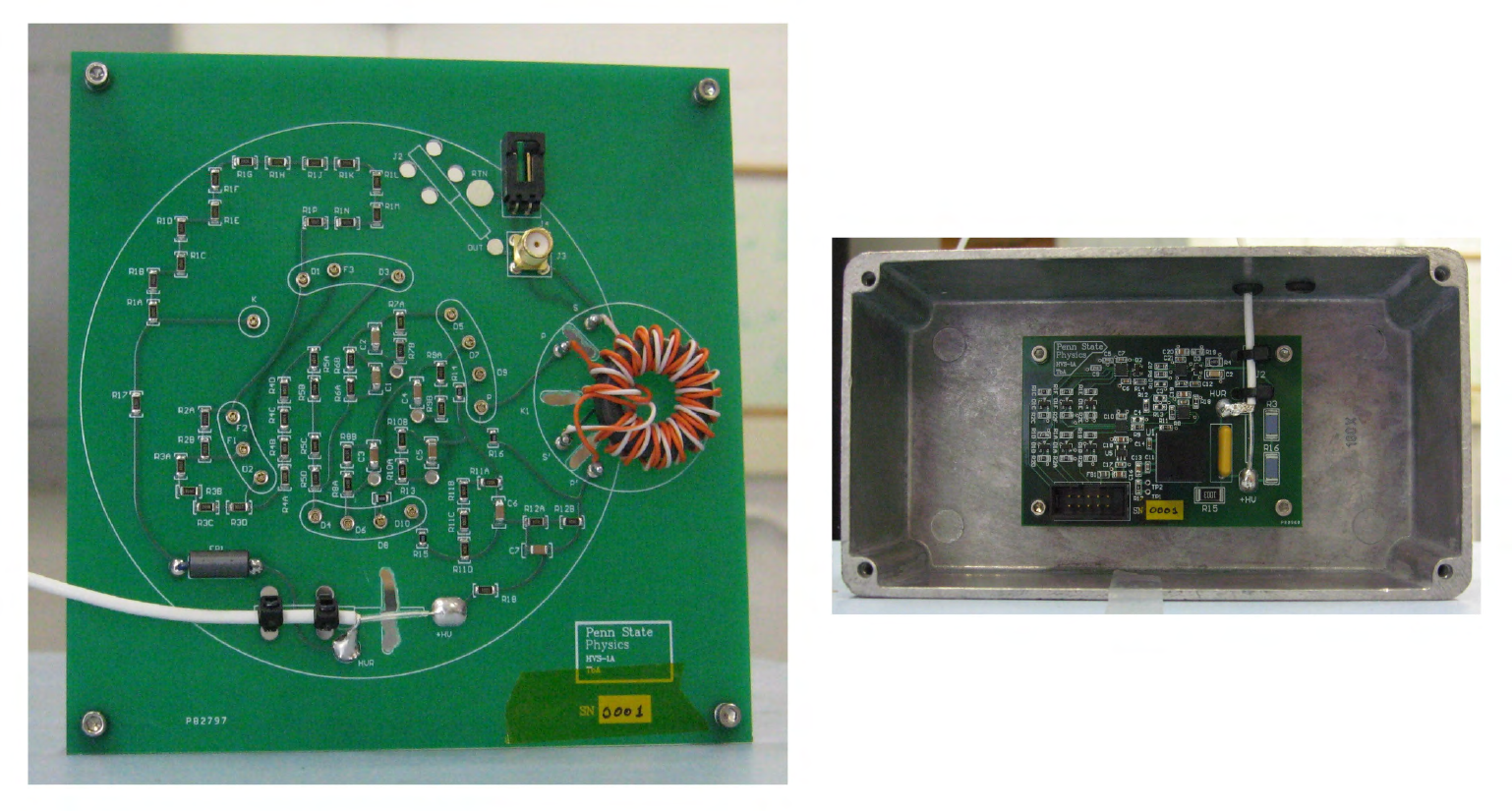}
  \caption{The prototype Gen2-DOM HVD (left) and HVS (right) boards.}
  \label{fig:proto_gen2dom_hv}
\end{figure}

Figure~\ref{fig:proto_gen2dom_hv} shows first-revision hardware for
the PDOM high voltage system. The High Voltage Divider (HVD) board provides
the high-voltage resistive bias and output signal coupling networks
for the PMT, while the High Voltage Supply (HVS) board provides
a simple analog interface for controlling and monitoring the system's
high voltage setting. The design utilizes a miniaturized high voltage DC-DC
converter module, which is capable of supplying a maximum 2~kV output
voltage while consuming only 125~mW under the HVD's nominal
130~M$\Omega$ load. The total ripple injected into the PMT's output
signal by the high-voltage system has been measured as
$<$10~$\mu$V. These new designs hew closely to the proven high-voltage
design strategies of operating IceCube DOMs, while also employing
recent advances in passive component technology to increase voltage
de-ratings and improve manufacturability.

\begin{figure}
  \centering
  \includegraphics[width=1.\textwidth]{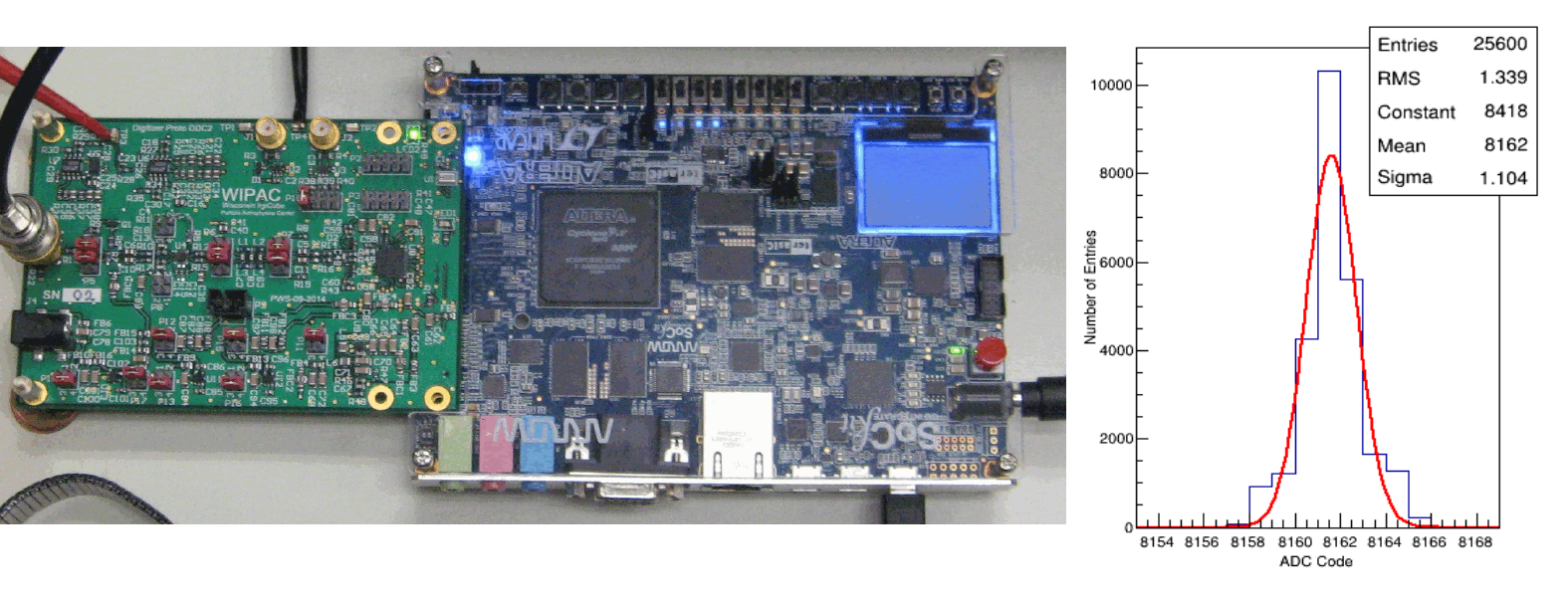}
  \caption{The DDC and SoCKit (left) and a measured baseline
    distribution (right).}
  \label{fig:proto_ddc_and_result}
\end{figure}

A prototype design for the PDOM's on-board digitizer, called the
Digitizing Daughter Card (DDC), is shown in
Fig.~\ref{fig:proto_ddc_and_result}. At the heart of the DDC is the
Texas Instruments ADS4149, a 14-bit, differential ADC, configured to
sample at its maximum rate of 250~Msps. The DDC also employs an Analog
Devices ADA4930 fully differential amplifier for the dual purposes of
single-ended to differential signal conversion, and signal shaping for
optimal identification of single photoelectron pulses. Various other
bits of circuitry (\textit{e.g.,} an anti-aliasing filter, a baseline
offset adjustment circuit, an external trigger circuit, and digital
expansion headers) are also included on the
DDC. Figure~\ref{fig:proto_ddc_and_result} also shows a measurement of
the DDC's baseline distribution with its input terminated to
100~$\Omega$. The measured $\sim$1.4 RMS error in ADC counts is very
close to the manufacturer's $\sim$1 RMS error specification for
typical ADS4149 transition noise.

\begin{figure}
  \centering
  \includegraphics[width=1.0\textwidth]{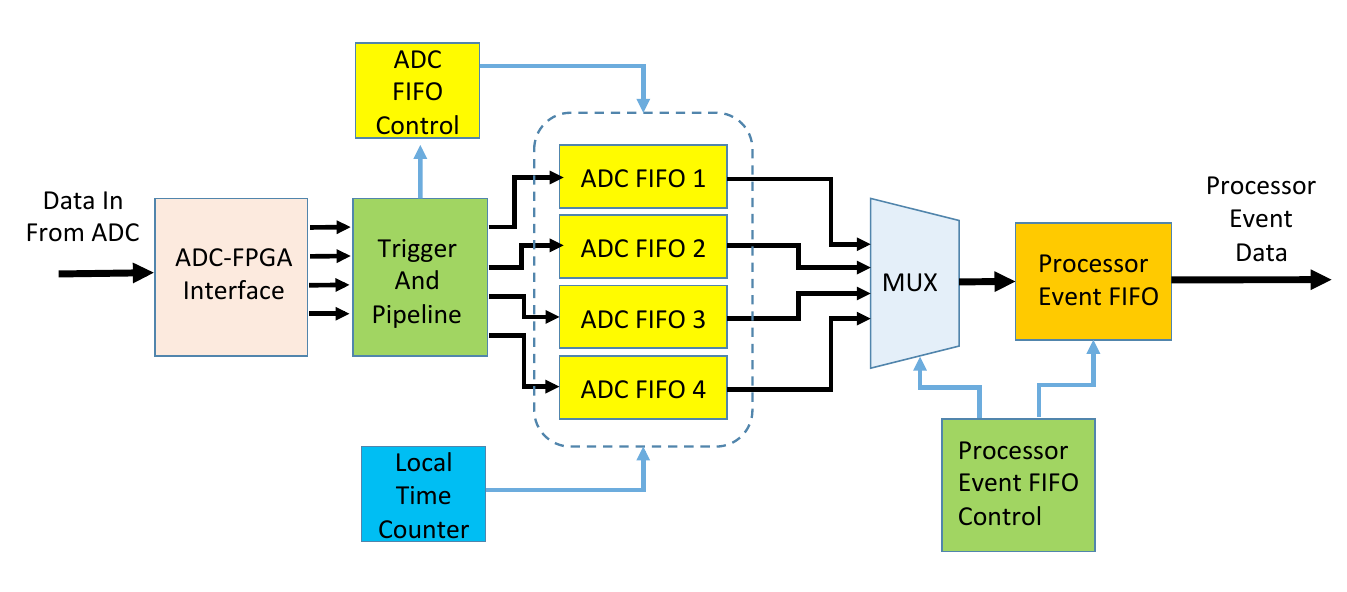}
  \caption{The prototype firmware block diagram.}
  \label{fig:proto_fw_block_diagram_simple}
\end{figure}

The DDC is controlled and read out by the Terasic
Inc. System-on-a-Chip Development Kit (SoCKit). The SoCKit utilizes
Altera's Cyclone V SX-series FPGA, a 110K~logic element FPGA which
includes a built-in ARM Cortex-A9 hard processor. The DDC connects to
the SoCKit via Samtec's High Speed Mezzanine Connector
(HSMC). Figure~\ref{fig:proto_fw_block_diagram_simple} shows a block
diagram representation of the firmware which was developed to read out
the DDC. The ADC-FPGA Interface module divides the incoming Double
Data Rate (DDR) 250~Msps ADC data stream into four parallel Single
Data Rate (SDR) streams, each running at 62.5~Msps. Dividing the data
stream in this way allows the design to easily meet all synchronous
timing requirements, while making only modest tradeoffs in resource
utilization and design complexity. The Trigger and Pipeline module
examines the data stream for the occurrence of any of a number of
user-programmable trigger conditions, signalling the ADC FIFO Control
module to begin filling the ADC FIFOs with event data when the active
trigger condition is satisfied. A 48-bit timestamp generated by the
Local Time Counter module is also stored with each event. Data on the
output side of the ADC FIFOs are sequentially re-ordered by the
Processor Event FIFO Control module and presented to the FPGA's
processor for formatting and transmission to the user.

\begin{figure}
  \centering
  \includegraphics[width=0.9\textwidth]{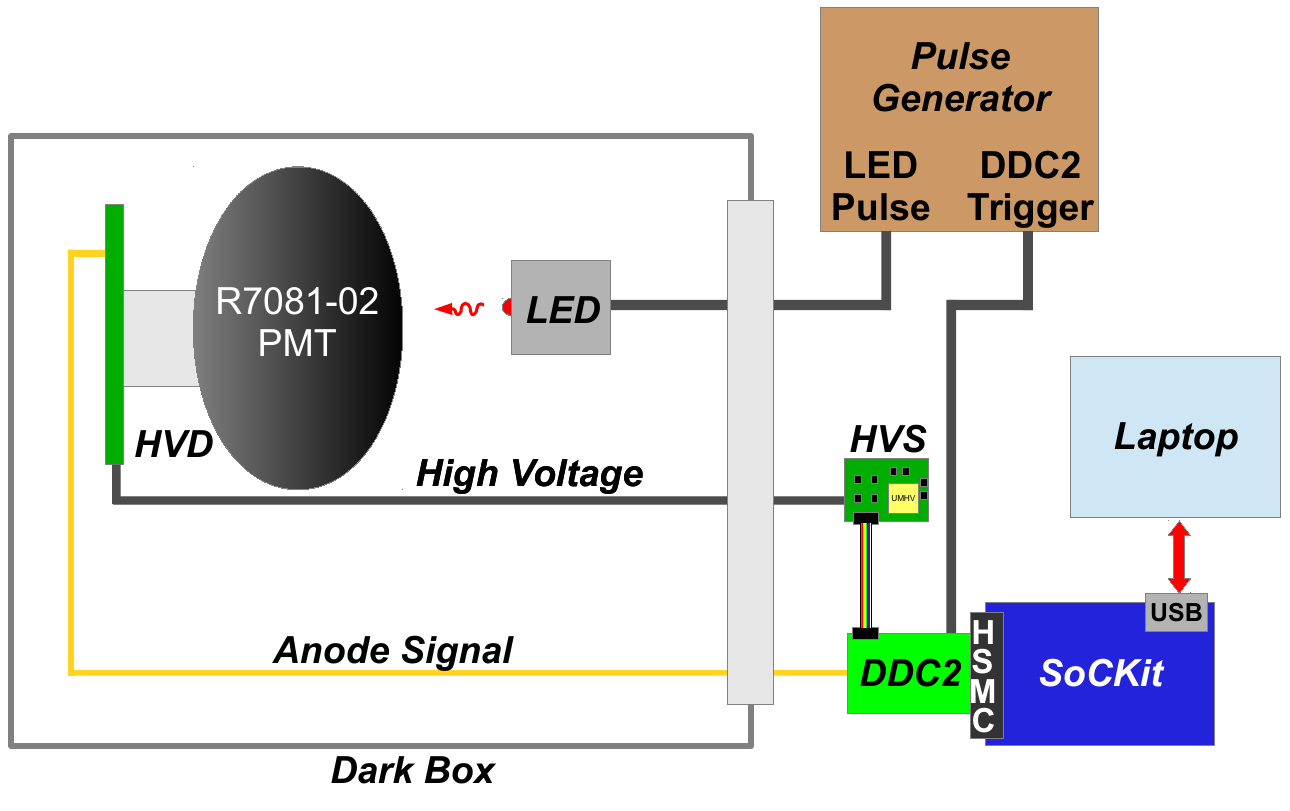}
  \caption{Block diagram of the test setup for the prototype Gen2-DOM
    hardware.}
  \label{fig:proto_gen2dom_test_setup}
\end{figure}

The prototype hardware described above was combined to make an early
test setup, as shown in the block diagram of
Fig.~\ref{fig:proto_gen2dom_test_setup}. This test setup was used in
order to exercise all prototype electronics elements according to their expected
end use, and also to begin to characterize overall system
performance. An IceCube R7081-02 Hamamatsu 10-inch PMT was mounted on
the HVD, and the resulting assembly was placed in a dark box. The
PMT's anode signal was read out by the DDC. High voltage was supplied
to the PMT assembly from the HVS, which was controlled by the SoCKit
with all control signals routed through a DDC expansion header. The
high voltage setting was tuned for a PMT gain of $\sim$10$^7$. Single
photoelectron pulses were stimulated in the PMT by drving an LED with
a pulse generator whose amplitude and width settings were adjusted
such that only $\sim$1/50 pulse generator triggers resulted in a
signal at the PMT's output. A second channel of the pulse generator
was connected to the DDC's external trigger input in order to force
coincidental triggering and readout.

\begin{figure}
  \centering
  \includegraphics[width=1.\textwidth]{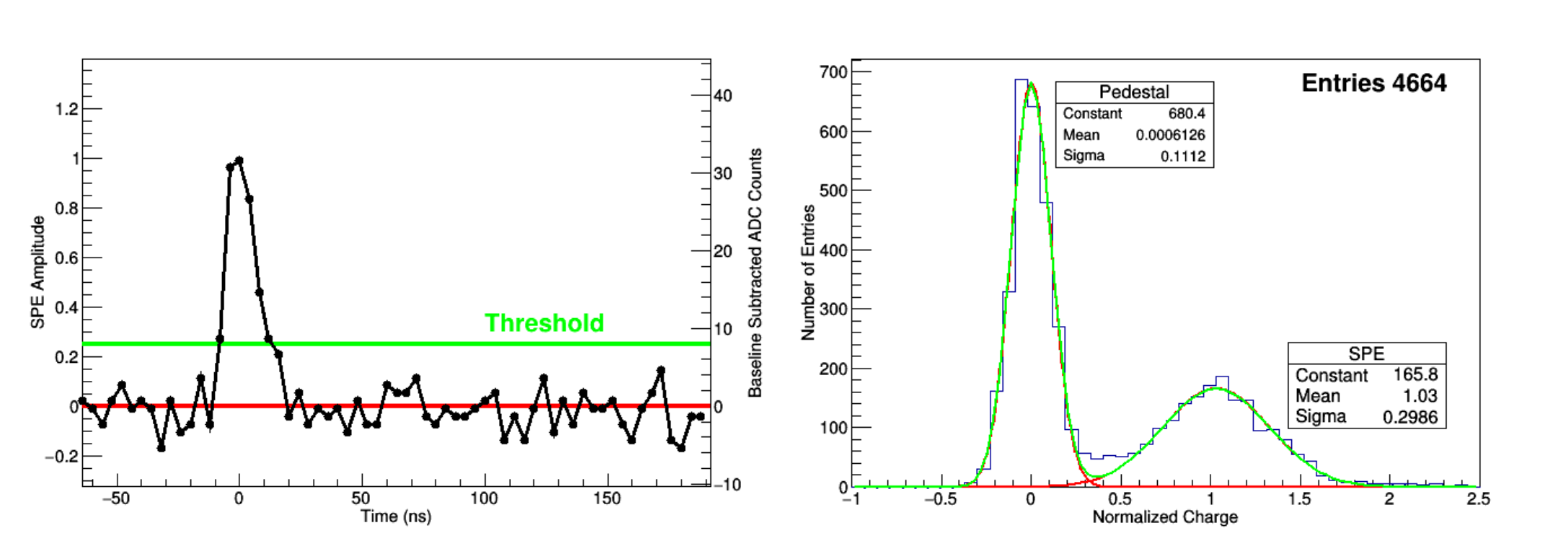}
  \caption{A typical single photoelectron waveform (left) and the
    resulting charge distribution for the run (right).}
  \label{fig:proto_gen2_test_results}
\end{figure}

Figure~\ref{fig:proto_gen2_test_results} shows a typical single
photoelectron waveform and the resulting charge distribution of all
events for which a 0.25~PE amplitude threshold was surpassed. The
typical SPE is seen to be easily distinguishable from the baseline noise,
and the measured charge distribution exhibits a relative width and
peak-to-valley ratio similar to the characteristics of
typical DOMs being used by IceCube.

\IfFileExists{NewCommands.tex}       {}       {}
\IfFileExists{../NewCommands.tex}    {}    {}
\IfFileExists{../../NewCommands.tex} {} {}

\graphicspath{{figures/}{RandD/figures/}}

\subsection{New Photon Detection Technologies}
\label{sec:NewPhotonDetectionTechnologies}

In parallel to the PDOM design described in
Sec.~\ref{sec:OptSensDesign}, alternative optical modules are also
under active development. The focus of this development is to increase
the photon collection area, minimize the noise, and improve the time
and directional resolution of the optical module.  Full detector
simulations for PINGU and other detectors using the new types of
optical modules are currently being developed, focusing on the
following two key areas:

\textbf{Photon Collection Area:} The energy and zenith
resolutions of the detector will improve with the count statistics of
measured photons. To a good approximation, the total number of emitted
photons scales linearly with the neutrino energy. 

\begin{figure}
  \centering
  \includegraphics[width=0.75\textwidth]{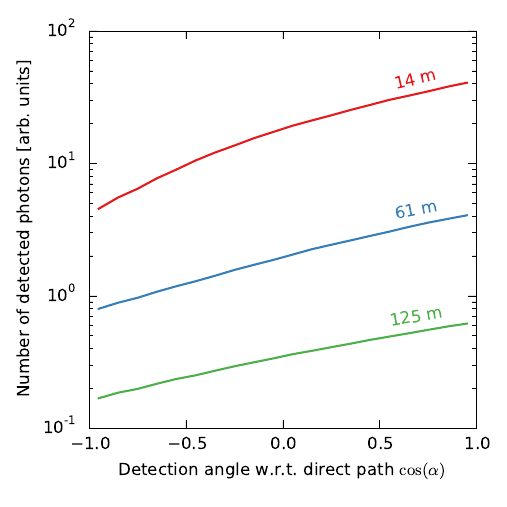}
  \caption{Number of \Cerenkov photons from a bare muon track
    arriving at an optical module with a given angle with respect to
    the direct, unscattered light path for different distances between
    the emitter and optical module.}
  \label{fig:impact_angle}
\end{figure}

\textbf{Directional Sensitivity:} Even though scattering dominates
over absorption in the Antarctic ice, the average scattering angle is
small. The distribution of detected \Cerenkov photons therefore
partially maintains its directional characteristics present at
emission, carrying important information about the direction of
the emitting particle. This is especially true for distances that are
comparable to the scattering length, as in PINGU.
Figure~\ref{fig:impact_angle} shows the number of photons as a
function of the angle between the direct, unscattered Cherenkov path,
and the impact direction on the optical module.  For an
emitter-receiver distance of 14~m, typical for these events, it has
been seen that the signal is dominated by direct light. Thus optical
modules with intrinsic directional resolution can deliver additional
input for event reconstruction. Even at a distance of 125~m some of
the directionality is preserved, in particular when considering the 
arrival time with respect to that of a unscattered photon.

The installation of PINGU may allow for deployment of new photon detection instruments that will improve the overall sensitivity to photons as well as potentially providing additional information about the events themselves.
Taking advantage of the properties of Antarctic ice (low temperature,
low radioactivity, and little absorption above UV wavelengths of
200~nm) the R\&D efforts so far focus on two major directions:

\begin{itemize}
\item multi-PMT optical modules ($e.g.$ mDOMs and D-Eggs, discussed below)
  with increased photodetection sensitivity and directional
  information and
\item low noise, UV-sensitive, wavelength-shifting modules ($e.g.$ WOMs, also
  discussed below) with large area but relaxed timing requirements.
\end{itemize}

While both approaches provide an enhanced photodetection sensitivity,
the directional information and precise timing of the mDOM and D-Egg
may prove particularly beneficial for accurate event reconstruction. The mDOM in particular may provide strong advantages in the
PINGU energy regime.  The low noise and UV-sensitivity of the WOM
approach mostly promises a reduced energy threshold and improved
energy resolution.  Beyond PINGU, these technologies may eventually be
combined to exploit their complementarity.

\IfFileExists{NewCommands.tex}       {}       {}
\IfFileExists{../NewCommands.tex}    {}    {}
\IfFileExists{../../NewCommands.tex} {} {}

\graphicspath{{figures/}{RandD/mDOMs/figures/}}

\subsubsection{Multi-PMT Optical Modules (mDOMs)}
\label{sec:mDOMs}

\begin{figure}
\centering
\includegraphics[width=0.95\linewidth]{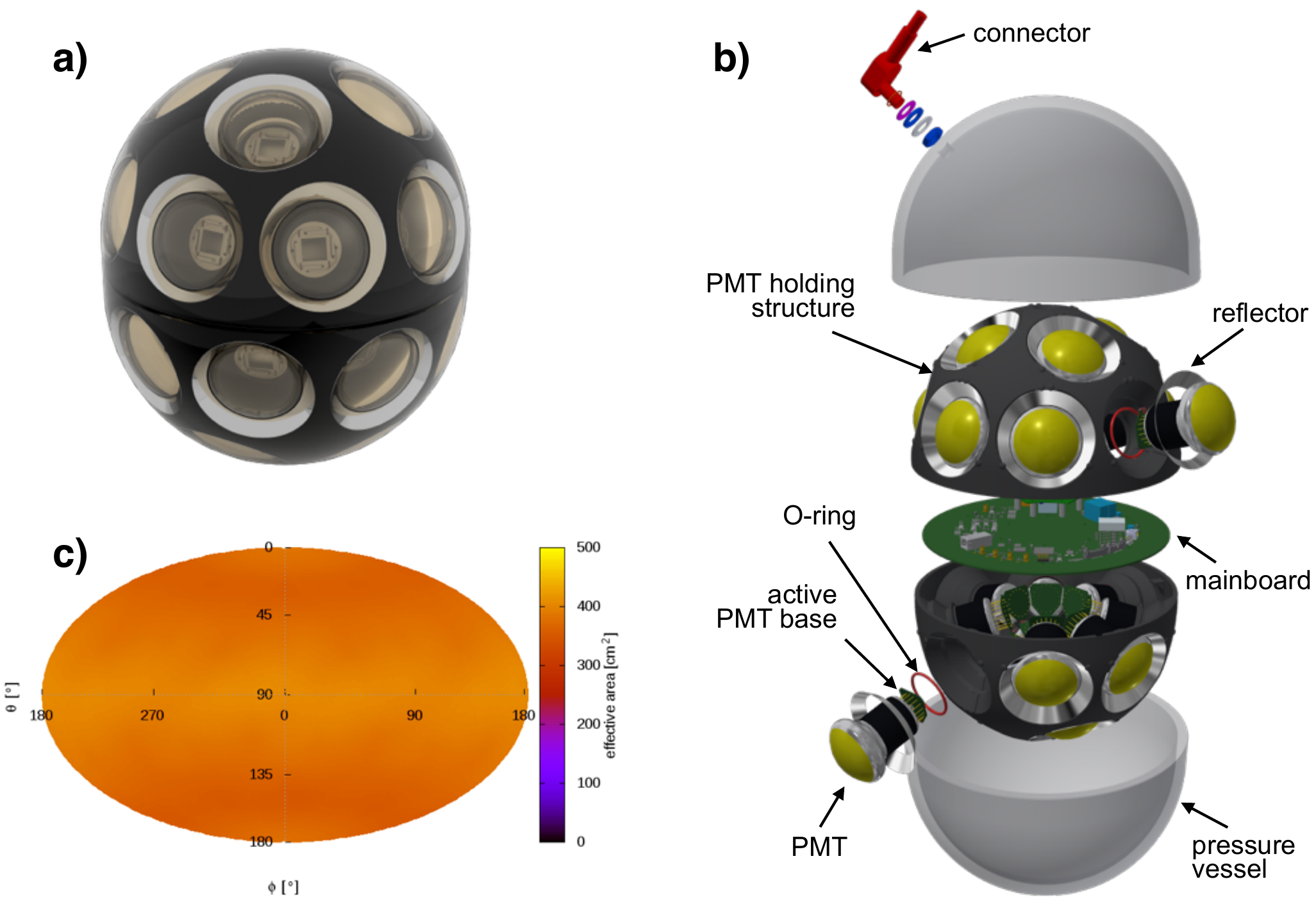}
\caption{ Rendering (a) and exploded view (b) of current mDOM design. (c) mDOM effective area for a plane wave
  front as a function of its zenith and azimuth angle simulated at $470$\,nm, quantum efficiency not included.}
\label{fig:mDOM}
\end{figure}

As discussed previously, the IceCube neutrino telescope uses optical
modules consisting of a spherical glass pressure vessel containing a
single large-diameter PMT, as well as front-end and digitization
electronics. In contrast, multi-PMT optical
modules~\cite{Adrian-Martinez:2014vja}, such as those developed for
the KM3NeT neutrino telescope in the Mediterranean Sea, house arrays
of many small PMTs and their read-out electronics. Such modules
offer a number of attractive advantages compared to the PDOM:
\begin{itemize}
\item increased photocathode area with almost uniform 4$\pi$
  acceptance,
\item improved event reconstruction and background suppression due to
  directional sensitivity,
\item improved photon counting due to segmentation of the photocathode
  area,
\item local coincidence recognition for improved signal selection and
  background suppression, and
\item superior timing without the need for magnetic shielding.
\end{itemize}
We adapted the KM3NeT multi-PMT concept by incorporating 24 three-inch
PMTs in a 14-inch spherical glass housing with a short cylindrical
extension suitable for deployment in holes achievable with the IceCube
drill equipment (Fig.~\ref{fig:mDOM}a,b). The goal is to achieve an overall price per cathode
area at least as low as in conventional optical modules. The status of
the R\&D and prototyping efforts is discussed below.

\noindent {\bf Housing:} Unlike full glass spheres, spherical pressure
vessels with a cylindrical extension, required by the mDOM to
accommodate the main electronics board, are not commercially available
in the appropriate form factor and pressure rating. Cooperation with a
glass vessel manufacturer (Nautilus GmbH) has resulted in a technical
solution which, according to the manufacturer specifications, fulfills
the requirements. The pressure vessel consists of two glass
half-spheres which are extended at their equator with a cylinder of
$27.5$~mm height (see Fig.~\ref{fig:mDOM}b). The two segments
will be joined centrally using beveled glass surfaces and application
of low pressure. Each hemisphere is able to house 12 three-inch PMTs
providing an almost uniform $4\pi$ sensitivity in the full module (see
Fig.~\ref{fig:mDOM}c).

\noindent {\bf PMTs:} At the time of this writing, three companies
(Hamamatsu, ET Enterprises, HZC) provide three-inch PMT prototypes
that have been optimized for KM3NeT, including length ($<
12$~cm), and transit time spread (RMS $< 2$~ns). Their
mushroom-shaped entrance windows offer superior timing and 10 dynodes
provide the required gain. For the first mDOM prototype, the Hamamatsu
R12199-02 model has been selected due to its better form factor with
respect to the dimensions of the pressure vessel.

\noindent {\bf Internal PMT Support structure:} The PMTs (and possibly
calibration devices) will be suspended using a support structure that
will also accommodate reflector rings around each PMT. The reflector
rings increase the sensitivity of the PMTs in the central
$\pm50^\circ$ direction by about 30\%, thereby enhancing the
directional sensitivity of the module significantly. The support
structure has to provide temperature resistance, limited shrinkage due
to pressure and temperature, negligible chemical reactivity with PMT
glass and optical silicon gel, low intrinsic radioactivity, and
durability throughout the lifetime of the experiment. The first
prototypes of such a holding structure (see Fig.~\ref{fig:mDOM}b) 
have been produced by rapid prototyping (``three-dimensional
printing'') by an external company. For mass production, alternative,
more cost-effective production methods are being investigated.

\noindent {\bf Readout electronics:} The front-end readout electronics
will be based on an extended KM3NeT layout. The high voltage is
generated on the PMT base with Cockcroft-Walton circuitry designed by
Nikhef \cite{Timmer:2010zz}. For the digitization of the analogue waveform produced by the
PMT, two concepts are currently under investigation. In the baseline
concept, the waveform is split after amplification and passed to four
discriminators with different thresholds. The time-over-threshold
signal from each discriminator is routed to an FPGA on the main board
where the leading and trailing edge times for each channel are
determined. In a more ambitious approach, the waveform is fed into a
custom Application-Specific Integrated Circuit (ASIC) currently under
development, which incorporates 63 thresholds. The ASIC contains a
$2^N - 1$ to $N$ encoder which reduces the data to a 6-bit output
word. Processing of the output corresponds to that of the baseline
concept. In the ASIC scheme, the larger number of thresholds allows
for a detailed sampling of even complex waveforms. The target for the
total energy consumption of the mDOM is $<3$\,W.

\noindent {\bf Current project status:} The design of the pressure
vessels has been finalized. First prototypes have been delivered and
will undergo pressure tests. Three-inch PMT prototypes from Hamamatsu
are on hand and will be incorporated in the first mDOM prototype. A
climate chamber is available to measure properties crucial for the use
in deep ice, such as the dark rate, under realistic conditions (down
to $-50^\circ$C). A first design of the PMT support structure has been
developed and a prototype manufactured. Work on the design of the PMT
base incorporating the pre-amplifier and time-over-threshold circuitry
(both discrete and ASIC) are progressing. Simulations are currently
underway to address the optical properties of the module including the
signature of optical background from radioactive processes in the
glass and the impact of the utilization of mDOMs on the physics
performance of PINGU.

\IfFileExists{NewCommands.tex}       {}       {}
\IfFileExists{../NewCommands.tex}    {}    {}
\IfFileExists{../../NewCommands.tex} {} {}

\graphicspath{{figures/}{RandD/D-Egg/figures/}}

\subsubsection{Dual optical sensors in an Ellipsoid Glass (D-Egg)}
\label{sec:d-egg}

The existing IceCube DOMs have a proven record of very high
stability. Each DOM carries a 10-inch PMT (Hamamatsu R7081-2) which
looks downwards into the ice. The D-Egg design is based on the
successful technology of the current IceCube DOM but housing two
8-inch PMTs (Hamamatsu R5912-100) with one looking up and one looking
down. The improved aspects of this setup will be detailed here.

\begin{figure}
  \centering
  \includegraphics[height=4.6cm]{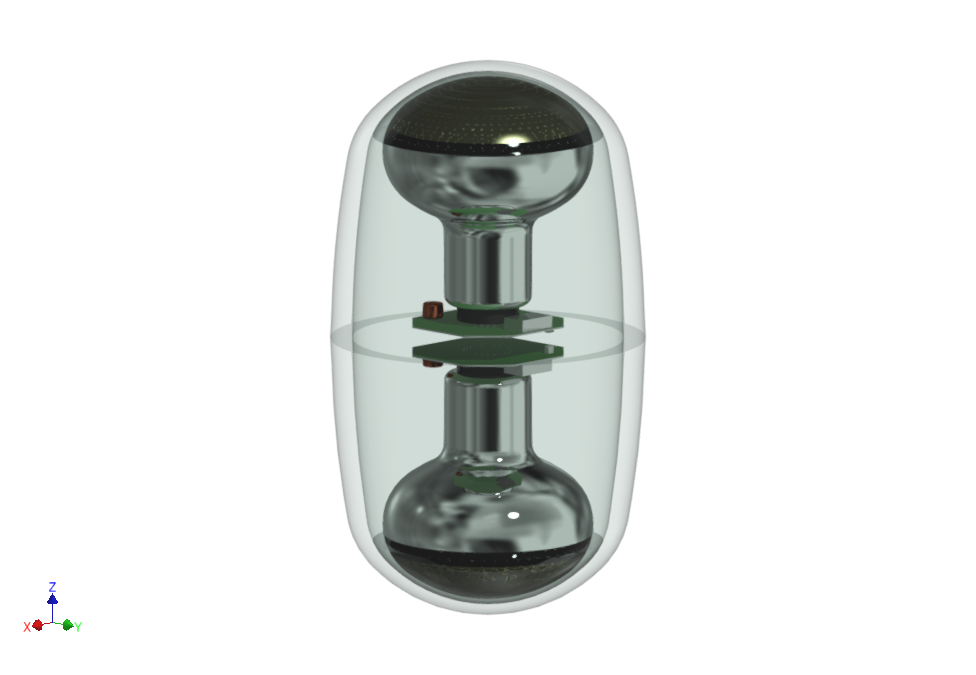}
  \caption{ The D-Egg schematic: the height of the glass vessel is
    $534$~mm and its diameter at the equator is $300$~mm.}
  \label{fig:DEgg}
\end{figure}

\noindent \textbf{Housing:} The glass housing of the D-Egg has been
optimized to ensure the best optical performance while preserving the
mechanical strength necessary to withstand the pressure stresses to
which it will be exposed.  The mechanical strength of the glass
housing has been tested up to 70~MPa, which is
higher than the greatest pressure observed during the freeze-in
process of the current IceCube strings.

The thickness of the glass varies with latitude, being thicker at the
housing's equator and thinner at its poles. The diameter of the D-Egg
at the equator is $300$~mm, which is $30.2$~mm less than the
current IceCube DOM. This will reduce the drilling costs by allowing for
holes with smaller diameter (and therefore less fuel). Figure~\ref{fig:DEgg} shows
the existing components of the D-Egg. The electronics and digital
outputs are under development and are based on modifications of the PDOM
designs.

\begin{figure}
  \centering
  \includegraphics[height=6cm]{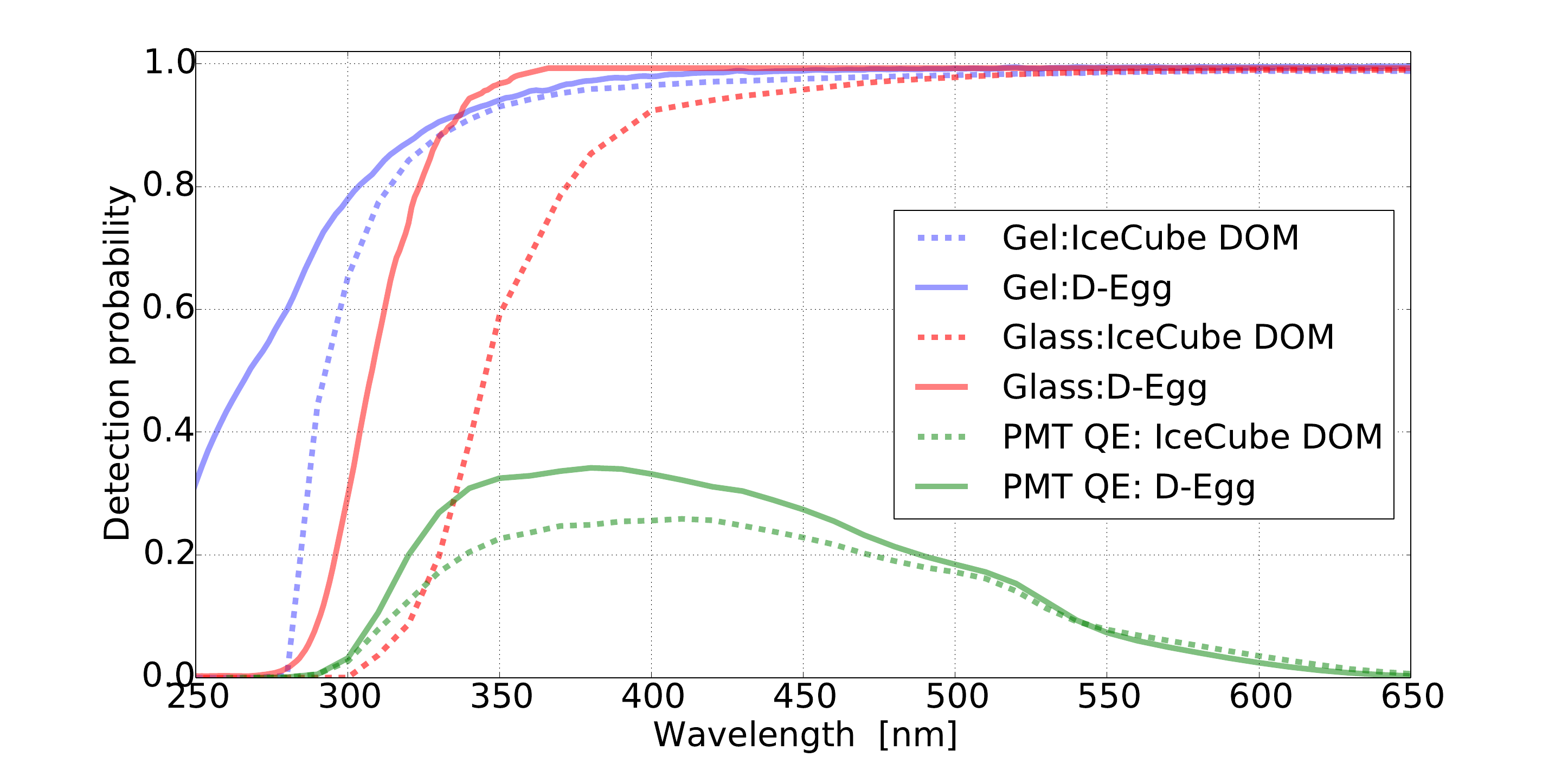}
  \caption{The photon detection efficiency is determined by the glass,
    gel, and PMT. The transmittance of the glass and gel and the QE of
    the PMTs for the D-Egg and IceCube DOMs are shown as a function of
    wavelength.}
  \label{fig:PhotonEfficiency}
\end{figure}

\noindent \textbf{Improved sensitivity to UV photons:} The
transmittance of the glass for UV photons has been improved compared
to the current IceCube glass. For instance $40\%$ of $350$~nm photons
are absorbed by the current IceCube glass while over $95\%$ of
vertical photons pass through the D-Egg glass at the pole. A
comparison of the transmittance of the glass and the gel is shown in
Fig.~\ref{fig:PhotonEfficiency}.

\begin{figure}
  \centering
  \includegraphics[height=6cm]{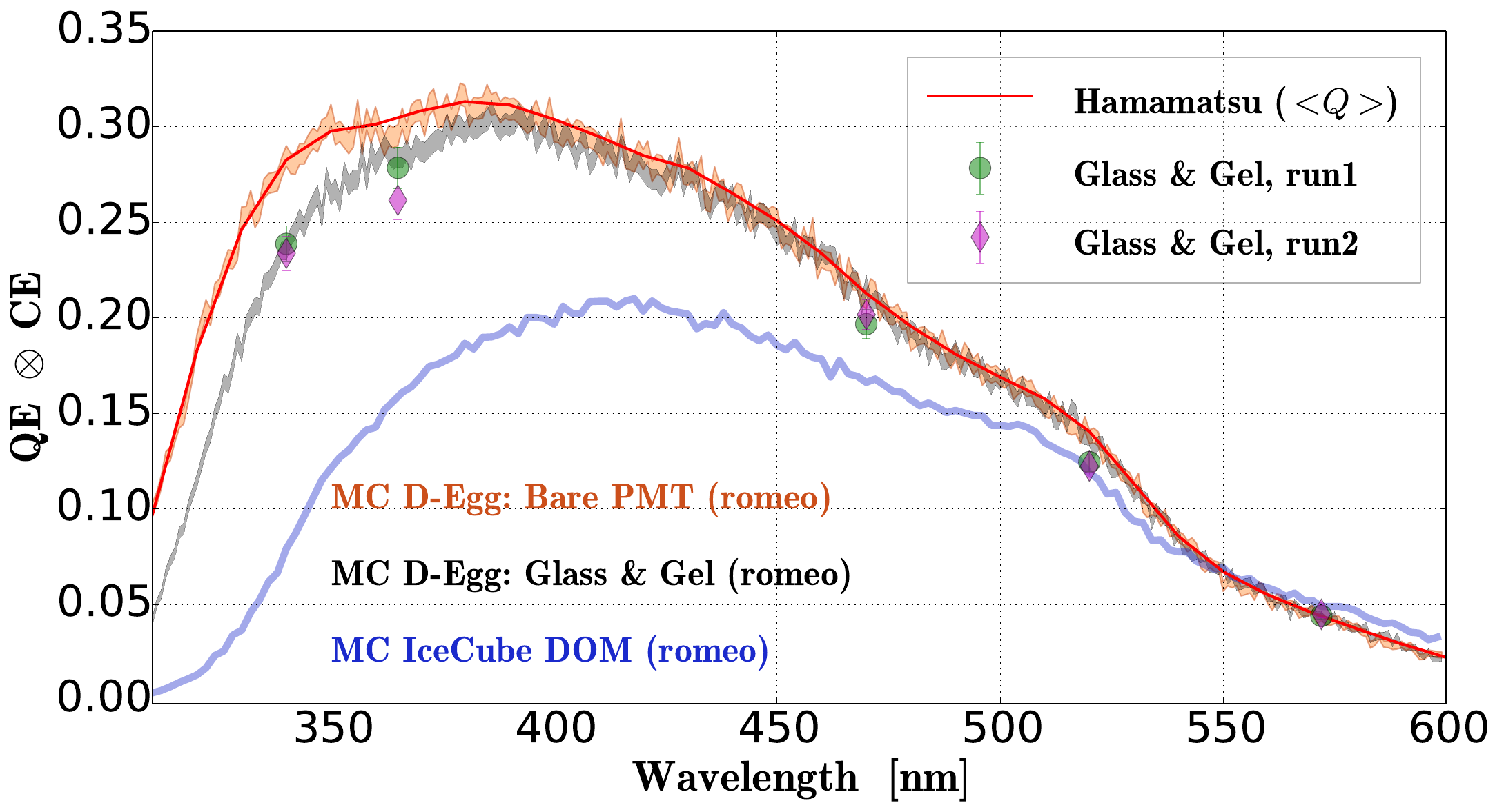}
  \caption{Comparison between the quantum efficiency measured in the
    lab and generated in the simulations, based on GEANT4 and Romeo.  The bare-PMT
    simulation is compared to Hamamatsu measurements which have been
    corrected for the charge response. The blue curve shows the
    simulated efficiency when photons hit the centre of the
    photocathode of an IceCube DOM.}
  \label{fig:QEMeasurements}
\end{figure}

The quantum efficiency (QE) of the PMT has also been improved and is
shown together with the transmittance of the glass in
Fig.~\ref{fig:PhotonEfficiency}. At $\lambda = 380$~nm, the QE of the
8-inch D-Egg PMT is increased from $25\%$ to $\sim 33\%$ compared to
a standard IceCube PMT. It can be inferred that the acceptance of UV
photons at $\lambda = 320$~nm is limited by the quantum efficiency of
the PMTs. The photon detection probability of the D-Egg (with glass
and gel) has been measured in a freezer with five LEDs of wavelengths
$340$, $365$, $470$, $520$, and $572$~nm. The result is compared to
GEANT4 simulations which take into account the absorption of the glass
and the gel, the geometry of D-Egg and the properties of the PMTs
(Fig.~\ref{fig:QEMeasurements}).

The QE measured by Hamamatsu has been corrected by the charge response
of $\sim 90\%$ and compared to simulations of the bare PMT. More
measurements are planned to verify the simulations, especially at UV
wavelengths and at different photon-incident angles. Compared to
the current IceCube DOM, the detection efficiency of photons at
$350$~nm has been increased by almost a factor of two.

\noindent \textbf{4$\pi$-coverage of acceptance:} The D-Egg has two
PMTs covering $4\pi$ solid angle of the detection volume. This is
important in the case of vetoing down-going muons. It also provides
directional information of the incoming photons and potentially
could help reconstructing events which are cascade-like. Detailed
studies are in progress.

\begin{figure}
  \begin{subfigure}[]
    {
      \includegraphics[trim={1cm 0 0.cm 0},clip,height=5.2cm]{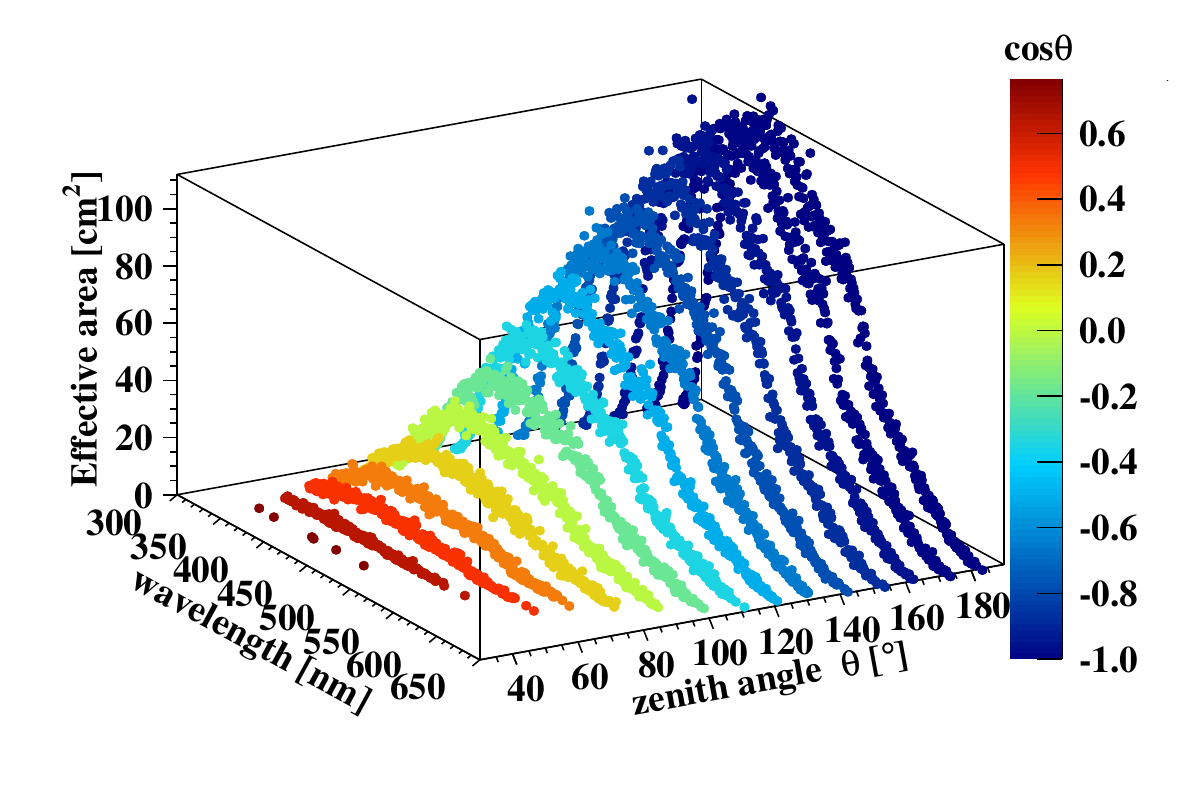}
      \label{fig:EffectiveAreaSphereDOM}
    }
  \end{subfigure}
  \hfill
  \begin{subfigure}[]
    {
      \includegraphics[trim={1cm 0 0.cm 0},clip,height=5.2cm]{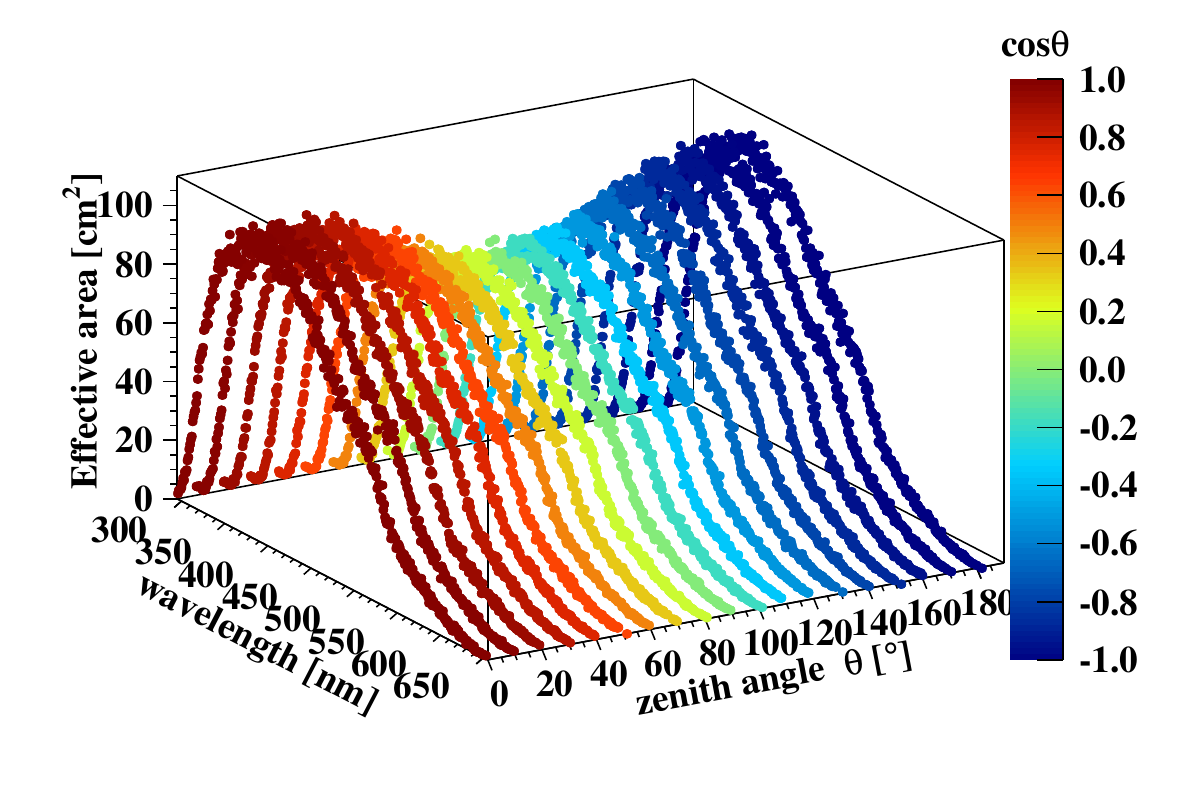}
      \label{fig:EffectiveAreaDEgg}
    }
  \end{subfigure}
  \caption{The effective area of the current IceCube DOM (a)
    and the D-Egg (b). The optical properties of the glass and gel have been
    implemented in GEANT4 simulations. The PMT simulations include the
    charge response correction and the quantum and collection efficiency.
    Each point scales with the probability of a photon of a certain
    wavelength and zenith angle being converted to a photoelectron.}
  \label{fig:EffectiveArea}
\end{figure}

The geometric and optical properties of the D-Egg have been implemented in
GEANT4~\cite{Agostinelli:2002hh}. Photons of wavelengths $300$~nm to
$650$~nm are simulated in circular beams with a diameter of $0.534$~m,
which is the height of the D-Egg. The quantum efficiency of the PMTs has
also been included in the simulation. Photon beams are then
injected and rotated from the bottom to the top of the optical module.
The effective areas of the D-Egg and the IceCube DOM at different
wavelengths and photon angles are shown in
Fig.~\ref{fig:EffectiveArea}.

The IceCube DOM has greater sensitivity to photons arriving from below
while the D-Egg has nearly isotropic sensitivity to photons from all
angles. At $340$~nm, the effective area of the D-Egg integrated over
all angles is $\sim 4$ times the effective area of the IceCube DOM.

The assembly of a complete D-Egg is expected in 2017
including the final version of the HV divider board and signal readouts.
The performance of 8-inch PMTs with transformer-based or capacitor-based
HV boards is being tested in a freezer at $-45^{\circ}$C.

\IfFileExists{NewCommands.tex}       {}       {}
\IfFileExists{../NewCommands.tex}    {}    {}
\IfFileExists{../../NewCommands.tex} {} {}

\graphicspath{{figures/}{RandD/WOMs/figures/}}

\subsubsection{Wavelength-Shifting Optical Modules (WOMs)}
\label{sec:WOMs}

One way to improve the photo-collection efficiency of the
photodetectors is through the use of wavelength-shifting and
light-guiding materials to concentrate the light from a large area
onto one or a few PMTs with small photocathode area. The goal of this
R\&D project is to provide modules with the following properties:
\begin{itemize}
\item very low noise rates on the order of 10 Hz,
\item high UV sensitivity,
\item large geometric acceptance and module sensitivity,
\item long term stability,
\item no necessity for magnetic shielding, and
\item adequate timing resolution.
\end{itemize}

\begin{figure} [h!]
   \begin{center}
      \includegraphics[height=5.0cm, angle=0]{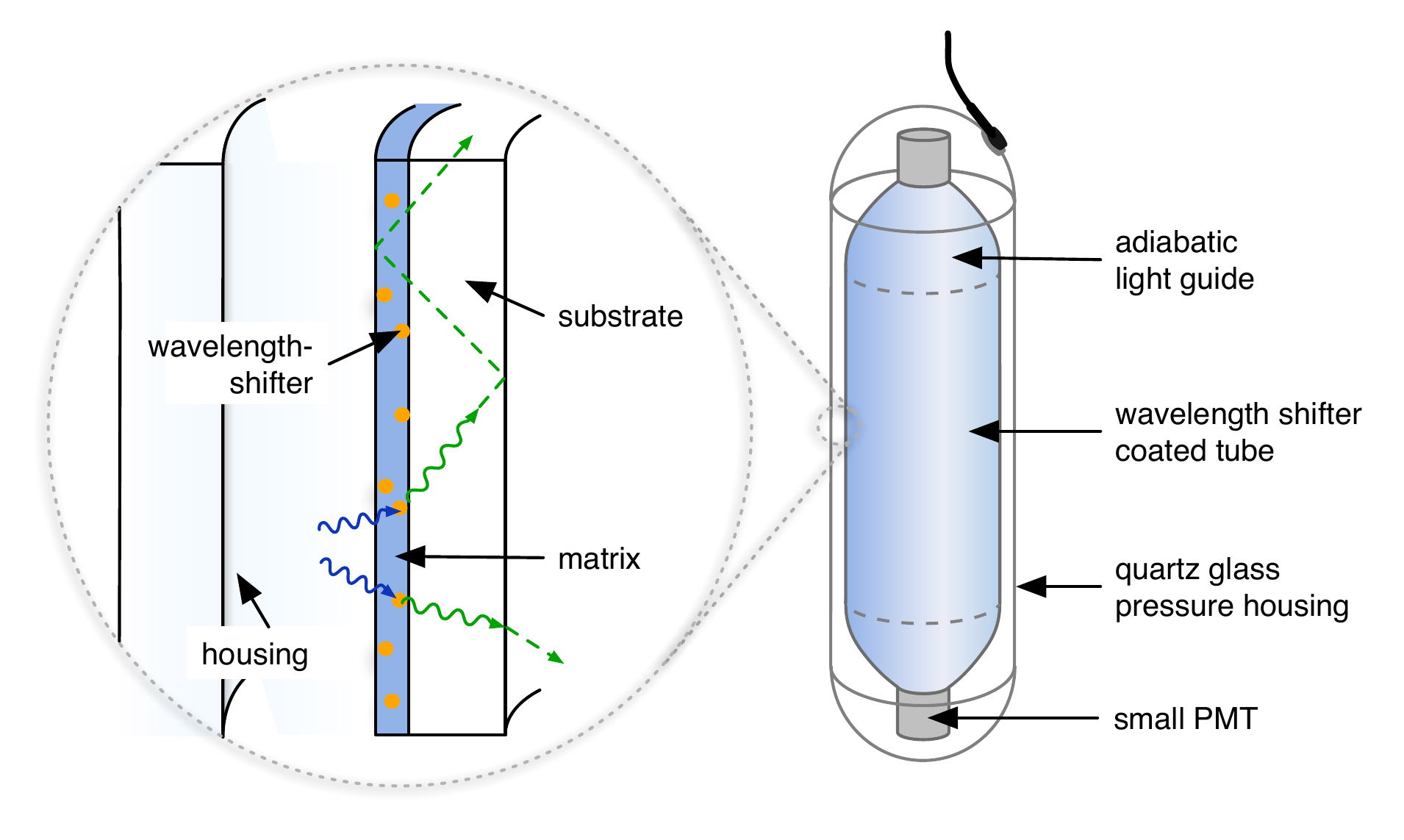}
      \includegraphics[height=5.0cm, angle=0]{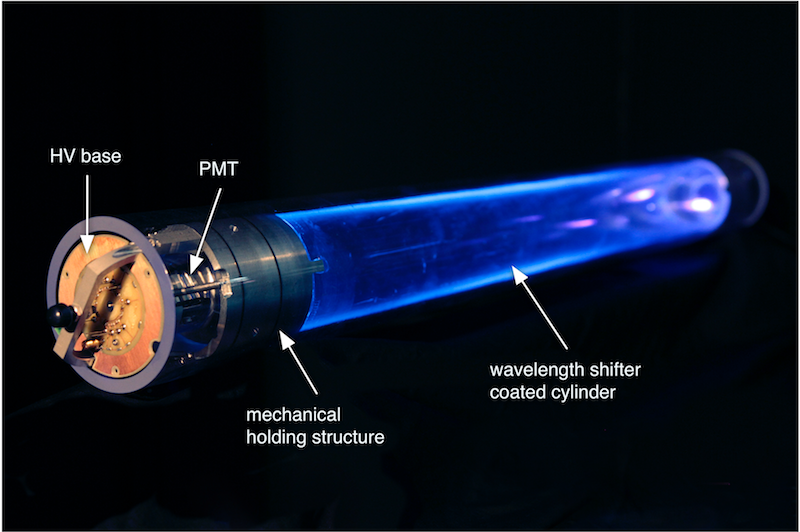}
   \end{center}
   \caption{Schematic drawing of the Wavelength-shifting Optical
   Module (left) and a prototype illuminated under UV light (right). Converted light
   guided in the cylinder is focused onto the PMT by an adiabatic light guide
 (within the holding structure).}
   \label{fig:WOM}
\end{figure}

The essential idea of the WOM is to increase the sensitive area of a
PMT by using passive components that act as light collectors and
concentrators, as illustrated in Fig.~\ref{fig:WOM}. The use of cigar-shaped
modules maximizes the photon collection area that can be deployed, for modules
produced at a price per photosensitive area that is lower than that of standard
optical modules. The dimensions are mainly limited by the requirement to safely
withstand high pressure and be handled easily. With a larger photo-detection
efficiency, the resulting geometry can provide a more elongated module that has
a significantly smaller diameter, allowing a decrease in the hole size and thus
yielding another significant cost reduction.  Except for the readout
electronics, all components for a first prototype have
been produced and are currently evaluated with the results discussed below.

\noindent {\bf Housing:} In order to provide sensitivity at
wavelengths below 300~nm and suppress noise arising from radioactivity
in the glass, quartz tubes of suitable dimensions have been
investigated.  An R\&D project with the glass vessel manufacturer
Nautilus GmbH resulted in a technical solution designed to withstand
pressures of 68~MPa, significantly larger than what is expected during
the freeze-in process.  The final design specs of 9~mm thickness,
114~mm diameter, and 1200~mm length (with hemispherical borosilicate
end caps) are a compromise with weight, achievable dimensional
accuracy and cost taken into consideration.

In total six modules were fabricated. One module has been successfully
pressure-cycled multiple times to a pressure above 34~MPa. The optical
transparency was measured on smaller samples from the same production
batch and shows excellent transmittivity for wavelengths larger than
290~nm, and $\gtrsim20\%$ transmittivity down to a wavelength of
215~nm.

\noindent {\bf Wavelength-shifting cylinder:} A wavelength shifting
and light-guiding cylindrical tube, with a diameter to allow for only
a small air gap between the tube and the pressure vessel, provides the
optimal geometry for light collection and concentration. In order to enhance the
performance towards the UV, transparent tubes are coated with a thin
film containing fluorescent dye. \Cerenkov photons, peaking in the UV,
are absorbed and re-emitted isotropically in the dye at a larger wavelength. Of
these photons, a large fraction (up to 75\%) are guided towards the ends via 
total internal reflection. The dye is applied as a thin (few $\mu{}m$) film
on cast PMMA tubes through a dip-coating technique, resulting in high
light yield, durability and long-term stability.

\noindent {\bf PMT and readout:} On both ends of the tube, the light
is guided to the spherical surface of a PMT through adiabatic light
guides. In principle, the minimum diameter of the PMT is determined by the wall thickness
of the light-guiding tube, however the selection is limited by the availability
of small PMTs with sufficient gain to detect single photo-electrons. For the
current prototype, we use the same 3~inch Hamamatsu PMT
(R12199-02) with a peak quantum efficiency of ~25\% at 390~nm that is
evaluated for the mDOM concept (c.f.~\S\ref{sec:mDOMs}). The dark-noise rate
of this PMT has been measured to be $\lesssim 30$~Hz over the full temperature
range of interest from $-50^\circ$C to  $-10^\circ$C. The readout electronics
for the PMT will be based on those developed for the PDOM
(c.f.~\S\ref{sec:OptSensDesign}), but altered to match the smaller dimensions
of the housing.

\begin{figure} [h!]
   \begin{center}
     \includegraphics[height=6.7cm, angle=0]{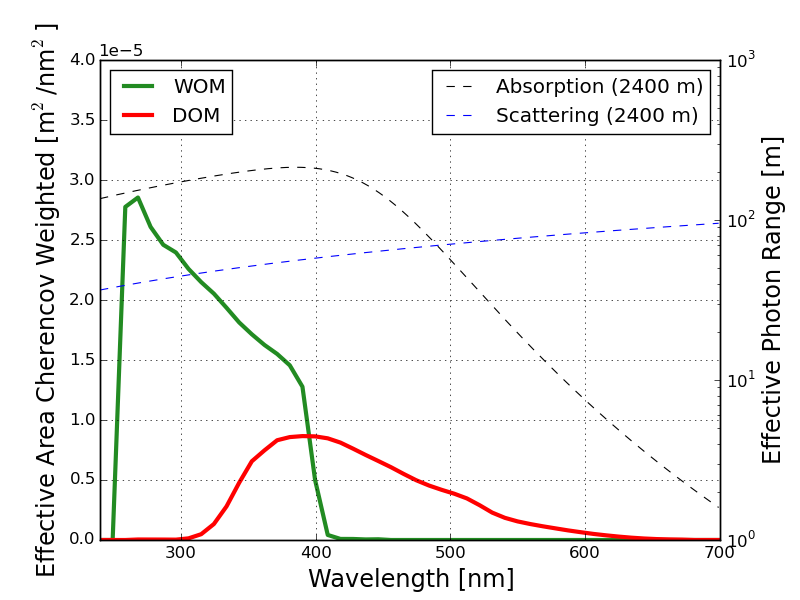}
     \includegraphics[height=6.5cm, angle=0]{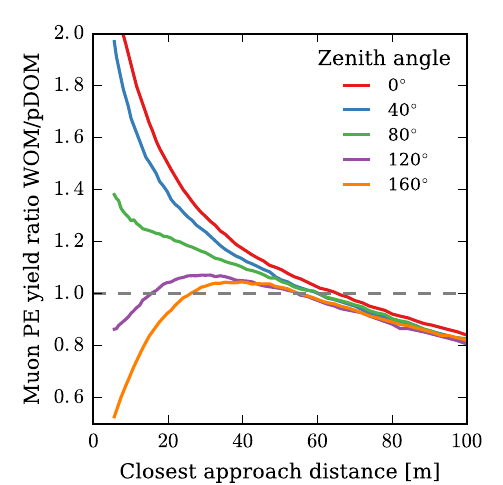}
   \end{center}
   \caption{Left: \Cerenkov spectrum weighted effective photo-detecion area as a
     function of incident wavelength for the Wavelength-shifting
     Optical Module (WOM) and an IceCube high-QE module (DOM). Absorption and
     scattering length in the ice a depth of 2400m are shown for reference.
     Right: Average relative photoelectron yield from minimum ionizing muons compared to the
     PDOM as a function of distance and zenith angle.} \label{fig:WOM-efficiency}
\end{figure}

\noindent {\bf System performance:}
Different wavelength-shifting materials have been tested in lab
measurements as candidates for use in such a sensor, and their photon
capture efficiency (including conversion and light-guiding) has been
measured. Best results have been obtained with a dye that is made from a mixture
of different wavelength shifters in the same polymer matrix. For the prototype,
we use a PEMA-matrix (Paraloid B72) with an admixture of 0.6$\mathrm\%_w$ of
1,4-Bis(2-methylstyryl)benzol (Bis-MSB) and 1.2$\mathrm\%_w$ p-Terphenyl (pT) as
active ingredients. Photon capture and transport efficiencies well above
20\% and as high as 50\% have been achieved in this configuration, depending on
the dye concentration, surface quality of the coating and other factors.
Figure~\ref{fig:WOM-efficiency} shows the \Cerenkov spectrum-weighted
effective photosensitive area of the prototype module, including the
transparency of the pressure housing, the measured capture and transport
efficiency and the quantum efficiency, of the currently used PMT. In particular
for short wavelengths, the performance of the prototype built with readily
available technology exceeds that of modules currently used in IceCube. Stronger
absorption and scattering of the UV-light reduces the effect, resulting in the
most notable gain for close-by horizontal events. For a geometry with a 24~m
string spacing, the average distance to the closest sensor is 7.2~m.

With the current design already outperforming the existing modules, in
particular for dense sensor geometries, future improvements are to be expected
by increasing the light capture and transport efficiency and a more optimal
choice of PMT.

\subsection{Cable}

Sets of PINGU sensors will be deployed along a single cable similar to
those used for IceCube.  These cables consist of multiple twisted wire
quads, each of which serves power and communications for several sensors,
a polymer core for structural support, and a robust outer sheath.  The
PINGU sensor spacing motivates some changes to the cable attachment
configuration, but the electrical and mechanical specifications will
be essentially the same as for previous IceCube cables.  This allows
reuse of a proven engineering solution, leveraging the IceCube project
experience with procurement and verification of this critical
component.

\begin{figure}[htbp]
\begin{center}
\includegraphics[width=5in]{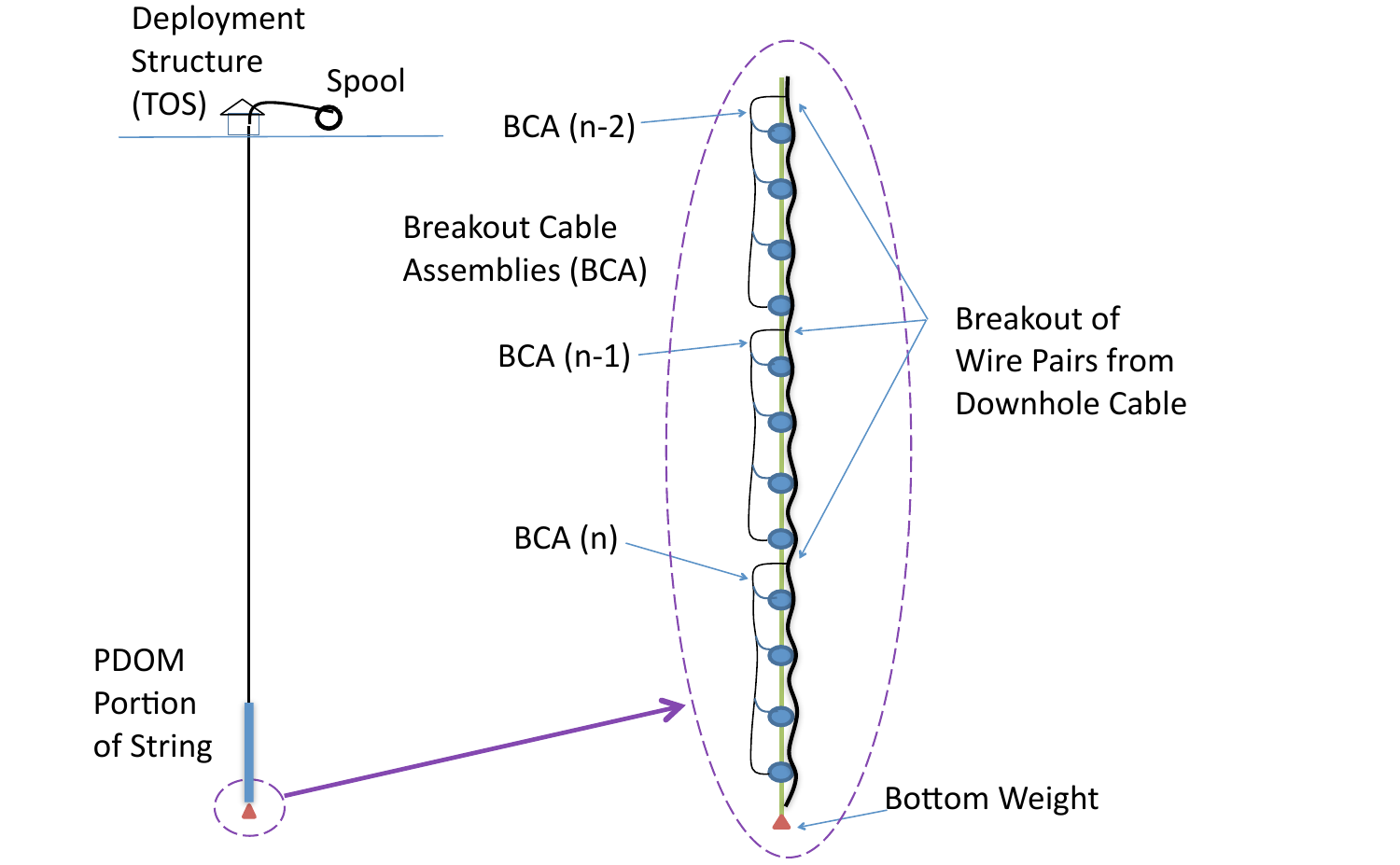}
\vskip0.5in
\includegraphics[width=5in]{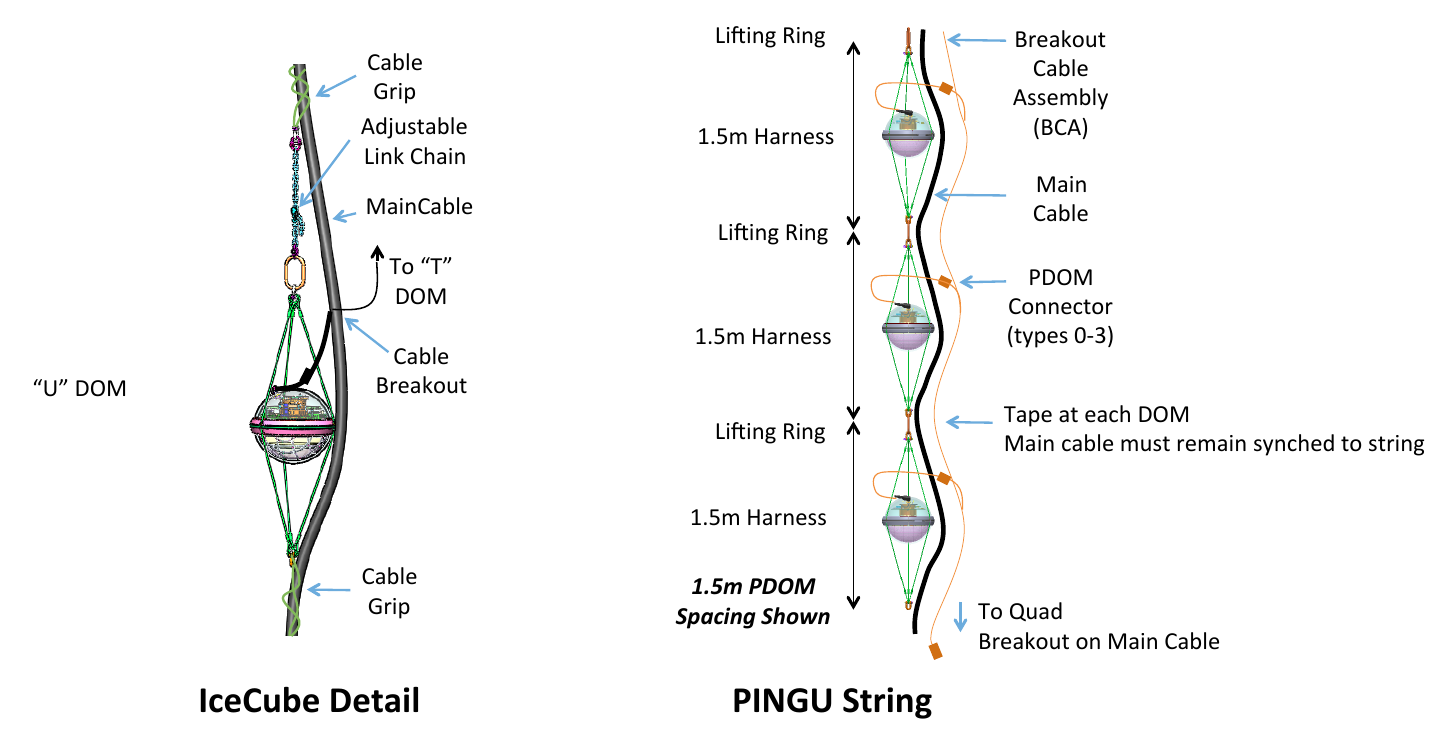}
\caption{Attachment and support of PDOMs along a PINGU string.  The
  Seacon XSJJ connectors are of the same type used in IceCube.}
\label{fig:StringCable}
\end{center}
\end{figure}

Figure~\ref{fig:StringCable} shows how PDOM sensors could be attached
to the down-hole cable.  The attachment of other sensor designs, such
as the mDOM, would be similar.  The cable itself carries the load from
the top of the PDOM chain, 2150~m up to the surface of the ice.  In
the instrumented depth range, the load is carried alternately by short
segments of steel wire rope and the PDOM harnesses, with electrical
cables secured alongside.  The main load is transferred from the
down-hole cable to the steel rope by grips above the topmost PDOM and
at the bottom of the chain.  These cable grips are like those used in
IceCube above and below every DOM, but the closer PINGU module spacing
allows most of them to be replaced by the simpler and lower cost steel
rope hardware.  The strings are connected to the data acquisition
computers via surface cables that run to the IceCube laboratory
building.

\begin{figure}[h!]
\begin{center}
\includegraphics[width=5in]{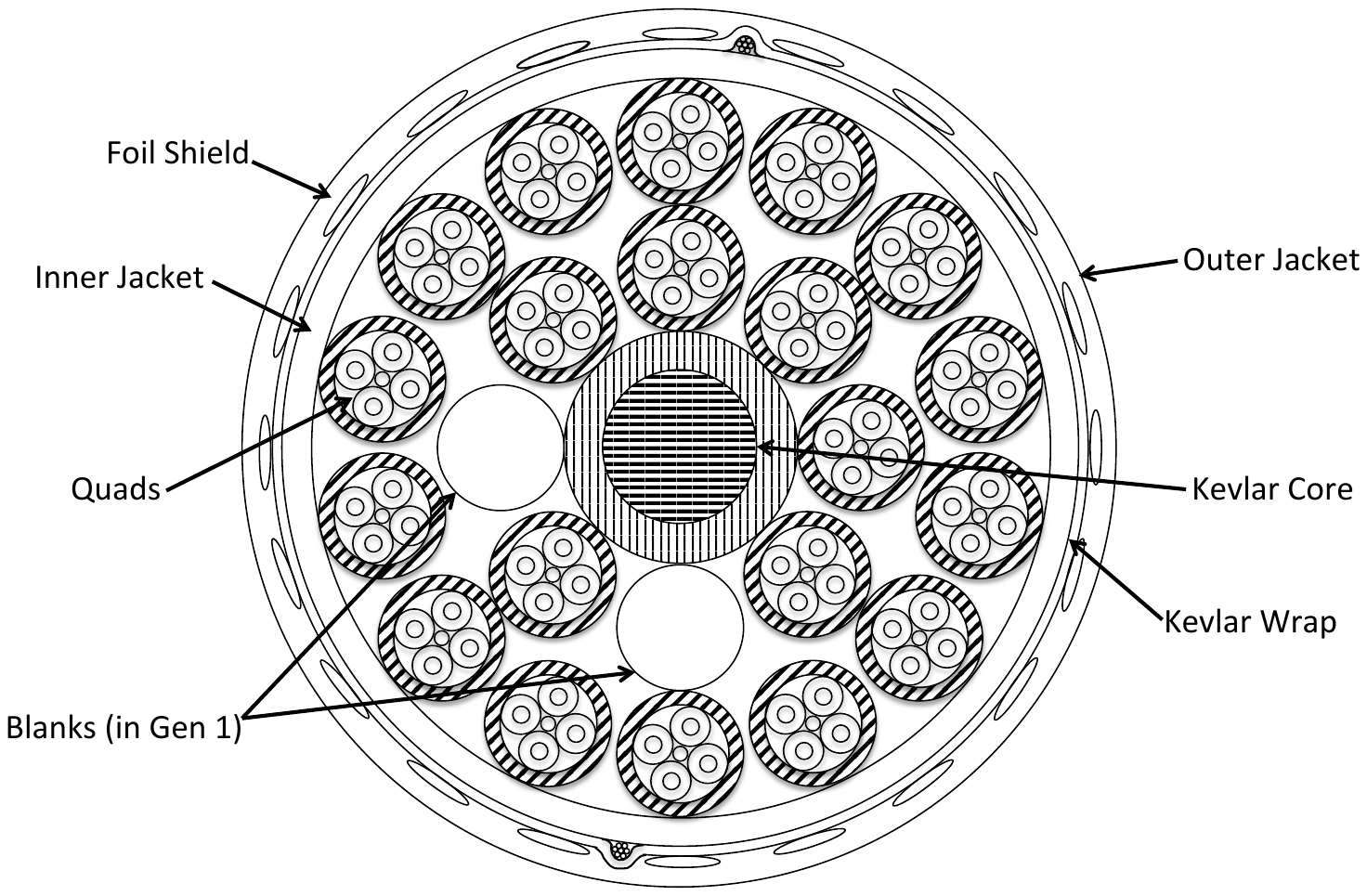}
\caption{Cross section of an Ericsson cable designed for a PINGU
  string.  The outer diameter is 36~mm.}
\label{fig:CableCrossSection}
\end{center}
\end{figure}

With a lower power design and improved data compression, more sensors
can be serviced by each wire pair than in IceCube.  As shown in
Fig.~\ref{fig:StringCable}, each group of four sensors is serviced by
one wire pair broken out from the main cable.  This is compared with
only two DOMs per wire pair in IceCube.  Another simplification of the
cable design comes from the omission of extra wires previously
dedicated to detecting local coincidences between modules hit, so that
each breakout construction is much more straightforward.  These
changes allow considerably more sensors to be deployed on each cable,
reducing cable and drilling costs and avoiding the need for redesign
and attendant risk.  The same advantages apply to other sensor
designs, such as the mDOM, D-Egg, and WOM.



\subsection{Drilling and Deployment}
\label{sec:drilling_deployment}
PINGU strings will be deployed in holes drilled with an upgraded
IceCube Enhanced Hot Water Drill.  Each hole is approximately 55~cm
diameter and 2500~m depth, the same as the original IceCube holes.
The drill provides 5~MW of thermal power in the form of a
high-pressure (1000~psi) and high flow (200~gallons/minute)
$90^\circ$C water jet which will melt ice to a depth of 2500~m in
30~hours.  After drilling each hole, a string of modules is deployed
over a period of 10~hours and will be fully frozen-in after 3~weeks.

During the construction of IceCube, a highly trained crew of 30 drillers
were able to drill up to 20 holes in each South Pole construction
season (Table~\ref{table:ICStringsPerSeason}).  Key personnel from the
IceCube project and the South Pole support network remain active in
hot water drilling and will still be available for PINGU, 
leveraging their extensive experience to optimize the required
drilling time and effort.  Existing equipment and procedures can be
utilized, including necessary refurbishment and replacement of some
components, to complete the construction of the 26-string PINGU within
$1+2$ seasons.  As shown in Table~\ref{table:PINGUStringsPerSeason},
the first season will be used for reassembly, staging and testing of
the drill equipment, followed by two for drilling and deployment.

\begin{table}[ht]
\renewcommand{\arraystretch}{1.05}
\renewcommand{\tabcolsep}{0.2cm}
\sffamily
\begin{center}
\begin{tabular}{|c|c|} 
\hline
Season&Strings deployed\\
\hline\hline
2004-2005&1\\
\hline
2005-2006&8\\
\hline
2006-2007&13\\
\hline
2007-2008&18\\
\hline
2008-2009&19\\
\hline
2009-2010&20\\
\hline
2010-2011&7\\
\hline
\end{tabular}
\caption{IceCube yearly records for drilling and deployment.}
\label{table:ICStringsPerSeason}
\end{center}
\end{table}

\begin{table}[ht]
\renewcommand{\arraystretch}{1.05}
\renewcommand{\tabcolsep}{0.2cm}
\sffamily
\begin{center}
\begin{tabular}{|c|l|c|} 
\hline
Season&Special activities&Strings Deployed\\
\hline\hline
1&Drill Rebuilding and Staging, Firn Pre-drilling&\\
\hline
2&Recommission Drill&6-10\\
\hline
3&Decommission Drill&10-16\\
\hline
\end{tabular}
\caption{PINGU drilling and deployment schedule.}
\label{table:PINGUStringsPerSeason}
\end{center}
\end{table}

\begin{figure}[htb]
\begin{center}
\includegraphics[width=6in]{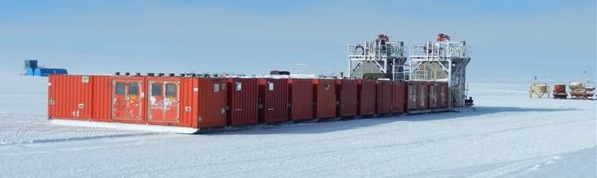}
\vskip0.2in
\includegraphics[width=6in]{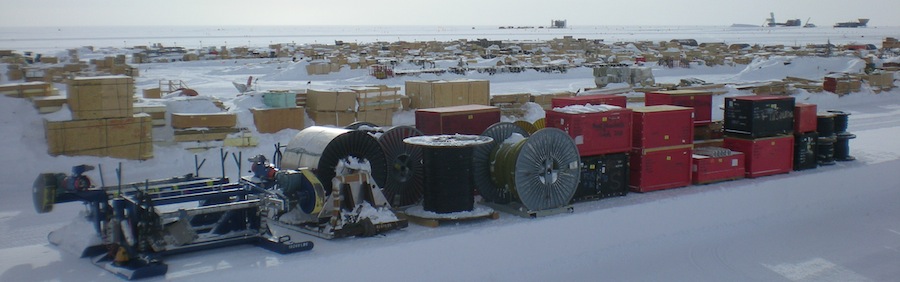}
\caption{IceCube drill components stored at South Pole, 2011.}
\label{fig:EHWD-photos}
\end{center}
\end{figure}

The IceCube drill is a complex system with many components, with
complete written and video documentation of its operation.  Following
the last deployments in 2010--2011, the system was winterized for
storage and to facilitate reuse (Fig.~\ref{fig:EHWD-photos}).  Major
components remain in IceCube/NSF stewardship, such as the hose reel,
drill towers, and associated structures.  These represent significant
engineering solutions and assembly efforts that can be reused by
PINGU.  Heating plants, high pressure pumps, and water tanks remain
similarly available.  Other components were allocated for use by other
projects, of which only some are still at the South Pole.  A
preliminary inventory of drill-related equipment needing replacement
has been made for PINGU cost estimation (see
Sec.~\ref{sec:ScheduleAndCost}).  To recommission the drill system,
PINGU will likely need to replace the main drill hose, cable reels,
electrical generators and distribution, the independent firn drill,
and the Rodwell system that supplies make-up water from melted snow.

\begin{figure}[htb]
\begin{center}
\includegraphics[width=3in]{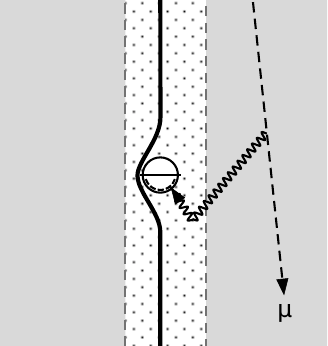}
\caption{Scattering of light by bubbles in hole ice near DOMs in
  IceCube.  The photocathode is on the bottom side of each DOM,
  indicated with a dashed curve.}  
\label{fig:HoleIceScattering}
\end{center}
\end{figure}

Certain aspects of the drilling system will be upgraded for PINGU.  In
particular, improvements are planned that will increase optical
clarity of the refrozen ice near the deployed sensors.  In AMANDA and
IceCube, it was found that light scattering is enhanced in such
refrozen ice~\cite{Lundberg:2007mf} (or ``hole ice'') modifying the
DOMs' angular response function and complicating calibration and
systematic error studies.  The effect is most apparent for
near-vertical background muons, where downward traveling light can be
scattered in the upward direction and more easily strike the PMT
photocathode (Fig.~\ref{fig:HoleIceScattering}).  Study of the number
and timing of photons in such events has resulted in a quantitative
hole ice model that is routinely used for simulation and analysis of
neutrino events in IceCube.  Subsequently, a camera system was
deployed in one IceCube hole and confirmed the scattering
visually~\cite{Rongen:2016sbk}.

The hole ice scattering is believed to be caused by dissolved gas that
is released from the solution as very small bubbles during the refreezing
process.  Glacial ice incorporates air entrained with the snow from
which it was slowly formed over thousands of years, but in a clathrate
solid form (the air is trapped within the crystal structure) that is
optically clear.  The hole-refreezing process takes place over a much
shorter time scale, apparently causing the observed reduction in hole
ice clarity.  Supporting this view is the amount of air measured in
the original ice, about 600~mg/kg, and the solubility of air in cold
water, only 20~mg/kg at 1~bar.      Drill circulation passes about
two-thirds of the meltwater through equipment at the surface, where
much of the gas can escape, but the remaining one-third still
contributes 200~mg/kg to the final mix.  During refreezing, the inward
growing ice phase can exclude this dissolved gas into the remaining
liquid phase along the axis, where the concentration thus increases
until it reaches saturation even at the high pressures in the hole
(200~bar).  At this point bubbles can be expected to form a reduced-transparency core, consistent with the camera observations.

For the surface tanks of the IceCube array (IceTop), a
special degassing system was used to ensure clear ice, but this
degassing was not performed for water in the deep holes of AMANDA and
IceCube.  For PINGU, the drilling operation will be modified so that the
instrumented region of the hole contains only water that has been
brought to the surface and degassed.  Standard membrane degassing
equipment will remove 85\% of air dissolved at atmospheric pressure,
leaving only 3~mg/kg for return to the hole.  By slowing the drilling
in the bottom 350~m of depth and then raising the drill head slowly
while injecting cold degassed water, gas-laden water will be displaced
above the PINGU fiducial volume while adding only a few hours to the
drilling time.  Additional filtering will also
prevent the introduction of mineral impurities from the heating equipment
and their possible precipitation.  Possible ways to reduce ice
fracturing through pressure reduction during the refreeze are also under
consideration.  In this way the PINGU hole ice is expected to be made
effectively clear.  Remaining light scattering near PINGU
modules, as well as in the bulk ice away from the holes, will be
precisely calibrated by studies with LED beacons and downgoing
background muons (Sec.~\ref{sec:Calibration}).

Deployment of sensors into PINGU holes will be similar to IceCube
deployment, with some modification for the use of steel wire rope
between modules (Fig.~\ref{fig:StringCable}).  One AMANDA string was
successfully deployed with such steel rope, but the standard IceCube
procedure involved attachment of harnessed DOMs directly to the main
electrical cable, with installation of special grips for transferring
the vertical load between harness and cable.  In PINGU, the chain of
modules will be assembled and lowered by repeatedly adding
harnessed modules and steel rope links.  This is done in an iterative
process that supports the already deployed section on a tethered hook
while the next module or rope segment is suspended above it from a chain
hoist.  After attachment of the new component, the whole load is
hoisted up slightly to allow detachment of the tether and then lowered
a few meters to allow attaching the new top to the tether.  In
synchrony, the main electrical/support cable is lowered and
secured appropriately to the module chain, along with its breakout cable
assemblies that connect to the modules.  After the final module link is
deployed, support is transferred to a grip on the main cable and the
assembly is then lowered 2150~m into the hole.  The equipment for
handling the main cable, the enclosing structure, and procedural
safeguards can all be reused from IceCube, giving confidence in
achieving a similarly successful operation.

\clearpage


\resetlinenumber

\IfFileExists{NewCommands.tex}       {}       {}
\IfFileExists{../NewCommands.tex}    {}    {}
\IfFileExists{../../NewCommands.tex} {} {}

\graphicspath{{figures/}{DAQ/figures/}}

\section{The Data Acquisition System}

The IceCube data acquisition system (DAQ) is now a mature and
well-understood composition of firmware and software components
maintained by a team of IceCube scientists and software professionals.
In order to leverage the existing infrastructure, both the optical modules
will be designed in such a way as to be fully compatible with the
IceCube DAQ system.

\subsection{Firmware and Software}
As described in Sec.~\ref{sec:OptSensDesign}, PINGU modules, like
their IceCube counterparts, trigger and digitize PMT pulses
autonomously.  The digitization process is foreseen to be continuous
and the trigger formed digitally with logic in the FPGA.  Triggered
pulses will then be processed by a deconvolution algorithm existing in
firmware, software, or some combination of the two running within the
module. The algorithm will extract the underlying photon pulse charge
and local time stamp from the sampled waveforms and output the
resulting list of (charge, time) pairs to a memory buffer. The data
volume of this representation of the pulses is approximately ten times
more compact than the uncompressed waveforms, thereby dramatically
increasing the effective buffer depth of the module and decreasing the
bandwidth requirements on the module-to-surface communications
channel. The information content of these data objects will be made
compatible with the stream output by the IceCube DOM; the minimum
needed to process the objects is an identifier of the originating
channel and a timestamp. Further information, such as charge, is
optional and may be used if available. To support these anticipated
changes in the module hardware design, the following new features will
be implemented:
\begin{itemize}
\item module-specific digitizer hardware, including firmware-based
  trigger,
\item firmware and software to perform in-module pulse deconvolution
  (under development), and
\item firmware to support the higher density of channels per wire pair
  relative to the IceCube DOM (under development).
\end{itemize}

\subsection{Integration of PINGU Channels into IceCube Surface DAQ}
\begin{figure}[htbp]
\begin{center}
\includegraphics[width=0.8\textwidth]{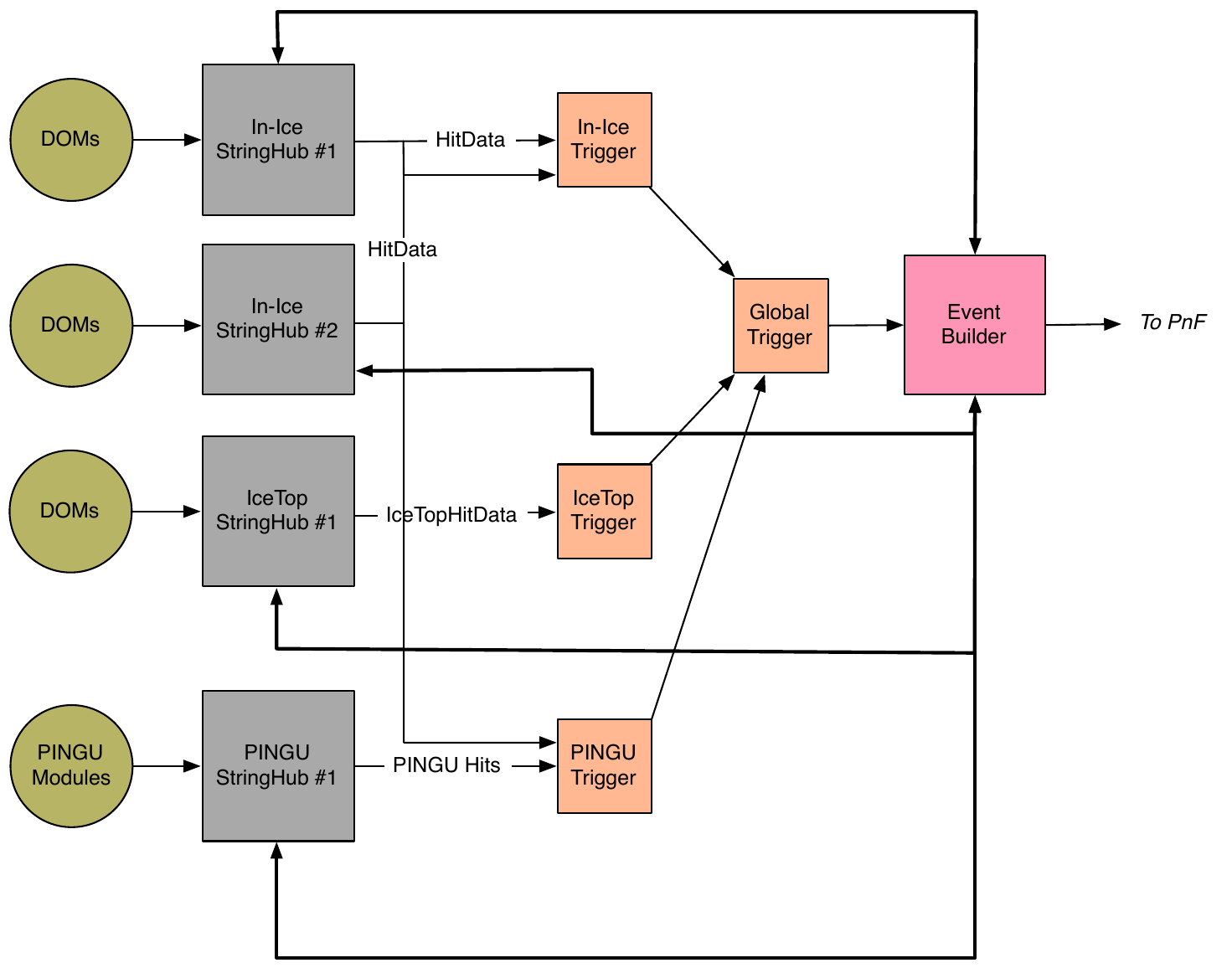}
\caption{Schematic representation of the integration of the PINGU
  modules into the IceCube data acquisition system.  At left are the
  modules (circles), an entire string of which are connected to data
  collection processes running on the readout computers (grey boxes).
  In IceCube there are 86 such entities corresponding to the deep ice
  modules and another 11 entities for the IceTop surface channels.
  From software there is no limitation to the number of readout hosts
  in the system.  The StringHub readout processors forward 
  module hit summaries to the trigger processors which may, upon
  detection of an interesting hit pattern, request a readout of all
  hits in the detector in a given time window around the trigger time.
  Trigger processors can  operate on hits from both PINGU and the existing IceCube and DeepCore
  DOMs in combination.  Each global trigger results in an event being
  built by the Event Builder.  These events are then sent to the
  Processing and Filtering (PnF) farm at the South Pole for initial
  reconstruction and data reduction.}
\label{fig:daqint}
\end{center}
\end{figure}

PINGU channels will be integrated into the IceCube data acquisition
system as shown in Fig.~\ref{fig:daqint}. The PINGU readout hub
(StringHub) components are each connected to a string of modules. They
extract hit summaries (channel and time information, currently) from
the stream of data buffers sent by the modules and send them to a trigger
unit.  Each trigger unit examines the stream of hits and identifies
patterns of interest by sending a trigger to the Global Trigger unit.
The Global Trigger unit merges triggers from the sub-detectors which
overlap in time.  A global trigger is then sent to an Event Builder
processor which requests the full readout data, including waveform
information buffered in the hubs, in time windows around the triggers,
packaging this data into an event data structure on disk.

The individual components, and their changes relative to IceCube
entities are discussed next.

\subsubsection{Communications System and Readout Electronics}
It is expected that between 4 and 8 modules will share a common copper
pair connection to the surface in order to reduce the cost and size of
the surface-to-DOM cables relative to the IceCube in-ice cable
system. This is a two- to four-fold increase in the channel
aggregation and will require upgraded communications firmware. At the
same time it will be necessary to increase the density of the readout
electronics in order to accommodate additional PINGU channels in the
rack space available in the IceCube Computing Laboratory. An improved
readout system is currently being designed to meet these new
requirements of PINGU. The communications must be made compatible with
existing IceCube channels to allow for a uniform deployment of readout
hardware across all channels. Despite this requirement, it will be
possible, via advanced protocol-discovery during the link negotiation
phase, to encompass the improved communications protocol of the PINGU
channels using more robust phase-shift keying modulation techniques
while still supporting the baseband signalling currently used in
IceCube.

In addition to data communication, the IceCube and PINGU data
transmission protocols are required to implement a mechanism for time
translation between the local timestamps in the IceCube and PINGU
modules and the globally referenced time system UTC.  In IceCube this
mechanism is known as RAPCal and is fully documented
in~\cite{Abbasi:2009domdaq}.  The corresponding solution for PINGU
DOMs is currently under design with requirements for the same or
better time resolution with respect to RAPCal (less than 5~ns) which
loosens the cable crosstalk specification.  With the above-mentioned
phase-shift keying modulation schemes, pseudo clock recovery can be
implemented.

At the application level, the PINGU StringHubs should function
nearly identically to those of IceCube: each hub collects and buffers
data, sending pulse hit summaries to the trigger processors. The
notable extension to the functionality concerns the tagging of hits
that are sent to the triggers. Due to past computing limitations in
sending the full 2.5~MHz rate of hits to the triggers, only hits
tagged with ``local coincidence'' in the ice are forwarded.  This
reduces the rate of hits resulting in triggers to approximately 50~kHz.
Recent improvements in both the computing hardware and the IceCube DAQ
will support hit sorting and triggering at the expected aggregate
PINGU hit rate, also roughly 2.5~MHz.

\subsubsection{Triggers}
The DAQ layout depicted in Fig.~\ref{fig:daqint} explicitly separates
the PINGU trigger processor from the IceCube trigger processor,
providing a degree of isolation between the two. Development of
another trigger processor specific to PINGU is the only infrastructure
development foreseen at the DAQ level. These modifications are minimal
and only involve adding additional bookkeeping data structures to the
trigger code. While existing DAQ trigger algorithms can be
re-parameterized and used in PINGU, new trigger algorithms will also
likely be developed to fully exploit the capabilities of the new
subdetector. It is also possible that the PINGU trigger could operate
as one or more algorithms within the In-Ice Trigger unit. This latter
configuration has the advantage that a single trigger can be formed
from the combination of information from IceCube and PINGU channels.


\subsection{Event Builder}
The Event Builder interacts with the readout computers through a rather
simple network interface.  The PINGU StringHub will implement
this interface and thereby participate directly in the event assembly
process.  As such, there are no customizations needed in the Event
Builder for PINGU deployment.

\clearpage

\resetlinenumber

\IfFileExists{NewCommands.tex}       {}       {}
\IfFileExists{../NewCommands.tex}    {}    {}
\IfFileExists{../../NewCommands.tex} {} {}

\graphicspath{{figures/}{Calibration/figures/}}

\section{Calibration}
\label{sec:Calibration}

\subsection{Overview}
\label{subsec:Overview}

PINGU event reconstruction uses the timing, location, and amount of
deposited light to reconstruct the properties of the
incident particle: position, time, direction and energy. Accurate
reconstruction relies on calibration of the timing (arrival time of
photons at the modules), geometry (module positions), photon detection
efficiency (photosensor sensitivity), and a model for the optical
properties of the ice. The latter determines how \Cerenkov photons
propagate through the ice from their creation until they are absorbed
or reach a module. Uncertainties in any of these measurements
constitute the detector-based systematic uncertainties in PINGU
physics measurements. Other systematic errors come from uncertainties
in the atmospheric neutrino flux, including the energy spectrum and
zenith angle distribution, the atmospheric oscillation parameters
$\dmatm$, $\thTT$ and $\thOT$, the neutrino cross-section and
branching ratios to specific final states, and the light output from a
given neutrino interaction final state. As shown in
Sec.~\ref{sec:NMHAnalysisSystematics}, the largest impact in the
neutrino mass ordering measurement is from uncertainty in the
atmospheric oscillation parameters.

Here we review the impact of detector-based uncertainties on PINGU
physics measurements. We then discuss the methods for detector
calibration in IceCube and PINGU, and planned calibration devices for
PINGU.  In this
section, the NMO analysis is used as a proxy for all PINGU analyses,
as it is one of the most sensitive analyses and is susceptible to the
impact of most of systematic uncertainties discussed in
Sec.~\ref{sec:NeutrinoOscillations}.

Note that these calibration devices could equally well be deployed and
used in the proposed Phase~1 detector (\ref{sec:gen2phase1}), with the
same benefits to low energy oscillation analyses.  In~\ref{sec:gen2phase1}
we also describe the advantages that would accrue to high energy
IceCube neutrino analyses as well.

\subsection{Impact of Detector Systematics in PINGU}
\label{subsec:detsys}
A complete study of the impact of detector-based uncertainties on
PINGU's sensitivity to the NMO requires more simulation statistics
than are currently available. At present, the only detector-based
systematics incorporated in the NMO study are the optical properties
of the ice and module efficiency. Conservatively, the uncertainties on
these quantities are taken to be the current IceCube values, and these
values are sufficient to measure the NMO.

The default model of the optical properties of the ice is the SPICE
Mie model, based on LED flasher data from
IceCube~\cite{Aartsen:2013rt}. This model is compared to the more
recent SPICE Lea model~\cite{Chirkin:icrc2013}, which has a similar
basis but includes an observed azimuthal anisotropy in the scattering
length. The effect of unsimulated anisotropy in SPICE Mie appeared as
a spread in the difference between the predicted and observed charge
deposited in IceCube DOMs by the LED flashers. The
difference between SPICE Mie and SPICE Lea is used therefore as a proxy for
unsimulated, as yet unknown, effects in the ice model. The DOM
efficiency is known to within 10\% in IceCube; a conservative estimate
of the uncertainty in PINGU uses this number. In order to evaluate the
impact of these parameters in the existing simulations, the mean PINGU
module efficiency has been shifted by 10\%, and the individual module
efficiency has been smeared by sampling from a Gaussian distribution
with a width of 10\%.  This method of including the efficiency shift
is applicable to all modules used in PINGU, and the results are not
expected to change with the module type.

\begin{figure}[h]
\begin{center}
\includegraphics[width=14cm, angle=0]{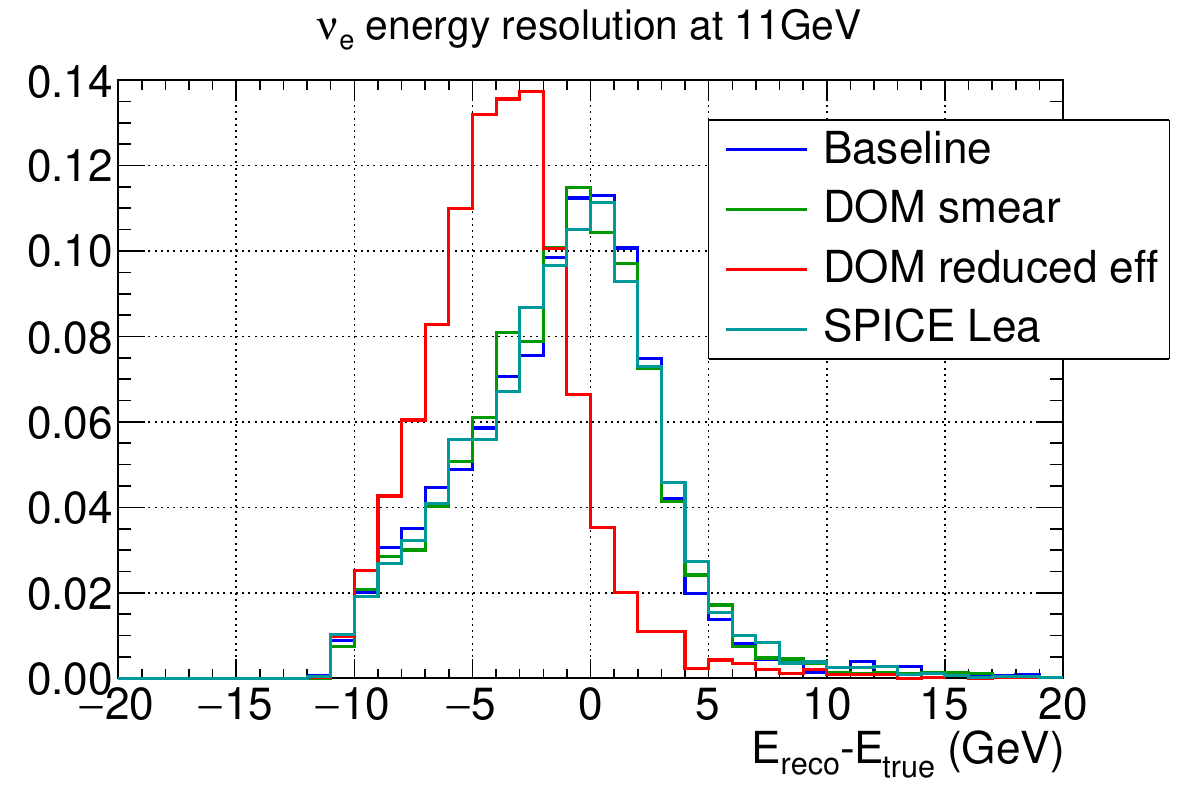}
\caption{Energy resolution for 11~GeV electron neutrinos in PINGU
  (charged current only). Blue: baseline geometry and SPICE Mie ice
  model. Red: Mean module efficiency reduced to 70\% of nominal; this is
  larger than our 10\% uncertainty in order to see the impact more
  clearly. Green: Individual module efficiency smeared by
  10\%. Blue-green: SPICE Lea ice model. The largest shift is seen for
  the change in the mean module efficiency; the peak of the reconstructed
  energy distribution is reduced by the same amount as the
  efficiency.}
\label{fig:energyres}
\vspace{-1.\baselineskip}
\end{center}
\end{figure}

Of all detector systematics evaluated thus far in the simulation, the
overall shift in the mean value of the module efficiency has the largest
impact on the energy scale in PINGU, as shown in
Fig.~\ref{fig:energyres}. The shift in the energy scale corresponds to
a shift in the mean value of the module efficiency. Therefore, the
uncertainty in the energy scale is taken to be 10\% in the NMO
systematics study in Sec.~\ref{sec:NMHAnalysisSystematics},
corresponding to the current uncertainty in the IceCube DOM
efficiency.

\subsection{Detector Calibration}

IceCube {\it in-situ} detector calibration employs reference
electronics onboard the DOMs, LED flashers co-located with each DOM,
in-ice calibration lasers and low energy muons in the ice. Most of
these methods will be used again in PINGU.

\subsubsection{Geometry}

\label{sssec:geometry}
The positions of IceCube DOMs have been measured to within 1~m with
several methods, which we will also use in PINGU. During module
deployment, the position of strings will be surveyed and the absolute
depth of the modules will be estimated using data from the drilling
process and pressure sensors attached to the bottom of each
string. The vertical distance between modules will be measured with a
laser ranger or similar device during deployment, and will be
corrected for possible cable stretching. Corrections to these
measurements will be determined using {\it in situ} light sources to
triangulate the final position of modules in the ice. Given the PINGU
interstring spacing of about 20~m, an uncertainty of 1~m in the interstring
spacing corresponds to a less than 3$^{\circ}$ error in direction
resolution. Studies are underway to determine how much the geometry
measurement may be improved in PINGU.

IceCube also includes several DOMs with onboard inclinometers to
measure a possible differential ice flow that would cause
time-dependent changes in the detector geometry.  Significant changes
have not yet been observed, but similar devices may be installed in
PINGU.

In addition to the position of the modules, the azimuthal orientation of
the modules after freeze-in will be measured by LED flashers and/or
onboard compasses. In IceCube, the azimuthal orientations of the DOMs
are measured with LED flashers to within
10$^{\circ}$~\cite{Aartsen:2013rt}. 

\subsubsection{Timing}
\label{sssec:gain}

The IceCube time calibration system uses reciprocal pulses sent
between the free-running clock in each DOM and the master clock on the
surface, resulting in a time synchronization across the detector of 3
ns or better~\cite{Achterberg:2006md}.  PINGU will use a similar
method, but with a continuous clock signal that is integrated into
the communications system as opposed to isolated discrete pulses.
Small transit time offsets in each PMT will be measured using pulses
from integrated LEDs.  Calibration light sources with independent
timing will be used during commissioning to verify that overall time
offsets among modules are accurately tracked throughout the hardware
and data processing chain.

\subsubsection{Module Gain and Sensitivity}
\label{sssec:gainsens}
Similar to IceCube, PINGU reconstruction will make use of PMT output
waveforms based on the time and number of photons arriving at each
module.  From observed single photon responses, PMT high voltage
settings will be chosen for gain close to 10$^7$.  In order to control
for small offsets or drifts from the nominal single photoelectron
(SPE) response of 1.6~pC, PINGU will continuously monitor the
waveforms for SPEs in triggered events.  The result is a measured SPE
charge for each module, including any effects due to PMT or electronic
gains.  This will then serve as the fundamental calibration quantity
when making precise correspondences between PMT waveforms and photon
counts.  In IceCube, such an SPE calibration has been shown to be stable at
the level of 1\% or better.

Linear response of the module to pulses with multiple photons will be
verified by in-ice LED flashers capable of operating at different
brightnesses.  As previously done in IceCube~\cite{Aartsen:2013vja},
ratios between pulse brightness can be gauged by counting single
photon responses in sufficiently distant modules, and then used to map
 any nonlinearities in near modules.

Independent measurements of module optical sensitivity will be performed
in the lab before shipping and deployment in the ice. In particular, the
sensitivity of a selected subset of modules will be thoroughly
characterized as a function of wavelength and angle. To
transfer this detailed characterization to the full population, all
modules will undergo a time-efficient testing process similar to that used in IceCube~\cite{Achterberg:2006md}, but with increased
emphasis on angular and wavelength dependences. Similar measurements
of IceCube DOMs~\cite{Tosi:2015ica} suggest that a systematic error of
2\% can be achieved in the laboratory.

The efficiency in ice includes the effect of the local ice properties,
and is measured in IceCube using low energy
muons~\cite{Aartsen:2013vja}, with an uncertainty of 10\%. This
measurement may be improved by better reconstruction of the position
and direction of low energy muons.

\subsubsection{Ice Properties}
\label{sssec:iceprop}
The ice into which the modules are deployed consists of two components:
the undisturbed bulk ice between strings, and the melted and refrozen
hole ice in which strings are deployed. The optical properties of the
bulk ice depend on local dust concentrations, which were measured in
IceCube with dust-logging devices during deployment and with LED
flashers after deployment~\cite{Aartsen:2013rt}. IceCube measurements
have shown that the scattering and absorption properties of the bulk
ice between the holes are highly location-dependent. Concentrations of
dust in the ice vary with depth, creating approximately horizontal
layers. These dust layers are not perfectly horizontal, and the
concentration of dust at a given depth for one string can therefore differ from
the concentration of dust at the same depth for another string. Both
the absorption and scattering lengths of the ice track the dust
concentrations. Additionally, the strength of scattering in the ice
appears to be anisotropic with respect to the azimuthal angle of the
photon direction~\cite{Chirkin:icrc2013}.

At the depth where PINGU will be deployed (between 2150 and 2450~m
below the surface of the ice), typical scattering and absorption
lengths are greater than 25~m and greater than 100~m,
respectively~\cite{Aartsen:2013rt}. The scattering length is therefore
comparable to or longer than the PINGU inter-string spacing of about
20~m.  Currently the scattering and absorption lengths in IceCube are
measured to within 10\% using LED flashers co-located with IceCube
DOMs~\cite{Aartsen:2013rt}.  In addition to the 10\% uncertainty on
the optical properties of the ice, there is additional uncertainty due
to unsimulated effects, which appears as a difference in the simulated
and actual charge deposited by LED flashers. For example, if the
anisotropy is not simulated, this introduces a 30\% spread in the
difference between the simulated and actual
charge~\cite{Aartsen:2013rt}. Improvements to the LED flashers in
PINGU are under study (see Sec.~\ref{subsec:Devices}) in order to
better measure the bulk ice properties.

The hole ice includes bubbles from the refreezing process in the
columns surrounding the sensors, as discussed in
Sec.~\ref{sec:drilling_deployment}.  These bubbles modify the angular
sensitivity of the module in ice, especially in the forward direction, as measured by flashers and muons in
ice~\cite{Aartsen:2013rt}. The degassing process planned for PINGU
deployments (see Sec.~\ref{sec:Instrumentation}) is intended to reduce
or eliminate the formation of bubbles in the hole ice. Instruments
discussed in Sec.~\ref{subsec:Devices} will investigate any residual
effects of the hole ice.

\subsection{Calibration Devices}
\label{subsec:Devices}
As the studies of calibration requirements progress, we are looking at
cost-effective improvements to the calibration devices used in IceCube.  In this
section, we will discuss several new or improved calibration devices
designed to significantly improve our understanding of the detector. A
summary of calibration devices and their purposes is shown in
Table~\ref{tab:calibdevices}.

\begin{table}
\caption{Summary of proposed PINGU calibration devices and their purposes\label{tab:calibdevices}.}
\centering
\begin{tabular}{l c c c c c c}\hline \hline
& LED flashers& POCAM& Cameras& MTOMs& Compass &Inclinometer\\\hline
Energy scale& \checkmark & \checkmark& & & &\\\hline
Bulk ice& \checkmark& \checkmark& & & &\\\hline
Hole ice& \checkmark& \checkmark & \checkmark& & &\\\hline
Sensitivity &\checkmark & \checkmark& &\checkmark & &\\\hline
Geometry&\checkmark & &\checkmark & &\checkmark &\checkmark\\\hline
Timing& \checkmark& & & & &\\\hline
Azimuth & \checkmark& &\checkmark & & \checkmark&\checkmark\\\hline
Ice motion& \checkmark& & & & &\checkmark\\\hline
Cable shadow& & & \checkmark& & &\\\hline
\hline \hline
\end{tabular}
\end{table}

\subsubsection{LED Flashers}
\label{sec:calib_led}
\begin{figure}[h]
\begin{center}
\includegraphics[width=10cm, angle=0]{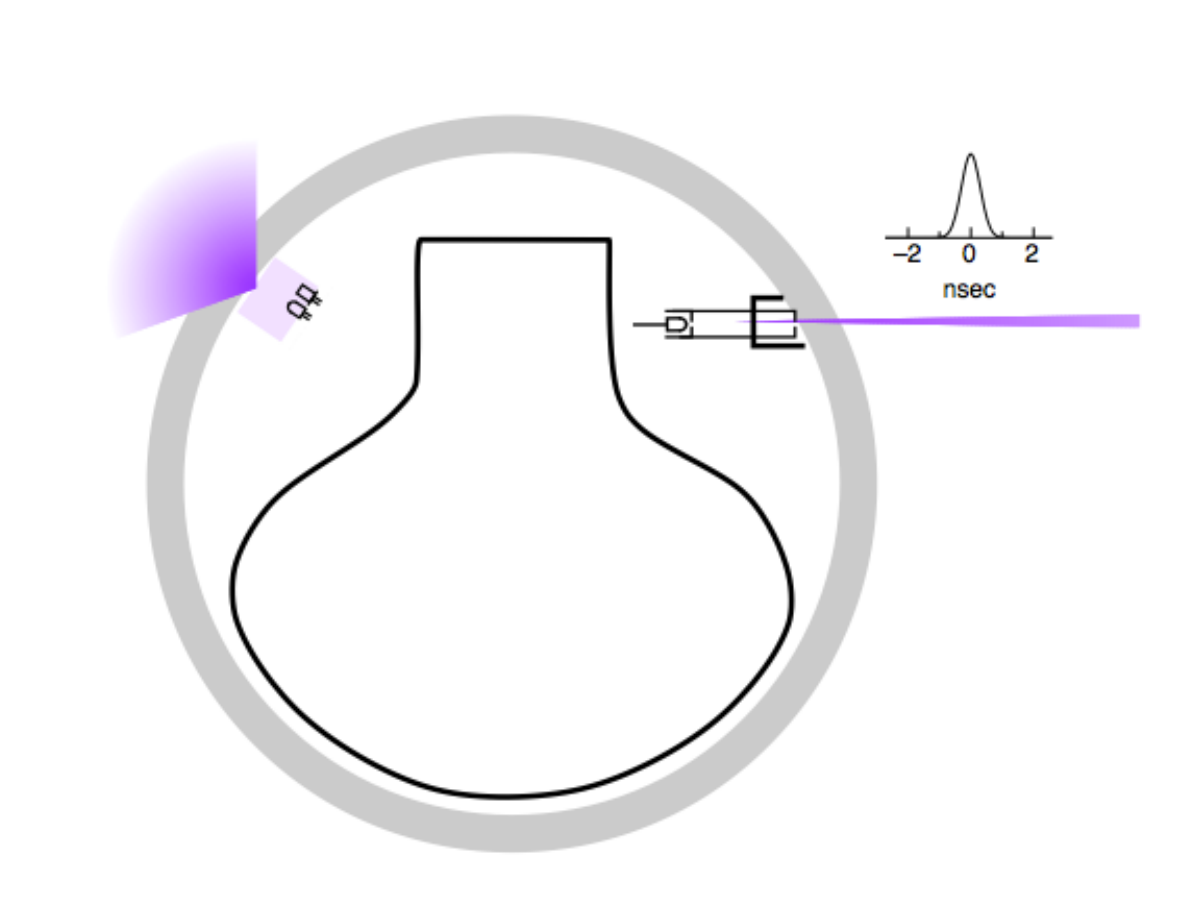}
\caption{An upgraded PINGU module with flashers.
\label{fig:pinguflashers}}
\vspace{-1.\baselineskip}
\end{center}
\end{figure}

LED flashers are installed with IceCube/DeepCore
DOMs~\cite{IceCube:2007aa} and will be co-located in all PINGU
modules.  Flashers are versatile devices with a range of settings
which have been used in IceCube to measure ice properties, module
sensitivity, timing, orientation and the coordinates of deployed
modules. Measuring the highly location-dependent variation in the
optical properties of the ice requires calibration sources at multiple
locations within the PINGU instrumented volume. Several improvements
are planned for both the design and characterization of PINGU light
sources in order to achieve the desired calibration goals. A
conceptual sketch of a PDOM with upgraded PINGU flashers is shown in
Fig.~\ref{fig:pinguflashers}. An extensive simulation effort is
currently underway to optimize the properties of the flashers. 
\begin{itemize}
\item {\bf LED light output}: Photodiode monitoring of the LEDs on the
  control board will increase confidence in the laboratory
  characterization and will allow for better control of the brightness
  settings, especially at low light levels, which will be necessary in
  order to measure the ice properties across the short distances
  between PINGU modules.
\item{\bf LED pulse timing profile}: In IceCube, the minimum LED pulse
  width is 7~ns. In order to measure the scattering function more
  precisely over short distances, the LED pulse width should be
  reduced below 2~ns so that the time behavior of the received light
  is almost entirely due to scattering rather than the source pulse
  shape. R\&D efforts are underway to determine the feasibility of
  producing a 1~ns pulse in the ice. Pulse widths of 3~ns have been
  produced in the lab using IceCube LEDs, and it is expected that
  widths of 1-2~ns can be achieved with an updated driver circuit.  A
  very simple fast driver, proposed by Kapustinsky {\it et
    al.}~\cite{Kapustinsky1985612}, has been successfully used by a
  variety of experiments for calibration.
\item {\bf LED orientation}: IceCube LEDs are positioned at one of two
  zenith angle orientations: pointing horizontally outward
  (perpendicular to the cable) and pointing upwards at an angle of
  45$^{\circ}$ above horizontal. Other angles are being investigated
  for PINGU flashers, such as flashers pointed straight upward in
  order to be maximally sensitive to the optical properties of the
  hole ice. The mDOM sensor design allows the most flexibility in
  placement of LEDs.
\item {\bf LED direction}: The orientation of the IceCube LEDs within
  a DOM is only known to within $\pm 5^{\circ}$. A housing for the LED
  in the PINGU module would control the direction to within
  1$^{\circ}$, which is better than needed given the spacing in
  PINGU. The azimuthal LED orientations will be determined {\it in
    situ} using the observed light pattern in surrounding modules. 
\item {\bf LED angular emission profile}: IceCube LEDs have a beam
  width of 10$^{\circ}$ in ice. Both wider and narrower beam widths
  are under investigation; a wide beam can be achieved by coupling the
  LED to a diffuser, and a narrow beam can be achieved by using a
  collimator. The LED emission pattern will be measured in the lab.
\item {\bf LED wavelength}: Most IceCube LEDs have a wavelength of
  405~nm, with a few modules containing LEDs with wavelengths of 505,
  450, 370 and 340~nm. These LEDs have been used to validate the
  wavelength dependence of the scattering and absorption lengths in
  the ice. PINGU modules may contain different wavelength LEDs in
  order to further study the wavelength dependence of the ice
  properties and the module response in ice.
\end{itemize}

\subsubsection{POCAM}
\label{sec:calib_pocam}

A diffuse light source called the Precision Optical CAlibration Module
(POCAM) is under development to complement the
LED flashers~\cite{Jurkovic:2016kxn}.  The POCAM is a light source that will not be located
inside a module, but rather on the cable alongside the modules.  The main
goal of the POCAM is isotropic illumination of a large part of the
PINGU volume by building a light source that is not shadowed by a
PMT.  Similar to the LED flashers, we plan to use monochromatic LEDs
with several different wavelengths as a primary light source. A
single LED per wavelength is used to achieve isotropic illumination.  The
POCAM will be self-calibrating and the light output will be
continuously monitored.

The PINGU baseline geometry shows an overall symmetry in the azimuthal
direction. However, for the PDOM has an up-down asymmetry due to the
module's asymmetric construction (see
Fig.~\ref{fig:PDOMAppearance}). In addition, the orientation of any
module (sensor or POCAM) will be random in azimuth along the string axis
during deployment and will be fixed after hole ice refreezing. Given
that, we aim to reach an ice illumination in the azimuthal direction
with a homogeneity on the order of $1\,$\% and relax the requirement
on homogeneity in the zenith direction to $< 10\,$\%. Overall, we aim
to measure the POCAM's emission profile with a precision better than
$1\,$\%.

Initial POCAM studies used the well-tested IceCube pressure sphere as the housing, and we identified three major
components that challenged the isotropic-emission requirement. These
included (a) the cable penetrator, (b) the main cable itself and
(c) the waistband and harness, which causes a certain amount of
shadowing, interrupting the homogeneity of the illumination.



\begin{figure}[hbt]
   \centering
   \begin{tabular}{c@{\hspace{0.5in}}c}
     \includegraphics[width=4.762213in]{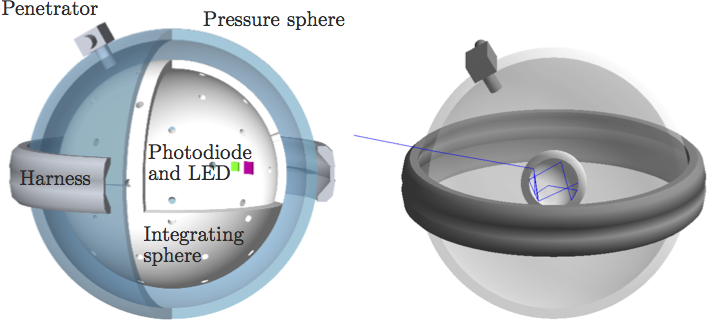} & \
   \includegraphics[width=1in]{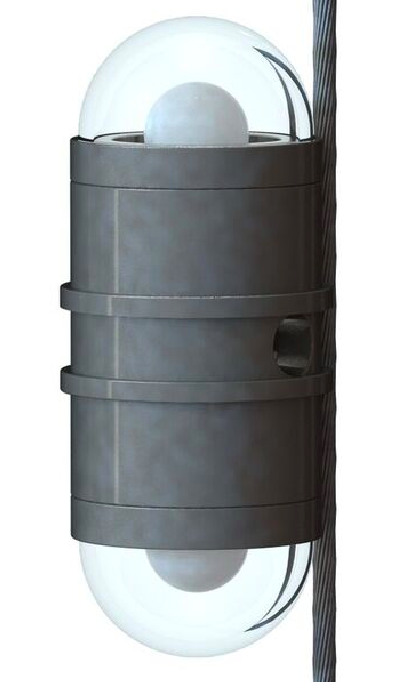} \\
   \end{tabular}
  \caption{Left: A multi-port POCAM placed in the IceCube pressure
    sphere.  Middle: A diffusing semi-transparent setup of the
    POCAM. The blue line shows a sample simulation of a photon
    emitted from a LED and undergoing several reflections before
    leaving POCAM. Right: cylindrical POCAM design.}
    \label{fig:pocam}
\end{figure}

We  developed a dedicated Geant4~\cite{Agostinelli:2002hh}
simulation to study and optimize the light emission of the
POCAM~\cite{Krings:icrc2015}. We investigated the POCAM's performance
for two configurations: (a) a multi-port (see Fig.~\ref{fig:pocam},
left) and (b) a diffusing semi-transparent layer setup (shown in
Fig.~\ref{fig:pocam}, middle). In the first configuration we
investigated the light homogeneity and timing depending on the
integrating sphere diameter, the number of ports, the position and the
radius. To investigate the properties of the second
configuration we modified only the radius of the semi-transparent
sphere and its thickness and thus the probability of photons being
reflected, transmitted or absorbed through the sphere's
wall. Investigation of shadowing, and inhomogeneities which limited
light emission in the spherical design, led to an updated pill-shaped
design (see Figure~\ref{fig:pocam}, right). This design allows
light emission over 4$\pi$ from two separated hemispheres. The
simulation results shown below are for the spherical POCAM but have
informed the development of the pill-shaped POCAM.

The performance of both configurations was studied for two diameters,
$24\:$cm and $6\:$cm. The larger diameter is limited by the size of
the IceCube pressure sphere leaving some space for electronics and
cables. The smaller diameter was chosen empirically considering
handling during construction. The settings of those four
configurations are summarized in
Table~\ref{tab:pocam_geant4_geometry_configurations}.

\begin{table}[htb]
  \caption{POCAM configurations: integrating sphere diameter $D$, number of
    ports $n$, port opening angle $\alpha$, effective photon transparency $T$,
    photon absorption $A$, and inhomogeneity for three different zenith angle
    zones $\theta$.}
  \centering
  \resizebox{\textwidth}{!} {
    \begin{tabular}[c]{cccccccccc}
      \hline\noalign{\smallskip}
      Config. & $D/\rm cm$ & $n$ & $\alpha/^{\circ}$ & $T/\%$ & $A/\%$ &
      \multicolumn{3}{c}{Inhomogeneity/\%} & $\tau/ns$ \\
      \noalign{\smallskip}\cline{7-9}\noalign{\smallskip}
        &   &   &   &   &   &
        $\theta < 60^{\circ}$ & $60^{\circ} \leq \theta \leq 120^{\circ}$ &
        $\theta > 120^{\circ}$ & \\
      \noalign{\smallskip}\hline\hline\noalign{\smallskip}
      $C1$ & 24 & 768 & 1  & $1.5$ & $61.0$ & $11.9$ & $18.5$ & $11.6$ & $35.1$ \\
      $C2$ & 6  & 768 & 1  & $1.5$ & $74.4$ & $14.7$ & $56.4$ & $10.3$ & $11.7$ \\
      $C3$ & 24 & -   & -  & $2.5$ & $50.3$ & $12.4$ & $14.5$ & $12.4$ & $13.1$ \\
      $C4$ & 6  & -   & -  & $2.5$ & $50.7$ & $14.3$ & $61.3$ & $10.3$ & $3.1$  \\
      \noalign{\smallskip}\hline
    \end{tabular}
  }
  \label{tab:pocam_geant4_geometry_configurations}
\end{table}

\begin{figure}[bt]
   \centering
   \includegraphics[width=0.8\textwidth]{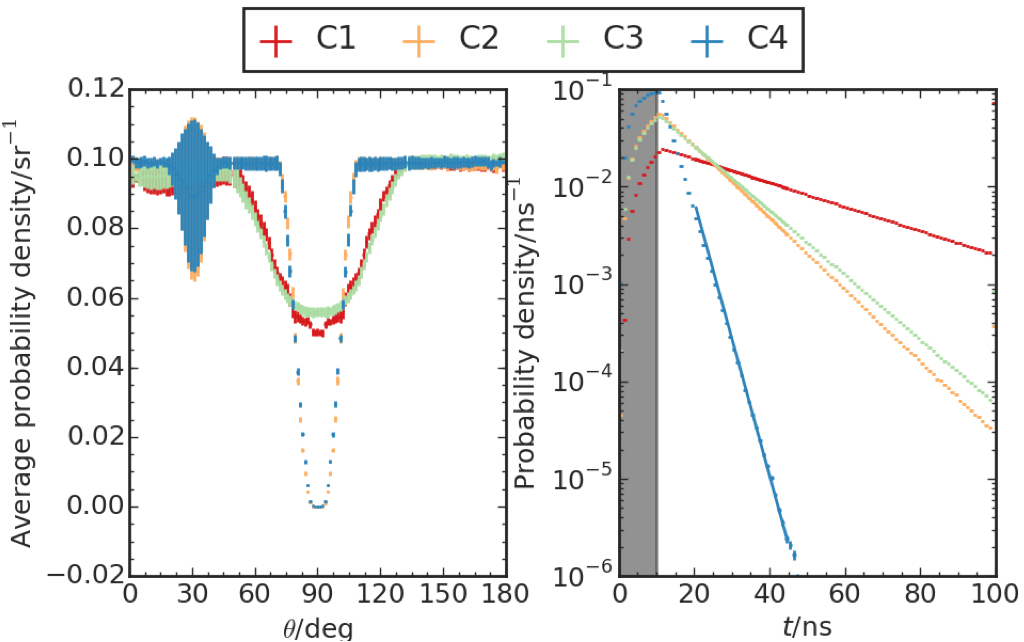}
   \caption{Left: Average photon emission direction in azimuth
     direction for different zenith angle bands. The error bars show
     the standard deviation.  Right: Emission time profile. The shaded
     area refers to the rectangular LED time profile with a width of
     $10\:\rm ns$. An exponential distribution is fitted to each time
     profile for $t > 20\:\rm ns$ and shown as a solid line. The decay
     times are listed in
     Table~\ref{tab:pocam_geant4_geometry_configurations}.}
   \label{fig:pocam_results}
\end{figure}

The resulting inhomogeneities for three zenith angle bands, together
with the decay constants of the time response of the integrating
spheres, are also listed in the
Table~\ref{tab:pocam_geant4_geometry_configurations}. In
Fig.~\ref{fig:pocam_results} we show the average photon emission
direction in azimuth for $1^{\circ}$ zenith angle bands (left) and the
emission time profile (right) of the studied integrating spheres.

Generally, the smaller spheres for both versions of the setup show
better homogeneity, reaching $\sim 2\:$\% inhomogeneity in azimuth
direction inside narrow zenith bands in the regions not affected by
shadowing. On the other hand the diffuse semi-transparent
configurations show significantly shorter response compared to the
multi-port version of the same diameter. Overall we can conclude that
a POCAM consisting of a diffuse semi-transparent integrating sphere
with a diameter of $6\:$cm shows the smallest inhomogeneities while
giving the shortest response time. The pill-shaped POCAM design uses
glass hemispheres with an external diameter of 4.5~inches, with the diffuse semi-transparent layer made of polytetrafluoroethylene (PTFE).

We also plan to investigate the feasibility of using a completely
different approach for a diffusing light source based on the concept
of a diffusing ball described in~\cite{Moffat2005255}.  This concept
provides a simple and inexpensive solution for the POCAM.

For the POCAM's light source we are investigating individual
monochromatic fast-switched (few nanosecond) LEDs and a fast LED
driver circuit. The Kapustinsky driver is able to drive LEDs with
pulses of $\Delta t = 1-2\,$ns (FWHM), and the LED provides light
pulses with up to $10^9$~photons per pulse. This photon yield
corresponds to an electromagnetic shower with an energy of
$\sim6\,$TeV. Light intensity can be controlled directly by the
applied supply voltage ($V_{CC} < 24\,$V).  Assuming the dynamic range
of the modules is large enough, and that they do not saturate, the
energy resolution can reach $1\,$\% in the case of an ideal detector.

An important feature of the POCAM is the continuous pulse-by-pulse
monitoring of the light emission with a precision of
$\sim1\,$\%.  Here we are considering fast solid-state detectors such
as photodiodes or silicon photomultipliers. In the recent years the
technology of the latter sensors shows a continuous improvement in
performance, which makes it an interesting alternative to photodiodes.

\subsubsection{Cameras}
\label{sec:calib_cameras}
A high-resolution camera similar to the one already deployed in
IceCube~\cite{Rongen:2016sbk} will be deployed with at least one
string to photograph the hole ice during and after refreezing.
Comparison of PINGU hole ice photographs with photographs taken by
IceCube will provide early data on the efficiency of the degassing and
filtering program.

\begin{figure}[h]
\begin{center}
\includegraphics [width=0.6\textwidth,height=0.40\textheight]{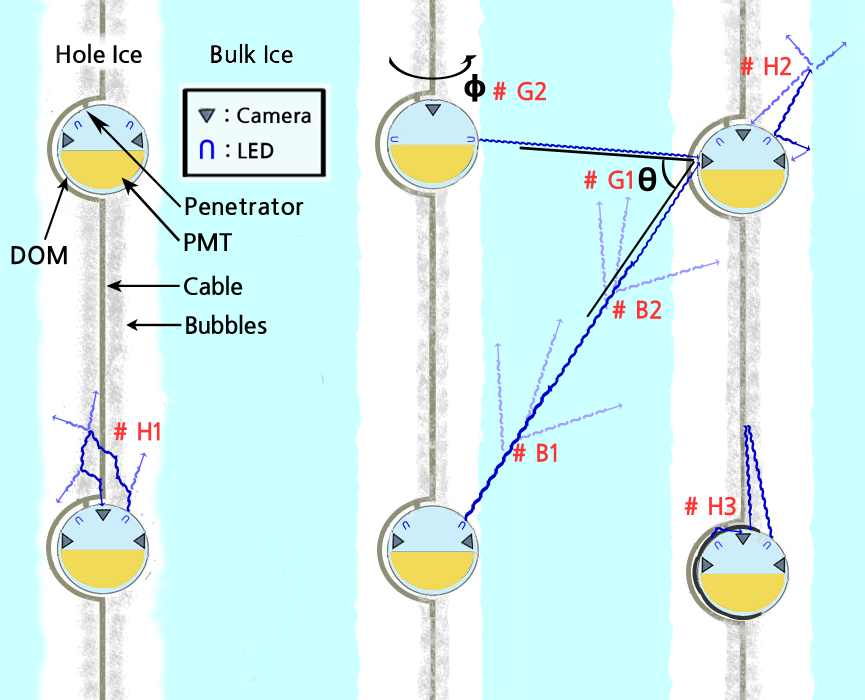}
\includegraphics [width=0.35\textwidth,height=0.20\textheight]{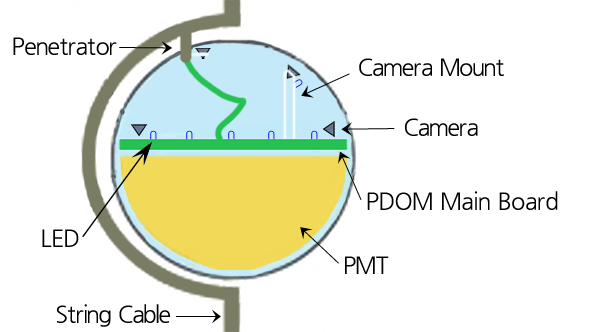}
\caption{Left: Schematic drawing to illustrate potential camera
  measurements: (H1) Hole ice survey; (H2) Mapping of hole shape; (H3)
  Cable position and orientation; (B1) Light transmission and
  scattering at hole ice - bulk ice interface; (B2) Light attenuation
  and scattering in the bulk ice; (G1) Orientation of the camera in
  the module; (G2) module geometry. Right: Potential camera positions
  in the PDOM.}
\label{fig:camera1}
\vspace{-1.\baselineskip}
\end{center}
\end{figure}

We are also planning a more
comprehensive set of visual hole ice data via deployment of low cost
cameras~\cite{Bose:icrc2015}. These on-board cameras, located on all or
a large fraction of PINGU modules, could provide quantitive data and
qualitative information about the local environment of each module,
valuable to the interpretation of other calibration
measurements. Together with the cameras, one or more bright LEDs are
required as light sources for the system.  The goals of the
camera measurements can be categorised into five groups: (1)
Hole-ice measurements, (2) Bulk-ice measurements, (3) Geometry
calibration, (4) Freeze-in observations, (5) General survey of the
module environment. Figure~\ref{fig:camera1} illustrates the potential
camera measurements.

The main goal of the camera system is to better understand the
hole ice. It could be utilised to observe where individual modules are
located with respect to the interface between the bulk ice and the
hole ice. The camera would detect bubbles in the refrozen hole, dust
or large contaminants. The camera system could also determine the module
position with respect to the cable and thereby reliably
measure the effect of cable shadowing on the module sensitivity.

Measurements of bulk ice properties are possible by observing the
scattered light from LEDs on adjacent strings to the observing
camera. A variety of geometry measurements that compliment the standard
IceCube-like geometry calibration are also possible. These measurements
could range from measuring the module $\phi$-orientation to measuring
interstring distances. Observations during and after freeze-in allow
for a better understanding of the fundamental hole-ice properties and ice
formation process. Another virtue of the camera system is the ability
to survey changes in the module environment.

\begin{figure}[h]
\begin{center}
\includegraphics [width=1.0\textwidth] {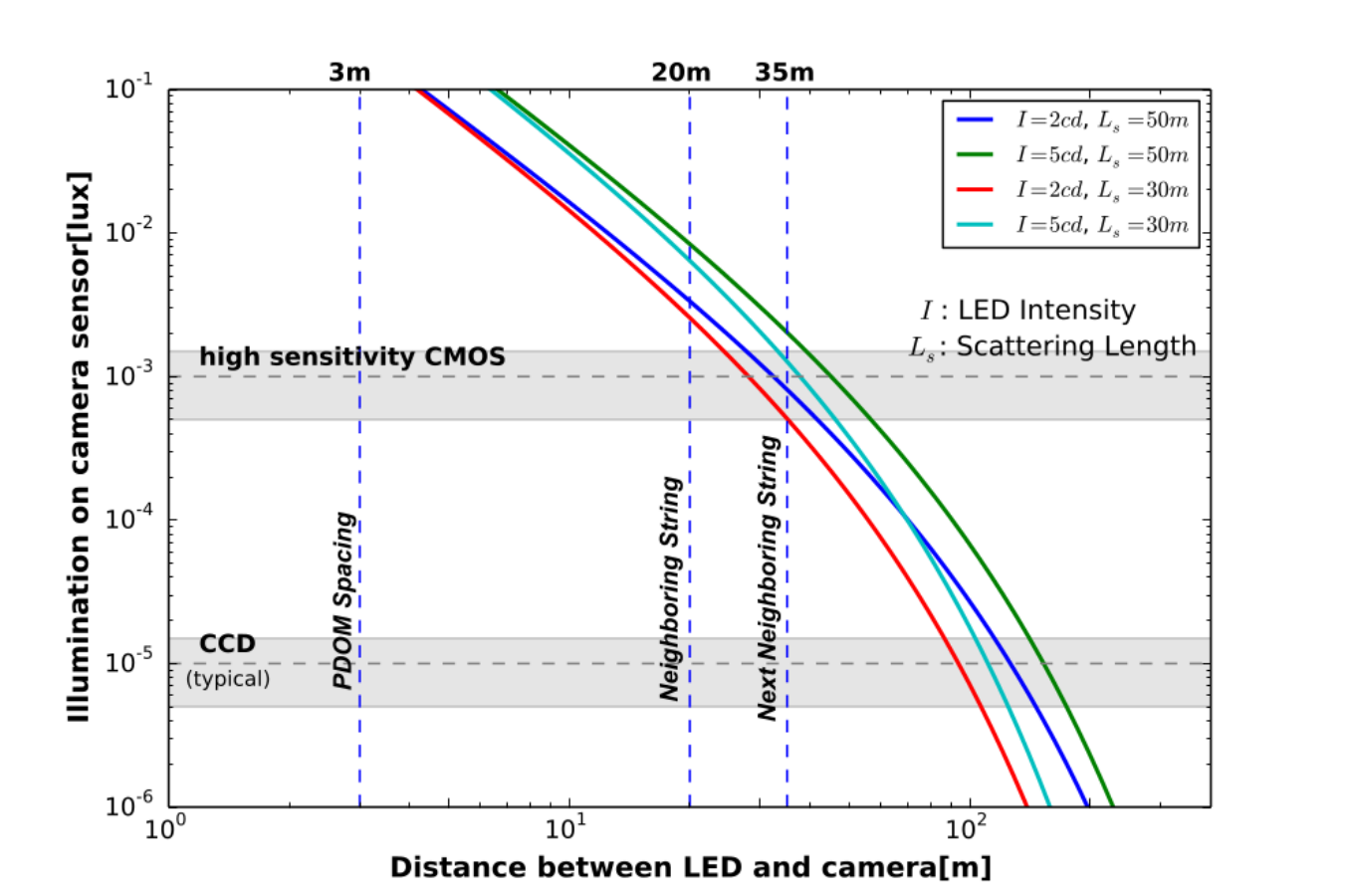}
\caption{The minimum illuminance required for a camera to see an LED
  located in the neighboring module on the same string, and on an adjacent
  string, are shown. An exposure time of 1s is assumed. The gray area
  at the top is that of a typical CMOS camera sensitivity. The gray area
  at the bottom represents the typical sensitivity of a CCD
  camera.}
\label{fig:camera2}
\vspace{-1.\baselineskip}
\end{center}
\end{figure}

The camera system would use small inexpensive fixed focus CMOS or CCD
cameras mounted on the main board or attached to penetrator of the
pressure sphere. For increased light sensitivity and reduced data
volume, black and white cameras would be used. Colored images could
still be obtained by illuminating with RGB-LEDs. CMOS cameras combine low power usage ($\sim 0.3$~W), wide field of
view, high resolutions, and good light sensitivity. CCD cameras have a
significantly higher light sensitivity compared to CMOS cameras, but
consume more power ($\sim$ 1W) and have lower
resolutions. Figure~\ref{fig:camera2} compares light sensitivities in
the PINGU geometry.  Long-exposure still
images would be taken that are only limited by the intrinsic camera
noise.

 As an example, to see a typically bright LED
($\sim$ 2--3 candela) located in a neighboring module in the adjacent
string at 20~m distance, the camera should have sensitivity $\sim$
0.001~lux (typical sensitivity of a CMOS camera).  A fish-eye lens
might be used to increase the field of view of the camera. Precise
knowledge of the camera and LED position and orientation on each module is
required to use the system for geometry calibration.

IceCube data transfer rates even allow for large raw images to be
transferred on time scales of several minutes, keeping calibration
measurements short. Images would be analysed offline, corrected for
image distortions from the pressure sphere, and used to study ice
properties and hole ice conditions.

The camera system would be mounted separately on a small camera board,
and communicate with the module mainboard via an SPI (Serial Peripheral
Interface) or similar interface. The camera board (approximately
$\unit[35]{mm}\times\unit[35]{mm}$) would contain a MCU (Micro
Controller Unit) next to the camera for image processing. Most of the time this system
operates in power-saving mode.

\subsubsection{Muon Tagging Optical Module (mTOM)}

We are investigating the possibility of deploying scintillator
detectors called Muon Tagging Optical Modules (mTOMs) in PINGU to tag the position of muons with high precision (to within a few
cm). This approach is inspired by a recent study using the DM-Ice17 dark matter
detector, that consists of two 5-inch NaI crystals
deployed in the IceCube volume~\cite{Cherwinka:2014xta}. While the primary
purpose of the DM-Ice17 detector is the direct detection of dark matter, the crystals detect 1-2 muons per day that also trigger
the IceCube array. 

A study of the impact in energy, direction, and position uncertainty of muons using the DM-Ice17 detector in conjunction with IceCube has been performed~\cite{HubbardPhD}. This has shown a reduction in mis-reconstructed events, when tagging the muon using the NaI crystals, as well as an improvement in the track reconstructed position. Given that the mTOM's size is 10 times smaller than the DM-Ice17 crystal, the track can be better localized.

The mTOMs in PINGU would be constructed of solid, 5~cm$\times$5~cm$\times$1~cm plastic scintillator block placed within a light-tight reflective aluminium casing. A prototype of this design has been developed in the context of a stand alone desktop muon counter~\cite{mtom}. The optical read-out is provided via a silicon photomultiplier (SiPM). These Modules have a thin profile, so that they require very little space within the module, making it possible to have four modules per PDOM. The detector would be hardwired into the standard PINGU DAQ.

If four mTOMs were deployed in each PINGU DOM, we expect to detect approximately $10^6$ downgoing cosmic ray muon interactions per year. These identified muons could then be used for detailed calibration of PINGU.

\clearpage

\resetlinenumber

\IfFileExists{NewCommands.tex}       {}       {}
\IfFileExists{../NewCommands.tex}    {}    {}
\IfFileExists{../../NewCommands.tex} {} {}

\graphicspath{{figures/}{Software/figures/}}

\section{Monte Carlo Simulation Software}
\label{sec:MCSoftware}

Simulations of the PINGU detector have been carried out using the
Monte Carlo tools developed for the IceCube experiment.  The strong
similarities between the IceCube and PINGU hardware permit capitalizing on the extensive efforts that have gone into refining the
IceCube simulation model of both the Antarctic ice and also the
hardware response.  However, some modifications to the standard
IceCube tools were necessary to produce reliable simulations of
neutrino interactions in the range of 1--20~GeV, and there have been modifications to the IceCube simulation tools in order to more
correctly simulate the detector.  All of these elements are detailed in
this section.
\subsection{Neutrino Event Generator}

Although most high-energy IceCube neutrino simulations rely on the
NuGen event generator (based on ANIS~\cite{Gazizov:2004va}),
simulations of neutrinos interacting in the PINGU detector are based
on the GENIE Neutrino Monte Carlo generator~\cite{Andreopoulos:2009rq}
version 2.8.6, that is used extensively in the accelerator-based
neutrino physics community.  GENIE conducts a full simulation of the
neutrino-nucleon interaction, and produces a list of secondary
particles exiting from the interaction vertex, modeling intranuclear
interactions where necessary.  All neutrino interaction channels are
modeled by GENIE, although the bulk of the events detected by PINGU
arise from deep inelastic scattering (DIS).

A wrapper was created to embed GENIE within the IceCube
software framework.  Events are generated isotropically from a
user-defined power law distribution (normally $E^{-1}$) and need to be
reweighted to reproduce the atmospheric neutrino
spectrum~\cite{Honda:2006qj}. A newer atmospheric model was
specifically developed for use at the South Pole~\cite{Athar:2012it}
in order to accurately incorporate the Earth's geomagnetic effects on
the arrival directions of cosmic rays. The asymmetry in the resultant neutrino flux as a
function of azimuth and zenith starts at $\sim10$~GeV and becomes more
pronounced at lower energies. While a rewrite of the weighting
software to incorporate the asymmetry is underway, an average flux
over the full azimuth range was used to produce an atmospheric
neutrino flux value.  The spectrum currently used for generation
($E^{-1}$) is very different from the expected spectrum so that the
higher energy events can be sampled more efficiently.  This is
necessary for parameterizing the response at all energies as explained
in Secs.~\ref{sec:EventReconstruction} and~\ref{sec:NMOAnalysisTechnique}.

In addition to considering the atmospheric weighting, this reweighting
procedure permits rapid assessment of the impact of uncertainties in
the true flux and spectrum of the atmospheric neutrinos. GENIE enables
extensive modeling of the uncertainties for neutrino
interaction physics \cite{Andreopoulos:2009rq} be
taken into account in the simulation.  Specifically, the uncertainties
that are studied address the neutrino interactions for the major
types of events that are simulated for use in the analyses: deep
inelastic scattering (DIS), resonance production (RES) and
quasi-elastic (QEL) interactions.  These interactions are
each addressed separately.

Deep inelastic scattering events will be the most prevalent type in the
PINGU low-energy dataset.  In the GENIE simulation code employed for
PINGU simulations, the model used is that provided by Bodek and Yang
\cite{BodekYang:2003} which accounts for higher twist and corrections
to the target mass.  The parameters specifically used in the PINGU
analysis for the DIS events deal with the twist parameterization
(AhtBY and BhtBY) and the correction to the $u$ valence quark in the
GRV98 PDF (CV1uBY and CV2uBY).

The resonance production events are generated based on the Rein-Sehgal
model \cite{Rein:1980wg} and have the associated uncertainties
included in the production calculations for the axial mass parameter
for charged current (MaCCRES) and neutral current (MaNCRES)
interactions.  Finally, the QE events are also considered and the
axial mass is also taken into account for charged current interactions
(MaCCQE).  In all these cases, the effects of variations of these
parameters up to $\pm2\sigma$ are computed and parameterized for use
in the analysis.

To improve performance, neutrino interactions were simulated inside a
cylinder of 200~m radius and 500~m height centered on PINGU.  Because
the atmospheric muon veto is extremely effective at rejecting
particles originating outside of the PINGU fiducial volume, the
reduction in neutrino rate at the analysis level introduced by this
approximation should be negligible.

\subsection{Atmospheric Muon Event Generators}

Computational resource limitations do not permit generation of event
samples comparable to the expected atmospheric muon rates, so we rely
primarily on the demonstrated muon rejection performance of the
IceCube DeepCore detector to support our estimates of muon background
rates.  Since the extra instrumentation available in PINGU will
improve the currently available reconstructions in the DeepCore
volume, the existing algorithms that reject atmospheric muons
will be further improved and, therefore, this estimate should be
strongly conservative.  Accordingly, our neutrino efficiencies are
calculated by applying slightly modified DeepCore event selection
routines to our simulated neutrino events, and the atmospheric muon
samples produced for PINGU are primarily intended to confirm that
these event selections behave as expected.

Two methods were used to produce samples of the atmospheric muon
background.  The first utilized full Corsika~\cite{Heck:1998vt}
simulation of air showers produced by cosmic rays incident on the
upper atmosphere, tracking the muons produced in such air showers to
the PINGU detector.  The second, known as the ``muon gun,''
parametrizes the angular and energy distributions of the muons
produced in the Corsika simulations and injected muons from a
hemisphere extending just outside the IceCube detector volume.

\subsection{Particle Propagation}

The relativistic particles produced at a GENIE neutrino interaction
vertex were tracked by GEANT4~\cite{Agostinelli:2002hh} until they,
and any daughter particles produced, fell below the \Cerenkov
threshold.  All of the \Cerenkov photons produced were stored for propagation through the ice by a separate program (see below).  This
provided a more detailed model of the light emission from the neutrino
interaction vertex than the standard IceCube tools, which relies
instead on parametrized descriptions of prototypical hadronic and
electromagnetc showers.  Atmospheric muons traveling through the
detector were modeled using the standard IceCube particle propagator
PROPOSAL~\cite{PROPOSAL}, instead of GEANT4.

\subsection{Light Propagation}
\label{sec:SoftwareLightPropagation}

Once the relativistic particles were propagated through the detector
volume by GEANT4 or PROPOSAL, the \Cerenkov photons produced were tracked
by the IceCube software tool ``clsim.''  This is a parallelized,
GPU-based software package that permits full treatment of photon
propagation throughout the extremely large volume occupied by IceCube.
The depth-dependent optical properties of the Antarctic ice have been
extensively studied by the IceCube Collaboration, and the full details
of this optical model~\cite{Aartsen:2013rt} were included.  At each
propagation step, photon scattering is modeled by numerical
approximations to the Mie scattering function, as developed and tuned
by IceCube.

\subsection{Detector Response}

The main optical elements of the planned PINGU DOMs are identical to
those used in DeepCore, so the detector response modeling of the DOMs
was used without modification (except that an average calibration
function was used for each DOM, rather than the individualized DOM
models used in IceCube simulations).  Recent studies of dark noise in
IceCube have shown that there is a significant non-Poissonian
component to the DOM noise, possibly arising from scintillation in the
pressure housing glass or the PMT glass at very low temperatures; this effect
was included in the detector simulation.  Improvements to this noise
model~\cite{LarsonMasters} were some of the main updates included in
the simulation since the first version of this letter.

Although the digitization electronics used in PINGU will differ from
the IceCube and DeepCore design, no modifications were made to this
aspect of the simulation software.  The dynamic range of PINGU may be
slightly narrower, but given the much lower energy range of the
neutrinos of interest, saturation of the PINGU electronics should be
rare for events in the desired energy range.  The timing resolution of
the PINGU electronics will be similar to that of IceCube, so the
photon timing and pulse resolution should be similar.  The PINGU
electronics will provide this timing resolution over the full event
time range, whereas the IceCube electronics occasionally need to fall
back on a slower FADC with coarser timing resolution.

In IceCube and DeepCore, data rates from individual DOMs are reduced
by imposition of a local coincidence in hardware.  If neighboring DOMs
also detect light, a full data record is transmitted to the surface;
otherwise, only a brief summary record is sent.  No dedicated local
coincidence circuitry is planned for PINGU, but we anticipate that
onboard photoelectron pulse extraction in FPGAs, and string-level
software coincidence logic at the surface, will allow us to fit PINGU
data comfortably within the available cable bandwidth.  Again, to the
extent that the PINGU hardware differs from the model used in
simulation, the additional flexibility in software-based coincidence
logic will only improve PINGU performance over the model used in
simulation.  We therefore have a high degree of confidence in our
simulation of PINGU.

The trigger rate in DeepCore is quite low, demanding only three
close-neighbor DOMs with locally-coincident hits within a time window
of a 2.5 microseconds.  Although this threshold may not be feasible
given the much higher number of DOMs in PINGU, very few simulated
neutrino events near the trigger threshold survive the relatively
strict event selection applied to neutrino events used in the
performance studies presented in this document.  We anticipate that
the inclusion of spatial, as well as temporal, information into the
PINGU trigger will permit use of trigger algorithms which would record
all of the events used in these studies.

\clearpage

\resetlinenumber

\IfFileExists{NewCommands.tex}       {}       {}
\IfFileExists{../NewCommands.tex}    {}    {}
\IfFileExists{../../NewCommands.tex} {} {}

\graphicspath{{figures/}{ScheduleCost/figures/}}

\section{Schedule and Cost}
\label{sec:ScheduleAndCost}

The proposed schedule and cost for PINGU is largely based on the
successful construction of IceCube. IceCube was completed on time and
on budget over a 10~year period that included eight seasons of
construction and installation at the South Pole starting in the
austral summer of 2003/2004. IceCube construction was completed in
January 2011. The plan for the construction schedule of PINGU
incorporates knowledge obtained from IceCube along with considerations
related to the length of time between IceCube completion and the start
of PINGU. The schedule is driven primarily by funding availability and
is not technically limited. 

We anticipate that two years will be required for refurbishment and
improvement of the IceCube hot water drill.  Optical module assembly and
transportation to the South Pole would occur in parallel with this effort.  Once the
drill and optical modules are available at the South Pole, the full
PINGU array can be deployed in two seasons.  Some
preparatory activity (snow compacting, firn drilling) would be
required in the prior South Pole season to enable a prompt start to
deployment once the drill arrives.  A summary of the envisaged schedule is shown
in Table~\ref{Tab:Schedule}.
\begin{table}
  \begin{center}
    \begin{tabular}{c|c} 
      Project Year   & Activities \\ \hline \hline
      1                  & Drill engineering and refurbishment,\\
                     & Optical module production   \\ \hline
      2                  & Drill engineering and refurbishment, \\
                     & Firn drill transport, \\
                     & Optical Module production and transport\\ \hline
      3                  & Drill transport, Site preparation,  \\
                     &Optical module production and transport  \\ \hline
      4                   & Deployment, optical module transport\\ \hline
      5                   & Deployment           \\
    \end{tabular}
    \caption{Summary schedule for construction of PINGU.\label{Tab:Schedule}}
  \end{center}
\end{table}



\noindent In summary, PINGU will take approximately five Project Years
to build and will require two austral summer deployment seasons at the
South Pole. The schedule does not currently take into account possible
funding constraints related to federal budget years that begin each
October. If a Phase~1 proposal is approved, all of the work related
to the drill and module development will be completed in advance of a
full PINGU deployment, corresponding to approximately the first three
years of a full PINGU construction timeline.  There would also be a
commensurate reduction in the cost for completion of PINGU, since
development, transportation and testing of the drill, as well as
module development, will have already been completed.  Similarly,
Phase~1 will retire virtually all of the associated risk for PINGU.

The start of module production is planned for PY1 with an expected
integration and test rate of 24 modules per week at each production
site. One such site will be the Physical Sciences Laboratory (PSL) in
Stoughton, Wisconsin. PSL successfully integrated and tested
approximately 3500 Digital Optical Modules for IceCube from 2004 to
2009. Physical infrastructure including important Dark Freezer Labs,
along with key personnel, remain at PSL making this a natural choice
for one of the several production facilities.

The first scheduled module deployment season for PINGU will be in
PY4. It is planned that 6--8 holes would be drilled in this initial
season. To accommodate a potentially faster drilling schedule, 10
complete strings of instrumentation will ship from production sites
and the associated cable system production site to meet Polar
Program logistics requirements. Instrumentation production would take
place continuously to accommodate the deployment of the remaining
additional strings of detectors in PY5. Final string commissioning and
IceCube integration would occur late in PY5 when PINGU would be
complete.

As with the schedule estimates described above, the cost estimates for
PINGU are largely based on extensive experience obtained from IceCube
construction.  We have created a detailed Work Breakdown Structure (WBS) with the following top
level task areas:
\begin{enumerate}
   \item Project Office
   \item Drilling
   \item Detector Modules
   \item Cable System
   \item Power, Communications, and Timing System
   \item Calibration System
   \item IceCube Integration
   \item Polar Operations (except drilling)
   \item Antarctic Support Contractor
\end{enumerate}

A detailed bottom-up estimate that extends three levels into the WBS,
including inflation but not contingency, has been performed for the
40-string, 96 module/string PINGU geometry~\cite{LoI}. This earlier
estimate leveraged some cost savings with PINGU being a part of a
large-scale IceCube high energy extension of order 100 strings.
However, with PINGU built in two seasons with many fewer strings,
its overall cost decreases even when budgeted as a standalone project.
With fewer holes drilled in fewer seasons, but more modules per
string, project costs are driven less by drilling, cable and
deployment expenses, and more by module procurement and production
expenses.  Experience gained from the IceCube deployment indicates
that installing up to 20 strings in a single season is feasible.  The
anticipated schedule involves the installation of 8 strings in the
first season (allowing time for restarting the procedure) and 18 in
the second for a total of 26.  A pessimistic scenario is also
presented in which 6 strings are installed in the first season and 14
in the second for a total of 20.  In each of these scenarios, the
number of modules per string has been increased to 192.  The rough costs
of these two scenarios are given in
Table~\ref{tab:ScheduleCostTwoScenarios}.

\begin{table}
\begin{centering}
       \begin{tabular}{c|c|c}
        & Cost (20 Strings) & Cost (26 Strings)\\ \hline\hline
        Drill refurbishment & \TwentyStringDrillCost & \TwentySixStringDrillCost \\
        Deployment (labor) & \TwentyStringDeploymentCost & \TwentySixStringDeploymentCost \\
        Instrumentation & \TwentyStringInstrumentationCost & \TwentySixStringInstrumentationCost \\
        Management \& other costs & \TwentyStringManagementCost & \TwentySixStringManagementCost \\ \hline
        Total & \TwentyStringTotalCost & \TwentySixStringTotalCost \\
        Fuel & \TwentyStringFuel & \TwentySixStringFuel \\
   \end{tabular}
   \caption{ Summary costs for construction of PINGU in U.S. dollars.
   Deployment labor includes the scientists and engineers associated
   with the hot water drill and string installation effort.  Fuel requirements
   for the hot water drill are provided as volumes due to uncertainties in the
   price of oil and the impact of the overland traverse on transport costs;
   recent costs are approximately \$20/gal.}
     \label{tab:ScheduleCostTwoScenarios}	
     \end{centering}
\end{table}

The potential for the use of a new optical module has also been
considered with minimal impact on the total budget.  The cost for an
IceCube-like PDOM is \$5.2k while the cost for the mDOM-like module
(see Sec.~\ref{sec:mDOMs}) has been estimated to be \$8k.  (An mDOM
has more than double the photocathode area of a PDOM.)  The use of 125
mDOMs per string would then result in a net savings while increasing
the overall number of photons collected.


\resetlinenumber

\IfFileExists{NewCommands.tex}       {}       {}
\IfFileExists{../NewCommands.tex}    {}    {}
\IfFileExists{../../NewCommands.tex} {} {}

\graphicspath{{figures/}{Facility/figures/}}

\section{Antarctic South Pole Facility and Logistics}

At the Amundsen-Scott South Pole Station, operated as a scientific
facility on behalf of the National Science Foundation (NSF),
astrophysical observations have been underway for several decades, and
to date represent the majority of ground-based astrophysical work in
Antarctica.  Relatively recent overviews of neutrino, cosmic ray and astronomy
programs at the South Pole have been presented at a workshop in 2011
in Washington~\cite{AstroAnt} and at the Symposium 288, {\it
  Astrophysics from Antarctica}, at the International Astronomical
Union General Assembly~\cite{IAU288}.

\begin{figure}[tb]
\begin{center}
    \includegraphics[width=0.8\textwidth]{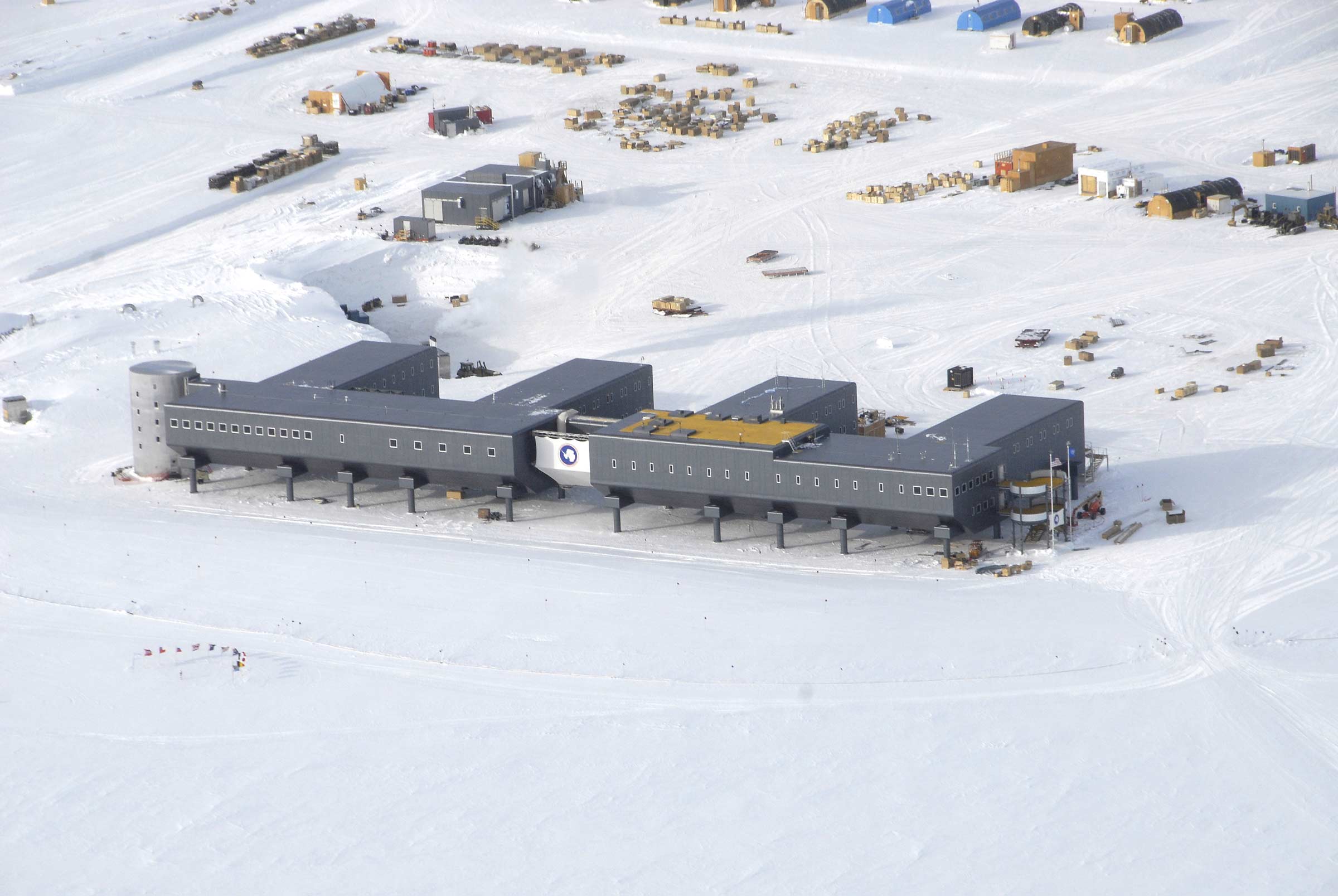}
   \caption{The Amundsen-Scott South Pole Station.}
   \label{fig:station}
\end{center}
\end{figure}

The Amundsen-Scott station provides excellent infrastructure for
scientific activities at the South Pole, including the IceCube
Laboratory (see Fig.~\ref{fig:ICL}) building that houses power,
communications, and data acquisition systems for IceCube.  The 
South Pole Station facility, shown in Fig.~\ref{fig:station}, was completed in 2008, prior to the completion of IceCube.  The realization of IceCube established the South Pole
ice cap as an underground laboratory that is available as a resource
for scientific endeavors beyond IceCube.

The station is accessible by aircraft from approximately November
through mid-February, supporting a summer population of approximately
150, including both scientists and support personnel.  During the
remainder of the year, the station is occupied by approximately 40
winter-over personnel.  Scientists typically make up around 25--30\% of
the personnel on station.

\begin{figure}[tb]
\begin{center}
    \includegraphics[width=0.8\textwidth]{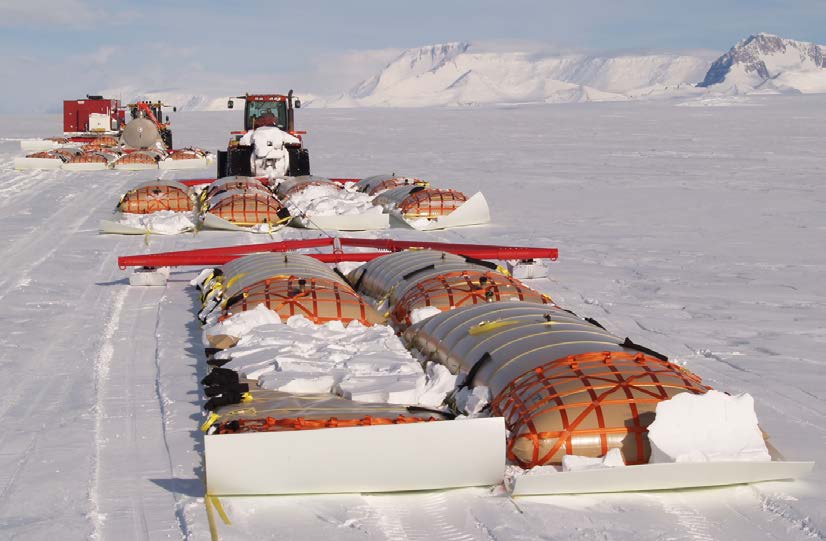}
    \caption{From {\it Economic Analysis of the South Pole
      Traverse}~\cite{SPoT}: ``Three SPoT sled trains heading up the
      Leverett Glacier en route to the South Pole during the 2008--09
      season. The front-most train includes the living and generator
      modules (red). Next in line is an MT865 towing a steel fuel sled
      (grey) and eight 3000 gal. bladders on flexible sleds (tan). The
      red Quadtrac is towing two coupled sets of bladder sleds,
      initially consisting of twelve 3000 gal. bladders but
      subsequently reduced by fleet consumption.''}
   \label{fig:SPoT}
\end{center}
\end{figure}

Whereas during the construction of IceCube all Antarctic
intercontinental passenger and cargo transport was via ski-equipped
LC-130 aircraft operated by the New York Air National Guard, the US
Antarctic Program has been transitioning fuel and cargo transport to
tractor traverses between the McMurdo and Amundsen-Scott stations (see
Fig.~\ref{fig:SPoT}). The cargo capacity of a single traverse, 350
metric tons, is thirty fold that of an LC-130 mission.  A number of forty-five day
round trip traverses have been achieved in recent austral seasons.  This has 
dramatically reduced the program's dependency on the limited LC-130
fleet, and cargo and fuel costs per unit of load delivered to South
Pole are more cost efficient by \$11.50 per kg compared to an LC-130 airlift.

The PINGU effort is expected to rely heavily on the overland traverse to deliver
many of its large-scale components, including elements of the drill,
fuel, down-hole cables and modules.  This will dramatically reduce the
logistical footprint of PINGU construction at the South Pole, as will
the significantly shorter active drilling and deployment duration of
two seasons.
 
IceCube uses 60~kW of power and has access to a bandwidth of more than
100~GB/day for data transmission and detector control.  Although PINGU
and IceCube will have a comparable number of deployed sensors, PINGU
is being designed to require roughly 1/5 the power and 1/10 the
bandwidth of IceCube.  PINGU is also being designed to integrate
seamlessly with the IceCube data acquisition cyberinfrastructure at
the South Pole, leading to economies of scale that reduce the
additional computational power needed on the surface.  Electrical
power is provided by the station generators and is highly reliable with
the IceCube detector operating with an uptime of approximately
99\%.   The primary means of data transmission from the South Pole is
via a TDRS satellite link maintained by the NSF.

\resetlinenumber
\section{Acknowledgements}

IceCube acknowledges the support from the following agencies: U.S. National
Science Foundation-Office of Polar Programs, U.S. National Science
Foundation-Physics Division, University of Wisconsin Alumni Research
Foundation, the Grid Laboratory Of Wisconsin (GLOW) grid
infrastructure at the University of Wisconsin - Madison, the Open
Science Grid (OSG) grid infrastructure; U.S. Department of Energy, and
National Energy Research Scientific Computing Center, the Louisiana
Optical Network Initiative (LONI) grid computing resources; Natural
Sciences and Engineering Research Council of Canada, WestGrid and
Compute/Calcul Canada; Swedish Research Council, Swedish Polar
Research Secretariat, Swedish National Infrastructure for Computing
(SNIC), and Knut and Alice Wallenberg Foundation, Sweden; German
Ministry for Education and Research (BMBF), Deutsche
Forschungsgemeinschaft (DFG), Helmholtz Alliance for Astroparticle
Physics (HAP), Research Department of Plasmas with Complex
Interactions (Bochum), Germany; Fund for Scientific Research
(FNRS-FWO), FWO Odysseus programme, Flanders Institute to encourage
scientific and technological research in industry (IWT), Belgian
Federal Science Policy Office (Belspo); Science and Technology Facilities Council (STFC) and University of Oxford, United
Kingdom; Marsden Fund, New Zealand; Australian Research Council; Japan
Society for Promotion of Science (JSPS); the Swiss National Science
Foundation (SNSF), Switzerland; National Research Foundation of Korea
(NRF); Villum Fonden, Danish National Research Foundation (DNRF), Denmark

\clearpage
\resetlinenumber

\IfFileExists{NewCommands.tex}       {}       {}
\IfFileExists{../NewCommands.tex}    {}    {}
\IfFileExists{../../NewCommands.tex} {} {}

\graphicspath{{figures/}{Appendices/figures/}}

\appendix
\addcontentsline{toc}{section}{Appendix}
\addtocontents{toc}{\protect\setcounter{tocdepth}{-1}}
\section{Mass Ordering Analysis Techniques}
\label{sec:NMOAnalysisTechnique}

\subsection{Simulation of experimental Outcome}
\begin{figure}[h]
  \centering
  \includegraphics[scale=.45]{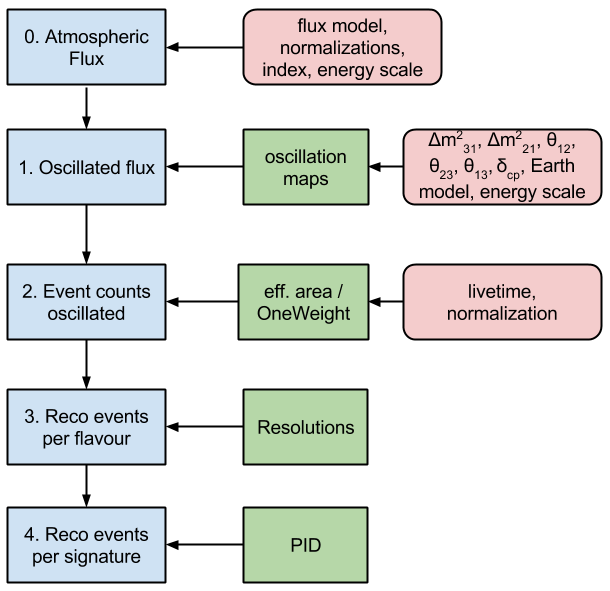}
  \caption{Template simulation scheme. See text for details.}
  \label{fig:templategen}
\end{figure}

All statistical analysis methods presented here require an
experimental expectation for the number of neutrinos, which is created
in the simulation step (see Sec.~\ref{sec:MCSoftware}).  The same code
is used to obtain this distribution of events in the $(E_\nu, \cos
\theta_\nu)$ or ``template'' for any given set of physical and
systematic parameter values.  Within the template the event
information is binned in 20 equally spaced linear bins in the cosine
of the zenith angle for the range $-1 < \cos \theta_\nu < 0$ and 39
logarithmically spaced bins in energy for the range $1\,\rm{GeV} <
\Enu < 80\,\rm{GeV}$. The bins are chosen to be small enough to not
artificially degrade the resolution, yet large enough to keep the
computational effort affordable and retain sufficient statistics in
each bin. To generate the template, we employ a staged approach, in
which we assume that the effect modeled by each step is independent of
any of the other steps.  Figure~\ref{fig:templategen} shows the
generation scheme that yields the final level templates. The
individual detector response functions are generated using a set of
simulated events, thus fully taking into account both the physics
modelled by GENIE, GEANT4, and the propagation of \Cerenkov photons
within a realistic ice model (see Sec.~\ref{sec:MCSoftware}), as well
as the effects of the detector resolution predicted by the full
reconstruction and event selection pipeline
(Sec.~\ref{sec:EventReconstruction}).

The simulation starts off at the flux stage, where two-dimensional
maps of the expected primary atmospheric neutrino flux as a function
of $(\Enu, \cos \theta_\nu)$ are produced for $\nue$, $\numu$ and the
corresponding anti-particles. Here, we allow for uncertainties in the
neutrino-to-antineutrino ratio, the muon-to-electron flavor ratio and
the spectral index.

In the second step, the oscillation probabilities are computed on a
fine grid for a given set of 3-flavor mixing parameters, taking into
account the varying electron densities from
Ref.~\cite{Dziewonski:1981xy} as the neutrinos traverse the Earth's
mantle or core. Since this is a task that is highly parallelizable for
different trajectories through the Earth, we have adapted the existing
GPU version of the \texttt{Prob3++} neutrino oscillation
library~\cite{prob3++} from Ref.~\cite{Calland:2013vaa} to be able to
take into account layers of alternating matter density. Calculating
the probabilities on a GPU instead of a CPU reduces the template
generation time by up to a factor of 30, from around 3~s down to
around 0.1~s. The four histograms containing the electron and muon
(anti)neutrino fluxes are then each weighted with the appropriate
survival and appearance probabilities of the various oscillation
channels (averaged over the bin) to get the corresponding histograms
for the oscillated flux, of which there are twelve in
total.

Oscillated fluxes are converted into event counts by multiplying with
the desired detector livetime, and folding in the detector's effective
areas for the various interaction channels.  The effective areas
follow from the event selection procedure laid out in
Sec.~\ref{sec:atmo_event_sel}.  This is the stage at which the
simulated detector response is included into this process.

Both the energy and angular resolutions of the detector are accounted
for by multiplying the 2D histogram of detected events in the true
parameters ($\Enu^{\rm true},\cos \theta_\nu^{\rm true}$) with the 3D
detector response matrix $\mathcal{R}:E_\nu^{\rm true} \to
\left(\Enu^{\rm reco},\cos \theta_\nu^{\rm reco}\right)$.  The matrix
maps the true parameters on the reconstructed parameters ($\Enu^{\rm
  reco},\cos \theta_\nu^{\rm reco}$) as a function of energy, thereby
properly accounting for events that are lost due to being
mis-reconstructed outside of the considered energy or zenith ranges.

In order to model the reconstruction resolutions as accurately as
possible in each bin of ($\Enu^{\rm true},\cos \theta_\nu^{\rm
  true}$), we employ a variable-bandwidth kernel density estimation
(VBW KDE) algorithm, that smoothes the reconstructed
neutrino energy and direction histograms obtained directly from MC
simulations. This is necessary in order to mitigate the effect of the
limited amount of MC statistics available, which would otherwise
artificially bias the analysis results. As mentioned in
Sec.~\ref{sec:EventReconstruction}, our algorithm is the fusion of two
previously-published methods: first, a fixed-bandwidth KDE is made
\cite{botev2010}, with a modified version of the Sheather-Jones
bandwidth selection criteria~\cite{10.2307/2345597} that does not
assume that the underlying distribution is normal. The individual
kernels' bandwidths for the full VBW KDE are then derived from the
pilot density estimate \cite{abramson1982}. A Gaussian kernel shape is
used for both the fixed- and variable-bandwidth parts of the
algorithm. At this step, a distinction is not currently being made
between interactions of neutrinos and antineutrinos of the same
flavor, nor do we assume different resolutions for NC interactions of
different-flavor neutrinos.

Application of the event classification probabilities (as depicted in
Fig.~\ref{fig:PID} and described in Sec.~\ref{sec:ParticleID}), but as
a function of reconstructed neutrino energy, leads to the two final
level templates, with event counts classified as either track-like or
cascade-like using the particle identification.

\subsection{Log-Likelihood Ratio Analysis} 
\label{sec:LLRAnalysis}
The most statistically detailed analysis we employ---the
log-likelihood ratio (LLR) analysis---entails two concepts: that of a
{\it template} (introduced previously) and that of a {\it pseudo-experiment}, which is
generated by calculating a bin-wise Poisson variation of the
template. 

The log-likelihood of observing the pseudo-data given a certain
realization of physical and systematic parameters (the hypothesis used
to generate the template)
\begin{equation} \mathcal{L} \equiv \ln L =
  \sum_{i,j} n_{ij} \ln \mu_{ij} - \mu_{ij} - \ln n_{ij}!\;,
\label{Eq:loglikelihood}
\end{equation}
where $n_{ij}$ is the content of the $(i,j)$-th bin in the pseudo-data
and $\mu_{ij}$ is the expectation for that bin, taken from the
template corresponding to the hypothesis. Gaussian priors are included
via an additional sum of the form
$\mathcal{L}_\mathrm{prior}=-\sum_{k}(\Delta p_k)^2/(2\,\sigma^2_k)$,
where $\Delta p_k$ is the deviation of the k-th systematic parameter
from its central value, and $\sigma_k$ is the width of its prior
distribution.

Compared to the original version of this document~\cite{LoI}, we are
now able to include a larger number of systematics through the use of
a numerical minimizer that maximizes the log-likelihood $\mathcal{L}$
over all parameters, instead of performing a scan of the
multi-dimensional space. The CPU time required to perform a single fit
to a given pseudo-experiment depends on the dimension of the parameter
space and on the rigorousness of the minimizer, but is of the order of
one to two minutes for our default settings in the eight dimensional
space. \footnote{Note that whenever the mixing angle $\theta_{23}$ is
  a free parameter the same pseudo-experiment is fit twice under the
  same hypothesis, with the minimizer being started in a different
  octant each time. This ensures that the global minimum is found,
  even when the likelihood landscape exhibits a strong octant
  degeneracy.}

\begin{figure}[h]
  \centering
  \includegraphics[scale=.5]{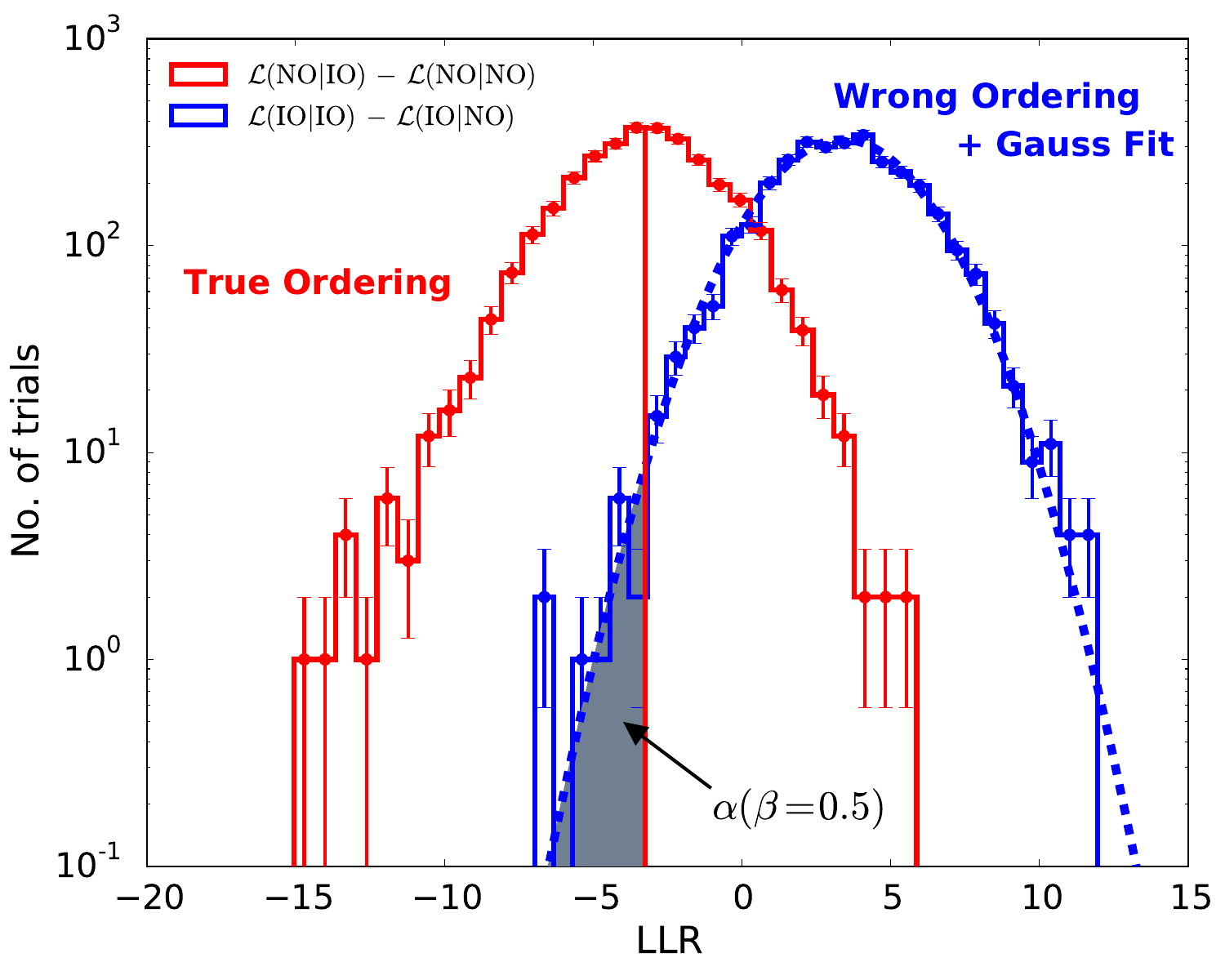}
  \caption{Calculation of the median NMO sensitivity using the
    log-likelihood ratio (LLR) method. Shown are the LLR distributions
    based on a particular set of pseudo-experiments for true NO. The
    vertical red line indicates the median LLR for the underlying NO
    fiducial model.  From a Gaussian fit to the wrong ordering
    distribution (here: IO), we determine the probability
    $\alpha(\beta=0.5)$ to misidentify the ordering in the ``average"
    experiment. See text for details.}
  \label{fig:teststatistics}
\end{figure}

The LLR test statistic is used in the NMO analysis to determine how
much more one ordering (either the Inverted Ordering (IO) or the
Normal Ordering (NO)) is favored by the pseudo-experiment.  For any
given pseudo-experiment, the two log-likelihoods
$\mathcal{L}($~pseudo-experiment~$|$~$\mathrm{IO}$~$)$ and
$\mathcal{L}($~pseudo-experiment~$|$~$\mathrm{NO}$~$)$ are calculated
by finding the respective maxima of $\mathcal{L}$. Since the
pseudo-experiment can be generated assuming either the IO or the NO is
correct, this leads to the four possible combinations
\begin{equation}
  \mathcal{L}(\mathrm{IO}|\mathrm{IO})\quad {\rm and} \quad
  \mathcal{L}(\mathrm{IO}|\mathrm{NO}), \quad {\rm or }\quad
  \mathcal{L}(\mathrm{NO}|\mathrm{IO}) \quad {\rm and} \quad
  \mathcal{L}(\mathrm{NO}|\mathrm{NO})\;
  .  
  \label{Eq:FourLLHs} 
\end{equation} 
These, in turn, are used to create the LLRs\footnote{Logarithm of the
  likelihood ratios, equivalent to the difference between the
  log-likelihoods.}
  \begin{equation}
  \mathcal{L}(\mathrm{IO}|\mathrm{IO})-\mathcal{L}(\mathrm{IO}|\mathrm{NO})
\quad {\rm and} \quad
\mathcal{L}(\mathrm{NO}|\mathrm{IO})-\mathcal{L}(\mathrm{NO}|\mathrm{NO})\; ,
\label{Eq:TwoLLRs} 
\end{equation}
respectively. By choice of this convention higher values of this test
statistic imply that the data is more consistent with the IO, while
lower values suggest better agreement with the NO. By producing a
large number of pseudo-experiments for each ordering, in which the individual
outcomes each are subject to random fluctuations, one can build up the
two LLR distributions, and quantify the agreement and disagreement
between the underlying hypothesis and both hierarchies on a
statistical basis. An exemplary case of the LLR distribution is shown
in Fig.~\ref{fig:teststatistics}.

\subsubsection{Median Sensitivity}
\label{sec:MedSensLLR} 
Due to its use throughout the literature, we haven chosen to adopt the
median sensitivity $\alpha(\beta=0.5)$, which yields the confidence
level $1-\alpha$ at which an experiment will reject the wrong ordering
hypothesis with $\beta=50\%$
probability~\cite{PDG_review}. Figure~\ref{fig:teststatistics}
demonstrates how the median sensitivity is obtained from two LLR
distributions $(\ref{Eq:TwoLLRs})$: we employ a one-sided test,
i.e. $\alpha$ is given by the fraction of cases in which the
log-likelihood ratio of a wrong ordering experiment exceeds (true IO)
or falls below (true NO) the median log-likelihood ratio of a true
ordering experiment. The double-sided convention is applied to convert
it into a corresponding number of Gaussian standard deviations,
\begin{equation} 
   n_\sigma = \sqrt{2}\,\mathrm{erfc}^{-1}(\alpha)\;.
\label{eq:DoubleSidedSignificance}
\end{equation}

To mitigate statistical fluctuations on the determination of $\alpha$,
we fit a Gaussian model to the test statistic distribution for the
wrong/opposite hypothesis with a large number of trials
(see~\ref{sec:estimate_median}).  Now the question arises of how to
calculate the median sensitivity; every ensemble of pseudo-experiments
we generate is based on one well-defined, fixed hypothesis, i.e. a
certain combination of the ordering and oscillation parameter values,
as well as the detector exposure time. Ideally in the next step, one
would select one pseudo-experiment within the ensemble at hand,
compute the best fitting solution within each ordering and produce one
ensemble of pseudo-experiments for each hypothesis, in order to build
up the LLR distributions ($\ref{Eq:TwoLLRs}$) and calculate
$\alpha(\beta=0.5)$. Repeating this for each pseudo-experiment in the
original ensemble would yield a distribution of sensitivities, the
median of which would constitute the confidence level at which the
average experiment would be able to exclude the wrong ordering, valid
for the hypothesis assumed to be true in the first place.  The problem
with this approach is the number of pseudo-experiments that would be
needed to be fit in order to adequately sample median sensitivities
and LLR distributions. It would be of the order of $(10^4)^2=10^8$,
for only one possible realization of the NMO, each requiring roughly
two minutes to fit the systematic parameters to the pseudo-experiment.

Instead, we take an approach that is still highly computationally
demanding, but much more feasible. Here, we also generate
pseudo-experiments for a certain fixed hypothesis, but adjust the
procedure laid out above in two aspects. First, only a single true
ordering (TO) LLR distribution is created by directly fitting to these
pseudo-experiments, from which the median LLR for the underlying
hypothesis is determined.  Second, the combination of the atmospheric
parameters ($\theta_{23}$ and $\Delta m^2_{31}$) are adjusted to
determine the realization of the wrong ordering (WO) most resembling
the true one on average.  The test statistic is then built based on
those true ordering and wrong ordering hypotheses only, reducing the
number of pseudo-experiments that are fitted to roughly $10^4$.

\begin{figure}[t]
  \centering
  \subfigure{
   	\includegraphics[scale=0.3]{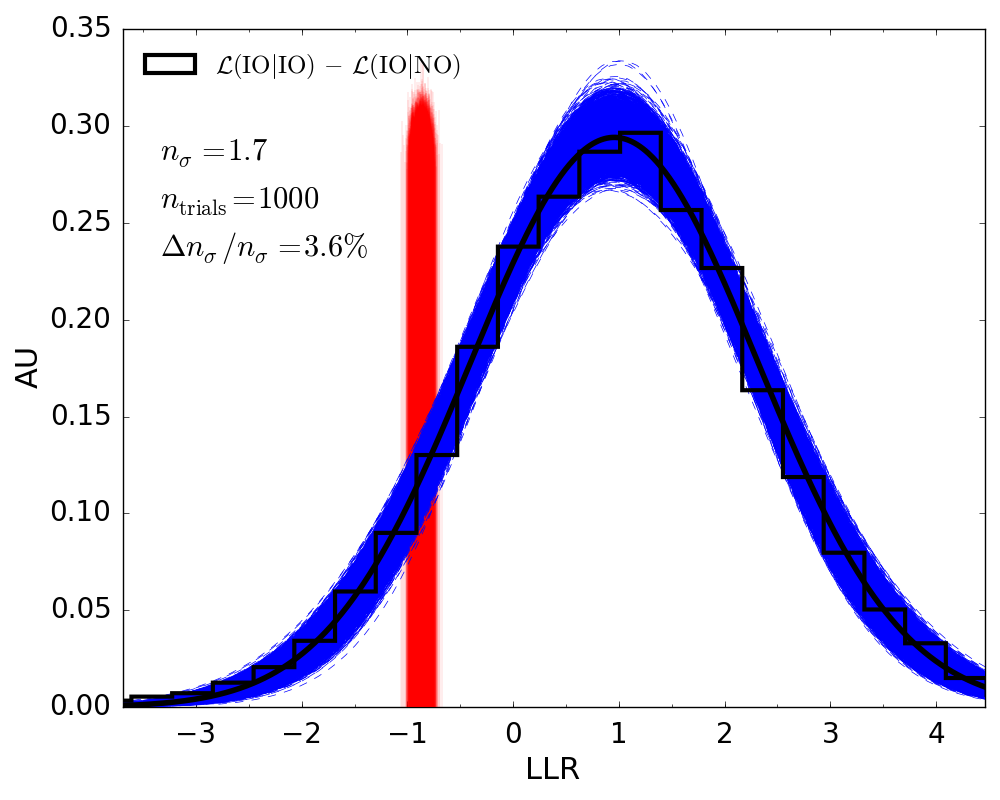}
        \label{fig:LLRSim1k}
  }
  \subfigure{
   	\includegraphics[scale=0.3]{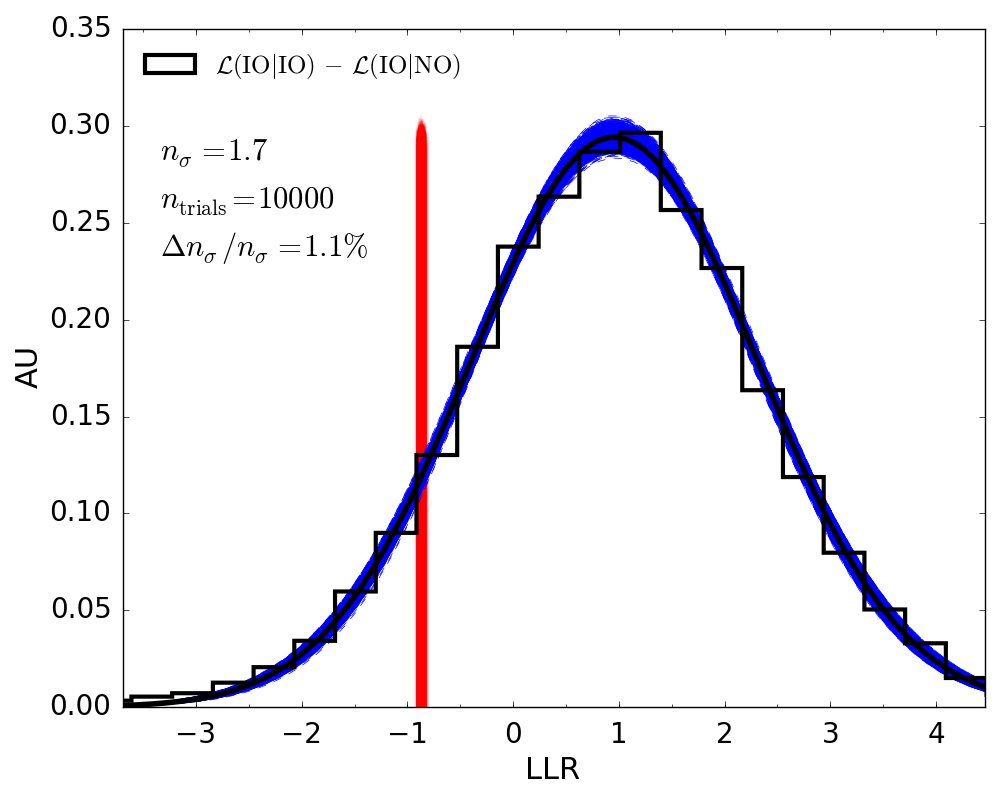}
    \label{fig:LLRSim10k}
  }  
  \caption{10\,000 Gaussian fits (dashed blue) and TO-LLR means (red) from
    randomly sampling the underlying (Gaussian) LLR distributions. The
    left panel shows the case of 1\,000 pseudo-experiments, while in the
    right panel 10\,000 pseudo-experiments were generated. See text for
    details.}
  \label{fig:LLRSim}
\end{figure}

\subsubsection{Estimate of Statistical Uncertainty of Median Sensitivity}
\label{sec:estimate_median}

The fact that the number of available pseudo-experiments is not
infinite means that the quoted median sensitivity is subject to
statistical fluctuations. In general, the statistical uncertainty will
depend mainly on the number of trials $n_\mathrm{trials}$, but also on
the value of the sensitivity itself. In order to ensure that the
relative statistical error is on the order of a few percent or less,
we have repeatedly simulated each LLR procedure $10\,000$ times by
randomly drawing $n_\mathrm{trials}$ samples from the Gaussian fit to
the LLR distribution, refitting those samples, and calculating the
deviations of the obtained sensitivities from the original one. This
allows us to give an estimate of the standard error on the (two-sided)
median sensitivity, given $n_\mathrm{trials}$. Here, we make the
assumption that we can determine the mean of the true-ordering LLR
distribution with an accuracy of
$\sigma_\mathrm{mean}=\sigma_\mathrm{TO}/\sqrt{n_\mathrm{trials}}$,
where $\sigma_\mathrm{TO}$ is the standard deviation of its Gaussian
fit.  Two examples depicting this procedure can be seen in
Fig.~\ref{fig:LLRSim}, where we compare the spreads in the TO-LLR mean
and the variations of the WO-LLR Gaussian fits for the 1-year NO dataset
for the (hypothetical) cases of 1\,000 and 10\,000
pseudo-experiments. The black histogram corresponds to the original
LLR distribution, while the Gaussian fit, assumed to represent the true
shape of the distribution, is also shown in black.  Here, the ten-fold
increase in the number of pseudo-experiments would reduce the relative
statistical uncertainty of the median sensitivity from $3.5\%$ to
$1.1\%$, i.e. by roughly a factor of $\sqrt{10}$ in accordance with
the analytical calculation of that uncertainty.

Tables \ref{tab:TrialsSigVsTime} and \ref{tab:TrialsSigVsTheta23} give
an overview of the number of pseudo-experiments we have processed for
different detector livetimes and true values of $\thTT$, and also show
the obtained median sensitivities including their relative statistical
uncertainties estimated via simulation.

\begin{table}[h]
\centering
\begin{tabular}{c c c c}\hline \hline
livetime (years) & \# Trials & $n_\sigma$ (NO/IO) & $\Delta n_\sigma / n_\sigma$ (NO/IO) (\%) \\\hline
$1$ & 16220 & 1.7/1.5 & 0.9/0.9\\
$4$ & 11120 & 2.9/2.6 & 1.1/1.1 \\
$5$ & 3240 & 3.1/2.8 & 1.9/1.9\\
$10$ & 16000  & 4.2/3.8 & 0.9/0.9\\
\hline \hline
\end{tabular}
\caption{Number of pseudo-experiments generated and fitted for
  different PINGU livetimes at the two NMO fiducial models, together
  with the obtained median NMO sensitivities and their relative
  statistical uncertainties based on the finite number of trials.  No prior is applied for $\theta_{13}$ or $\Delta m^2_{31}$.}
 \label{tab:TrialsSigVsTime}
\end{table}

\begin{table}[h]
\centering
\begin{tabular}{c c c c}\hline \hline
$\theta_{23}$ $(^\circ)$ & \# Trials & $n_\sigma$ (NO/IO) & $\Delta n_\sigma / n_\sigma$ (NO/IO) (\%) \\\hline
$39$ & 2000 & 2.6/2.8 & 2.4/2.4\\
$42$ & 1930 & 3.2 (IO)  & 2.5 (IO) \\
$42.3$ & 11120 & 2.9 (NO) & 1.1 (NO) \\
$45$ & 1500 & 4.4/3.3 & 2.8/2.8\\
$47$ & 2000 & 5.5/2.9 & 2.5/2.4\\
$49$ & 2000 & 6.7 (NO) & 2.4 (NO)\\
$49.5$ & 11120 & 2.6 (IO) & 1.1 (IO)\\
$52$ & 2000 & 7.2/3.0 & 2.5/2.4\\
\hline \hline
\end{tabular}
 \caption{Same as Tab.~\ref{tab:TrialsSigVsTime}, but here for a fixed
   livetime of four years and different true values of $\theta_{23}$
   in both NMOs (unless given otherwise).}
 \label{tab:TrialsSigVsTheta23}
\end{table}

\newpage

\subsection{Asimov Analysis} 
\label{sec:AsimovAnalysis}

Our second analysis method makes use of the $\deltaChiSq$ metric also
used extensively in the literature. It is based on the usual
definition of $\chi^2$ as the weighted sum of squared differences
between $n$ measurements (bins) $x_i$ drawn from the normal
distributions with standard deviation $\sigma_i$
and corresponding median outcomes
$\mu_i$,
\begin{equation} 
   \chi^2 = \sum_{i=1}^{n} \frac{\left(x_i - \mu_i
\right)^2}{\sigma_i^2}\; .  
\end{equation} 
When the model predictions $\mu_i$ are dependent on a set of $P$
parameters $\{p_i\}$, one usually considers the
minimum 
\begin{equation} 
  \chi^2_\mathrm{min} \equiv \min_{\{p_i\}}
  \chi^2
\end{equation} 
with respect to $p_i$. The quantity $\chi^2$ can be minimized with
respect to the (nuisance) parameters either using the NO or the IO
hypothesis, leading to the two minima
\begin{equation} 
  \chi^2_{\mathrm{NO(IO)}}\equiv
  \min_{\{p_i\}\, \in\, \mathrm{NO(IO)}} \chi^2\;
  .  
  \label{Eq:Chi2Min} 
\end{equation} 
The test statistic 
\begin{equation} 
  \Delta\chi^2\equiv\chi^2_{\mathrm{NO}}
  -\chi^2_{\mathrm{IO}} 
  \label{Eq:DeltaChi2} 
\end{equation} 
is then employed to determine which one of the two ordering hypotheses
constitutes the better fit to the data, with $\deltaChiSq<0$
$(\deltaChiSq>0)$ favouring the NO (IO) hypothesis, in close analogy
to the LLR method detailed in the previous section.

In the so-called ``Asimov"\footnote{This is a reference to the Isaac
  Asimov short story \textit{Franchise}, in which the single most
  typical voter replaces the full electorate.} approach, the randomly
sampled pseudo-data $x_i$ are replaced by the exact model predictions
$\mu_i$ at the nominal parameter values $p_0$ within the corresponding
true ordering:
\begin{equation} 
  \overline{\Delta\chi^2_\mathrm{TO}}(p_0) \equiv
  \min_{p\, \in\, \mathrm{WO}} \sum_{i=1}^n
  \frac{\left(\mu_i^{\mathrm{TO}}(p_0) -
    \mu_i^{\mathrm{WO}}(p)\right)^2}{\sigma^2_i}\;
  , 
  \label{Eq:AsimovChi2}
\end{equation} 
where the minimization is performed over the parameters $p$ within the
wrong ordering. Since by construction $\chi^2_\mathrm{TO}=0$ when
using the Asimov data set, $\overline{\Delta\chi^2_\mathrm{IO}}$ is
the equivalent of the test statistic defined in
Eq.~(\ref{Eq:DeltaChi2}), while $\overline{\Delta\chi^2_\mathrm{NO}}$
only differs by the sign.\footnote{This is the reason for denoting
  this quantity by ``$\deltaChiSqBar$", and not by
  ``$\deltaChiSq$". Also, be aware that the subscript denotes the
  assumed true ordering and not the one that is minimized over, as in
  Eq.~(\ref{Eq:Chi2Min}), since minimization over the parameters
  within the WO will always be implied.} The Gaussian prior penalty
term added to $\overline{\Delta \chi^2_\mathrm{TO}}$ is $-2\,
\mathcal{L}_\mathrm{prior}$.

Essentially, this technique assumes that statistical fluctuations in
the experimental data are as likely to reinforce as to obscure the
signature of the correct ordering, so that only the single data set
most likely to be observed for any given set of oscillation parameters
needs to be analysed.

Usually, $\overline{\Delta \chi^2}$ is transformed into an actual
sensitivity by evaluating it under the assumption that it is
$\chi^2$-distributed with 1 degree of freedom, so that the number of
Gaussian standard deviations at which a given ordering can be
identified is simply $n_\sigma = \sqrt{\overline{\Delta \chi^2}}$. As
detailed in Ref.~\cite{Ciuffoli:2013rza}, this approach is not well
justified due to the discrete nature of the neutrino mass ordering,
but it turns out that a precise frequentist definition of the
sensitivity for any desired rejection power is possible under the
Gaussian approximation for the distribution of the test statistic
$\Delta \chi^2$ \cite{Blennow:2013oma}. In these references it is
shown that $\chi^2_\mathrm{TO}$ is $\chi^2$ distributed with $n-P$ (P
being the number of parameters) d.o.f., while $\chi^2_\mathrm{WO}$
follows a non-central $\chi^2$ distribution, and as a result
\begin{equation} 
  \Delta \chi^2 = \mathcal{N}\left(\pm
  \overline{\Delta \chi^2},\,2\sqrt{\overline{\Delta
      \chi^2}}\right)\,, \textrm{ with + (-) for IO
    (NO)} 
  \label{Eq:DeltaChi2Gauss}
\end{equation} 
is Gaussian distributed, with mean $ \overline{\Delta \chi^2}$ and
standard deviation $2\sqrt{\overline{\Delta
    \chi^2}}$~\cite{Blennow:2013oma}. $ \overline{\Delta \chi^2}$
depends on the true parameter values within the given mass ordering,
especially the mixing angle $\theta_{23}$, which determines the
amplitude of the matter effect and thus significantly alters the event
rates.

Following Ref.~\cite{Blennow:2013oma}, in order to obtain an estimate
of the median sensitivity under the Gaussian assumption that is
consistent with our Monte Carlo based approach of generating
pseudo-experiments laid out in \ref{sec:MedSensLLR}, one has to start
off with the critical value $\Delta\chi^{2}_\alpha$, such that a
certain ordering hypothesis can be rejected at the Confidence Level CL$(1-\alpha)$. From
the definition (\ref{Eq:DeltaChi2}) it is clear that in the case of
true NO, we would decide to reject the IO hypothesis at CL($1-\alpha$)
if $ \Delta\chi^2<\Delta\chi^{2}_{\alpha}$, i.e.
$\alpha=P(\Delta\chi^2<\Delta\chi^{2}_{\alpha})$. Using
Eq.~(\ref{Eq:DeltaChi2Gauss}) one finds 
\begin{eqnarray} 
  \alpha =
  \frac{1}{2}\,\mathrm{erfc}\left(\frac{\overline{\Delta
      \chi^2_\mathrm{IO}}-\Delta
    \chi^2_\alpha}{\sqrt{8\overline{\Delta
        \chi^2_\mathrm{IO}}}}\right)\textrm{ ,}\\ \beta =
  \frac{1}{2}\,\mathrm{erfc}\left(\frac{\overline{\Delta
      \chi^2_\mathrm{NO}}+\Delta
    \chi^2_\alpha}{\sqrt{8\overline{\Delta
        \chi^2_\mathrm{NO}}}}\right)\textrm{ ,} 
\end{eqnarray} 
so that requiring $\beta=0.5$ implies $\Delta \chi^2_\alpha =
-\overline{\Delta \chi^2_\mathrm{NO}}$. Applying the two-sided
convention to convert $\alpha$ into a number of Gaussian standard
deviations, the median sensitivity is
\begin{equation} 
  n_\sigma =
  \sqrt{2}\,\mathrm{erfc}^{-1}\left[\frac{1}{2}\,\mathrm{erfc}\left(\frac{\overline{\Delta
  \chi^2_\mathrm{IO}}+\overline{\Delta
  \chi^2_\mathrm{NO}}}{\sqrt{8\overline{\Delta
  \chi^2_\mathrm{IO}}}}\right)\right] \label{Eq:AsimovMedianSensitivity}
\end{equation} 
for true normal ordering. The true inverted ordering case is
calculated analogously.

Both $\overline{\Delta \chi^2_\mathrm{IO}}$ as well as
$\overline{\Delta \chi^2_\mathrm{NO}}$ above will depend on the true
parameter values $p_0$ within the respective hypotheses. This means
that we actually need to perform two fits for any given true
hypothesis TO, with injected values $p^\mathrm{TO}_0$. Again sticking
to true NO for illustration purposes, first we determine
$\overline{\Delta\chi^2_\mathrm{NO}}(p^\mathrm{NO}_0)$
(Eq.~(\ref{Eq:AsimovChi2})). The resulting best fitting parameters
under the IO hypothesis then serve as the IO fiducial model
$p^\mathrm{IO}_0$ for which the expected experimental outcome under
the IO is computed, from which in turn
$\overline{\Delta\chi^2_\mathrm{IO}}(p^\mathrm{IO}_0)$ is
obtained. The median sensitivity then follows directly from
Eq.~(\ref{Eq:AsimovMedianSensitivity}).

\begin{figure}[t]
  \centering
  \subfigure{
   	\includegraphics[scale=0.42]{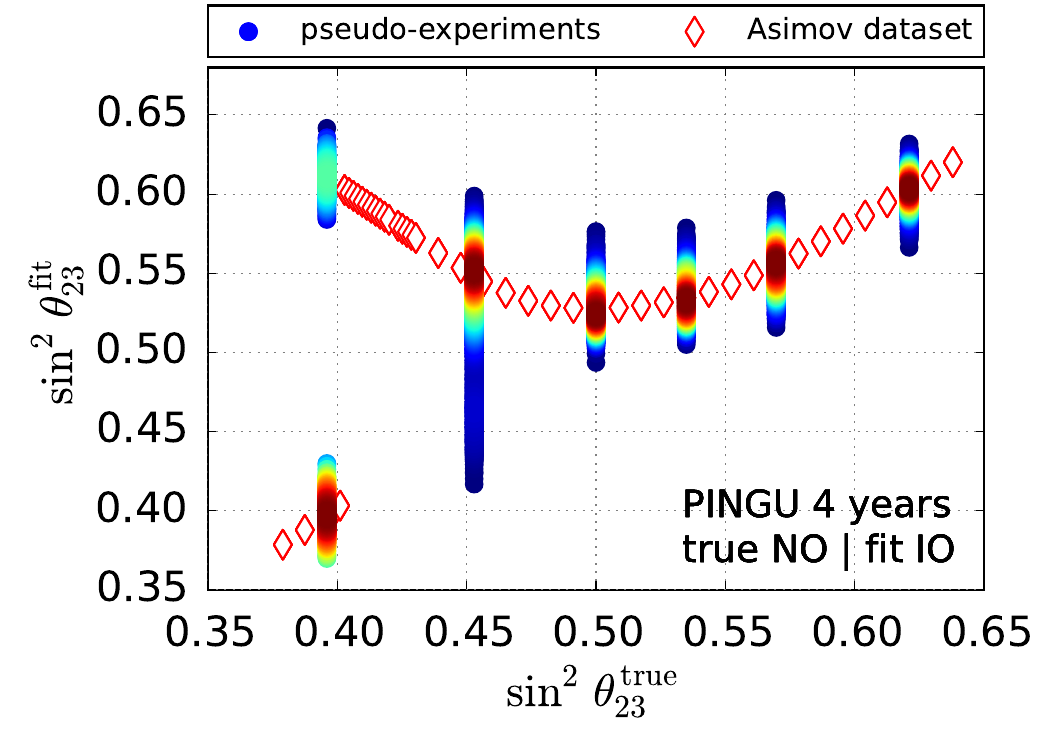}
        \label{fig:theta23trueNOhypoIO}
  }
  \subfigure{
   	\includegraphics[scale=0.42]{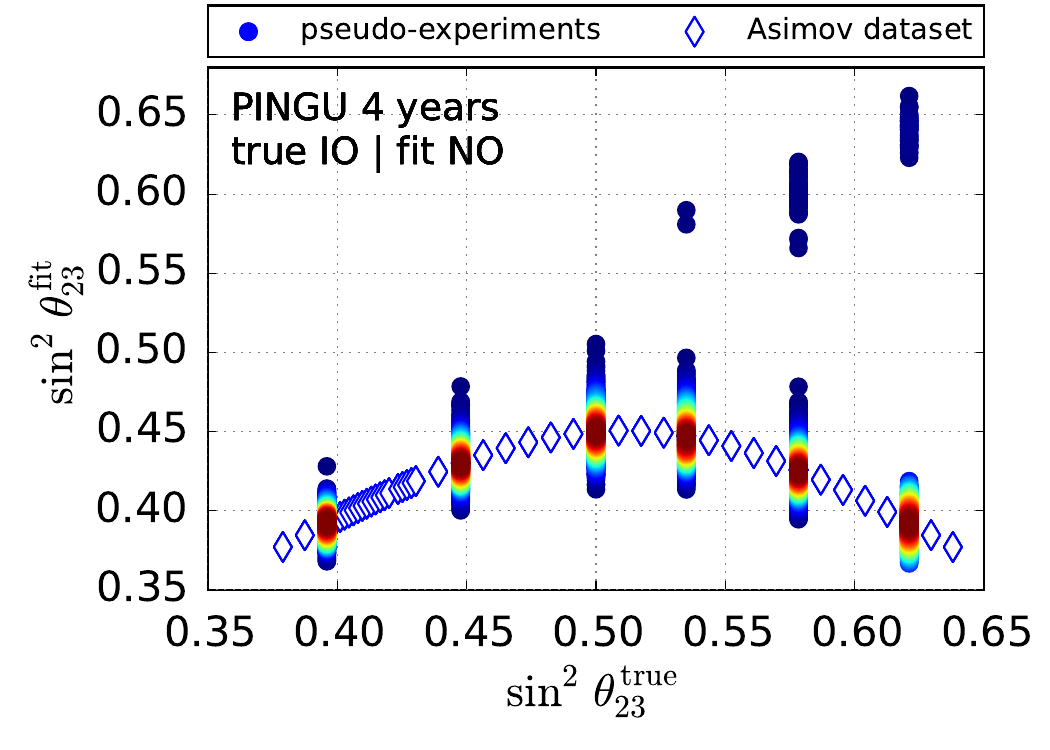}
    \label{fig:theta23trueIOhypoNO}
  }  
  \caption{Distribution of best fit values $\sin^2
    \theta^\mathrm{fit}_{23}$ as a function of the true value $\sin^2
    \theta^\mathrm{true}_{23}$, when the NMO is misidentified. Both
    the outcomes of invididual pseudo-experiments (point density
    color-coded) as well as fits to the corresponding Asimov datasets
    are shown, for true normal ordering in the left panel, and for
    true inverted ordering on the right.}
  \label{fig:theta23fitvstrue}
\end{figure}

The usual $\sqrt{\overline{\Delta \chi^2}}$ prescription frequently
encountered in the literature is recovered if the one-sided convention
is employed to convert from $\alpha$ to $n_\sigma$, and if the moduli
of the means of the two test statistic distributions are approximately
the same, i.e.  $\overline{\Delta \chi^2_\mathrm{IO}} \approx
\overline{\Delta \chi^2_\mathrm{NO}}$. We find that this approximation
does not hold for $\sin^2 \theta_{23}\gtrsim 0.5$ and the case in
which NO is true, where it actually leads to an underestimation of the
sensitivity. The reason is that while $\overline{\Delta
  \chi^2_\mathrm{NO}}$ continues growing in this region of parameter
space, $\overline{\Delta \chi^2_\mathrm{IO}}$ stays small due to the
ordering-octant degeneracy: the best fit $\theta_{23,\mathrm{IO}}$
lies in the second octant, which means that the expected experimental
outcome under the wrong (inverted) ordering hypothesis is reasonably
well fit with $\theta_{23,\mathrm{NO}}$ in the first octant. This
behavior is depicted in Fig.~\ref{fig:theta23fitvstrue}: here we
compare the fit results for $\sinsqTT$ based on the Asimov
dataset to the outcomes of the individual pseudo-experiments in the
LLR method.

\begin{figure}[t]
  \centering
  \subfigure{
   	\includegraphics[scale=0.37]{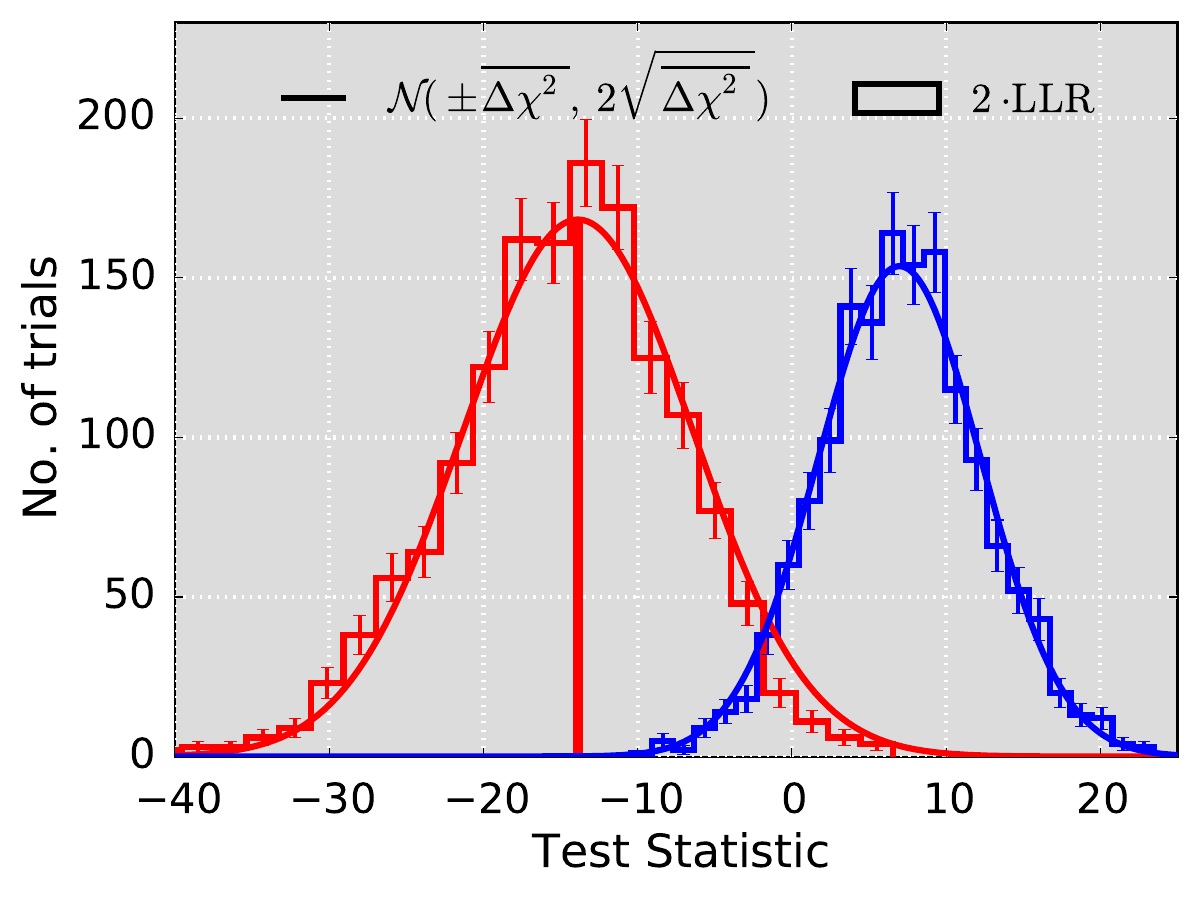}
        \label{fig:Chi2GaussVsLLRNOMaxMix}
  }
  \subfigure{
   	\includegraphics[scale=0.37]{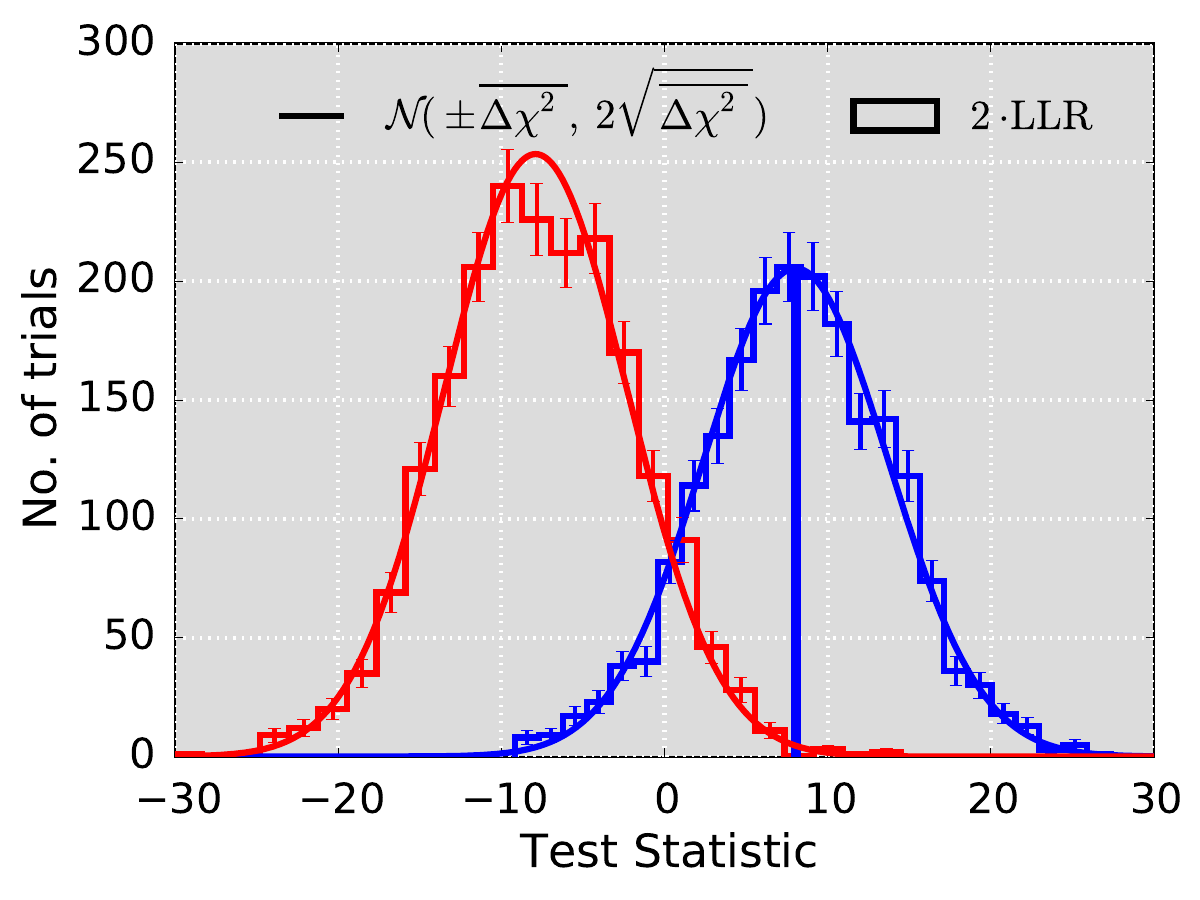}
    \label{fig:Chi2GaussVsLLRIOOscOnly}
  }  
  \caption{Gaussian approximations for the distribution of $\Delta
    \chi^2$, together with the corresponding LLR distributions (scaled
    by a factor of two) obtained from pseudo-experiments. The left
    panel shows a representative dataset assuming the NO, while the
    right panel shows a case where the IO is true. Only two systematic
    parameters are included, see text for details.}
  \label{fig:Chi2GaussVsLLR}
\end{figure}

In Fig.~\ref{fig:Chi2GaussVsLLR} we provide an example of how the
distribution of the LLR test statistic obtained from performing
several thousands of pseudo-experiments compares to the Gaussian
approximation (\ref{Eq:DeltaChi2Gauss}) of $\Delta \chi^2$. Here, we
show two representative datasets; the left panel corresponds to the
true NO case and maximal mixing, for four years of detector livetime,
while the right panel displays the distributions obtained for true IO,
with $\thTT=49.5^\circ$.  Only the atmospheric oscillation parameters
are included as systematics
(cf. Tab.~\ref{Tab:SystematicsImpact}). The vertical line in each
panel is drawn at the position of the mean of the $\Delta
\chi^2_\mathrm{TO}$ distribution; its absolute value corresponds to
the ``Asimov-$\Delta\chi^2$", i.e.  $\overline{\Delta
  \chi^2_\mathrm{TO}}$ defined in Eq.~(\ref{Eq:AsimovChi2}).

\subsubsection{Notes on the Validity of the Gaussian Approximation 
(\ref{Eq:DeltaChi2Gauss})}
In the scheme we describe above it is not guaranteed that all the
assumptions upon which Eq.~(\ref{Eq:DeltaChi2Gauss}) rests indeed
hold. This applies especially to the case in which there is a tight
prior penalty term on one or more parameters, when the WO fiducial
model is tested, with corresponding ``true" fiducial values
$p^\mathrm{TO}_0$.  Illustratively, the reason is that the injected
value of the i-th parameter $p^\mathrm{TO}_{0,i}$ does not necessarily
correspond to the central value of its prior (since it is obtained
from a fit to the alternative ordering), so that the minimum within
the same ordering will be pulled away from $p^\mathrm{TO}_{0,i}$, and
be systematically biased.  As a result,
$\chi^2_\mathrm{TO}(p^\mathrm{TO}_0)$ will follow a non-central
$\chi^2$ distribution also, and the expression
(\ref{Eq:DeltaChi2Gauss}) is no longer valid. The modification of
$\Delta \chi^2$ is not as trivial as just a shift of its mean by
$\overline{\chi^2_\mathrm{TO}(p^\mathrm{TO}_0)}$ however, as it also
affects the higher percentiles. As a consequence, for studies using
the current NuFit prior on $\Delta m^2_{31}$ we only rely on
pseudo-experiments as in the LLR methodology to determine the
significance.

\begin{figure}[t]
  \centering
  \includegraphics[scale=.45]{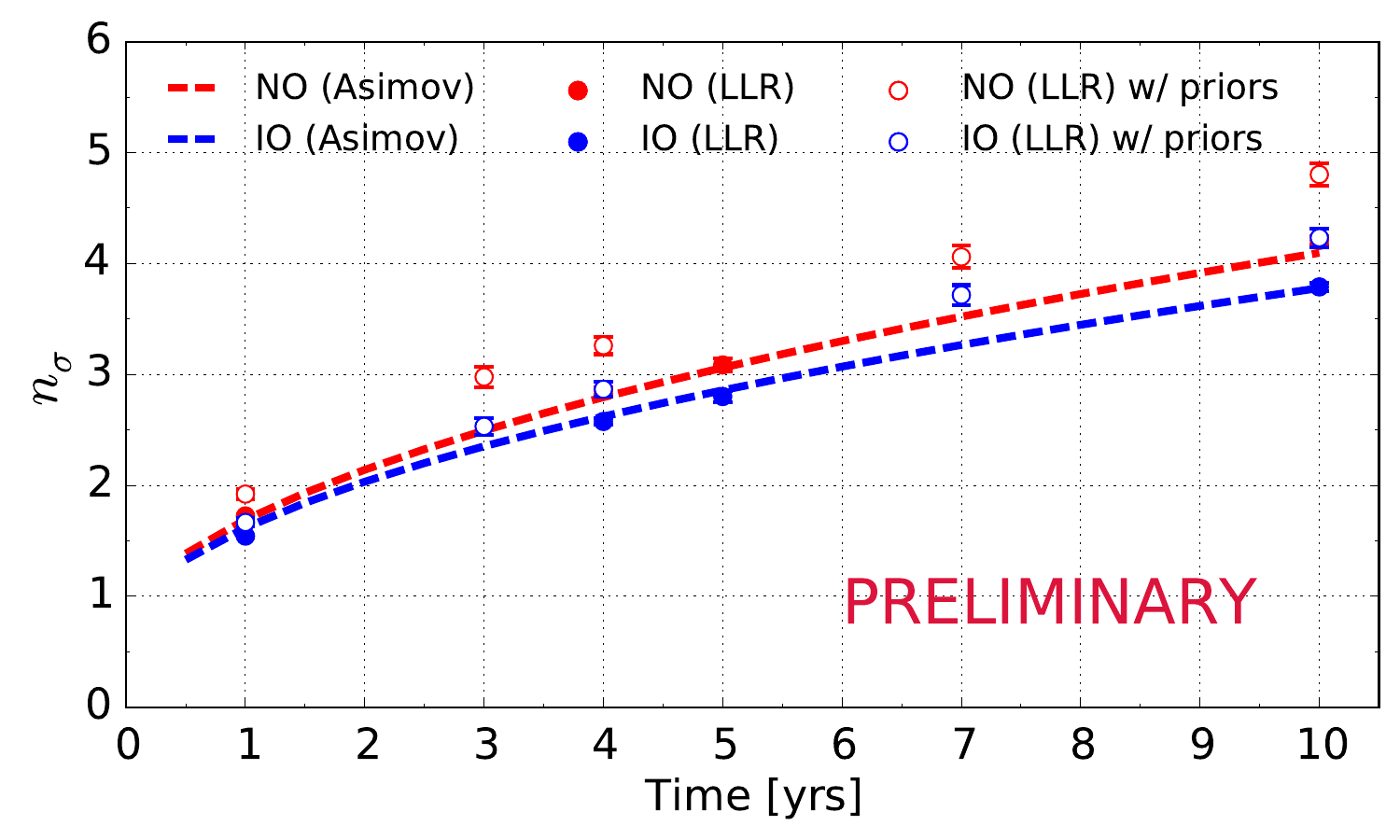}
  \caption{Projected evolution of median NMO sensitivity as a function
    of full detector livetime, in number of Gaussian standard
    deviations as defined in Eq.~(\ref{eq:DoubleSidedSignificance}),
    for both true normal (red) and true inverted (blue) ordering. For
    the case of no external constraints on $\thTT$ and $\Delta
    m^2_{31}$, results from the Asimov method are shown as dashed
    lines, while LLR results are shown as filled points. Circles
    represent the sensitivities obtained in the LLR method with
    NuFit~2.0~\cite{NuFIT20} constraints applied to the two
    oscillation parameters in addition.  See text for details.}
  \label{fig:SigmaVsTime}
\end{figure}

Fig.~\ref{fig:SigmaVsTime} shows the expected evolution of the NMO
sensitivity as a function of full detector livetime both without and
with including current knowledge of the oscillation parameters $\thTT$
and $\Delta m^2_{31}$. In the former case we compare the significances
yielded by the Gaussian approximation of the $\Delta \chi^2$ test
statistic to those obtained from fitting Poisson-fluctuated
pseudo-experiments in the LLR approach (cf.
Tab.~\ref{tab:TrialsSigVsTime}). In the latter, however, only results
from the LLR method are shown due to the limitations of the Gaussian
approximation (\ref{Eq:DeltaChi2Gauss}) pointed out above. The prior
on $\dmTO$ is assumed to be Gaussian, with central value and standard
deviation taken from \cite{NuFIT20}, and chosen according to the NMO
assumption of the fit. The (non-Gaussian) constraints placed on
$\thTT$ are shown in the lower panel of Fig.~\ref{fig:SigmaVsTheta23}.
We apply the according prior to $\thTT$ whenever it is fit within a
given NMO, but remove the overall penalty of about one unit in $\Delta
\chi^2$ when NO is assumed in the fit. Even though one would not
expect different shapes for NO and IO if current oscillation data were
not sensitive to the ordering, this procedure allows us to capture the
impact of the current constraints on $\thTT$ while retaining the two
global best fits as central values and without introducing an external
preference for the IO in the NMO determination.

Both priors are implemented as penalty terms added to
Eq.~\ref{Eq:loglikelihood}, and are applied when fitting both NMO
assumptions to pseudo-experiments, and when searching for the
wrong-ordering hypothesis that mimics the average true ordering
experiment the most. The observed increase in NMO sensitivity with
respect to the case of no external constraints on $\dmTO$ and $\thTT$
is on the order of $10-20\%$.

\subsection{Atmospheric Oscillation Parameter Sensitivity}
In Sec.~\ref{sec:MuonNeutrinoDisappearance} we determine regions of
various two-dimensional confidence levels for the atmospheric
oscillation parameters $(\Delta m^2_{31}, \sin^2 \thTT)$ by generating
the mean experimental outcome under the ``true" model and evaluating
$\chi^2_\mathrm{TO}$ defined in Eq.~(\ref{Eq:Chi2Min}) on a fine grid
in the plane of the two parameters, assuming the correct NMO in the
fit. Under the assumption that $\chi^2_\mathrm{TO}$ follows a $\chi^2$
distribution with 2 d.o.f., each iso-contour is easily converted into
a corresponding C.L. at which an experiment will yield a combination
of $(\Delta m^2_{31}, \sin^2 \thTT)$ contained within.

Since it is not guaranteed that the distribution follows that of a
$\chi^2$ with a given number of d.o.f., we have compared the
C.L. associated with different iso-contours to their true
coverage. This can be done by determining the outcomes of a large
number of pseudo-experiments, as mentioned in
Sec.~\ref{sec:NumuDisappearance:Results} and depicted in
Fig.~\ref{fig:PseudoExpVsScanResults}, and then finding the fraction
of outcomes a given contour of constant $\chi^2_\mathrm{TO}$ actually
contains.

\begin{figure}[t]
  \centering
  \includegraphics[scale=.45]{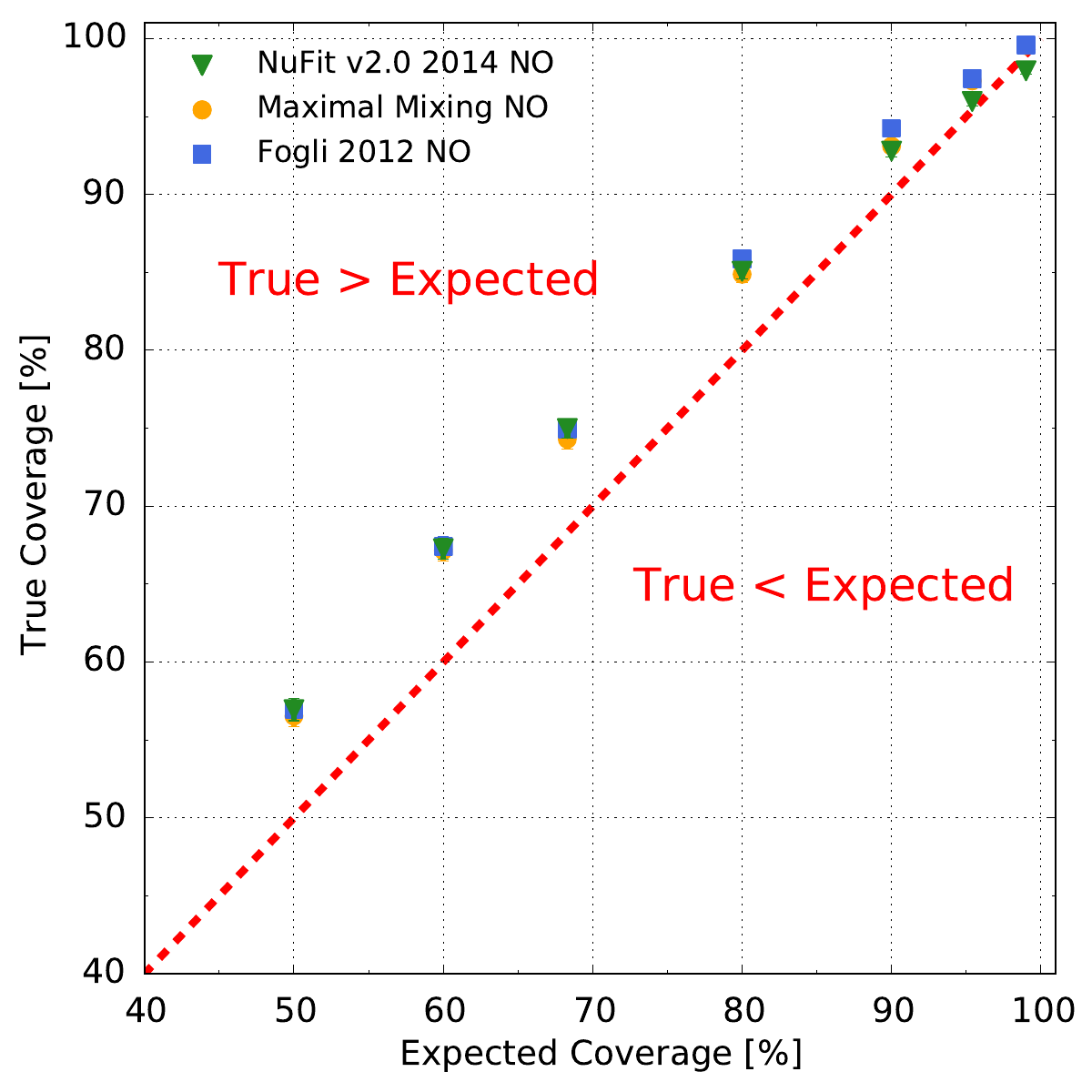}
  \caption{True coverage of various joint confidence regions for the
    atmospheric oscillation parameters $(\Delta m^2_{31}, \sin^2
    \thTT)$ from the Asimov approach, as a function of the expected
    confidence level obtained according to a $\chi^2$ distribution
    with 2 d.o.f. See text for details.}
  \label{fig:contourcoverage}
\end{figure}

The result of this study is shown in Fig.~\ref{fig:contourcoverage},
for the same three injected values of $\sin^2 \thTT$ used in
Sec.~\ref{sec:MuonNeutrinoDisappearance}, and true NO. As one can see
from the plot, the fractions of pseudo-experimental outcomes found
inside the iso-contours exceed the expectations based on assuming a
$\chi^2$ distribution with 2 d.o.f. up to about $95\%$ (expected)
C.L., implying that the extents of the respective contours are
conservative. For true IO, we do not anticipate a qualitatively
different picture.

\clearpage
\resetlinenumber

\IfFileExists{NewCommands.tex}       {}       {}
\IfFileExists{../NewCommands.tex}    {}    {}
\IfFileExists{../../NewCommands.tex} {} {}

\graphicspath{{figures/}{Appendices/figures/}}

\addcontentsline{toc}{section}{Appendix}
\addtocontents{toc}{\protect\setcounter{tocdepth}{-1}}

\section{Neutrino Cross-section and Model Uncertainties} \label{sec:GENIESysts}

Neutrino interaction models play an important role in oscillation analyses; the oscillation probability is inferred from the event rate, which is a convolution of the flux, cross section, efficiency and oscillation probability. In the PINGU mass ordering analysis, as an example, the relative difference in cos(zenith) vs. E space for neutrinos and antineutrinos is used to infer the mass ordering.  Flavor separation of charged current (CC) $\numu$ interactions and $\nue$, $\nutau$ into track and cascade samples also introduces dependence on the relative population of these interactions, as predicted by the interaction model and efficiency. In addition, neutral current (NC) interactions contribute to the backgrounds in cascade samples.  The uncertainties on these models therefore play an important role in the determination of the physics results obtained from the neutrino analyses.

\subsection{Neutrino interaction processes from 1-100 GeV}

PINGU is sensitive to neutrino and antineutrino interactions above a few GeV, as shown by Fig.~\ref{fig:pingu_effvol} which shows the effective areas divided into neutrinos and antineutrinos.  Both of these interactions are then important to consider in the final analysis of the event rates.

%
%
\begin{figure}
\begin{center}
\includegraphics[width=0.7\columnwidth]{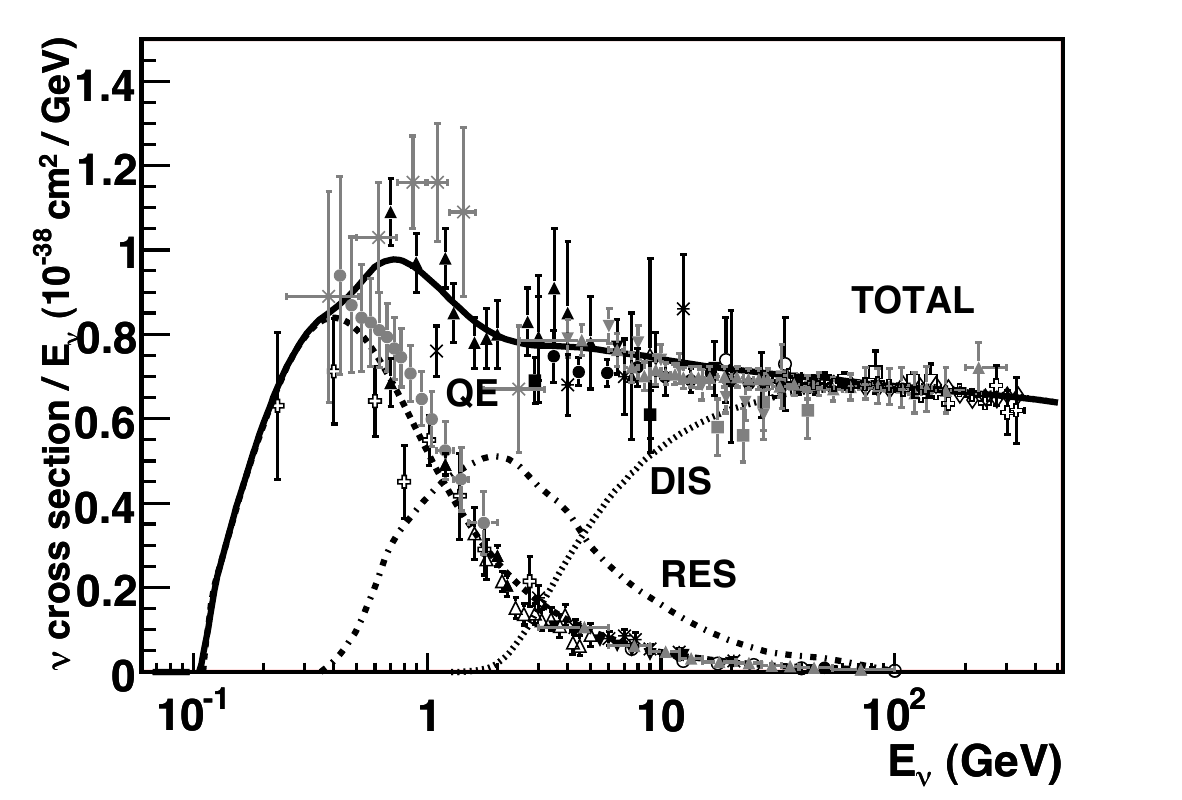} 
\includegraphics[width=0.7\columnwidth]{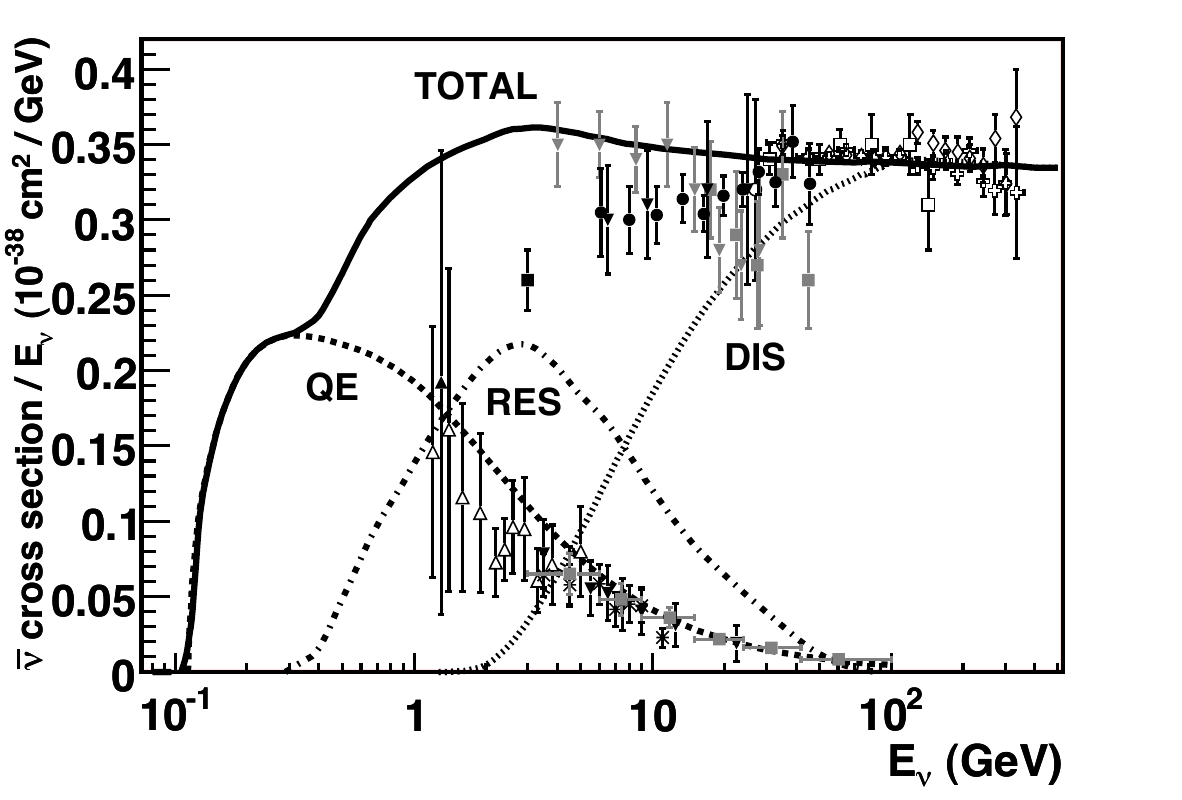} 
\end{center}
\caption{Total neutrino and antineutrino per nucleon CC cross sections  
      (for an isoscalar target) divided by neutrino energy and 
      plotted as a function of energy.  Contributions include quasi-elastic scattering (dashed), resonance 
      production (dot-dash), and deep inelastic scattering (dotted). 
      Example predictions for each are provided by the NUANCE 
      generator~\cite{Casper:2002sd}. Note that the quasi-elastic scattering
      data and predictions have been averaged over neutron and proton
      targets and hence have been divided by a factor of two for the
      purposes of this plot.}
    \label{fig:totalxsec}
\end{figure}

Figure~\ref{fig:totalxsec}, copied from Ref~\cite{RevModPhys.84.1307}, shows the neutrino and antineutrino cross section processes as a function of neutrino (or antineutrino) energy.  The dominant processes in the PINGU event sample, as shown later, are CC and NC deep inelastic scattering (DIS) interactions, in which a neutrino ($\nu_{\ell}$) strikes a quark in a nucleon ($N$), which produces a hadronic shower, $X$, and the corresponding flavor lepton partner ($\ell=e,\mu,\tau$) in the final state:
\begin{linenomath*}
\begin{equation}
\nu_{\ell} + N \rightarrow \ell^- + X
\end{equation}
\end{linenomath*}

In the case of antineutrino interactions $\bar{\nu}$, the final state lepton will be positively charged ($\ell^+$); we note PINGU is  insensitive to the charge of the final state lepton and separates antineutrino from neutrino interactions through the neutrino and antineutrino cross sections and relative initial flux.

At energies of a few GeV the most significant contribution to the total cross-section is from CC resonant (RES) interactions, in which the neutrino interacts with a nucleon,  exciting a resonance, which decays to a meson, and predominantly to a pion: 
\begin{linenomath*}
\begin{equation}
\nu_{\ell} + N \rightarrow \ell + N^{\prime} + \pi,
\end{equation}
\end{linenomath*}
where $N^{\prime}$ represents the final state nucleon. 

Below 1 GeV, the dominant process is CC quasi-elastic scattering (CCQE):
\begin{linenomath*}
\begin{equation}
\nu_{\ell} + N \rightarrow \ell + N^{\prime},
\end{equation}
\end{linenomath*}

There are also NC analogs of each of these processes, which result in a neutrino in the final state instead of the charged lepton.  Table ~\ref{tab:eventrate} lists the contributions to the total event rate for the default simulation, separated by flavor and assuming the Honda flux model.

\subsection{The GENIE event generator}

GENIE is a software package~\cite{Andreopoulos:2009rq} which simulates neutrino interaction; GENIE stands for Generates Events for Neutrino Interaction Experiments. For the PINGU analysis studies discussed here, the version of GENIE  used was 2.8.6.  
That this version of GENIE includes a bug fix to the GRV98 PDF, having $\approx1\%$ effect on the total cross-section.

 Table~\ref{tab:geniesysttab} summarizes the errors on the parameters in GENIE believed to be the most significant for PINGU. These are taken directly from the GENIE Users Manual 
 \cite{Andreopoulos:2015wxa}). No justification is provided in the GENIE documentation, however in the following sections we provide additional information to explain why these parameters are the most significant and why others may be disregarded at this time. 

\begin{table}
\centering
\begin{tabular}{l c c c} \hline \hline
Interaction & $\numu$\&$\numubar$ & $\nue$\&$\nuebar$ & $\nutau$\&$\nutbar$ \\ \hline
CC & 83.8 & 84.9 & 61.9 \\
\hspace{0.5cm}QE & 9.8 & 9.7 & 6.5 \\
\hspace{0.5cm}RES & 13.7 & 13.3 & 9.8 \\
\hspace{0.5cm}COH & 0.7 & 0.8 & 0.03 \\
\hspace{0.5cm}DIS & 59.5 & 61.1 & 45.6 \\
\hspace{0.5cm}Other & 0.0 & 0.0 & 0.0 \\
NC & 16.2 & 15.1 & 38.1 \\
\hspace{0.5cm}QE & 0.004 & 0.007 & 0.04 \\
\hspace{0.5cm}RES & 0.3 & 0.3 & 1.0 \\
\hspace{0.5cm}COH & 0.2 & 0.2 & 0.5 \\
\hspace{0.5cm}DIS & 15.8 & 14.6 & 36.6 \\
\hspace{0.5cm}Other & 0.02 & 0.0 & 0.0 \\
\hline \hline
\end{tabular}
\caption{The breakdown of the event rate, in percentage, using the default GENIE simulation for each neutrino flavor.  This table was made using the Honda flux model.} \label{tab:eventrate}
\end{table}


\begin{table}
\centering
\begin{tabular}{c c c}\hline \hline
Name & nominal value  & uncertainty (\%) \\\hline
$M_A^{CCQE}$    & 0.99 & $-15,+25$ \\
$M_A^{RES}$    & 1.120 & $\pm 20$ \\ 
$A^{BY}_{HT}$    & 0.538 &  $\pm 25$ \\ 
$B^{BY}_{HT}$    & 0.305 &  $\pm 25$ \\ 
$C^{BY}_{V1u}$    & 0.291 &  $\pm 30$ \\ 
$C^{BY}_{V2u}$    & 0.189 &  $\pm 30$ \\ 
\hline \hline
\end{tabular}
\caption{List of parameters and their associated uncertainties based on GENIE. Nominal values of parameters from Appendix B of the GENIE User Manual, Section 8.1.} \label{tab:geniesysttab}
\end{table}


\subsection{Uncertainties not considered}

Table~\ref{tab:eventrate} shows that several of the potential process do not contribute significantly to the overall event rate and therefore need not be considered in the analysis.  These include the coherent production channel for both CC and NC events (contributing 0.7 and 0.2\% respectively) and elastic scattering for NC events (totalling 4$\times$10$^{-3}$\%).  Grouped into the ``Other'' category are single kaon, $\eta$, and $\Lambda$ production as well as inverse muon decay.  These events may be considered for a future analysis and can be either neglected or included with a large uncertainty.



The total DIS cross-section is known at approximately the 3-5\% level.  We do not consider this as a source of systematic uncertainty because a floating normalization on the dominant channel would be degenerate with the neutrino or antineutrino normalization.  Therefore, for this analysis we focus on aspects which may affect the relative difference between neutrino or antineutrino event rates, including:
\begin{itemize}
\item Possible variations in the cross section model which may affect the shape of the predicted event spectrum with energy
\item Differences between channels (CC vs. NC)
\item Differences between flavors 
\end{itemize}

Due to the technical implementation of the cross section systematics into the analysis (via effective area) it is not possible to study the effect of uncertainties on the hadronic final state. However, systematic errors from the hadronic system are rather minor in this analysis~\cite{Katori:2014fxa}.

While we recognize the importance of CC vs. NC uncertainties, we do not consider any additional uncertainties on this ratio for DIS, RES or QE. According to Table~\ref{tab:eventrate}, the only significant contribution to the analysis comes from NC DIS interactions and the Bodek-Yang model parameters, discussed in later sections, affect both CC and NC cross sections as implemented in GENIE.

This analysis is also constrained by the error propagation method used in GENIE,
a technique often called ``re-weighting''.   This technique is applied in GENIE
for a set of physics parameters $p$.  For each $p$ that can be reweighted, there
is a corresponding systematic parameter $x_p$.  Tweaking this systematic
parameter modifies the physics parameter $p$ as:


\begin{equation}
  p \rightarrow p(x_p) = p\cdot(1+x_p \frac{\delta p}{p})
\end{equation}
where $\delta p$ is the estimated standard deviation of $p$. After tweaking $p$ by some value, a new cross section is calculated.  This allows one to see the effect on the cross section as compared to the nominal cross section~\cite{Andreopoulos:2009rq}. 
Because the new cross-sections are weighted to the nominal cross-section with different weights, there are no effects on the reconstructed variables such as energy.

The only uncertainties considered on the neutrino cross-section models are those on the QE, RES, and DIS models, which will be addressed in the following sections.

\subsection{Uncertainties on QE and RES model}
For the greater neutrino community, most experiments are at lower energies than the region of interest for PINGU (less than a few GeV) where cross section uncertainties are very important.  In those experiments QE and resonant events dominate the sample and the uncertainties on these types of interactions are relatively high.   Although PINGU will be dominated by DIS events, we have addressed the uncertainties associated with the QE interactions.

The largest effect on the overall shape of CCQE events is driven by uncertainties on the axial mass in the axial form factor, $M_A^{CCQE}$. It is suspected that $M_A^{CCQE}$ is merely an effective parameter; it reproduces the right differential cross section at low energy but is masking nuclear physics processes which are not included in the current GENIE model. The two extreme effects on the analysis are either a raw increase to the CCQE cross section, achieved by $M_A^{CCQE}$, or the presence of ``multi-nucleon effects''\footnote{These are sometimes referred to two particle, two hole (2p2h) effects or meson exchange current (MEC).} where the neutrino interacts on a correlated pair of nucleons. Multi-nucleon interactions, as a high momentum transfer process which mimics CCQE, can create a bias in the neutrino energy reconstruction for low energy experiments, however generous uncertainties on the RES channel, which produce a similar feed down, can mitigate the effect.


\subsection{Uncertainties on the DIS model}

The dominant contribution to the total cross-section at the energy ranges of interest with PINGU comes from DIS interactions. In GENIE, the DIS model uses the Bodek-Yang correction (BY)~\cite{BodekYang:2003}; the uncertainties in this model are encapsulated in the parameters, $A^{BY}_{HT}$, $B^{BY}_{HT}$, $C^{BY}_{V1u}$, $C^{BY}_{V2u}$.
The parameter $A$ is used to account for the higher order QCD terms and dynamic higher twist in the form of an enhanced target mass while the parameter $B$ is used to account for the initial state quark transverse momentum.  
The $C_{V1u}$ and $C_{V2u}$ parameters are used by vector $K$ factors so the GRV98 PDFs have the correct form in the low $Q^2$ photo-production limit.  There is an additional $K$ factor that is used for sea quarks but at neutrino energies less than 50 GeV, the charm sea contribution is very small and can be neglected~\cite{Bodek:2010km}.


At sufficiently high energies, the masses of the leptons can be neglected in the cross-section calculations. However, at energies near the production threshold, one must consider differences between the $\nue$,$\numu$ and $\nutau$ cross-sections. 
This effect is assumed to be negligible at PINGU's energies.

\subsection{Implementation\label{sec:xsecimplementation}}
The effects of the neutrino cross section systematics on the neutrino mass ordering study were studying using the $\deltaChiSqBar$ method. Apart from the addition of the GENIE systematics, the same systematics were used in the other $\deltaChiSqBar$ results as described in Table~\ref{Tab:Systematics}.

The neutrino cross section systematics were included with the assumption that they would only change the effective areas and not the reconstructed variables.   Indeed this is true when GENIE is only used to propagate cross-section systematics to the effective area where the error is propagated as a weight on an event. 
Because of this, we are currently not able to investigate the effects of these systematics on the energy or zenith reconstruction and this will be investigated in the future.
In practice, the GENIE systematic weights are multiplied with the standard weights and the new effective area is calculated.  This is done at the -2~$\sigma$, -1~$\sigma$, 1~$\sigma$, and 2~$\sigma$ values for each of the six parameters.  These files are then used in a line fit to obtain slopes $\Delta A_{i}$.  The modified effective area is calculated as
\begin{equation}
  A^{\prime}_{\textrm{eff}}(E) = A_{\textrm{eff}}\left(1+\sum_ip_i\Delta A_i\right).
\end{equation}


\subsection{Results}

By allowing each of the six GENIE systematics errors to be fitted individually, the total effect of each parameter can be determined.

\begin{figure}
\begin{center}
\includegraphics[width=0.9\columnwidth]{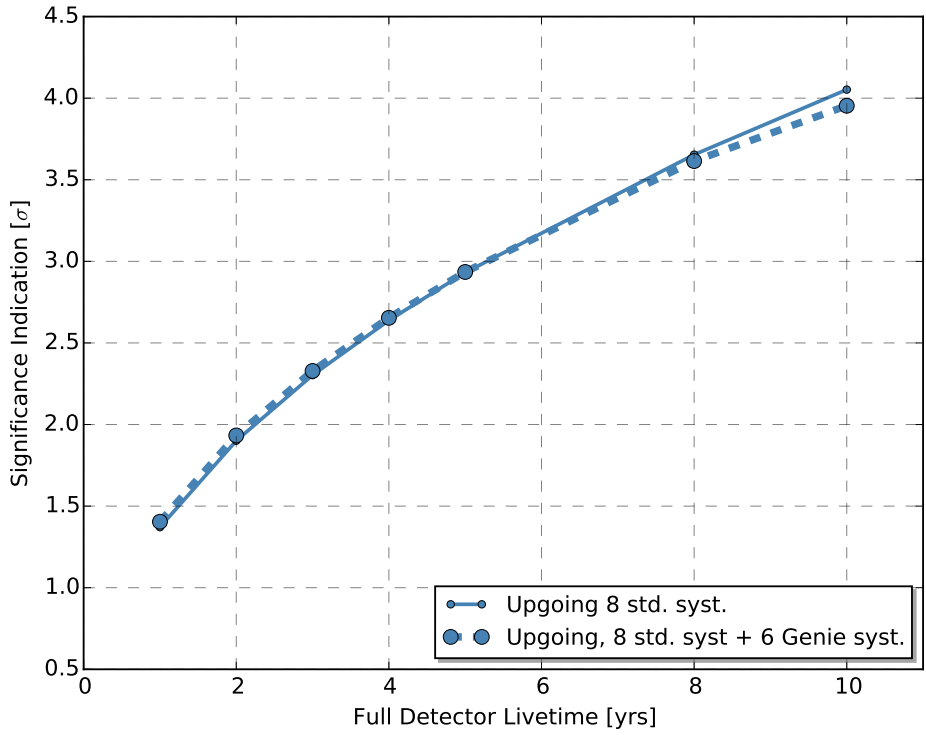}
\caption{The NMO significance shown with (dashed line) and without (solid line) the GENIE systematics added.} \label{fig:significance} 
\end{center}
\end{figure}

The total impact on the NMO significance over time can been seen in Fig.~\ref{fig:significance}.  
This graph shows that the significance when including the six extra systematics is consistent to within 1\% of the standard systematics case for livetimes below 5 years and still give small to  negigible impacts at livetimes up to 10 years.


We suspect that the $\ahtby$ and $\bhtby$ along with the $\cvouby$ and $\cvtuby$ may be correlated with each other.  
We have looked into this by fully correlating the higher twist parameters and the valence quark parameters and by having each parameter independent.
It was found that when a prior is included, all results are compatible with each other.

\clearpage
\resetlinenumber

\IfFileExists{NewCommands.tex}       {}       {}
\IfFileExists{../NewCommands.tex}    {}    {}
\IfFileExists{../../NewCommands.tex} {} {}

\graphicspath{{figures/}{Appendices/figures/}}

\addcontentsline{toc}{section}{Appendix}
\addtocontents{toc}{\protect\setcounter{tocdepth}{-1}}

\section{Neutrino flux uncertainties} \label{sec:FluxSysts}

Atmospheric neutrinos are generated in hadronic showers induced by the
interactions of primary cosmic rays in the Earth's atmosphere. Their flux
therefore depends on the properties of the primary flux (spectrum and
composition) and of the atmosphere (in particular its density profile) as
well as the hadron shower physics. At cosmic ray energies
$E_{\rm CR}\lesssim 20\rm{GeV}$, the Earth's magnetic field additionally affects the primary
cosmic ray flux, effectively shielding against low-energy charged
particles and thus providing a {\it geomagnetic cutoff}. Detailed Monte-Carlo
simulations are employed to predict the resulting neutrino flux at various
places on Earth~\cite{Honda:2015fha,Barr:2004br}, which on the one hand enhances the precision
of the calculation, but simultaneously limits the exploration of systematic
effects to a few selected scenarios~\cite{Honda:2015fha,Athar:2012it}. While
faster calculation methods based on one-dimensional cascade equations are
available~\cite{Fedynitch:2015zbe}, the collinear approximation in the shower
development limits their application to $E_\nu > 100\rm{GeV}$. 

In the following we will briefly discuss the various sources of uncertainty in
the calculations of the atmospheric neutrino flux and their effects on the
determination of the neutrino mass ordering with PINGU.

\subsection{Primary flux uncertainties}
To study modifications of the flux by the geomagnetic cutoff effect, the primary
cosmic ray spectrum is measured in balloon-borne, satellite or space-station
experiments
\cite{Abe:2015mga,Kopenkin:2009zz,Christ:1998zz,Aguilar:2015ooa,Adriani:2015aps,Yoon:2011aa}.
While most magnetic spectrometers provide high statistics measurements of the flux in
the range of $1 - 200\rm{GeV}$ per nucleon, calorimetric approaches typically
have a larger energy threshold of $\sim1\rm{TeV/n}$. This leaves a gap, 
particularly in the intermediate energy range, that has only recently been
populated by data from the AMS-02~\cite{Aguilar:2015ooa},
PAMELA~\cite{Adriani:2015aps} spectrometer and CREAM~\cite{Yoon:2011aa}
calorimeter experiments. 

In the absence of this data, the so called GSHL power-law approximation~\cite{Gaisser:2001jw}
\begin{equation}
  \Phi(E_p) = a\left[E_p + b\exp\left(c\sqrt E_p\right)\right]^{-d}
  \label{eqn:GSHL}
\end{equation}
has been adopted in earlier calculations which introduces four parameters a, b, c, and d that are obtained from fits to the data. In a study by Barr et
al.~\cite{Barr2006}, the impact of varying these parameters within their
model~\cite{Barr:2004br} is performed, assuming the uncertainties listed in
Tab.~\ref{Tab:PrimFluxUncertainties}.

\begin{table}
\begin{center}
  \begin{tabular}{l>{$}c<{$}>{$}c<{$}}
    Parameter & \text{Proton fluxes} & \text{Nuclear fluxes} \\ \hline
    $a$ (normalization) &   1.49 \pm 0.10  &  0.060 \pm 0.004 \\
    $b$ (shape)         &   2.15 \pm 0.025 &  1.25 \pm 0.03 \\
    $c$ (shape)         &  −2.21 \pm 0.02  & −0.14 \pm 0.02 \\
    $d$ (index) $> 200\rm{GeV/n}$ & 2.74 \pm 0.01 & 2.64 \pm 0.02\\
    $d$ (index) $< 200\rm{GeV/n}$ &2.74 \pm 0.03 & 2.64 \pm 0.04\\
\end{tabular}
\caption{Parameter values and uncertainties for the GSHL parameterization of the primary
cosmic ray flux, taken from~\cite{Barr2006}\label{Tab:PrimFluxUncertainties}}
\end{center}
\end{table}

\begin{figure}
  \begin{center}
    \includegraphics[width=0.54\columnwidth]{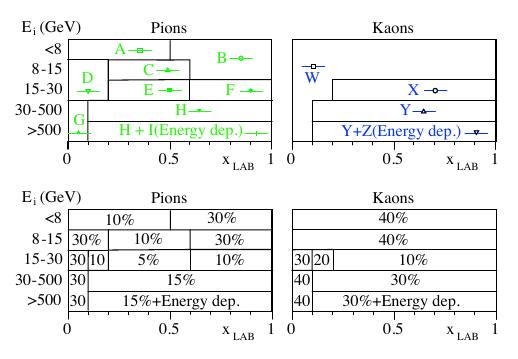} 
    \includegraphics[width=0.45\columnwidth]{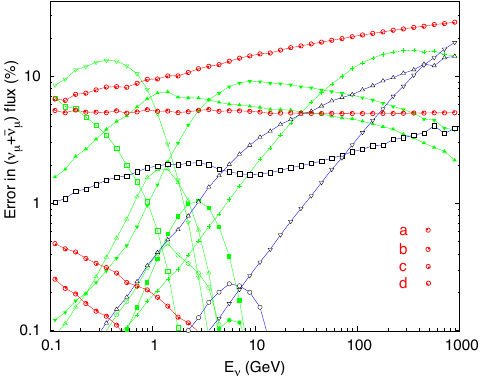} 
  \end{center}
  \caption{Uncertainties of the atmospheric neutrino flux from variations of the
    primary flux parameterization (a-d, red) as well as from uncertainties in various kinematic
    ranges of the hadronic interactions of pions (A-I, green) and kaons
    (W-Z, blue).  Figure
  adapted from~\cite{Barr2006}\label{fig:BarrNuFluxUncertainties}}
\end{figure}

The red curves in Figure~\ref{fig:BarrNuFluxUncertainties} shows the ensuing uncertainty on the
atmospheric neutrino flux as a function of neutrino energy, from each parameter respectively. The
spectral index uncertainty clearly dominates and ranges from $10-20\%$ in
the neutrino energy range of interest. The uncertainty in normalization is
independent of energy in the resulting neutrino flux at a value of
$\sim5\%$, while the {\it shape} parameters $a$ and $b$ provide a negligible
impact at low energies. This is consistent with a more recent investigation
\cite{HondaANW} using newly available data to find a deviation from the
power-law approach of $<8\%$ over the whole energy range.

To incorporate these uncertainties in our calculation,
the spectral index of the atmospheric neutrino flux is allowed to vary in a
range of $\pm 0.05$, while the overall normalization is a free parameter (c.f.
Table~\ref{Tab:Systematics}). It should be noted that this is a very
conservative approach, as the newly available data on the primary cosmic ray
flux is incorporated. While it reveals additional features in the
spectrum~\cite{Aguilar:2015ooa,Adriani:2015aps,Yoon:2011aa} that can not be
modelled with the simple GSHL parameterization given in Eq.~\ref{eqn:GSHL}, the
significantly improved event statistics constrain the primary flux much more
tightly. Detailed Monte Carlo studies will be necessary to assess the remaining
uncertainty.

\subsection{Magnetic fields}
Today, not only the spatial but also the the temporal evolution of the Earth's
magnetic field are known with high precision~\cite{Thebault2015}. The geomagnetic
cutoff, which effectively shields the Earth from cosmic rays at energies below $E_{\rm
CR}\lesssim 20\rm{GeV}$ and affects the neutrino flux at energies below
$E_\nu \lesssim 5\rm{GeV}$ can therefore be calculated with sufficient
accuracy~\cite{Honda:2011nf}.

However, another complication which is dealt with in modern Monte Carlo
calculations is the lateral spreading of the cosmic ray showers both from
transverse momentum acquired in the interactions and decay of mesons and from
bending of muons in the Earth's magnetic field. The field is not
symmetric enough to allow for a simplification of the problem. This makes the
computation challenging as it requires the generation of particles in all
directions at all points on the Earth.

As a consequence, the neutrino flux has to be calculated individually for each
detector location. A commonly used speed-up trick is to
extend the size of the detector to cover a large area surrounding it~\cite{Barr:2004br,Athar:2012it}. Through the geomagnetic effects
described above, this generates a systematic bias in the calculated flux that
is corrected for by studying how the flux changes as a function of the detection
radius~\cite{Honda:2011nf,Honda:2015fha}. The resulting uncertainty on the
atmospheric flux depends on the Monte Carlo statistics employed in the
calculation, but is typically $\lesssim 1\%$~\cite{Honda:2011nf}.

We therefore do not include any systematic effects related to the geometry or
magnitude of the magnetic field.

\subsection{Seasonal variation}
The atmosphere of the Earth has been studied in great detail and a plethora of
atmospheric data has been available for a long time. For simplicity, most atmospheric neutrino flux
calculations~\cite{Battistoni:2002ew,Barr:2004br,Honda:2011nf} employ the time-averaged atmosphere
model {\it US-standard '76} presented in \cite{USStd:1976}. In a
recent calculation~\cite{Honda:2015fha}, the newer {\it NRMLSISE-00} model~\cite{JGRA:JGRA16630}
has been adopted since it provides a temporally and spatially changing atmospheric
model.

\begin{figure}
\begin{center}
\includegraphics[width=0.4\columnwidth]{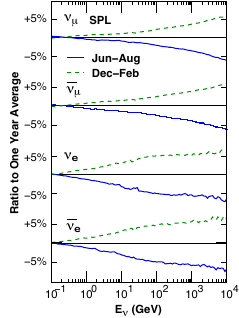} 
\includegraphics[width=0.4\columnwidth]{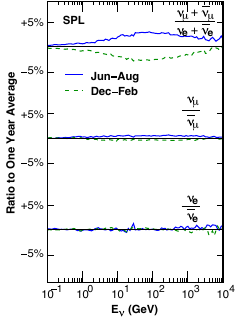} 
\end{center}
\caption{Temporal variation of atmospheric neutrino fluxes at the South Pole for
  different flavors (left) and of flavor ratios(right). Figure taken
from~\cite{Honda:2015fha}
    \label{fig:flux_yearly_variation}}
\end{figure}

Comparing the flux calculated with the {\it US-standard '76} and
the time-averaged flux calculated with the {\it NRMLSISE-00} model results in
differences of $\lesssim 2\%$, significantly smaller than the variability
introduced by seasonal effects. The most important parameter in these
calculations is the atmospheric density, where the primary effect on the
 neutrino flux is the concurrency of interaction and decay of the
generating mesons in the atmosphere. The resulting variation over the course of
the year -- shown in Figure~\ref{fig:flux_yearly_variation} -- is well below 1\%
at a neutrino energy of $E_\nu \simeq 1\rm{GeV}$, but increases to $\pm 5\%$ at
$E_\nu \simeq 1\rm{TeV}$, where the pions and kaons experience stronger
time-dilatation effects that suppress their decay. While this effect works the
same on all neutrino flavors, the muons originating in the mesons decay lose
more energy in a denser atmosphere before they decay. Therefore the variability
of $\nue$ which are predominantly generated from the decay of these muons is
enhanced at $E_\nu \simeq 10-100\rm{GeV}$ with respect to the variability of the
$\numu$ flux. 

Despite the large effective volume of the PINGU detector, event
statistics comprising several years of data will have to be collected
in order to determine the neutrino mass ordering
(c.f. Sec.~\ref{sec:NeutrinoMassHierarchy}
and~\ref{sec:NMOAnalysisTechnique}). It is therefore expected that
atmospheric variations largely cancel out in the resulting dataset. We
do not incorporate any systematic effect to account for this
uncertainty. Detailed atmospheric data is available~\cite{AIRS} that allows to
correct for temporal variations should it become necessary.

We note however, that the matter effects giving rise to the mass ordering sensitivity
arises from the asymmetry in the $\nu/\bar\nu$ flux and interaction ratios. The sensitivity is
therefore affected less by uncertainties in the absolute flux, and more by
uncertainties in the ratios of $\numu/\numubar$ and $\nue/\nuebar$. Low energy muons leave short
tracks in the detector, leading to poor flavour identification at those
energies, as the muon neutrinos cannot be disentangled 
from the rest of the sample with high purity. Thus  the ratio of $(\numu+\numubar)/(\nue+\nuebar)$ is of interest for the neutrino mass ordering
determination. Figure~\ref{fig:flux_yearly_variation} shows the variation of
these ratios with the seasonal changes in the atmosphere. Both the
$\numu/\numubar$ and $\nue/\nuebar$ ratios stay constant throughout the year,
while the electron neutrino fluxes are suppressed in summer as a result of the
decreased cosmic ray interaction height causing more muons to hit the ground
before they decay. An uncertainty of $\pm 3\%$ on the ratio of
$(\numu+\numubar)/(\nue+\nuebar)$ is adopted in our calculation, and of $\pm
10\%$ on the ratios of $\numu/\numubar$ and $\nue/\nuebar$. Both are chosen
significantly larger than the atmospheric variability would suggest, as they
additionally incorporate uncertainties in the hadronic interaction models (see~\ref{ssec:HadronicInteract}) 
and neutrino cross-sections.

\subsection{Hadronic interaction uncertainties} \label{ssec:HadronicInteract}
A dominant source of uncertainty in the atmospheric neutrino flux stems from the
not well determined hadronic production rates for mesons generated in $pN \to
\pi^\pm X$ and $pN \to K^\pm X$ reactions, where $p$ denotes an incoming cosmic
ray nucleus, $N$ is a target nucleus in the atmosphere and $X$ represents the
rest of the interaction products. In the absence of comprehensive experimental
results and with the computational challenge involved in a detailed exploration
of the allowed ranges in the underlying nuclear physics models, early analyses
have constrained themselves to a comparison of a number of selected hadronic
interaction models~\cite{Battistoni:2002ew}.

In contrast, Barr et al.~\cite{Barr2006} have made an assessment of the
uncertainties based on a collection of available experimental data.
The parameter space is defined in the kinematic variables $E_i$ and $x_{\rm
lab}$, where $E_i$ is the
energy of the incident projectile, $E_s$ is the energy of the secondary meson
and $x_{\rm lab} = E_s/E_i$.
Figure~\ref{fig:BarrNuFluxUncertainties} shows the uncertainty in the hadronic
meson production assigned in~\cite{Barr2006} to each region in the $(E_i,x_{\rm
lab})$-space. To proceed with the calculation of the resulting uncertainties in
the atmospheric neutrino flux, a number of zones (denoted A-I for pions and W-Z
for kaons) is defined.  Within each zone, the meson production rates are varied
in the calculation assuming $\pm 1\sigma$ ranges as given by the hadronic meson
production uncertainties, with the assumption that these uncertainties are 100\%
correlated within each zone, and fully uncorrelated between zones. From these
variations, uncertainties on the atmospheric neutrino flux are determined as
shown in figure~\ref{fig:BarrNuFluxUncertainties}.

While (through the higher kaon mass) kaon production uncertainties only become relevant at higher
energies, uncertainties in the pion production (regions D,C,H) dominate the overall flux uncertainties
below $E_\nu < 1\rm{GeV}$, and are second largest (after the spectral index) in
the remaining energy range. Yet, in contrast to the spectral index variations which provide a
a smooth power-law modification of flux as a function of energy, uncertainties
in the hadronic production may be confined to a narrow range in energy.
As the neutrino mass ordering effect also reveals itself in atmospheric neutrino
data through enhancements or suppressions of the flux in well defined energy
ranges, hadronic production uncertainties will provide a larger impact on the
NMO sensitivity.

\begin{figure}
\begin{center}
\includegraphics[width=0.8\columnwidth]{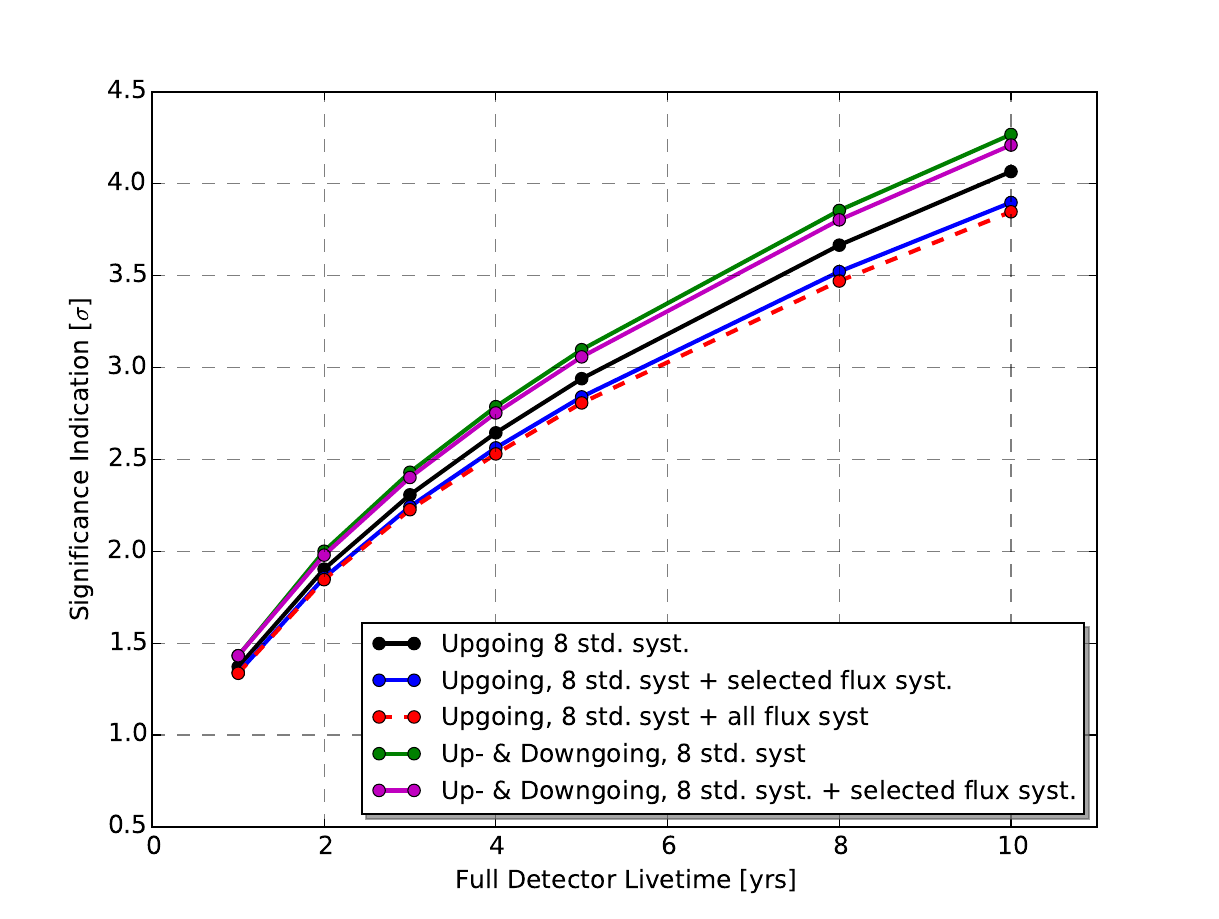} 
\end{center}
\caption{Neutrino mass ordering significance estimator vs. detector lifetime using
  up-going events and the standard set of systematic uncertainties listed in
  Tab.~\ref{Tab:Systematics} (black) as well as including 13 additional flux systematic
  uncertainties described in the text (red dashed) or just the 4 most relevant of those (blue). 
  Extending the analysis to the down-going region (green) enhances the sensitivity due to  earth-penetrating neutrinos that are
  mis-reconstructed above the horizon. Together with better constraints on the
  uncertainties from measuring the down-going flux, this over-compensates some
  of the sensitivity loss induced by the hadronic production uncertainties (purple). }
    \label{fig:flux_uncert_signifcances}
\end{figure}

In order to determine this impact we continue the strategy described
in~\cite{Barr2006}, assuming that each of the zones described above provides the
source of a systematic uncertainties with the resulting flux variation being
100\% correlated within the zone and fully uncorrelated between zones. In
analogy to the approach in \ref{sec:xsecimplementation}, we then
treat the flux variations $\Delta\Phi(E)$ shown in
Fig.~\ref{fig:BarrNuFluxUncertainties} as $\pm 1\sigma$ variation of these
uncertainties, to calculate the modified flux $\Phi(E)^\prime$ from the original
flux $\Phi(E)$ as

\begin{equation}
  \Phi(E)^\prime = \Phi(E)\left(1 + \sum\limits_j p_j\Delta\Phi_j(E)\right)
\end{equation}

with the uncertainties parameters $p_j$ in units of $\sigma$ of the associated
hadronic production zone and Gaussian priors of width $\sigma(p_j) = 1$.
Figure~\ref{fig:flux_uncert_signifcances} shows the resulting loss in
significance, which amounts to about 10\% in the neutrino mass
ordering estimator (red line). However, this result is overly conservative as
we are treating the zones as fully uncorrelated, while in reality a
smooth transition is expected between them. This conservative approach allows the parameters
$p_j$ to take e.g. extreme positive and negative value in adjacent zones,
therefore providing stronger shape modifications than expected.

Consistent with
this expectation, we find that variations in zones $H,Y,E$ and $I$ which
generate shape modifications in the same energy range as the neutrino
mass ordering effect provide the largest impact on the significance, while
variations in all other zones can be neglected (compare blue and red dashed line in
figure~\ref{fig:flux_uncert_signifcances}).  Furthermore, there is more
experimental data available now (and more will be available in the
future), which will more tightly constrain the hadron production rates and
therefore decrease the uncertainties. Finally, we note that our calculation
only uses neutrino events reconstructed as up-going, while the PINGU detector
will at the same time measure the down-going neutrino flux, allowing the
constraint on uncertainties in the atmospheric neutrino flux. 

In a first exploration of this option we extend our calculation to
incorporate the full zenith range of $-1 < \cos(\theta_z) < 1$, optimistically
assuming the same effective area for the up- and down-going flux. Even without
taking into account the additional constraints on the hadron production
uncertainties $p_j$, this results in a significant boost in sensitivity to the
neutrino mass ordering (c.f.~green line in 
Figure~\ref{fig:flux_uncert_signifcances}) due to events the have penetrated the
Earth and therefore are affected by matter effects, but are mis-reconstructed
as down-going events in PINGU. Adding back the relevant
hadronic production uncertainties (purple line in
Figure~\ref{fig:flux_uncert_signifcances}), there is still a net gain in the
sensitivity to the neutrino mass ordering provided by the additional constraints
on the atmospheric neutrino flux uncertainties. A more detailed study also
taking into account potentially larger remaining atmospheric muon background in
the down-going region will be required to fully assess the benefit of a $2\pi$
measurement.

\clearpage
\resetlinenumber

\IfFileExists{NewCommands.tex}       {}       {}
\IfFileExists{../NewCommands.tex}    {}    {}
\IfFileExists{../../NewCommands.tex} {} {}

\graphicspath{{figures/}{Appendices/figures/}}

\addcontentsline{toc}{section}{Appendix}
\addtocontents{toc}{\protect\setcounter{tocdepth}{-1}}

\section{IceCube--Gen2 Phase 1} 
\label{sec:gen2phase1}

As a first step toward a future IceCube--Gen2 detector, the
IceCube--Gen2 Collaboration has proposed to install a Phase~1 upgrade
of the IceCube detector.  The Phase~1 upgrade will consist of seven
new strings of photosensors and optical calibration devices, installed
in the center of the existing detector within the volume proposed for
PINGU.

The full scientific program of the Phase~1 upgrade, including the
impact on high energy neutrino astrophysics, will be described in
detail elsewhere.  In this Appendix we discuss the reach of Phase~1
with regard to the PINGU science program.  Phase~1 will enable a
world-leading measurement of tau neutrino appearance and provide the
most stringent test of unitarity in the tau sector of the PMNS matrix
(detailed in Sec.~\ref{sec:NeutrinoOscillations}). Phase~1 will also
enable a world-class measurement of atmospheric $\numu$ disappearance
and extend IceCube's indirect search for dark
matter~\cite{Aartsen:2012kia} down to WIMP masses of a few GeV.

The calibration devices will greatly improve our understanding of the
complex optics of the glacial ice, not only in the central region but
throughout the IceCube volume.  This will permit the retroactive
reanalysis of over a decade of archived IceCube data, providing
substantial improvements in IceCube's measurements of high energy
astrophysical neutrinos, including angular resolution (especially for
high energy cascades), energy resolution, and tau neutrino
identification.  The new photosensors will fill part of the proposed
PINGU volume, and would greatly enhance IceCube's performance in the
\unit[10]{GeV} energy regime.  The Phase~1 design is such that more strings
can be added in a later deployment, to attain a complete PINGU
detector.

\begin{figure}
	\begin{center}
		\includegraphics[width=0.9\columnwidth]{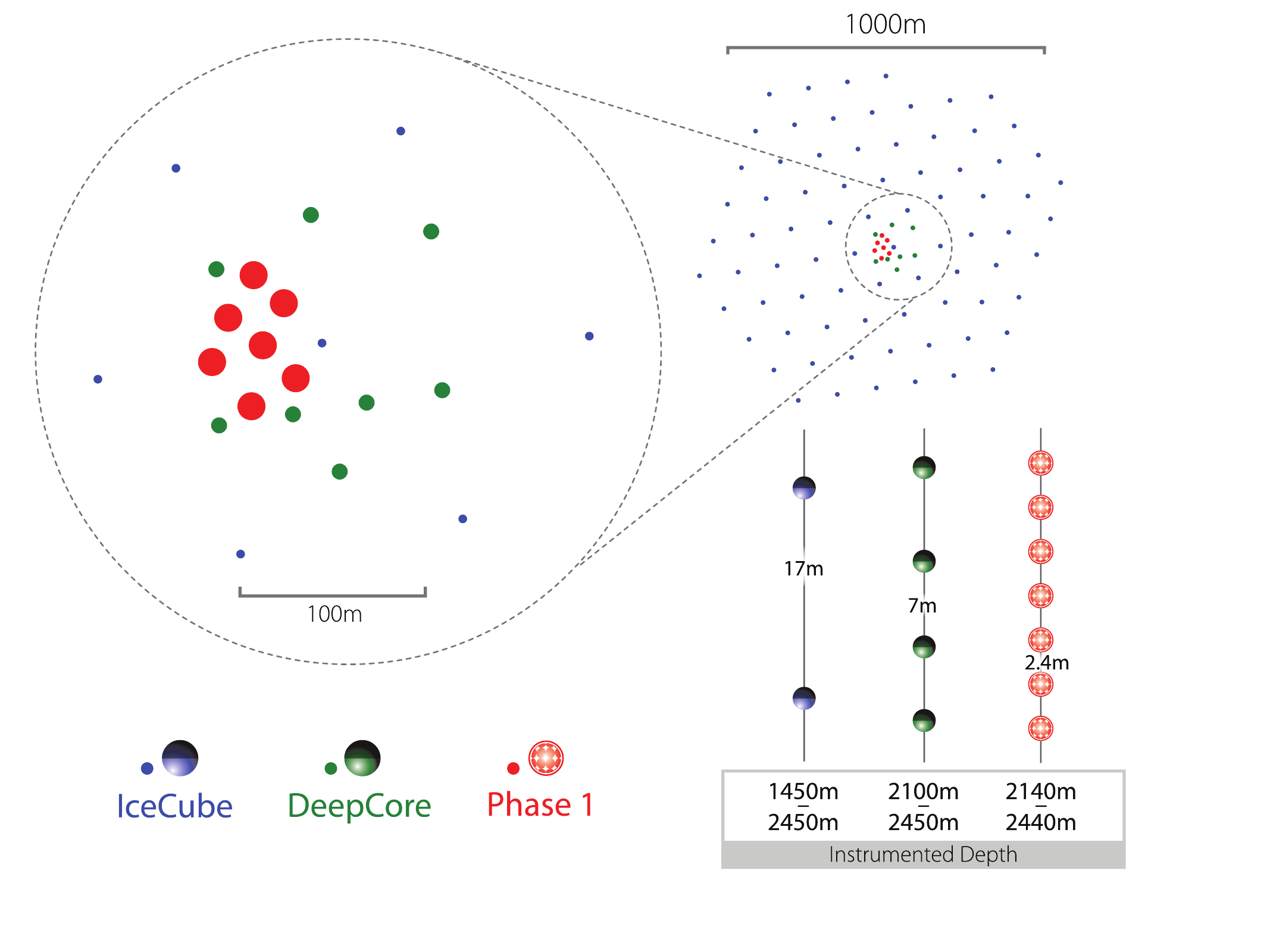}
		\caption{The layout of the IceCube--Gen2 Phase~1
                  strings, shown in red.  The locations of the existing
                  IceCube and DeepCore strings are also shown in blue
                  and green respectively.  The areas of the circles in
                  the zoomed-in top-view are proportional to the
                  relative photocathode density on each string.  This
                  is illustrated further in the lower right corner showing the
                  spacings on each of the strings progressing from
                  IceCube (\unit[17]{m}) to DeepCore (\unit[7]{m}) to Phase~1
                  (\unit[2.4]{m}).}
		\label{fig:phase1_geo} 
	\end{center}
\end{figure}

The new strings are envisioned to have roughly 125 multi-PMT Digital
Optical Modules (mDOMs), as described in detail in
Sec.~\ref{sec:mDOMs}.  The mDOM will provide more photocathode area
per module, at a lower cost per unit photocathode area and with
improved determination of photon directionality, than the traditional
IceCube DOM.  The layout of these strings in the IceCube detector is
shown in Fig.~\ref{fig:phase1_geo}.

\subsection{Science Objectives}

\subsubsection*{$\nutau$ Appearance and the Unitarity of the PMNS Matrix}

\begin{figure}
	\begin{center}
		\includegraphics[width=0.9\columnwidth]{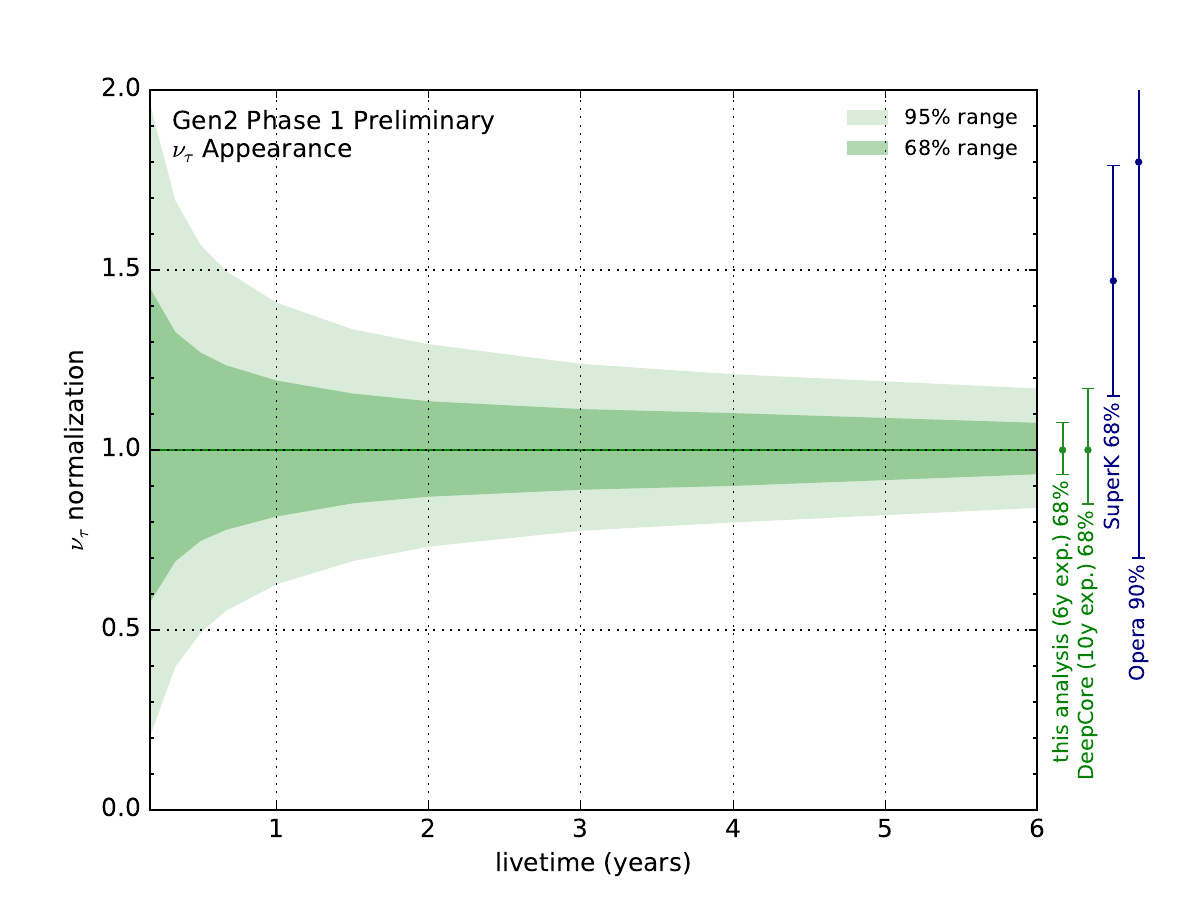}
		\caption{ Predicted measurement of the tau neutrino
                  normalization vs.\ time for Phase~1, with results
                  from Super-K~\cite{Abe:2012jj} and
                  OPERA~\cite{Agafonova:2015jxn}, and predicted results
                  from 10~yrs of DeepCore, shown for comparison. The
                  Phase~1 prediction uses fully reconstructed
                  simulated data (but without taking advantage of
                  better reconstruction predicted with mDOMs) and has
                  a full suite of systematic errors applied. With the
                  Phase~1 strings we improve the worldwide precision
                  by a factor of two with less than 2~yrs of
                  livetime, and reach roughly 10\% precision on this
                  key measurement with about 5~yrs of
                  livetime. The combined DeepCore+Phase~1 is much more
                  sensitive than DeepCore alone, which plateaus at
                  roughly 15\% precision. Including 10~yrs of
                  DeepCore data, the livetime that will have been accumulated when Phase~1
                  starts taking data, will slightly improve the
                  overall precision beyond what is shown here. (Note: The OPERA 90\% range is
                  only shown to the top of the plot scale but in fact
                  extends to 3.6.)}
		\label{fig:nutau_norm_phase1} 
	\end{center}
\end{figure}

The chief physics goal of the Phase~1 detector is to make a
world-leading measurement of $\nutau$ appearance and exploit the
resulting sensitivity to potential deviations from PMNS matrix
unitarity.  This topic has been discussed in detail in
Sec.~\ref{sec:TauNeutrinoAppearance} for the full PINGU detector.  The
projected results for Phase~1 are shown in
Fig.~\ref{fig:nutau_norm_phase1}.

The relatively high mass of the tau lepton suppresses $\nutau$
interactions in accelerator-based experiments, but at the higher
atmospheric neutrino energies available to the Phase~1 detector, the
event rate from $\nutau$ appearance is high.  IceCube's existing
DeepCore extension already detects about 800~$\nutau$ CC+NC
interactions per year; the addition of Phase~1 strings would increase
that to 2500. (For comparison, the Super-K experiment accumulated an
estimated 180 atmospheric $\nutau$ events with a livetime of
2806~days~\cite{Abe:2012jj}.)  Since insignificant numbers of $\nutau$
are produced directly in cosmic ray air showers, the $\nutau$ observed
in DeepCore and Phase~1 arise from $\numu \rightarrow \nutau$
oscillations at specific $L/E$ (the ratio of the neutrino's path length
through the Earth to its energy).  For these path lengths, the
oscillation maximum is at roughly $\Enu = 10\textrm{--}30$~GeV, comfortably
above the DeepCore and Phase~1 detector thresholds.

These $\nutau$ events can be distinguished on a statistical basis from
the background of $\nue$ and $\numu$ CC and NC events by their
characteristic angular distribution and energy spectrum. This allows
Phase~1 to measure the rate of $\nutau$ appearance with a
precision of better than 10\% with about 5~yrs of data, as shown
in Fig.~\ref{fig:nutau_norm_phase1}, providing a significantly deeper
probe of PMNS matrix elements dependent on the $\numu$ and $\nutau$
flavors than current state-of-the-art
experiments~\cite{Abe:2012jj,Agafonova:2015jxn}. The measurement will
either strengthen the 3-flavor model and the underlying unitarity of
its corresponding mixing, or point us in the direction of new physics
due to sterile neutrinos, non-standard interactions, or other effects.

\subsubsection*{$\numu$ Disappearance; Sterile Neutrino Searches; Dark Matter}

\begin{figure}
	\begin{center}
		\includegraphics[width=0.49\columnwidth]{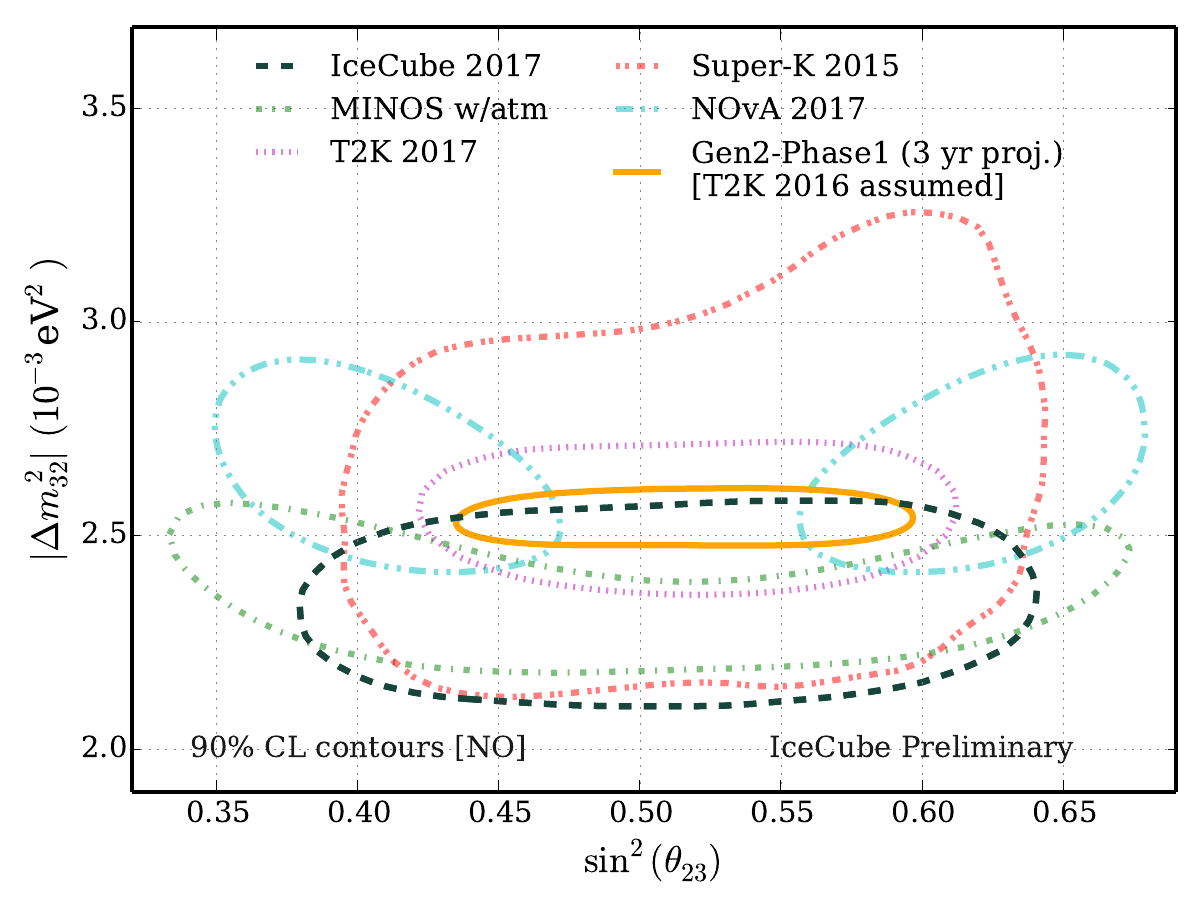}
		\includegraphics[width=0.49\columnwidth]{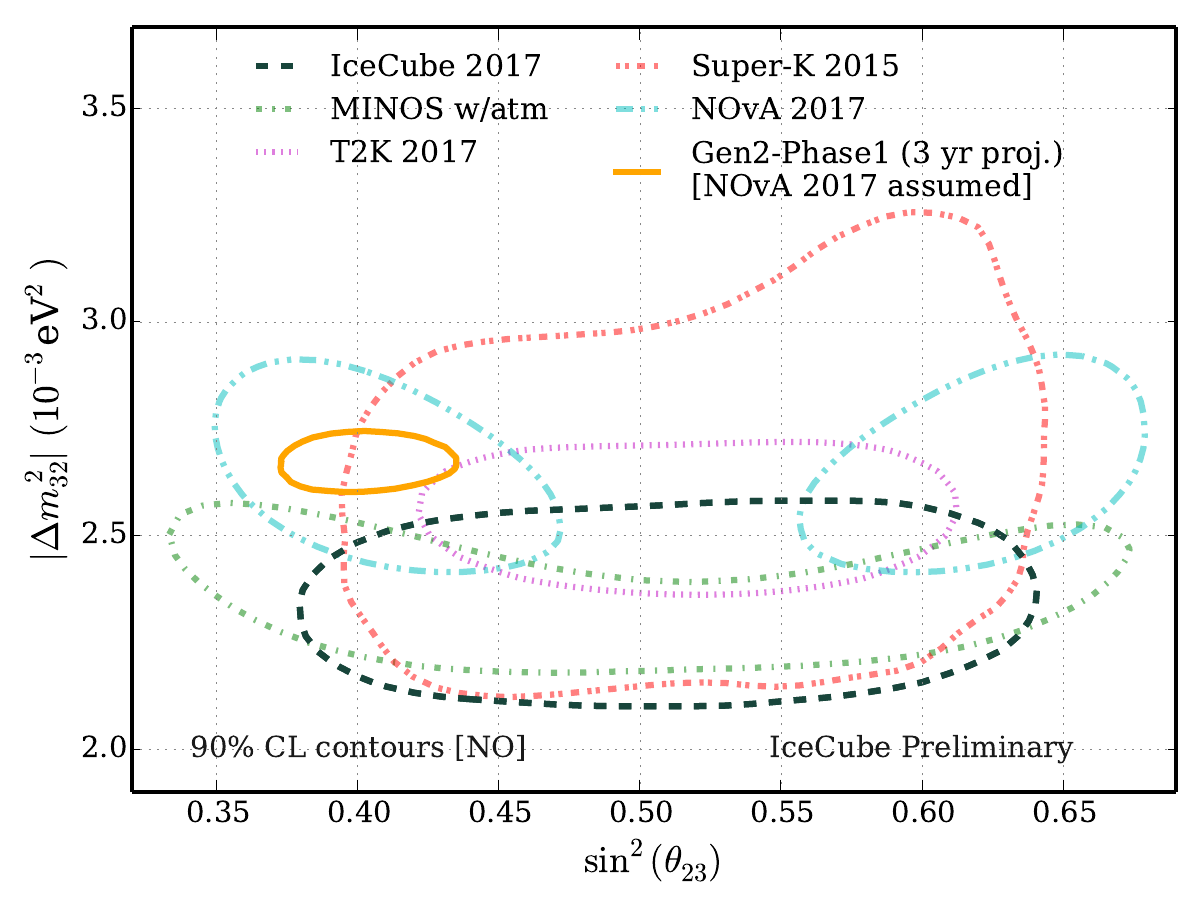}
		\caption{The orange lines show the sensitivity of the
                  IceCube--Gen2 Phase 1 detector to atmospheric
                  $\numu$ disappearance parameters after 3~yrs of
                  livetime. The sensitivity in the left-hand graph
                  assumes the true values of the parameters to be
                  $\Delta m^{2}_{32}=\unit[2.51\times
                  10^{-3}]{eV^{2}}$ and $\theta_{23}=45^{\circ}$,
                  which are the T2K best-fit point from
                  reference~\cite{ref:T2K2016PRL}. The sensitivity in
                  the right-hand graph assumes the true values of the
                  parameters to be $\Delta m^{2}_{32}=\unit[2.67\times
                  10^{-3}]{eV^{2}}$ and $\sin^{2}\theta_{23}=0.404$,
                  which are the lower-octant NOvA best-fit point
                  (statistically degenerate with their upper-octant
                  best-fit point) from
                  reference~\cite{NovaOctantPRL}. The graphs also show
                  the most recent measurements from
                  IceCube-DeepCore~\cite{ref:IC2017Contour},
                  MINOS~\cite{Adamson:2013},
                  T2K~\cite{ref:T2K2017Oscillations},
                  Super-Kamiokande~\cite{SuperKIV} and
                  NOvA~\cite{NovaOctantPRL}.}
		\label{fig:Contours_w_gen2phase1} 
	\end{center}
\end{figure}

The Phase~1 detector will have improved sensitivity to additional key
atmospheric neutrino oscillation parameters, as well as to sterile
neutrinos and to neutrinos from dark matter annihilations,
supplementing its highly important $\nutau$ appearance measurement.
Figure~\ref{fig:Contours_w_gen2phase1} shows the reach of Phase~1 for
the atmospheric neutrino oscillation parameters.  The two graphs in
the figure show the predicted Phase~1 result with 3~yrs of livetime,
assuming maximal and non-maximal true values of $\theta_{23}$. The
Phase~1 measurement of the atmospheric mixing parameters will reach a
precision competitive with that predicted from leading
accelerator-based experiments such as NOvA~\cite{NovaOctantPRL}
and T2K~\cite{ref:T2K2017Oscillations}, but with very different energies and
systematics, strengthening aggregate understanding of neutrino
oscillations.  In addition, Phase~1 will have sensitivity to the
$\thTT$ octant and maximal mixing, of particular current interest due
to the mild tension in recent measurements, with T2K~\cite{ref:T2K2017Oscillations}
consistent with maximal mixing and NOvA~\cite{NovaOctantPRL}
excluding it at $2.5\sigma$.  Phase~1 will be able to exclude maximal
mixing at $3\sigma$ in roughly 3~yrs if the true value of $\thTT$ is
at the current NOvA best-fit value.  Under conservative
assumptions, Phase~1 will also be able to exclude the wrong neutrino
mass ordering (NMO) at 1.5--2$\sigma$ with 3~yrs of data (with
favorable $\thTT$ and normal ordering, it could reach 3$\sigma$ in the
same time frame).

The greater event rate at low neutrino energies will also enable
Phase~1 to improve IceCube's world-leading sensitivity to sterile
neutrinos~\cite{ref:IceCubeHighEnergySteriles,ref:IceCubeLowEnergySteriles}.
Phase~1 can improve on the currently limits on $|U_{\tau 4}|^{2}$ from DeepCore and Super-K by about a factor
of three. This greater event rate will enable Phase~1 to test for non-standard neutrino
physics directly, complementing its ability to do so indirectly via
$\nutau$ appearance.

\begin{figure}
	\begin{center}
		\includegraphics[width=0.49\columnwidth]{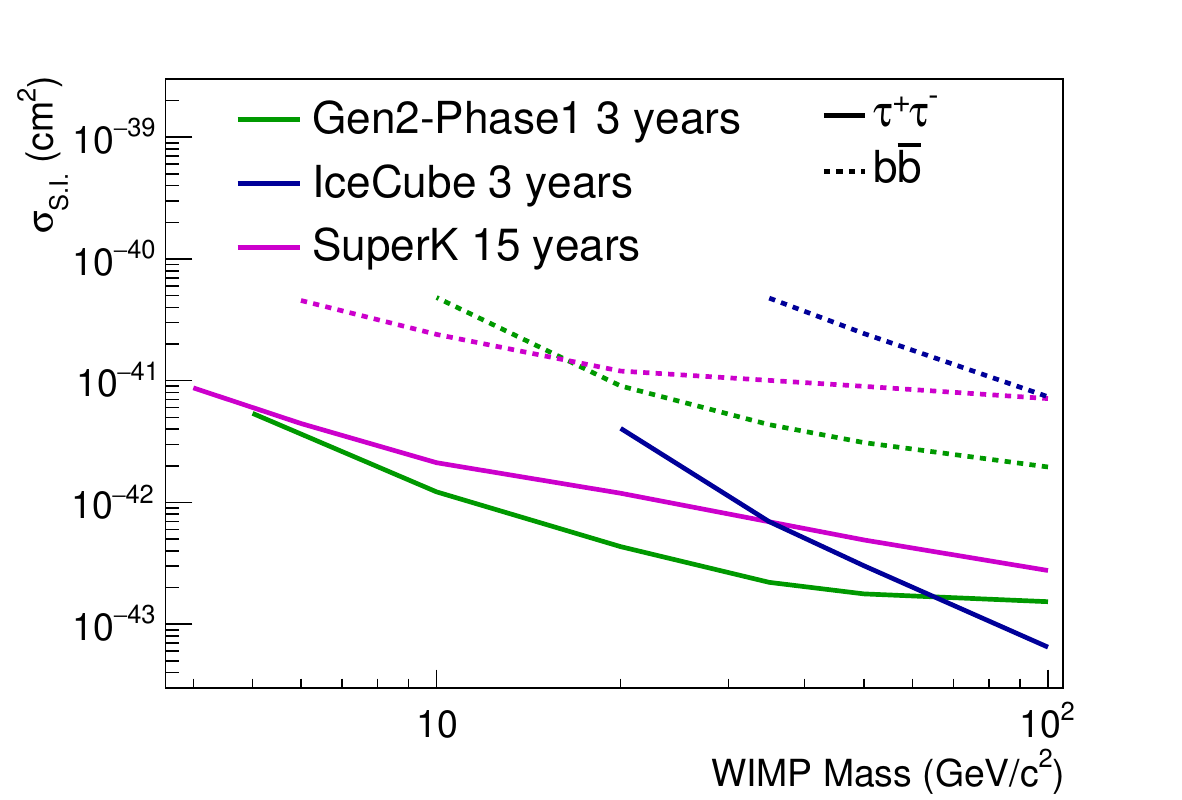}
		\includegraphics[width=0.49\columnwidth]{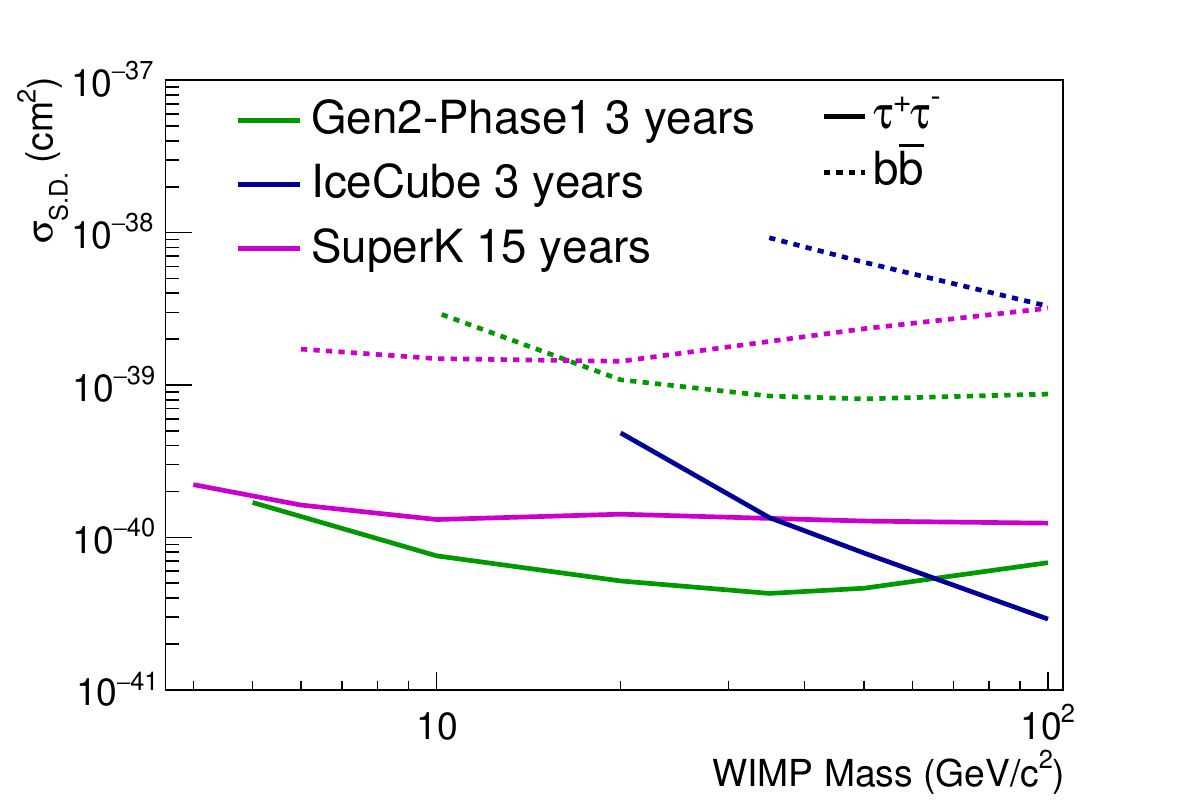}
		\caption{The IceCube--Gen2 Phase 1 sensitivity to
                  neutrinos from solar WIMP annihilations for
                  spin-independent (left) and spin-dependent (right)
                  models. This sensitivity is compared to existing
                  limits from Super-Kamiokande~\cite{Choi:2015ara} and
                  IceCube~\cite{ref:IceCubeWIMPLimit}.}
		\label{fig:phase1-wimps} 
	\end{center}
\end{figure}

Finally, with Phase~1 we will be able to extend our search for
neutrinos from solar WIMP annihilations down to WIMP masses as low as
5~GeV (see Fig.~\ref{fig:phase1-wimps}), building on IceCube's highly
competitive results on indirect dark matter
detection~\cite{Aartsen:2012kia}.

\subsubsection*{Calibration}

Phase~1 provides an opportunity to deploy new calibration sources, which are
intended to build on the lessons learned from IceCube and to improve
our knowledge of the ice, leading to corresponding improvements in
many IceCube analyses, including those at energies relevant for
neutrino oscillation analyses.  The majority of the calibrations in
IceCube are performed using LED flashers, co-located with the IceCube
DOMs.  These have been used to measure the detector geometry, validate
time calibration and, most importantly, to measure the optical
properties of the ice (the unaltered ``bulk ice'' between the strings,
and the refrozen ``hole ice'' in which the strings are deployed).

The IceCube LED brightness levels and directions are known to only
20\% and 10~degrees, respectively~\cite{Aartsen:2016nxy}.  There are
also no vertically oriented LEDs that can directly probe the hole ice.
The proposed Phase~1 devices are listed below.
\begin{itemize}
\item The Precision Optical CAlibration Module
  (POCAM)~\cite{Jurkovic:2016kxn,Krings:icrc2015}: An isotropic light
  source with a well understood light output to measure {\it in situ}
  the optical efficiency of the DOMs to within 3\% (see
  Sec.~\ref{sec:calib_pocam}).
\item LED flashers in the mDOMs: Housed in the internal support
  structure for the PMTs for precisely measured custom placement and
  orientation.  With horizontal distances as low as \unit[20]{m} between
  Phase~1 mDOMs, multiple flashers can be coordinated to probe
  ``double bang'' signatures at the \unit[20]{m} baseline at which IceCube is
  most sensitive to high energy $\nutau$ observation (see
  Sec.~\ref{sec:calib_led}).
\item Video cameras: Deployed with the mDOMs or in dedicated
  camera modules in order to monitor the hole ice, determine if the
  degassing process is successful and measure light from LEDs in order
  to further probe the absorption and scattering properties of the
  bulk ice (see Sec.~\ref{sec:calib_cameras}).
\end{itemize}

\subsection{Additional Science Benefits}

Beyond exploiting the distinct capabilities of an enhanced IceCube
in-fill array to achieve the core science mission of Phase~1, neutrino
oscillation measurements by a combined DeepCore and Phase~1 will
complement accelerator and reactor neutrino experiments, as the
different set of systematic uncertainties confronting it and the weak
impact of $\dcp$ on its measurements will increase the robustness of
global neutrino oscillation fits.  Comparison of DeepCore+Phase~1
observations to those made by both currently running experiments such
as T2K~\cite{Abe:2014ugx} and NOvA~\cite{Adamson:2016tbq} and
planned experiments such as DUNE~\cite{Acciarri:2015uup},
Hyper-Kamiokande~\cite{HyperK}, and JUNO~\cite{An:2015jdp} will
therefore also provide broad and model-independent potential for
discovery of new physics.

Moreover, a restart of deep ice drilling at the South Pole enables
future particle astrophysics at this unique site. The deep ice at the
South Pole is an optimal location for deployment of a direct dark
matter detector with scintillator crystals (DM-Ice~\cite{Cherwinka:2011ij})
to confirm or reject the annual modulation of WIMP dark matter claimed
by DAMA~\cite{Bernabei:2010mq}.  With surrounding IceCube strings used
as a veto shield, the South Pole site offers a highly stable, nearly
background-free environment. Deep ice drilling also allows for the
possibility of deploying next-generation optical sensor technology
prototypes for ultimate use in a fully-realized IceCube--Gen2 Neutrino
Observatory. The opportunity to retire risk at an early stage of
development would shorten the design phase of this formidable endeavor
and provide a pathway for utilization of novel and potentially
game-changing photodetection technology.

\clearpage

%
%

\resetlinenumber
\bibliography{biblio}

\providecommand{\href}[2]{#2}\begingroup\raggedright\begin{thebibliography}{100}

\bibitem{Aartsen:2013rt}
{\bf IceCube} Collaboration, M.~Aartsen {\em et~al.}, ``{Measurement of South
  Pole ice transparency with the IceCube LED calibration system},'' {\em
  Nucl.~Instrum.~Meth.} {\bf A711} (2013) 73--89,
  \href{http://xxx.lanl.gov/abs/arXiv:1301.5361}{{\tt arXiv:1301.5361}}.

\bibitem{Cherwinka:2011ij}
J.~Cherwinka, R.~Co, D.~Cowen, D.~Grant, F.~Halzen, {\em et~al.}, ``{A Search
  for the Dark Matter Annual Modulation in South Pole Ice},'' {\em
  Astropart.~Phys.} {\bf 35} (2012) 749--754,
  \href{http://xxx.lanl.gov/abs/arXiv:1106.1156}{{\tt arXiv:1106.1156}}.

\bibitem{Aartsen:2016nxy}
{\bf IceCube} Collaboration, M.~G. Aartsen {\em et~al.}, ``{The IceCube
  Neutrino Observatory: Instrumentation and Online Systems},'' {\em JINST} {\bf
  12} (2017), no.~03 P03012, \href{http://xxx.lanl.gov/abs/1612.05093}{{\tt
  1612.05093}}.

\bibitem{Mei:2005gm}
D.~Mei and A.~Hime, ``{Muon-induced background study for underground
  laboratories},'' {\em Phys. Rev.} {\bf D73} (2006) 053004,
  \href{http://xxx.lanl.gov/abs/astro-ph/0512125}{{\tt astro-ph/0512125}}.

\bibitem{Aartsen:2013pza}
{\bf IceCube} Collaboration, M.~G. Aartsen {\em et~al.}, ``{Evidence for
  High-Energy Extraterrestrial Neutrinos at the IceCube Detector},'' {\em
  Science} {\bf 342} (2013) 1242856,
  \href{http://xxx.lanl.gov/abs/arXiv:1311.5238}{{\tt arXiv:1311.5238}}.

\bibitem{Aartsen:2013jza}
{\bf IceCube} Collaboration, M.~Aartsen {\em et~al.}, ``{Measurement of
  Atmospheric Neutrino Oscillations with IceCube},'' {\em Phys.~Rev.~Lett.}
  {\bf 111} (2013) 081801, \href{http://xxx.lanl.gov/abs/arXiv:1305.3909}{{\tt
  arXiv:1305.3909}}.

\bibitem{NumuDisappearanceICRC2013}
{\bf IceCube} Collaboration, S.~Euler, L.~Gladstone, C.~Wiebusch, {\em et~al.},
  ``{Measurement of atmospheric neutrino oscillations with IceCube/DeepCore in
  its 79-string configuration},'' in {\em Proceedings of the 33rd International
  Cosmic Ray Conference (ICRC2013)}, 2013.
\newblock \href{http://xxx.lanl.gov/abs/arXiv:1304.0735}{{\tt
  arXiv:1304.0735}}.

\bibitem{Aartsen:2014yll}
{\bf IceCube} Collaboration, M.~Aartsen {\em et~al.}, ``{Determining neutrino
  oscillation parameters from atmospheric muon neutrino disappearance with
  three years of IceCube DeepCore data},'' {\em Phys. Rev.} {\bf D91} (2015)
  072004, \href{http://xxx.lanl.gov/abs/arXiv:1410.7227}{{\tt
  arXiv:1410.7227}}.

\bibitem{Aartsen:2014njl}
{\bf IceCube} Collaboration, M.~G. Aartsen {\em et~al.}, ``{IceCube-Gen2: A
  Vision for the Future of Neutrino Astronomy in Antarctica},''
  \href{http://xxx.lanl.gov/abs/arXiv:1412.5106}{{\tt arXiv:1412.5106}}.

\bibitem{MSW-W}
L.~Wolfenstein, ``Neutrino oscillations in matter,'' {\em Phys.~Rev.} {\bf D17}
  (1978) 2369--2374.

\bibitem{MSW-MS}
S.~Mikheyev and A.~Y. Smirnov, ``Resonant neutrino oscillations in matter,''
  {\em Prog.~Part.~Nucl.~Phys.} {\bf 23} (1989) 41--136.

\bibitem{Akhmedov}
E.~K. Akhmedov, A.~Dighe, P.~Lipari, and A.~Y. Smirnov, ``Atmospheric neutrinos
  at super-kamiokande and parametric resonance in neutrino oscillations,'' {\em
  Nucl.~Phys.} {\bf B542} (1999) 3--30,
  \href{http://xxx.lanl.gov/abs/arXiv:hep-ph/9808270}{{\tt
  arXiv:hep-ph/9808270}}.

\bibitem{Petcov-NOLR}
S.~T. Petcov, ``{Diffractive - like (or parametric resonance - like?)
  enhancement of the earth (day - night) effect for solar neutrinos crossing
  the earth core},'' {\em Phys. Lett.} {\bf B434} (1998) 321--332,
  \href{http://xxx.lanl.gov/abs/hep-ph/9805262}{{\tt hep-ph/9805262}}. See also
  \href{http://xxx.lanl.gov/abs/hep-ph/9811205}{{\tt hep-ph/9811205}}.

\bibitem{LoI}
{\bf IceCube PINGU} Collaboration, M.~G. Aartsen {\em et~al.}, ``{Letter of
  Intent: The Precision IceCube Next Generation Upgrade (PINGU)},''
  \href{http://xxx.lanl.gov/abs/arXiv:1401.2046}{{\tt arXiv:1401.2046}}.

\bibitem{Achterberg:2006md}
{\bf IceCube} Collaboration, A.~Achterberg {\em et~al.}, ``{First Year
  Performance of The IceCube Neutrino Telescope},'' {\em Astropart.~Phys.} {\bf
  26} (2006) 155--173,
  \href{http://xxx.lanl.gov/abs/arXiv:astro-ph/0604450}{{\tt
  arXiv:astro-ph/0604450}}.

\bibitem{Collaboration:2011ym}
{\bf IceCube} Collaboration, R.~Abbasi {\em et~al.}, ``{The Design and
  Performance of IceCube DeepCore},'' {\em Astropart.~Phys.} {\bf 35} (2012)
  615--624, \href{http://xxx.lanl.gov/abs/arXiv:1109.6096}{{\tt
  arXiv:1109.6096}}.

\bibitem{Benson:2014ag}
T.~Benson {\em et~al.}, ``{IceCube Enhanced Hot Water Drill functional
  description},'' {\em Ann.Glaciol.} {\bf 55} (2014), no.~68 105--114.

\bibitem{Andreopoulos:2009rq}
C.~Andreopoulos {\em et~al.}, ``{The GENIE Neutrino Monte Carlo Generator},''
  {\em Nucl. Instrum. Meth.} {\bf A614} (2010) 87--104,
  \href{http://xxx.lanl.gov/abs/arXiv:0905.2517}{{\tt arXiv:0905.2517}}.

\bibitem{Agostinelli:2002hh}
{\bf GEANT4} Collaboration, S.~Agostinelli {\em et~al.}, ``{GEANT4: A
  Simulation toolkit},'' {\em Nucl.~Instrum.~Meth.} {\bf A506} (2003) 250--303.

\bibitem{Chirkin:2013tma}
{\bf IceCube} Collaboration, D.~Chirkin, ``{Photon tracking with GPUs in
  IceCube},'' {\em Nucl.~Instrum.~Meth.} {\bf A725} (2013) 141--143.

\bibitem{LarsonMasters}
M.~Larson, ``{Simulation and Identification of Non-Poissonian Noise Triggers in
  the IceCube Neutrino Detector},'' {Master's Thesis}, University of Alabama,
  2013.

\bibitem{Aartsen:2013vja}
{\bf IceCube} Collaboration, M.~G. Aartsen {\em et~al.}, ``{Energy
  Reconstruction Methods in the IceCube Neutrino Telescope},'' {\em JINST} {\bf
  9} (2014) P03009, \href{http://xxx.lanl.gov/abs/arXiv:1311.4767}{{\tt
  arXiv:1311.4767}}.

\bibitem{Chirkin:icrc2013}
{\bf IceCube} Collaboration, D.~Chirkin, ``{Evidence of optical anisotropy of
  the South Pole ice},'' in {\em Proceedings of 33rd International Cosmic Ray
  Conference (ICRC2013)}, 2013.
\newblock \href{http://xxx.lanl.gov/abs/arXiv:1309.7010}{{\tt
  arXiv:1309.7010}}.

\bibitem{Ribordy:2013xea}
M.~Ribordy and A.~Y. Smirnov, ``{Improving the neutrino mass hierarchy
  identification with inelasticity measurement in PINGU and ORCA},'' {\em
  Phys.~Rev.} {\bf D87} (2013) 113007,
  \href{http://xxx.lanl.gov/abs/arXiv:1303.0758}{{\tt arXiv:1303.0758}}.

\bibitem{MultiNest_1}
F.~{Feroz}, M.~P. {Hobson}, and M.~{Bridges}, ``Multinest: an efficient and
  robust bayesian inference tool for cosmology and particle physics,'' {\em
  Mon.~Not.~Roy.~Astron.~Soc.} {\bf 398} (Oct., 2009) 1601--1614,
  \href{http://xxx.lanl.gov/abs/arXiv:0809.3437}{{\tt arXiv:0809.3437}}.

\bibitem{botev2010}
Z.~I. Botev, J.~F. Grotowski, and D.~P. Kroese, ``Kernel density estimation via
  diffusion,'' {\em Ann. Statist.} {\bf 38} (10, 2010) 2916--2957.

\bibitem{abramson1982}
I.~S. Abramson, ``On bandwidth variation in kernel estimates-a square root
  law,'' {\em Ann. Statist.} {\bf 10} (12, 1982) 1217--1223.

\bibitem{10.2307/2345597}
M.~C.~J. S.~J.~Sheather, ``A reliable data-based bandwidth selection method for
  kernel density estimation,'' {\em Journal of the Royal Statistical Society.
  Series B (Methodological)} {\bf 53} (1991), no.~3 683--690.

\bibitem{Aartsen:2012uu}
{\bf IceCube} Collaboration, M.~Aartsen {\em et~al.}, ``{Measurement of the
  Atmospheric $\nu_e$ flux in IceCube},'' {\em Phys.~Rev.~Lett.} {\bf 110}
  (2013) 151105, \href{http://xxx.lanl.gov/abs/arXiv:1212.4760}{{\tt
  arXiv:1212.4760}}.

\bibitem{ref:JasonNeutrinoProceedings}
J.~Koskinen, ``{Atmospheric neutrino results from IceCube/DeepCore and plans
  for PINGU},'' in {\em {Proceedings of the XXVII international conference on
  neutrino physics and astrophysics (Neutrino 2016)}}, 2016.

\bibitem{Hocker:2007ht}
A.~Hoecker, P.~Speckmayer, J.~Stelzer, J.~Therhaag, E.~von Toerne, and H.~Voss,
  ``{TMVA: Toolkit for Multivariate Data Analysis},'' {\em PoS} {\bf ACAT}
  (2007) 040, \href{http://xxx.lanl.gov/abs/arXiv:physics/0703039}{{\tt
  arXiv:physics/0703039}}.

\bibitem{PDG_review}
{\bf Particle Data Group} Collaboration, J.~Beringer {\em et~al.}, ``Review of
  particle physics,'' {\em Phys.~Rev.} {\bf D86} (Jul, 2012) 010001.

\bibitem{Honda:2006qj}
M.~Honda, T.~Kajita, K.~Kasahara, S.~Midorikawa, and T.~Sanuki, ``{Calculation
  of atmospheric neutrino flux using the interaction model calibrated with
  atmospheric muon data},'' {\em Phys. Rev.} {\bf D75} (2007) 043006,
  \href{http://xxx.lanl.gov/abs/astro-ph/0611418}{{\tt astro-ph/0611418}}.

\bibitem{Richard:2015aua}
{\bf Super-Kamiokande} Collaboration, E.~Richard {\em et~al.}, ``{Measurements
  of the atmospheric neutrino flux by Super-Kamiokande: energy spectra,
  geomagnetic effects, and solar modulation},''
  \href{http://xxx.lanl.gov/abs/1510.08127}{{\tt 1510.08127}}.

\bibitem{Daum:1994bf}
{\bf Frejus} Collaboration, K.~Daum {\em et~al.}, ``{Determination of the
  atmospheric neutrino spectra with the Frejus detector},'' {\em Z. Phys.} {\bf
  C66} (1995) 417--428.

\bibitem{Abbasi:2008ih}
{\bf IceCube} Collaboration, R.~Abbasi {\em et~al.}, ``{Search for Point
  Sources of High Energy Neutrinos with Final Data from AMANDA-II},'' {\em
  Phys. Rev.} {\bf D79} (2009) 062001,
  \href{http://xxx.lanl.gov/abs/0809.1646}{{\tt 0809.1646}}.

\bibitem{Abbasi:2010qv}
{\bf IceCube} Collaboration, R.~Abbasi {\em et~al.}, ``{The Energy Spectrum of
  Atmospheric Neutrinos between 2 and 200 TeV with the AMANDA-II Detector},''
  {\em Astropart. Phys.} {\bf 34} (2010) 48--58,
  \href{http://xxx.lanl.gov/abs/1004.2357}{{\tt 1004.2357}}.

\bibitem{Adrian-Martinez:2013bqq}
{\bf ANTARES} Collaboration, S.~Adrian-Martinez {\em et~al.}, ``{Measurement of
  the atmospheric $\nu_\mu$ energy spectrum from 100 GeV to 200 TeV with the
  ANTARES telescope},'' {\em Eur. Phys. J.} {\bf C73} (2013), no.~10 2606,
  \href{http://xxx.lanl.gov/abs/1308.1599}{{\tt 1308.1599}}. [Eur. Phys.
  J.C73,2606(2013)].

\bibitem{Aartsen:2014muf}
{\bf IceCube} Collaboration, M.~G. Aartsen {\em et~al.}, ``{Atmospheric and
  astrophysical neutrinos above 1 TeV interacting in IceCube},'' {\em Phys.
  Rev.} {\bf D91} (2015), no.~2 022001,
  \href{http://xxx.lanl.gov/abs/1410.1749}{{\tt 1410.1749}}.

\bibitem{Abbasi:2010ie}
{\bf IceCube} Collaboration, R.~Abbasi {\em et~al.}, ``{Measurement of the
  atmospheric neutrino energy spectrum from 100 GeV to 400 TeV with IceCube},''
  {\em Phys. Rev.} {\bf D83} (2011) 012001,
  \href{http://xxx.lanl.gov/abs/1010.3980}{{\tt 1010.3980}}.

\bibitem{Abbasi:2011ui}
{\bf IceCube} Collaboration, R.~Abbasi {\em et~al.}, ``{First search for
  atmospheric and extraterrestrial neutrino-induced cascades with the IceCube
  detector},'' {\em Phys. Rev.} {\bf D84} (2011) 072001,
  \href{http://xxx.lanl.gov/abs/1101.1692}{{\tt 1101.1692}}.

\bibitem{Honda:2011nf}
M.~Honda, T.~Kajita, K.~Kasahara, and S.~Midorikawa, ``{Improvement of low
  energy atmospheric neutrino flux calculation using the JAM nuclear
  interaction model},'' {\em Phys. Rev.} {\bf D83} (2011) 123001,
  \href{http://xxx.lanl.gov/abs/arXiv:1102.2688}{{\tt arXiv:1102.2688}}.

\bibitem{Dziewonski:1981xy}
A.~Dziewonski and D.~Anderson, ``{Preliminary reference earth model},'' {\em
  Phys.~Earth Planet.~In.} {\bf 25} (1981) 297--356.

\bibitem{Honda:2015fha}
M.~Honda, M.~S. Athar, T.~Kajita, K.~Kasahara, and S.~Midorikawa,
  ``{Atmospheric neutrino flux calculation using the NRLMSISE-00 atmospheric
  model},'' {\em Phys. Rev.} {\bf D92} (2015), no.~2 023004,
  \href{http://xxx.lanl.gov/abs/arXiv:1502.03916}{{\tt arXiv:1502.03916}}.

\bibitem{prob3++}
R.~Wendell, ``prob3++ software for computing three flavor neutrino oscillation
  probabilities..'' \url{http://www.phy.duke.edu/~raw22/public/Prob3++/}, 2012.

\bibitem{wilks1938}
S.~S. Wilks, ``The large-sample distribution of the likelihood ratio for
  testing composite hypotheses,'' {\em Ann. Math. Statist.} {\bf 9} (03, 1938)
  60--62.

\bibitem{NuFIT20}
M.~C. Gonzalez-Garcia, M.~Maltoni, and T.~Schwetz, ``{Updated fit to three
  neutrino mixing: status of leptonic CP violation},'' {\em JHEP} {\bf 11}
  (2014) 052, \href{http://xxx.lanl.gov/abs/arXiv:1409.5439}{{\tt
  arXiv:1409.5439}}.

\bibitem{PhysRevD.86.013012}
G.~L. Fogli, E.~Lisi, A.~Marrone, D.~Montanino, A.~Palazzo, and A.~M. Rotunno,
  ``Global analysis of neutrino masses, mixings, and phases: Entering the era
  of leptonic {CP} violation searchers,'' {\em Phys.~Rev.} {\bf D86} (Jul,
  2012) 013012.

\bibitem{NOvA}
``Nova plots and figures.''
  \url{http://www-nova.fnal.gov/plots\_and\_figures/plot\_and\_figures.html}.
\newblock Accessed: 2015.

\bibitem{Abe:2014tzr}
{\bf T2K} Collaboration, K.~Abe {\em et~al.}, ``{Neutrino oscillation physics
  potential of the T2K experiment},'' {\em PTEP} {\bf 4} (2015) 043C01,
  \href{http://xxx.lanl.gov/abs/arXiv:1409.7469}{{\tt arXiv:1409.7469}}.

\bibitem{ref:SNOPaper}
{\bf SNO} Collaboration, B.~Aharmim {\em et~al.}, ``{Electron energy spectra,
  fluxes, and day-night asymmetries of B-8 solar neutrinos from measurements
  with NaCl dissolved in the heavy-water detector at the Sudbury Neutrino
  Observatory},'' {\em Phys. Rev.} {\bf C72} (2005) 055502,
  \href{http://xxx.lanl.gov/abs/arXiv:nucl-ex/0502021}{{\tt
  arXiv:nucl-ex/0502021}}.

\bibitem{ARS}
E.~K. Akhmedov, S.~Razzaque, and A.~Y. Smirnov, ``{Mass hierarchy, 2-3 mixing
  and CP-phase with Huge Atmospheric Neutrino Detectors},'' {\em J.~High Energy
  Phys.} {\bf 1302} (2013) 082,
  \href{http://xxx.lanl.gov/abs/arXiv:1205.7071}{{\tt arXiv:1205.7071}}.

\bibitem{Barr2006}
G.~D. Barr, S.~Robbins, T.~K. Gaisser, and T.~Stanev, ``Uncertainties in
  atmospheric neutrino fluxes,'' {\em Phys. Rev. D} {\bf 74} (Nov, 2006)
  094009.

\bibitem{Capozzi:2015bxa}
F.~Capozzi, E.~Lisi, and A.~Marrone, ``{PINGU and the neutrino mass hierarchy:
  Statistical and systematic aspects},'' {\em Phys.Rev.} {\bf D91} (2015)
  073011, \href{http://xxx.lanl.gov/abs/arXiv:1503.01999}{{\tt
  arXiv:1503.01999}}.

\bibitem{Gonzalez-Garcia:2015qrr}
M.~C. Gonzalez-Garcia, M.~Maltoni, and T.~Schwetz, ``{Global Analyses of
  Neutrino Oscillation Experiments},'' {\em Nucl. Phys.} {\bf B908} (2016)
  199--217, \href{http://xxx.lanl.gov/abs/1512.06856}{{\tt 1512.06856}}.

\bibitem{Marrone:2015nip}
A.~Marrone, E.~Lisi, A.~Palazzo, D.~Montanino, and F.~Capozzi, ``{Global fits
  to neutrino oscillations: status and prospects},'' {\em PoS} {\bf
  EPS-HEP2015} (2015) 093.

\bibitem{Blennow:2013oma}
M.~Blennow, P.~Coloma, P.~Huber, and T.~Schwetz, ``{Quantifying the sensitivity
  of oscillation experiments to the neutrino mass ordering},'' {\em JHEP} {\bf
  03} (2014) 028, \href{http://xxx.lanl.gov/abs/arXiv:1311.1822}{{\tt
  arXiv:1311.1822}}.

\bibitem{Winter:2013ema}
W.~Winter, ``{Neutrino mass hierarchy determination with IceCube-PINGU},'' {\em
  Phys.~Rev.} {\bf D88} (2013) 013013,
  \href{http://xxx.lanl.gov/abs/arXiv:1305.5539}{{\tt arXiv:1305.5539}}.

\bibitem{Ge:2013zua}
S.-F. Ge, K.~Hagiwara, and C.~Rott, ``{A Novel Approach to Study Atmospheric
  Neutrino Oscillation},'' {\em JHEP} {\bf 06} (2014) 150,
  \href{http://xxx.lanl.gov/abs/1309.3176}{{\tt 1309.3176}}.

\bibitem{Ge:2013ffa}
S.-F. Ge and K.~Hagiwara, ``{Physics Reach of Atmospheric Neutrino Measurements
  at PINGU},'' {\em JHEP} {\bf 09} (2014) 024,
  \href{http://xxx.lanl.gov/abs/1312.0457}{{\tt 1312.0457}}.

\bibitem{Antusch:2006vwa}
S.~Antusch, C.~Biggio, E.~Fernandez-Martinez, M.~Gavela, and J.~Lopez-Pavon,
  ``{Unitarity of the Leptonic Mixing Matrix},'' {\em JHEP} {\bf 0610} (2006)
  084, \href{http://xxx.lanl.gov/abs/hep-ph/0607020}{{\tt hep-ph/0607020}}.

\bibitem{Antusch:2008tz}
S.~Antusch, J.~P. Baumann, and E.~Fernandez-Martinez, ``{Non-Standard Neutrino
  Interactions with Matter from Physics Beyond the Standard Model},'' {\em
  Nucl.Phys.} {\bf B810} (2009) 369--388,
  \href{http://xxx.lanl.gov/abs/0807.1003}{{\tt 0807.1003}}.

\bibitem{Meloni:2009cg}
D.~Meloni, T.~Ohlsson, W.~Winter, and H.~Zhang, ``{Non-standard interactions
  versus non-unitary lepton flavor mixing at a neutrino factory},'' {\em JHEP}
  {\bf 04} (2010) 041, \href{http://xxx.lanl.gov/abs/0912.2735}{{\tt
  0912.2735}}.

\bibitem{Abe:2012jj}
{\bf Super-Kamiokande Collaboration} Collaboration, K.~Abe {\em et~al.},
  ``{Evidence for the Appearance of Atmospheric Tau Neutrinos in
  Super-Kamiokande},'' {\em Phys.Rev.Lett.} {\bf 110} (2013), no.~18 181802,
  \href{http://xxx.lanl.gov/abs/arXiv:1206.0328}{{\tt arXiv:1206.0328}}.

\bibitem{MoriyamaNu2016}
S.~Moriyama, ``New atmospheric and solar results from super-kamiokande.''
  Presented at the XXVII International Conference on Neutrino Physics and
  Astrophysics (Neutrino2016), 2016.
\newblock
  \url{http://neutrino2016.iopconfs.org/IOP/media/uploaded/EVIOP/event_948/neutrino2016-moriyama-pub-2.pdf}.

\bibitem{Agafonova:2015jxn}
{\bf OPERA} Collaboration, N.~Agafonova {\em et~al.}, ``{Discovery of $\tau$
  Neutrino Appearance in the CNGS Neutrino Beam with the OPERA Experiment},''
  {\em Phys. Rev. Lett.} {\bf 115} (2015), no.~12 121802,
  \href{http://xxx.lanl.gov/abs/arXiv:1507.01417}{{\tt arXiv:1507.01417}}.

\bibitem{Gilbert1600}
W.~Gilbert, {\em {De magnete}}.
\newblock 1600.

\bibitem{Buffett16062000}
B.~A. Buffett, ``Earth's core and the geodynamo,'' {\em Science} {\bf 288}
  (2000), no.~5473 2007--2012,
  \href{http://xxx.lanl.gov/abs/http://www.sciencemag.org/content/288/5473/2007.full.pdf}{{\tt
  http://www.sciencemag.org/content/288/5473/2007.full.pdf}}.

\bibitem{Popov1999345}
Y.~A. Popov, S.~L. Pevzner, V.~P. Pimenov, and R.~A. Romushkevich, ``New
  geothermal data from the {K}ola superdeep well {SG}-3,'' {\em Tectonophysics}
  {\bf 306} (1999), no.~3--4 345 -- 366.

\bibitem{Hofmann1997}
A.~W. Hofmann, ``{Mantle geochemistry: the message from oceanic volcanism},''
  {\em Nature} {\bf 385} (Jan., 1997) 219--229.

\bibitem{2003TrGeo...2..547M}
W.~F. {McDonough}, ``{Compositional Model for the Earth's Core},'' {\em
  Treatise on Geochemistry} {\bf 2} (Dec., 2003) 547--568.

\bibitem{Oldham01021906}
R.~D. Oldham, ``The constitution of the interior of the earth, as revealed by
  earthquakes,'' {\em Q.~J.~Geol.~Soc.} {\bf 62} (1906), no.~1-4 456--475,
  \href{http://xxx.lanl.gov/abs/http://jgslegacy.lyellcollection.org/content/62/1-4/456.full.pdf+html}{{\tt
  http://jgslegacy.lyellcollection.org/content/62/1-4/456.full.pdf+html}}.

\bibitem{Resovsky2002}
J.~S. Resovsky and J.~Trampert, ``{Reliable mantle density error bars: an
  application of the neighbourhood algorithm to normal-mode and surface wave
  data},'' {\em Geophys.~J.~Int.} {\bf 150} (Sept., 2002) 665--672.

\bibitem{Smylie1992}
D.~E. Smylie, ``{The Inner Core Translational Triplet and the Density Near
  Earth's Center},'' {\em Science} {\bf 255} (Mar., 1992) 1678--82.

\bibitem{Winter:2015zwx}
W.~Winter, ``{Atmospheric Neutrino Oscillations for Earth Tomography},'' {\em
  Nucl. Phys.} {\bf B908} (2016) 250--267,
  \href{http://xxx.lanl.gov/abs/arXiv:1511.05154}{{\tt arXiv:1511.05154}}.

\bibitem{McDonough1995}
W.~McDonough and S.~Sun, ``{The composition of the Earth},'' {\em Chem.~Geol.}
  {\bf 120} (Mar., 1995) 223--253.

\bibitem{Birch1952}
F.~Birch, ``{Elasticity and constitution of the Earth's interior},'' {\em
  J.~Geophys.~Res.} {\bf 57} (June, 1952) 227--286.

\bibitem{Fearn1981}
D.~R. Fearn and D.~E. Loper, ``{Compositional convection and stratification of
  Earth's core},'' {\em Nature} {\bf 289} (Jan., 1981) 393--394.

\bibitem{Allegre1995}
C.~J. All\`{e}gre, J.-P. Poirier, E.~Humler, and A.~W. Hofmann, ``{The chemical
  composition of the Earth},'' {\em Earth Planet.~Sc.~Lett.} {\bf 134} (Sept.,
  1995) 515--526.

\bibitem{Li2007}
J.~Li and Y.~Fei, ``{Experimental Constraints on Core Composition},'' in {\em
  Treatise on geochemistry, Vol. 2: The Mantle and core} (R.~W. Carlson, ed.),
  vol.~2, ch.~2.14, pp.~1--31.
\newblock Elsevier, Amsterdam, 2007.

\bibitem{Narygina2011}
O.~Narygina, L.~S. Dubrovinsky, C.~a. McCammon, A.~Kurnosov, I.~Y. Kantor,
  V.~B. Prakapenka, and N.~a. Dubrovinskaia, ``{X-ray diffraction and
  M\"{o}ssbauer spectroscopy study of fcc iron hydride FeH at high pressures
  and implications for the composition of the Earth's core},'' {\em Earth
  Planet.~Sc.~Lett.} {\bf 307} (July, 2011) 409--414.

\bibitem{Allegre200149}
C.~All{\`e}gre, G.~Manh{\`e}s, and {\'E}.~Lewin, ``Chemical composition of the
  earth and the volatility control on planetary genetics,'' {\em Earth and
  Planetary Science Letters} {\bf 185} (2001), no.~1--2 49 -- 69.

\bibitem{Huang2011}
H.~Huang, Y.~Fei, L.~Cai, F.~Jing, X.~Hu, H.~Xie, L.~Zhang, and Z.~Gong,
  ``Evidence for an oxygen-depleted liquid outer core of the earth,'' {\em
  Nature} {\bf 479} (11, 2011) 513--516.

\bibitem{Rott:2015kwa}
C.~Rott, A.~Taketa, and D.~Bose, ``{Spectrometry of the Earth using Neutrino
  Oscillations},'' {\em Scientific Reports 5, 15225} (2015)
  \href{http://xxx.lanl.gov/abs/arXiv:1502.04930}{{\tt arXiv:1502.04930}}.

\bibitem{RevModPhys.84.1307}
J.~A. Formaggio and G.~P. Zeller, ``From {eV} to {EeV}: Neutrino cross sections
  across energy scales,'' {\em Rev.~Mod.~Phys.} {\bf 84} (Sep, 2012)
  1307--1341.

\bibitem{2011JPhCS.309a2029K}
L.~{K{\"o}pke} and {IceCube Collaboration}, ``{Supernova Neutrino Detection
  with IceCube},'' {\em Journal of Physics Conference Series} {\bf 309} (Aug.,
  2011) 012029, \href{http://xxx.lanl.gov/abs/arXiv:1106.6225}{{\tt
  arXiv:1106.6225}}.

\bibitem{ICRCSN}
{\bf IceCube} Collaboration, V.~Baum, L.~Demir\"ors, L.~K\"opke, and
  M.~Ribordy, ``{Supernova detection with IceCube and beyond},'' in {\em
  Proceedings of the 32nd International Cosmic Ray Conference (ICRC2011)},
  2011.

\bibitem{Gaisser}
T.~Gaisser, {\em Cosmic Rays and Particle Physics}.
\newblock Cambridge University Press, 1991.

\bibitem{Heck:1998vt}
D.~Heck, G.~Schatz, T.~Thouw, J.~Knapp, and J.~Capdevielle, ``{CORSIKA: A Monte
  Carlo code to simulate extensive air showers},'' {\em Forschungszentrum
  Karlsruhe Report FZKA 6019} (1998).

\bibitem{garching}
L.~{H{\"u}depohl}, B.~{M{\"u}ller}, H.-T. {Janka}, A.~{Marek}, and G.~G.
  {Raffelt}, ``{Neutrino Signal of Electron-Capture Supernovae from Core
  Collapse to Cooling},'' {\em Physical Review Letters} {\bf 104} (June, 2010)
  251101, \href{http://xxx.lanl.gov/abs/0912.0260}{{\tt 0912.0260}}.

\bibitem{Demiroers:2011am}
M.~Salathe, M.~Ribordy, and L.~Demirors, ``{Novel technique for supernova
  detection with IceCube},'' {\em Astropart. Phys.} {\bf 35} (2012) 485--494,
  \href{http://xxx.lanl.gov/abs/arXiv:1106.1937}{{\tt arXiv:1106.1937}}.

\bibitem{Varenna-Ribordy}
M.~Ribordy, ``{Methods and problems in neutrino observatories},'' {\em Proc.
  Int. Sch. Phys. Fermi} {\bf 182} (2012) 207--255,
  \href{http://xxx.lanl.gov/abs/1205.4965}{{\tt 1205.4965}}.

\bibitem{Bruijn:2013ibl}
{\bf IceCube} Collaboration, R.~Bruijn, ``{Supernova Detection in IceCube:
  Status and Future},'' in {\em {Proceedings of the Neutrino Oscillation
  Workshop (NOW 2012)}}, 2013.
\newblock \href{http://xxx.lanl.gov/abs/arXiv:1302.2040}{{\tt
  arXiv:1302.2040}}.

\bibitem{Keil}
M.~T. Keil, G.~G. Raffelt, and H.-T. Janka, ``Monte carlo study of supernova
  neutrino spectra formation,'' {\em The Astrophysical Journal} {\bf 590}
  (2003), no.~2 971.

\bibitem{Zwicky:1933gu}
F.~Zwicky, ``{Die Rotverschiebung von extragalaktischen Nebeln},'' {\em
  Helv.~Phys.~Acta} {\bf 6} (1933) 110--127.

\bibitem{Komatsu:2010fb}
{\bf WMAP} Collaboration, E.~Komatsu {\em et~al.}, ``{Seven-Year Wilkinson
  Microwave Anisotropy Probe (WMAP) Observations: Cosmological
  Interpretation},'' {\em Astrophys.~J.~Suppl.} {\bf 192} (2011) 18,
  \href{http://xxx.lanl.gov/abs/arXiv:1001.4538}{{\tt arXiv:1001.4538}}.

\bibitem{Bertone:2004pz}
G.~Bertone, D.~Hooper, and J.~Silk, ``{Particle dark matter: Evidence,
  candidates and constraints},'' {\em Phys.~Rep.} {\bf 405} (2005) 279--390,
  \href{http://xxx.lanl.gov/abs/arXiv:hep-ph/0404175}{{\tt
  arXiv:hep-ph/0404175}}.

\bibitem{Bertone:2010}
G.~Bertone, ed., {\em {Non-WIMP Candidates}}.
\newblock Particle Dark Matter: Observations, Models and Searches. 2010.

\bibitem{Kolb:2001}
I.~F.~M. Albuquerque, L.~Hui, and E.~W. Kolb, ``{High energy neutrinos from
  superheavy dark matter annihilation},'' {\em Phys. Rev. D} {\bf 64} (2001)
  083504.

\bibitem{Aartsen:2012kia}
{\bf IceCube} Collaboration, M.~Aartsen {\em et~al.}, ``{Search for dark matter
  annihilations in the Sun with the 79-string IceCube detector},'' {\em
  Phys.~Rev.~Lett.} {\bf 110} (2013) 131302,
  \href{http://xxx.lanl.gov/abs/arXiv:1212.4097}{{\tt arXiv:1212.4097}}.

\bibitem{Abbasi:2011eq}
{\bf IceCube} Collaboration, R.~Abbasi {\em et~al.}, ``{Search for Dark Matter
  from the Galactic Halo with the IceCube Neutrino Observatory},'' {\em
  Phys.~Rev.} {\bf D84} (2011) 022004,
  \href{http://xxx.lanl.gov/abs/arXiv:1101.3349}{{\tt arXiv:1101.3349}}.

\bibitem{Abbasi:2012ws}
{\bf IceCube} Collaboration, M.~Aartsen {\em et~al.}, ``{Search for Neutrinos
  from Dark Matter Self-Annihilations in the centre of the Milky Way with 3
  years of IceCube/DeepCore},''
  \href{http://xxx.lanl.gov/abs/arXiv:1705.08103}{{\tt arXiv:1705.08103}}.

\bibitem{Aartsen:2013dxa}
{\bf IceCube} Collaboration, M.~G. Aartsen {\em et~al.}, ``{IceCube Search for
  Dark Matter Annihilation in nearby Galaxies and Galaxy Clusters},'' {\em
  Phys. Rev.} {\bf D88} (2013) 122001,
  \href{http://xxx.lanl.gov/abs/arXiv:1307.3473}{{\tt arXiv:1307.3473}}.

\bibitem{IceCube:2011ae}
{\bf IceCube} Collaboration, R.~Abbasi {\em et~al.}, ``{The IceCube Neutrino
  Observatory IV: Searches for Dark Matter and Exotic Particles},''
  \href{http://xxx.lanl.gov/abs/arXiv:1111.2738}{{\tt arXiv:1111.2738}}.

\bibitem{Yuksel:2007ac}
H.~Yuksel, S.~Horiuchi, J.~F. Beacom, and S.~Ando, ``{Neutrino Constraints on
  the Dark Matter Total Annihilation Cross Section},'' {\em Phys.~Rev.} {\bf
  D76} (2007) 123506, \href{http://xxx.lanl.gov/abs/arXiv:0707.0196}{{\tt
  arXiv:0707.0196}}.

\bibitem{Dasgupta:2012bd}
B.~Dasgupta and R.~Laha, ``{Neutrinos in IceCube/KM3NeT as probes of Dark
  Matter Substructures in Galaxy Clusters},'' {\em Phys.~Rev.} {\bf D86} (2012)
  093001, \href{http://xxx.lanl.gov/abs/arXiv:1206.1322}{{\tt
  arXiv:1206.1322}}.

\bibitem{Savage:2008er}
C.~Savage, G.~Gelmini, P.~Gondolo, and K.~Freese, ``{Compatibility of
  DAMA/LIBRA dark matter detection with other searches},'' {\em
  J.~Cosmol.~Astropart.~P.} {\bf 0904} (2009) 010,
  \href{http://xxx.lanl.gov/abs/arXiv:0808.3607}{{\tt arXiv:0808.3607}}.

\bibitem{Feng:2011vu}
J.~L. Feng, J.~Kumar, D.~Marfatia, and D.~Sanford, ``{Isospin-Violating Dark
  Matter},'' {\em Phys.~Lett.} {\bf B703} (2011) 124--127,
  \href{http://xxx.lanl.gov/abs/arXiv:1102.4331}{{\tt arXiv:1102.4331}}.

\bibitem{Rott:2011fh}
C.~Rott, T.~Tanaka, and Y.~Itow, ``{Enhanced Sensitivity to Dark Matter
  Self-annihilations in the Sun using Neutrino Spectral Information},'' {\em
  J.~Cosmol.~Astropart.~P.} {\bf 1109} (2011) 029,
  \href{http://xxx.lanl.gov/abs/arXiv:1107.3182}{{\tt arXiv:1107.3182}}.

\bibitem{Choi:2015ara}
{\bf Super-Kamiokande} Collaboration, K.~Choi {\em et~al.}, ``{Search for
  neutrinos from annihilation of captured low-mass dark matter particles in the
  Sun by Super-Kamiokande},'' {\em Phys.Rev.Lett.} {\bf 114} (2015), no.~14
  141301, \href{http://xxx.lanl.gov/abs/arXiv:1503.04858}{{\tt
  arXiv:1503.04858}}.

\bibitem{Gondolo:2004sc}
P.~Gondolo {\em et~al.}, ``{DarkSUSY: Computing supersymmetric dark matter
  properties numerically},'' {\em J.~Cosmol.~Astropart.~P.} {\bf 0407} (2004)
  008, \href{http://xxx.lanl.gov/abs/arXiv:astro-ph/0406204}{{\tt
  arXiv:astro-ph/0406204}}.

\bibitem{Edsjoe:2013ws}
J.~Edsj\"{o}, ``{WimpSim Neutrino Monte Carlo}.''
  \url{http://www.fysik.su.se/~edsjo/wimpsim/}, 2013.

\bibitem{Rolke:2005tr}
W.~Rolke, A.~Lopez, J.~Conrad, and F.~James, ``{Limits and Confidence Intervals
  in presence of nuisance parameters},'' {\em Nucl.~Instrum.~Meth.} {\bf A551}
  (2005) 493--503, \href{http://xxx.lanl.gov/abs/arXiv:physics/0403059v5}{{\tt
  arXiv:physics/0403059v5}}.

\bibitem{Navarro:1995iw}
J.~F. Navarro, C.~S. Frenk, and S.~D.~M. White, ``{The Structure of Cold Dark
  Matter Halos},'' {\em Astrophys.~J.} {\bf 462} (1996) 563--575,
  \href{http://xxx.lanl.gov/abs/arXiv:astro-ph/9508025}{{\tt
  arXiv:astro-ph/9508025}}.

\bibitem{Aartsen:2013ae}
{\bf IceCube} Collaboration, M.~G. Aartsen {\em et~al.}, ``{The IceCube
  Neutrino Observatory Part IV: Searches for Dark Matter and Exotic
  Particles},'' in {\em {Proceedings, 33rd International Cosmic Ray Conference
  (ICRC2013)}}, 2013.
\newblock \href{http://xxx.lanl.gov/abs/arXiv:1309.7007}{{\tt
  arXiv:1309.7007}}.

\bibitem{Bernabei:2010mq}
{\bf DAMA/LIBRA} Collaboration, R.~Bernabei {\em et~al.}, ``{New results from
  DAMA/LIBRA},'' {\em Eur.~Phys.~J.} {\bf C67} (2010) 39--49,
  \href{http://xxx.lanl.gov/abs/arXiv:1002.1028}{{\tt arXiv:1002.1028}}.

\bibitem{Aalseth:2010vx}
{\bf CoGeNT} Collaboration, C.~Aalseth {\em et~al.}, ``{Results from a Search
  for Light-Mass Dark Matter with a P-type Point Contact Germanium Detector},''
  {\em Phys.~Rev.~Lett.} {\bf 106} (2011) 131301,
  \href{http://xxx.lanl.gov/abs/arXiv:1002.4703}{{\tt arXiv:1002.4703}}.

\bibitem{Aalseth:2011wp}
C.~Aalseth, P.~Barbeau, J.~Colaresi, J.~Collar, J.~Diaz~Leon, {\em et~al.},
  ``{Search for an Annual Modulation in a P-type Point Contact Germanium Dark
  Matter Detector},'' {\em Phys.~Rev.~Lett.} {\bf 107} (2011) 141301,
  \href{http://xxx.lanl.gov/abs/arXiv:1106.0650}{{\tt arXiv:1106.0650}}.

\bibitem{Agnese:2013rvf}
{\bf CDMS} Collaboration, R.~Agnese {\em et~al.}, ``{Silicon Detector Dark
  Matter Results from the Final Exposure of CDMS II},'' {\em Phys. Rev. Lett.}
  {\bf 111} (2013), no.~25 251301,
  \href{http://xxx.lanl.gov/abs/arXiv:1304.4279}{{\tt arXiv:1304.4279}}.

\bibitem{Abbasi:2009domdaq}
{\bf IceCube} Collaboration, R.~Abbasi {\em et~al.}, ``{The IceCube data
  acquisition system: Signal capture, digitization, and timestamping},'' {\em
  Nucl.~Instrum.~Meth.} {\bf A601} (2009) 294--316.

\bibitem{Adrian-Martinez:2014vja}
{\bf KM3NeT} Collaboration, S.~Adrian-Martinez {\em et~al.}, ``{Deep sea tests
  of a prototype of the KM3NeT digital optical module},'' {\em Eur.~Phys.~J.}
  {\bf C74} (2014) 3056.

\bibitem{Timmer:2010zz}
P.~Timmer, E.~Heine, and H.~Peek, ``{Very low power, high voltage base for a
  photo multiplier tube for the KM3NeT deep sea neutrino telescope},'' {\em
  JINST} {\bf 5} (2010) C12049.

\bibitem{Lundberg:2007mf}
J.~Lundberg, P.~Miocinovic, T.~Burgess, J.~Adams, S.~Hundertmark, P.~Desiati,
  K.~Woschnagg, and P.~Niessen, ``{Light tracking for glaciers and oceans:
  Scattering and absorption in heterogeneous media with Photonics},'' {\em
  Nucl. Instrum. Meth.} {\bf A581} (2007) 619--631,
  \href{http://xxx.lanl.gov/abs/astro-ph/0702108}{{\tt astro-ph/0702108}}.

\bibitem{Rongen:2016sbk}
{\bf IceCube} Collaboration, M.~Rongen, ``{Measuring the optical properties of
  IceCube drill holes},'' in {\em {{Proceedings of the 7th Very Large Volume
  Neutrino Telescope Workshop (VLVnT 2015)}}}, 2016.
\newblock
  \url{http://www.epj-conferences.org/articles/epjconf/pdf/2016/11/epjconf-VLVnT2015_06011.pdf}.

\bibitem{Tosi:2015ica}
{\bf IceCube} Collaboration, D.~Tosi and C.~Wendt, ``{Calibrating photon
  detection efficiency in IceCube},'' {\em PoS} {\bf TIPP2014} (2014) 157,
  \href{http://xxx.lanl.gov/abs/arXiv:1502.03102}{{\tt arXiv:1502.03102}}.

\bibitem{IceCube:2007aa}
{\bf IceCube} Collaboration, ``{The IceCube Collaboration: Contributions to the
  30th International Cosmic Ray Conference (ICRC 2007)},'' in {\em {Proceedings
  of the 30th International Cosmic Ray Conference (ICRC 2007)}}, 2007.
\newblock \href{http://xxx.lanl.gov/abs/arXiv:0711.0353}{{\tt
  arXiv:0711.0353}}.

\bibitem{Kapustinsky1985612}
J.~Kapustinsky {\em et~al.}, ``A fast timing light pulser for scintillation
  detectors,'' {\em Nucl.~Instrum.~Meth.} {\bf A241} (1985), no.~2--3 612 --
  613.

\bibitem{Jurkovic:2016kxn}
{\bf IceCube Gen2} Collaboration, M.~Jurkovič, K.~Abraham, K.~Holzapfel,
  K.~Krings, E.~Resconi, and J.~Veenkamp, ``{A Precision Optical Calibration
  Module (POCAM) for IceCube-Gen2},'' {\em EPJ Web Conf.} {\bf 116} (2016)
  06001.

\bibitem{Krings:icrc2015}
{\bf IceCube} Collaboration, K.~Krings, ``{A Precision Optical Calibration
  Module for IceCube-Gen2},'' in {\em Proceedings of 34rd International Cosmic
  Ray Conference (ICRC2015)}, 2015.
\newblock \href{http://xxx.lanl.gov/abs/arXiv:1510.05228}{{\tt
  arXiv:1510.05228}}.

\bibitem{Moffat2005255}
B.~Moffat {\em et~al.}, ``Optical calibration hardware for the sudbury neutrino
  observatory,'' {\em Nucl.~Instrum.~Meth.} {\bf A554} (2005), no.~1--3 255 --
  265.

\bibitem{Bose:icrc2015}
{\bf IceCube} Collaboration, D.~Bose, ``{PINGU Camera System to Study
  Properties of the Antarctic Ice},'' {\em Proceedings of 34rd International
  Cosmic Ray Conference (ICRC2015)} (2015)
  \href{http://xxx.lanl.gov/abs/arXiv:1510.05228}{{\tt arXiv:1510.05228}}.

\bibitem{Cherwinka:2014xta}
{\bf DM-Ice17} Collaboration, J.~Cherwinka {\em et~al.}, ``{First data from
  DM-Ice17},'' {\em Phys.Rev.} {\bf D90} (2014), no.~9 092005,
  \href{http://xxx.lanl.gov/abs/arXiv:1401.4804}{{\tt arXiv:1401.4804}}.

\bibitem{HubbardPhD}
A.~Hubbard, {\em {Muon-Induced Backgrounds in the DM-Ice17 NaI (Tl) Dark Matter
  Detector}}.
\newblock {PhD Dissertation}, University of Wisconsin -- Madison, 2015.

\bibitem{mtom}
S.~N. {Axani}, J.~M. {Conrad}, and C.~{Kirby}, ``{The Desktop Muon Detector: A
  simple, physics-motivated machine- and electronics-shop project for
  university students},'' {\em ArXiv e-prints} (June, 2016)
  \href{http://xxx.lanl.gov/abs/1606.01196}{{\tt 1606.01196}}.

\bibitem{Gazizov:2004va}
A.~Gazizov and M.~P. Kowalski, ``{ANIS: High energy neutrino generator for
  neutrino telescopes},'' {\em Comput.~Phys.~Commun.} {\bf 172} (2005)
  203--213, \href{http://xxx.lanl.gov/abs/arXiv:astro-ph/0406439}{{\tt
  arXiv:astro-ph/0406439}}.

\bibitem{Athar:2012it}
M.~Sajjad~Athar, M.~Honda, T.~Kajita, K.~Kasahara, and S.~Midorikawa,
  ``{Atmospheric neutrino flux at INO, South Pole and Pyh\"asalmi},'' {\em
  Phys.~Lett.} {\bf B718} (2013) 1375--1380,
  \href{http://xxx.lanl.gov/abs/arXiv:1210.5154}{{\tt arXiv:1210.5154}}.

\bibitem{BodekYang:2003}
A.~Bodek and U.~Yang, ``{Higher twist, $\xi_W$ scaling, and effective LO PDFs
  for lepton scattering in the few GeV region},'' {\em Journal of Physics G:
  Nuclear and Particle Physics} {\bf 29} (2003) 1899--1905,
  \href{http://xxx.lanl.gov/abs/arXiv:hep-ex/0210024v2}{{\tt
  arXiv:hep-ex/0210024v2}}.

\bibitem{Rein:1980wg}
D.~Rein and L.~M. Sehgal, ``{Neutrino Excitation of Baryon Resonances and
  Single Pion Production},'' {\em Annals Phys.} {\bf 133} (1981) 79--153.

\bibitem{PROPOSAL}
J.-H. Koehne, K.~Frantzen, M.~Schmitz, T.~Fuchs, W.~Rhode, D.~Chirkin, and
  J.~B. Tjus, ``Proposal: A tool for propagation of charged leptons,'' {\em
  Computer Physics Communications} {\bf 184} (2013), no.~9 2070 -- 2090.

\bibitem{AstroAnt}
C.~Pryke, A.~Karle, and C.~Kulesa, eds., {\em Report from the Workshop
  ``Astrophysics from the South Pole: Status and Future Prospects"}.
\newblock 2011.
\newblock \url{http://find.spa.umn.edu/~pryke/southpolemeeting/sp_ws_wp.pdf}.

\bibitem{IAU288}
M.~G. Burton, X.~Cui, and N.~F.~H. Tothill, eds., {\em Symposium S288
  (Astrophysics from Antarctica)}, vol.~8 of {\em Proceedings of the
  International Astronomical Union}.
\newblock 2012.
\newblock \url{http://www.phys.unsw.edu.au/IAUS288}.

\bibitem{SPoT}
J.~H. Lever and P.~Thur, ``{Economic Analysis of the South Pole Traverse,
  ERDC/CRREL TR-14-7}.''
  \url{http://acwc.sdp.sirsi.net/client/search/asset/1035041}.

\bibitem{Calland:2013vaa}
R.~G. Calland, A.~C. Kaboth, and D.~Payne, ``{Accelerated Event-by-Event
  Neutrino Oscillation Reweighting with Matter Effects on a GPU},'' {\em JINST}
  {\bf 9} (2014) P04016, \href{http://xxx.lanl.gov/abs/arXiv:1311.7579}{{\tt
  arXiv:1311.7579}}.

\bibitem{Ciuffoli:2013rza}
E.~Ciuffoli, J.~Evslin, and X.~Zhang, ``{Confidence in a neutrino mass
  hierarchy determination},'' {\em JHEP} {\bf 01} (2014) 095,
  \href{http://xxx.lanl.gov/abs/arXiv:1305.5150}{{\tt arXiv:1305.5150}}.

\bibitem{Casper:2002sd}
D.~Casper, ``{The Nuance neutrino physics simulation, and the future},'' in
  {\em {Proceedings of the 1st International Workshop on Neutrino-nucleus
  interactions in the few GeV region (NuInt 01)}}, 2002.
\newblock \href{http://xxx.lanl.gov/abs/hep-ph/0208030}{{\tt hep-ph/0208030}}.

\bibitem{Andreopoulos:2015wxa}
C.~Andreopoulos, C.~Barry, S.~Dytman, H.~Gallagher, T.~Golan, R.~Hatcher,
  G.~Perdue, and J.~Yarba, ``{The GENIE Neutrino Monte Carlo Generator: Physics
  and User Manual},'' \href{http://xxx.lanl.gov/abs/arXiv:1510.05494}{{\tt
  arXiv:1510.05494}}.

\bibitem{Katori:2014fxa}
T.~Katori and S.~Mandalia, ``{PYTHIA hadronization process tuning in the GENIE
  neutrino interaction generator},'' {\em J. Phys.} {\bf G42} (2015) 115004,
  \href{http://xxx.lanl.gov/abs/arXiv:1412.4301}{{\tt arXiv:1412.4301}}.

\bibitem{Bodek:2010km}
A.~Bodek and U.-k. Yang, ``{Axial and Vector Structure Functions for Electron-
  and Neutrino- Nucleon Scattering Cross Sections at all $Q^2$ using Effective
  Leading order Parton Distribution Functions},''
  \href{http://xxx.lanl.gov/abs/arXiv:1011.6592}{{\tt arXiv:1011.6592}}.

\bibitem{Barr:2004br}
G.~D. Barr, T.~K. Gaisser, P.~Lipari, S.~Robbins, and T.~Stanev, ``{A Three -
  dimensional calculation of atmospheric neutrinos},'' {\em Phys. Rev.} {\bf
  D70} (2004) 023006, \href{http://xxx.lanl.gov/abs/astro-ph/0403630}{{\tt
  astro-ph/0403630}}.

\bibitem{Fedynitch:2015zbe}
A.~Fedynitch, R.~Engel, T.~K. Gaisser, F.~Riehn, and S.~Todor, ``{MCEQ -
  numerical code for inclusive lepton flux calculations},'' in {\em
  {Proceedings of the 34th International Cosmic Ray Conference (ICRC 2015)}},
  2015.
\newblock \href{http://xxx.lanl.gov/abs/arXiv:1503.00544}{{\tt
  arXiv:1503.00544}}.

\bibitem{Abe:2015mga}
K.~Abe {\em et~al.}, ``{Measurements of cosmic-ray proton and helium spectra
  from the BESS-Polar long-duration balloon flights over Antarctica},'' {\em
  Astrophys. J.} {\bf 822} (2016), no.~2 65,
  \href{http://xxx.lanl.gov/abs/arXiv:1506.01267}{{\tt arXiv:1506.01267}}.

\bibitem{Kopenkin:2009zz}
V.~Kopenkin and T.~Sinzi, ``{Cosmic ray primary composition in the energy range
  10-1000 TeV obtained by passive balloonborne detector: Reanalysis of the
  RUNJOB experiment},'' {\em Phys. Rev.} {\bf D79} (2009) 072011.

\bibitem{Christ:1998zz}
M.~J. Christ {\em et~al.}, ``{Cosmic-ray proton and helium spectra: Results
  from the JACEE Experiment},'' {\em Astrophys. J.} {\bf 502} (1998) 278--283.

\bibitem{Aguilar:2015ooa}
{\bf AMS} Collaboration, M.~Aguilar {\em et~al.}, ``{Precision Measurement of
  the Proton Flux in Primary Cosmic Rays from Rigidity 1 GV to 1.8 TV with the
  Alpha Magnetic Spectrometer on the International Space Station},'' {\em Phys.
  Rev. Lett.} {\bf 114} (2015) 171103.

\bibitem{Adriani:2015aps}
{\bf PAMELA} Collaboration, O.~Adriani {\em et~al.}, ``{Measurements of
  Cosmic-Ray Hydrogen and Helium Isotopes with the PAMELA experiment},'' {\em
  Astrophys. J.} {\bf 818} (2016), no.~1 68,
  \href{http://xxx.lanl.gov/abs/arXiv:1512.06535}{{\tt arXiv:1512.06535}}.

\bibitem{Yoon:2011aa}
Y.~S. Yoon {\em et~al.}, ``{Cosmic-Ray Proton and Helium Spectra from the First
  CREAM Flight},'' {\em Astrophys. J.} {\bf 728} (2011) 122,
  \href{http://xxx.lanl.gov/abs/arXiv:1102.2575}{{\tt arXiv:1102.2575}}.

\bibitem{Gaisser:2001jw}
T.~K. Gaisser, T.~Stanev, M.~Honda, and P.~Lipari, ``{Primary spectrum to 1-TeV
  and beyond},'' in {\em {Proceedings of the 27th International Cosmic Ray
  Conference (ICRC 2001)}}, pp.~1643--1646, 2001.
\newblock \url{http://www.copernicus.org/icrc/papers/ici6694_p.pdf}.

\bibitem{HondaANW}
M.~Honda, ``{HKKM} atmospheric neutrino fluxes.'' Presented at the 1st
  Atmospheric Neutrino Workshop, 2016.
\newblock
  \url{http://indico.universe-cluster.de/indico/conferenceDisplay.py?confId=3533}.

\bibitem{Thebault2015}
E.~Th{\'e}bault {\em et~al.}, ``{International Geomagnetic Reference Field}:
  the 12th generation,'' {\em Earth, Planets and Space} {\bf 67} (2015), no.~1
  1--19.

\bibitem{Battistoni:2002ew}
G.~Battistoni, A.~Ferrari, T.~Montaruli, and P.~R. Sala, ``{The FLUKA
  atmospheric neutrino flux calculation},'' {\em Astropart. Phys.} {\bf 19}
  (2003) 269--290, \href{http://xxx.lanl.gov/abs/hep-ph/0207035}{{\tt
  hep-ph/0207035}}. [Erratum: Astropart. Phys.19,291(2003)].

\bibitem{USStd:1976}
``{U.S. Standard Atmosphere}, 1976.'' U.S. Government Printing Office, 1976.

\bibitem{JGRA:JGRA16630}
J.~M. Picone, A.~E. Hedin, D.~P. Drob, and A.~C. Aikin, ``{NRLMSISE-00}
  empirical model of the atmosphere: Statistical comparisons and scientific
  issues,'' {\em Journal of Geophysical Research: Space Physics} {\bf 107}
  (2002), no.~A12 SIA 15--1 to SIA 15--16. 1468.

\bibitem{AIRS}
``Airs -- atmospheric infrared sounder.'' \url{http://airs.jpl.nasa.gov/}.

\bibitem{NovaOctantPRL}
{\bf NOvA} Collaboration, P.~Adamson {\em et~al.}, ``{Measurement of the
  neutrino mixing angle $\theta_{23}$ in NOvA},'' {\em Phys.\ Rev.\ Lett.} {\bf
  118} (2016) 151802, \href{http://xxx.lanl.gov/abs/arXiv:1701.05891}{{\tt
  arXiv:1701.05891}}.

\bibitem{ref:T2K2017Oscillations}
{\bf T2K} Collaboration, K.~Abe {\em et~al.}, ``{First combined analysis of
  neutrino and antineutrino oscillations at T2K},'' {\em Phys.\ Rev.\ Lett.}
  {\bf 118} (2017) 151801.

\bibitem{ref:T2K2016PRL}
{\bf T2K} Collaboration, K.~Abe {\em et~al.}, ``{Measurement of muon
  antineutrino oscillations with an accelerator-produced off-axis beam},'' {\em
  Phys.\ Rev.\ Lett.} {\bf 116} (2016) 181801.

\bibitem{ref:IC2017Contour}
{\bf IceCube} Collaboration, M.~G. Aartsen {\em et~al.}, ``{Measurement of
  atmospheric neutrino oscillations at 6--56 GeV with IceCube DeepCore}.''
  arXiv:1707.07081, 2017.

\bibitem{Adamson:2013}
{\bf MINOS} Collaboration, P.~Adamson {\em et~al.}, ``{Measurement of Neutrino
  and Antineutrino Oscillations Using Beam and Atmospheric Data in MINOS},''
  {\em Phys.~Rev.~Lett.} {\bf 110} (2013) 251801,
  \href{http://xxx.lanl.gov/abs/arXiv:1304.6335}{{\tt arXiv:1304.6335}}.

\bibitem{SuperKIV}
R.~Wendell, ``{Atmospheric neutrino oscillations at Super-Kamiokande},'' {\em
  Proc.\ of Science} {\bf ICRC 2015} (2015) 1062.

\bibitem{ref:IceCubeHighEnergySteriles}
{\bf IceCube} Collaboration, M.~G. Aartsen {\em et~al.}, ``{Searches for
  sterile neutrinos with the IceCube detector},'' {\em Phys.\ Rev.\ Lett.} {\bf
  117} (2016) 071801.

\bibitem{ref:IceCubeLowEnergySteriles}
{\bf IceCube} Collaboration, M.~G. Aartsen {\em et~al.}, ``{Search for sterile
  neutrino mixing using three years of IceCube DeepCore data},'' {\em Phys.\
  Rev.} {\bf D95} (2017) 112002.

\bibitem{ref:IceCubeWIMPLimit}
{\bf IceCube} Collaboration, M.~G. Aartsen {\em et~al.}, ``{First search for
  dark matter annihilations in the Earth with the IceCube detector},'' {\em
  Eur.\ Phys.\ J.} {\bf C77} (2017) 82.

\bibitem{Abe:2014ugx}
{\bf T2K} Collaboration, K.~Abe {\em et~al.}, ``{Precise Measurement of the
  Neutrino Mixing Parameter $\theta_{23}$ from Muon Neutrino Disappearance in
  an Off-axis Beam},'' {\em Phys.Rev.Lett.} {\bf 112} (2014) 181801,
  \href{http://xxx.lanl.gov/abs/1403.1532}{{\tt 1403.1532}}.

\bibitem{Adamson:2016tbq}
{\bf NOvA} Collaboration, P.~Adamson {\em et~al.}, ``{First measurement of
  electron neutrino appearance in NOvA},'' {\em Phys. Rev. Lett.} {\bf 116}
  (2016), no.~15 151806, \href{http://xxx.lanl.gov/abs/1601.05022}{{\tt
  1601.05022}}.

\bibitem{Acciarri:2015uup}
{\bf DUNE} Collaboration, R.~Acciarri {\em et~al.}, ``{Long-Baseline Neutrino
  Facility (LBNF) and Deep Underground Neutrino Experiment (DUNE)},''
  \href{http://xxx.lanl.gov/abs/1512.06148}{{\tt 1512.06148}}.

\bibitem{HyperK}
K.~Abe, T.~Abe, H.~Aihara, Y.~Fukuda, Y.~Hayato, {\em et~al.}, ``{Letter of
  Intent: The Hyper-Kamiokande Experiment --- Detector Design and Physics
  Potential ---},'' \href{http://xxx.lanl.gov/abs/arXiv:1109.3262}{{\tt
  arXiv:1109.3262}}.

\bibitem{An:2015jdp}
{\bf JUNO} Collaboration, F.~An {\em et~al.}, ``{Neutrino Physics with JUNO},''
  {\em J. Phys.} {\bf G43} (2016), no.~3 030401,
  \href{http://xxx.lanl.gov/abs/1507.05613}{{\tt 1507.05613}}.

\end{thebibliography}\endgroup

\end{document}